\documentclass[aps,prx,twocolumn,footinbib,superscriptaddress,floatfix]{revtex4-2}
\usepackage{amsmath}
\usepackage{graphicx}
\usepackage{indentfirst}
\usepackage{physics}
\usepackage{braket}
\usepackage{float}
\usepackage{mathtools}
\usepackage{epstopdf} 
\usepackage{footnote}
\usepackage{esint}
\usepackage{comment}
\usepackage{color}
\usepackage[T1]{fontenc}
\usepackage[caption=false,position=top]{subfig}
\usepackage{amsfonts}
\usepackage{footmisc}
\usepackage{scrextend}
\usepackage{multirow}
\usepackage[hyperfootnotes=false]{hyperref}
\usepackage[acronym]{glossaries}
\usepackage[english]{babel}
\usepackage{url}
\usepackage{bm}
\usepackage{algpseudocode}
\usepackage{tikz}
\usepackage{etoolbox}
\usepackage{enumerate}
\usepackage{soul}
\usepackage[normalem]{ulem}
\usepackage{amssymb}
\definecolor{darkblue}{rgb}{0,0,0.5}
\hypersetup{
    colorlinks=true,
    linkcolor=black,
    filecolor=blue,
    citecolor=darkblue,  
    urlcolor=black,
}

\apptocmd{\sloppy}{\hbadness 9999\relax}{}{}

\DeclareMathOperator{\E}{\mathbb{E}}

\newtheorem{theorem}{Theorem}

\newcommand{\calI}{{\cal I}}

\newcommand{\calP}{{\cal P}} 

\newcommand{\calT}{{\cal T}}

\newcommand{\1}{^{(1)}}

\newcommand{\state}[1]{\ketbra{#1}{#1}}

\newcommand{\bI}{\boldsymbol I}

\newcommand{\bfz}{\mathbf{z}}

\def\be{\begin{equation}}
\def\ee{\end{equation}}
\def\ba{\begin{eqnarray}}
\def\ea{\end{eqnarray}}

\newcommand{\QZ}[1]{{{\textcolor{black}{#1}}}}
\newcommand{\BZ}[1]{{{\textcolor{black}{#1}}}}

\begin{document}

\title{Scaling Laws of Quantum Information Lifetime in Monitored Quantum Dynamics }

\author{Bingzhi Zhang}
\email{bingzhiz@usc.edu}
\affiliation{
Ming Hsieh Department of Electrical and Computer Engineering, University of Southern California, Los
Angeles, California 90089, USA
}

\author{Fangjun Hu}
\affiliation{Department of Electrical and Computer Engineering, Princeton University, Princeton, NJ 08544}

\author{Runzhe Mo}
\affiliation{
Ming Hsieh Department of Electrical and Computer Engineering, University of Southern California, Los
Angeles, California 90089, USA
}

\author{Tianyang Chen}
\affiliation{Department of Electrical and Computer Engineering, Princeton University, Princeton, NJ 08544}

\author{Hakan E. T\"{u}reci}
\affiliation{Department of Electrical and Computer Engineering, Princeton University, Princeton, NJ 08544}

\author{Quntao Zhuang}
\email{qzhuang@usc.edu}
\affiliation{
Ming Hsieh Department of Electrical and Computer Engineering, University of Southern California, Los
Angeles, California 90089, USA
}
\affiliation{ Department of Physics and Astronomy, University of Southern California, Los
Angeles, California 90089, USA
}

\begin{abstract}
Quantum information is typically fragile under measurements and environmental coupling. 
Remarkably, we find that its lifetime can scale exponentially with system size when the environment is continuously monitored via mid-circuit measurements---regardless of bath size. Starting from a maximally entangled state with a reference, we analytically prove this exponential scaling for typical Haar random unitaries and confirm it through numerical simulations in both \BZ{random unitary circuits} and chaotic Hamiltonian systems. In the absence of bath monitoring, the lifetime exhibits a markedly different scaling: it grows at most linearly---or remains constant---with system size and decays inversely with the bath size. We further extend our findings numerically to a broad class of initial states. 
\BZ{In the intermediate regime of partial monitoring, we identify and prove a two-scale transition, where the QMI decays logarithmically at microscopic time scales but linearly at macroscopic time scales.} We discuss implications for {monitored quantum circuits in the weak measurement limit, quantum algorithms such as quantum diffusion models and quantum reservoir computing, and quantum communication.} Finally, we \BZ{experimentally verify the gap of persisted information on IBM Quantum hardwares}.

\end{abstract}

\maketitle


\section{Introduction}
Quantum measurement is a defining feature of quantum mechanics---introducing irreducible randomness and collapsing superpositions in a way fundamentally distinct from unitary evolution. Beyond its foundational role, measurement has emerged as a powerful resource for engineering quantum states and probing the structure of many-body wavefunctions. In particular, mid-circuit measurements offer pathways for understanding thermalization in quantum systems, {where subsystem measurements induce an emergent maximally random state ensemble in the complementary system~\cite{ho2022exact,cotler2023emergent}.}
The monitored quantum circuit, a brickwork unitary circuit probabilistically doped with projective measurements, exhibits the measurement-induced phase transition (MIPT): as the measurement probability increases, a system can transition from the volume-law to area-law entanglement~\cite{li2019measurement,skinner2019measurement}. These results have resonated broadly, stimulating research across quantum information science, statistical mechanics, and condensed matter physics.


In the era of quantum science and engineering, a detailed understanding of the role of quantum measurement is essential for realizing quantum error correction~\cite{calderbank1996good}, a cornerstone of fault-tolerant quantum computation. Experimental advances, particularly in superconducting qubit platforms, have demonstrated mid-circuit operations conditioned on measurement outcomes as a potential practical pathway toward scalable quantum computation~\cite{acharya_quantum_2025}. Extending the lifetime of encoded quantum information in this setting requires carefully designed codes that preserve coherence in the logical quantum space during syndrome extraction. Mid-circuit measurements have also found utilities in quantum algorithms beyond error correction. To begin with, they enable the reuse of physical qubits as fresh ancillae, thereby reducing the physical resources in quantum computation. This approach has recently been demonstrated in the holographic deep thermalization protocol \BZ{for quantum and classical random number generation~\cite{zhang2025holographic, zhang2025quantum}} and discussed for a range of other algorithms~\cite{corcoles2021exploiting,decross2023qubit,baumer2024efficient,piroli2024approximating,buhrman2024state,smith2024constant,iqbal2024non, cao2025measurement}. In these applications and beyond, the notion of quantum information lifetime plays an important role, and yet its connection to how mid-circuit measurements are performed is unclear, as we detail below.

\begin{figure*}[t]
    \centering
    \includegraphics[width=0.7\textwidth]{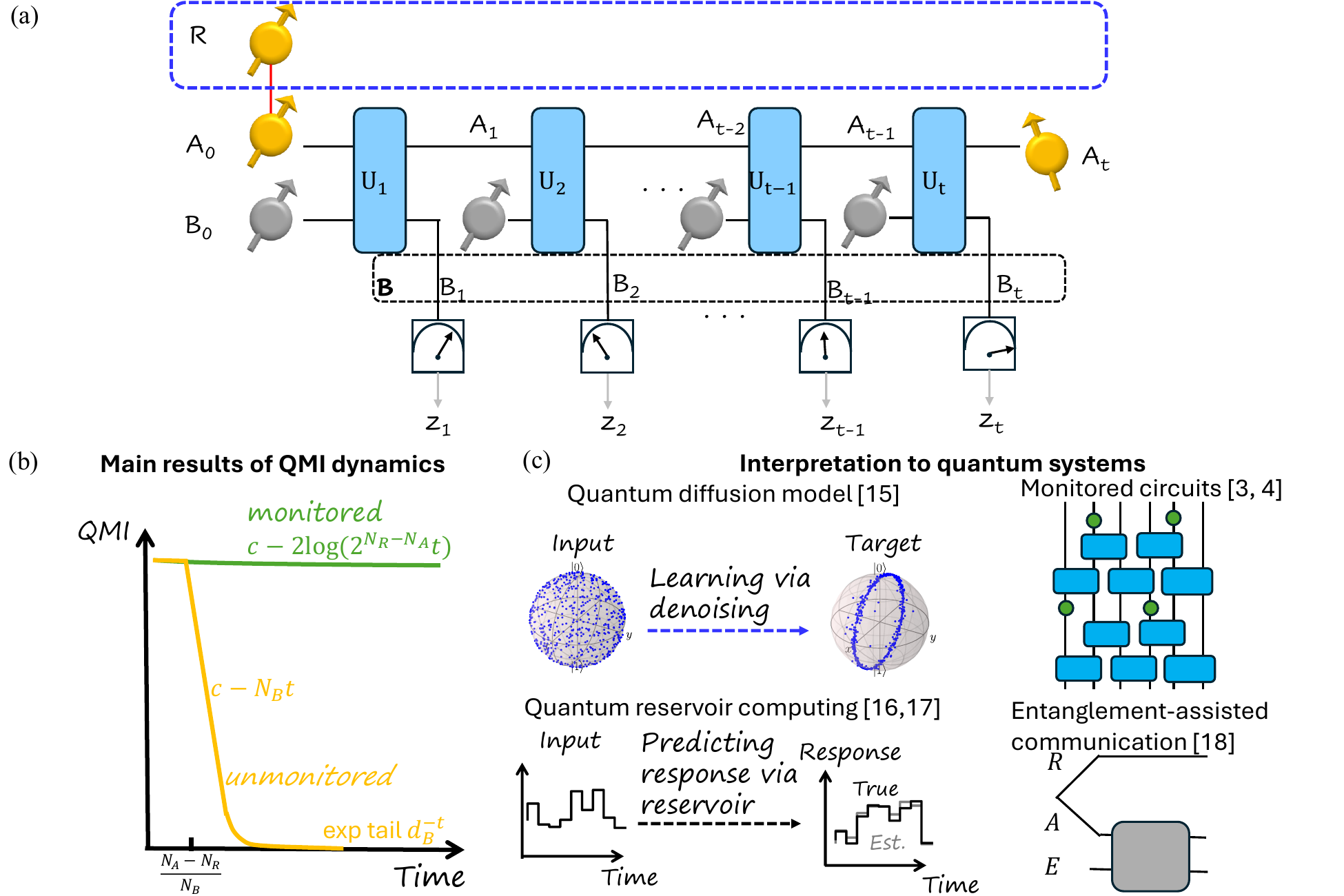}
    \caption{Schematic plot of quantum dynamics with mid-circuit measurements. In (a), initial input $A_0$ ($N_A$ qubits) goes through interactions with bath via unitaries $U_1,U_2,\cdots,U_t$ to produce the final output $A_t$. Between each unitary,  bath qubits $B$ ($N_B$ qubits) are measured and reset to $\ket{0}$'s in most of our discussions, except in Section~\ref{sec:reset}. Equivalently, one can regard the measurement as applied on bath qubits $B_1,B_2,\cdots, B_t$, each initialized in  $\ket{0}^{\otimes N_B}$. To track the evolution of quantum information, we introduce the reference system $R$ initially maximally entangled with $A_0$. The corresponding quantum mutual information (QMI) dynamics is sketched in (b), which decays {logarithmically in monitored dynamics but linearly in time} with \BZ{an initial perfect information protection  and a late exponential tail with unmonitored bath}. 
    Panel (c) highlights connections between the model above and other models and algorithmic frameworks, including quantum denoising diffusion probabilistic model (QuDDPM)~\cite{zhang2024generative}, quantum reservoir computing (QRC)~\cite{chen_temporal_2020, hu2024overcoming}, 
    {monitored quantum circuits~\cite{li2019measurement, skinner2019measurement}, and entanglement-assisted communication~\cite{bennett2002entanglement}}. The reference system surrounded by the blue dashed box in (a) indicates that it is not included in the original quantum circuit models except for communication setting. }
    \label{fig:scheme}
\end{figure*}

In generative quantum models such as the Quantum Denoising Diffusion Probabilistic Model (QuDDPM)~\cite{zhang2024generative}, of interest in the present work, mid-circuit measurement plays a central role.
{QuDDPM aims to learn a backward denoising process that maps a random state ensemble to the target ensemble, in a step-by-step fashion guided by a sequence of reference ensembles interpolating the target and initial ensembles, which are produced in a controlled forward scrambling process (see Fig.~\ref{fig:scheme}c). In each denoising step, mid-circuit measurement combined with a parameterized unitary is trained under a cost function.} In this regard, mid-circuit measurement serves a dual role: it supplies intrinsic quantum randomness for generative learning and acts as a Maxwell’s demon, denoising the system by steering transitions between quantum state ensembles. 
In this setting, the {\it quantum information lifetime} defined in this work serves as a proxy for the efficiency of state conversion and the rate of learning. 



Quantum Reservoir Computing (QRC)~\cite{nakajima2019boosting, fujii2017harnessing}, a general-purpose quantum machine learning paradigm for processing time-dependent data, is another setting where the scaling behavior of quantum information lifetime is a key design consideration.
An implementation of the QRC algorithm tailored for near-term quantum devices using mid-circuit measurements has been recently proposed and investigated~\cite{chen_temporal_2020, mujal_time-series_2023, yasuda_quantum_2023, martinez-pena_quantum_2023, hu2024overcoming}. In the noisy intermediate-scale QRC scheme (NISQRC)~\cite{hu2024overcoming} analyzed here, a subset of qubits, designated as readout qubits, is repeatedly measured and reset as the data is streamed into the circuit, to approximate a target function of past inputs (see Fig.~\ref{fig:scheme}c). The remaining qubits, referred to as memory qubits, remain unmeasured. A key quantity of interest is the {\it memory time}, the ability of the system state to retain information about its past inputs~\cite{jaeger2002short, ganguli_memory_2008}. In reservoir computing, memory time plays a subtle role and must be carefully balanced with the non-linearity the system introduces on its inputs -- a trade-off well recognized in classical reservoir computing~\cite{dambre_information_2012, verstraeten_memory_2010, inubushi_reservoir_2017}. It is therefore important to understand the scaling of memory time -- quantum information lifetime -- with the measurement rate and system size.


In this work, we derive scaling laws for the quantum information lifetime in dynamical settings with and without bath monitoring enabled by mid-circuit measurements. When the system interacts with a periodically monitored bath and the measurement trajectory is recorded, the information lifetime can scale exponentially with the system size (in the context of NISQRC, this corresponds to the size of the memory subsystem). While without the monitoring -- when the individual measurement trajectory is not recorded, the lifetime scales linearly or is a constant in system size. \BZ{In the presence of partial monitoring, we uncover a two-scale transition: QMI decays logarithmically on microscopic scales but linearly on macroscopic scales.}
We establish our results analytically for typical unitaries using quantum mutual information between the system and a maximally entangled reference. These findings are supported by numerical simulations under \BZ{Haar-random and random Clifford unitaries} and chaotic Hamiltonian dynamics across various initial states. For chaotic Hamiltonians, we identify atypical regimes where residual correlations persist for times that scale exponentially with system size.  Our analysis highlights the importance of training in QuDDPM for achieving efficient generative learning. In the context of monitored quantum circuits studied for the entanglement phase transitions, our results demonstrate an exponentially long quantum information lifetime in the weak measurement limit. We also interpret these results in the context of quantum reservoir computing, suggesting strategies for the design of the memory time in general dynamical systems. {In the quantum communication setting, our results have implications for entanglement-assisted protocols, both with and without environment assistance.} Finally, we propose a {hardware-friendly} protocol for verifying mutual information dynamics through \BZ{experiments on} the IBM Quantum hardware. 


\section{Overview}

As depicted in Fig.~\ref{fig:scheme}a, we study a quantum circuit, where system $A$ interacts unitarily with bath $B$ in a repeated manner. Between successive unitaries, mid-circuit measurements are applied on the bath $B$.
The central question we ask in this work is:
\begin{quote}
{\em 
    How does the initial quantum information in $A$ decay through unitary evolution and mid-circuit measurements? 
}
\end{quote}
To quantify the quantum information lost, we introduce a reference system $R$ initially correlated with $A_0$ (see Fig.~\ref{fig:scheme}a). We adopt the quantum mutual information (QMI) between $R$ and $A_t$ to quantify the amount of quantum information in the system $A_t$ for $t\ge 0$. The involvement of measurements means that access to measurement outcomes shapes the information dynamics, much like in the Wigner’s friend scenario~\cite{wigner1995remarks}. When the entire measurement trajectory is recorded, we refer to the bath as “monitored”, indicating complete knowledge of the measurement dynamics (Eq.~\eqref{state_RAt}); conversely, lack of the full information of measurement trajectory decreases our knowledge about the final quantum state. In the extreme case, when the entire measurement trajectory is discarded, we regard the bath as being not monitored (Eq.~\eqref{eq:state_uncond}). In this work, we \BZ{begin with} examination of the QMI dynamics under mid-circuit measurements with and without bath monitoring, and \BZ{extend to QMI dynamics with partial bath monitoring.}

We analyze the QMI dynamics by first deriving analytical scaling expressions with respect to system size, assuming the $R$–$A$ system starts in a maximally entangled Bell state \BZ{with $N_R \le N_A$ in general}. As sketched in Fig.~\ref{fig:scheme}b, we prove that for typical monitored dynamics, the QMI vanishes logarithmically with time steps, leading to a lifetime that scales exponentially with system size, $\tau\sim 2^{N_A}$ and independent of bath size; while for typical unmonitored dynamics, the QMI \BZ{initially experiences a perfect information protection plateau thanks to the dynamical quantum error correction (QEC) until $t \sim (N_A-N_R)/N_B$ and then} decays linearly with time \BZ{$2N_R/N_B$} before entering a long-tail residual region, leading to {a shorter lifetime with different scaling laws}: 
\BZ{$\tau \sim (N_A+N_R)/N_B$} or $\tau \sim 1/N_B$ depending on whether one focuses on the initial linear decay or the long-tail residual part. Our results are analytically established for typical dynamics under a maximally entangled initial state of the $R-A$ system and are further corroborated through numerical simulations of a broad class of correlated initial states.
\BZ{When erasure error is introduced as a partially monitored bath, the noisy conditional dynamics forms an interpolation between the monitored and unmonitored limits. We uncover a two-scale transition, with microscopic logarithmic decay accompanied by a macroscopic-scale linear decay.}

Besides the typical cases represented by dynamics generated by Haar unitaries, we also examine atypical cases of Hamiltonian dynamics without bath monitoring, which gives rise to a lifetime that scales exponentially.

\BZ{We also show that QMI dynamics remains unchanged under resetting to a fixed pure state or leaving the qubits in the measured pure state without reset.}

Our results have implications for monitored quantum circuits, QuDDPM and QRC algorithms employing partial mid-circuit measurements, and entanglement-assisted communication as indicated in Fig.~\ref{fig:scheme}c. Our results show that monitored quantum circuits in the weak measurement limit support exponentially long quantum information lifetime. In terms of QuDDPM, our results indicate that {focusing on a single measurement trajectory is inefficient in learning; it is indeed essential to take QuDDPM's approach of considering the states under different measurement trajectories.} For mid-circuit-measurement-based QRC algorithms~\cite{chen_temporal_2020, mujal_time-series_2023, yasuda_quantum_2023, martinez-pena_quantum_2023, hu2024overcoming}, our findings suggest that optimizing memory time for given resources may require going beyond the marginal distribution of measurements or employing non-generic Hamiltonian choices.

The paper is organized as follows. In Section~\ref{sec:preliminary}, we introduce the framework for information propagation in quantum dynamics with mid-circuit measurements, and the metric to quantify information lifetime. In Section~\ref{sec:exp}, we present results for quantum information dynamics with a monitored bath. In Section~\ref{sec:uncond}, we study the scenario without bath monitoring, including random unitaries and anomalous Hamiltonian dynamics. \BZ{In Section~\ref{sec:partial_monitor}, we investigate the information dynamics with a partially monitored bath.}  We interpret our results in {various scenarios} in Section~\ref{sec:interpretation}. In Section~\ref{sec:reset}, we evaluate the effect of reset strategies on the information lifetime. In Section~\ref{sec:experiment}, we examine the information dynamics from simulation of IBM quantum devices.


\section{Quantum system and mutual information lifetime}
\label{sec:preliminary}

\subsection{Quantum circuit setup}
We consider a general framework for quantum dynamics with mid-circuit measurements, as shown in Fig. \ref{fig:scheme}a. This framework underlies a broad class of algorithms including recent QRC algorithms~\cite{hu2024overcoming, kobayashi2024feedback} and the  QuDDPM~\cite{zhang2024generative}. The quantum system here consists of two subsystems: the data system $A$ and the bath system $B$ comprising $N_A$ and $N_B$ qubits respectively~\footnote{In the context of the NISQRC algorithm of Ref.~\cite{hu2024overcoming}, these are referred to as the Memory (M) and Readout (R) subsystems.}. For the convenience of description, we also add a subscript $t$ to $A$ and $B$ systems to denote the system at time step $t$. Starting from the step $t=0$, the data system $A_0$ takes an information-encoded quantum state, and the bath $B_0$ is initialized in a trivial product state, e.g. $\ket{\bm 0}_{B_0}\equiv \ket{0}^{\otimes N_B}$. At each time step $t$, the system undergoes a global unitary evolution $U_t$, which corresponds to either a chaotic evolution in the case of QRC or a parameterized quantum circuit in QuDDPM. This is followed by a projective measurement on the bath $B_t$, yielding a single-shot measurement outcome $z_t$. The resulting post-measurement state of the data subsystem $A$ is then used as the input for the subsequent time step $t+1$, where the process of unitary evolution and bath measurement is repeated. We first consider the case in which the bath subsystem $B$ is reset to the trivial product state $\ket{\bm{0}}_B$ at each time step. In Section~\ref{sec:reset}, we demonstrate that this reset procedure does not affect our main conclusions regarding lifetime scaling.

For simplicity, we take the reference and system to be initially in a pure state $\ket{\Phi}_{RA_0}$ for now.
When the bath is fully monitored in each step, in $t$-th step, the Kraus operator applied on $RA$ joint system corresponding to measurement outcome $z_t$ on bath is $K_t = {}_{B_t}\braket{z_t|(\bI_R \otimes U_t)|\bm 0}_{B_{t-1}}$, therefore the final quantum state of $RA_t$ conditioned on the measurement trajectory $\bfz =(z_1,\cdots,z_t)$ can be expressed as
\be
    \ket{\psi_{\bfz}}_{RA_t} = \frac{1}{\sqrt{P_{\bm U}(\bfz)}} K_t \cdots K_2 K_1 \ket{\Phi}_{RA_0}
    \label{state_RAt}
\ee
where $P_{\bm U}(\bfz)$ is the probability of obtaining the measurement outcome $\bfz$.


Meanwhile, when the bath is not monitored, we also consider the dynamics of measurement-unconditioned quantum state. The unconditioned state at step $t$ is the state averaged over all measurement trajectories as
\begin{align}
    \rho_{RA_t} &= 
    \sum_{\bfz} P_{\bm U}(\bfz) \state{\psi_{\bfz}}_{RA_t}.
    \label{eq:state_uncond}
\end{align}
Equivalently, one can regard the non-monitored bath being traced out in each time step.
In this regard, the input-output relation from $A_{t-1}$ to $A_t$, $\calP_t(\cdot)=\tr_{B_t}\left[ U_t \left(\cdot \otimes \state{\bm 0}_{{B_{t-1}}}\right) U_t^\dagger\right]$, defines the quantum channel $\calP_t$. Therefore, the unconditional joint state can be described by the overall quantum channel 
\be 
\rho_{RA_t}=\calI \otimes (\calP_t\circ \calP_{t-1}\circ \cdots \calP_{1}) (\rho_{RA_{0}}),
\label{unconditional_dynamics}
\ee 
where $\calI$ refers to the identity channel on the reference $R$.

\subsection{Connecting to quantum models}
Here we connect the quantum circuit dynamics under consideration to {several different quantum models}. The detailed interpretation can be found in Section~\ref{sec:interpretation}.


The trajectory-conditional state from bath monitoring in Eq.~\eqref{state_RAt} arises in the generative quantum machine learning models such as QuDDPM~\cite{zhang2024generative}. As shown in Fig.~\ref{fig:scheme}c, the goal of QuDDPM is to learn to sample a quantum state $\ket{\phi}$ from a certain distribution $P_\phi$ in a generative manner. The quantum circuit in Fig.~\ref{fig:scheme}a transports the initial random distribution towards the target distribution represented by the sample. Although the original QuDDPM does not require a reference system $R$, the dynamical transport of an ensemble of quantum states $\{\phi_k, P_{\phi_k}\}$ can be equivalently represented by the evolution of an entangled state between a reference $R$ and system $A$,
$
\ket{\psi}_{RA}=\sum_k \sqrt{P_{\phi_k}}\ket{k}_R \ket{\phi_k}_A,
$
where $\ket{k}$ is an orthonormal bases to denote the sample index.  
The induced quantum state ensemble structure change can generally be captured by the change in the quantum correlation between reference $R$ and $A_t$. 


On the other hand, the dynamics of the unconditional state in Eq.~\eqref{unconditional_dynamics} provides insight to the current approaches of QRC~\cite{fujii2017harnessing, nakajima2019boosting, hu2024overcoming, kobayashi2024feedback}. QRC aims to learn a complex functional mapping from a sequence of classical data to an output classical data sequence. 
{Ref.~\cite{hu2024overcoming} highlights the importance of balancing memory time and nonlinearity in QRC, using a Volterra-series-based analysis, and argues that the memory time should align with the timescale of the target learning task.}
In the NISQRC implementation,  classical data is encoded into the quantum system via parametric unitaries $U_1,\cdots,U_t$ in Fig.~\ref{fig:scheme}a and the output features are obtained from marginal distributions $P(z_1),\cdots P(z_t)$ of the measurement results (see details in Section~\ref{sec:qrc}), with dynamics characterized by Eq.~\eqref{unconditional_dynamics}. 
Although the original NISQRC framework does not involve a reference system $R$, introducing a classical-quantum (CQ) state over $RA$ provides an equivalent description of classical data input into the quantum system $A$ (for details, see Section~\ref{sec:qrc}. Therefore, memory time of classical input can be captured by the correlation between $R$ and $A_t$, in the unconditional state in Eq.~\eqref{unconditional_dynamics}.

{In the monitored circuit where a brickwork unitary circuit is probabilistically doped with projective measurements, as shown in Fig.~\ref{fig:scheme}c, both trajectory-conditional and unconditional states can be studied.
Our model considered in Fig.~\ref{fig:scheme}a can be regarded as a toy model of monitored circuits,
where the entire system is always fully scrambled between each round of subsystem projective measurements on a fixed number of $N_B$ qubits. Due to the fully scrambling unitary, we can fix the $N_B$ measured qubits as the bath and the rest as the system. While the entanglement dynamics is well-studied, the loss of quantum information due to measurement is unclear. To capture the quantum information, we can therefore introduce the reference system $R$ initially entangled with the data qubits $A$.
In this context, the conditional state in Eq.~\eqref{state_RAt} refers to the resulting state in a particular experiment, while Eq.~\eqref{eq:state_uncond} corresponds to the average state without post-selection. 
}

{The quantum circuit can also be interpreted in the scenario of entanglement-assisted communication through the channel $\calP_t \circ \cdots \calP_1$ specified in Eq.~\eqref{unconditional_dynamics}. In this case, the reference system $R$ provides the entanglement assistance in communication and the QMI quantity directly corresponds to entanglement-assisted communication capacity~\cite{bennett2002entanglement}. The case without environment assistance corresponds to the unmonitored dynamics described in Eq.~\eqref{unconditional_dynamics}. On the other hand, for monitored dynamics, in each experiment, the receiver will not only receive the conditional state $\ket{\psi_\bfz}_{RA_t}$ in Eq.~\eqref{state_RAt}, but as well as the measurement trajectory $\bfz$, which is a classical message about the environment bath status, therefore, we can interpret the monitored dynamics as environment-assisted communication.}

In all applications, we introduce the reference system $R$ to facilitate the study of the quantum information dynamics. 
\BZ{
We begin by considering the reference $R$ having the same size as the system $A$, $N_R = N_A$ and show the scaling of QMI lifetime,  a quantity widely examined in the contexts of communication and error correction. We then generalize to the case of a smaller reference $N_R < N_A$ in both monitored and unmonitored dynamics and highlight the resulting distinct behaviors of the QMI dynamics. Note that the case of $N_R> N_A$ is equivalent to $N_R = N_A$ since the maximal initial information is bounded by $2\min\{N_R, N_A\}$.}


\subsection{Quantum mutual information lifetime}

Taking into account the measurement-conditioned state $\ket{\psi_\bfz}_{RA_t}$ and unconditioned state $\rho_{RA_t}$ at time step $t$, we evaluate the information shared by parties $R$ and $A_t$ via QMI, \BZ{which quantifies the total correlation between $R$ and $A_t$ at time $t$.} 
Given a measurement trajectory $\bfz$, the quantum information retained in $A_t$ at time step $t$ is quantified by the {\it measurement-conditioned QMI} defined as
\be 
\overline{I(R:A_t|\bfz)} \equiv \E_{\bfz} I(R:A_t|\bfz),
\label{QMI_cond}
\ee 
where $\E_\bfz$ denotes averaging over measurement trajectories according to the outcome probabilities $P_{\bm U}(\bfz)$, and $I(R:A_t|\bfz)$ is the QMI of the conditional state $\ket{\psi_{\bfz}}_{RA_t}$ in Eq.~\eqref{state_RAt}. \BZ{Note that the measurement-conditioned QMI is equivalent to the entanglement entropy between $R$ and $A_t$ as the conditional state $\ket{\psi_\bfz}$ is pure.}
For the unconditional dynamics, the {\it measurement-unconditioned QMI} 
\be
    I(R:A_t) \equiv S(\rho_R) + S(\rho_{A_t}) - S(\rho_{RA_t}),
    \label{eq:uncond_MI}
\ee
is thus evaluated on $\rho_{RA_t}$ in Eq.~\eqref{unconditional_dynamics}, where $\rho_R, \rho_{A_t}$ are corresponding reduced states of reference and system $R, A_t$. 

The differing order of QMI evaluation and the ensemble average $\E_\bfz$ in the measurement-conditioned and unconditioned QMI reflects distinct physical scenarios. The measurement-conditioned QMI (Eq.\eqref{QMI_cond}) captures the information retained in a typical conditional state $\ket{\psi_\bfz}$ obtained from the corresponding measurement trajectory. In contrast, the unconditioned QMI is computed from the averaged state $\rho_{RA_t}$, which contains no information about any measurement outcome (see Eq.\eqref{eq:state_uncond}).
We further interpret the measurement-conditioned and unconditioned QMI introduced above in the setting of entanglement-assisted communication, as detailed in Section~\ref{sec:capacity}.

In both cases of monitored and unmonitored circuits, to quantify the preservation of quantum information, we define the $\epsilon$-lifetime $\tau_{\rm cond}$ and $\tau_{\rm uncond}$ of quantum information by
\be
\text{QMI}|_{t=\tau} = \epsilon \times \text{QMI}|_{t=0},
\label{epsilon_def}
\ee
where $\text{QMI}$ can be either $\overline{I(R:A_t|\bfz)}$ or $I(R:A_t)$ for conditional and unconditional dynamics, and $\epsilon<1$ is a free constant. Specifically, with Bell state as the initial state, the quantum communication capability measured by the coherent information $I(R \rangle A_t)$ starts with unity and decays to zero at the half-life of QMI (i.e. $\epsilon = 1/2$) in unmonitored dynamics, as detailed in Section~\ref{sec:capacity}.

To study the generic properties of quantum dynamics, we model each of the unitary $U_k$ as a fixed unitary sampled from Haar ensemble. This is a good approximation, e.g. when the quantum dynamics is chaotic or a deep random quantum circuit.  To characterize the typical case, we consider the Haar ensemble averaged conditional (unconditional) QMI. 
\BZ{To validate the asymptotic scaling of QMI dynamics in the large-system limit, we also perform the numerical simulations with random Clifford unitaries. Details of the stabilizer formalism are presented in Appendix~\ref{app:clifford}.}

To provide a concrete example, we also consider the Hamiltonian dynamics $U=e^{-iH t_H}$ generated by an all-to-all coupled Ising Hamiltonian~\cite{hu2024overcoming} through a relatively long time $t_H \gg 1$ 
\be 
H=\sum_{i<j} J_{ij} \sigma^z_j \sigma^z_j +\sum_{i=1}^L \eta_i^x \sigma^x + \sum_{i=1}^L \eta_i^z \sigma_i^z,
\label{H_Ising}
\ee 
where the coupling strength $J_{i,j}$, transverse x-field strength $\eta^x_i$ and longitudinal z-drive $\eta_i^z$ are randomly chosen, i.e. standard normal distributed (zero mean and unit variance). All unitaries $U_k = U = e^{-iH t_{H}}$ is identical throughout the various time steps ($t_H = 50$ is chosen throughout the paper).  Our numerical results, summarized in Appendix~\ref{app:indentical_or_not} indicate that our conclusions remain valid regardless of whether the same unitary is used at each time step or different unitaries are chosen independently. \BZ{We expect our results to extend to other chaotic Hamiltonians. As an additional illustration, in Appendix~\ref{app:MFIM} we analyze a mixed-field Ising model with only nearest-neighbor interactions.}

\section{Exponential QMI lifetime in monitored dynamics}
\label{sec:exp}

\subsection{\BZ{Equal reference and system, $N_R=N_A$}}
\label{sec:exp_equal}
With the framework for information propagation in place, we now examine the dynamics of quantum information under mid-circuit measurements and resets, focusing on the measurement-conditioned QMI in Eq.~\eqref{QMI_cond} proposed for monitored dynamics.
We begin with the maximally entangled state, $\ket{\Phi}_{RA_0} \propto \sum_{j} \ket{j}_R \ket{j}_{A_0}$, to obtain analytical results, and then we numerically extend the analytical results with less-entangled states including perturbed Haar-random states and classical-quantum (CQ) hybrid states. 
We define the perturbed Haar random states as $\ket{\phi^\delta}_{RA_0} \propto \ket{\phi_{\rm Haar}}_{RA_0} + \delta\ket{\bm 0}_{RA_0}$. Similarly, we define the family of CQ states to be  $(\rho^\delta)_{RA_0} \propto \sum \ketbra{j}{j}_R \otimes \ketbra{u_j^\delta}{u_j^\delta}_{A_0}$ with perturbed basis $\ket{u_j^\delta}_{A_0} \propto \ket{u_j}_{A_0} + \delta\ket{\bm 0}_{A_0}$, where $\ket{u_j}$ forms an orthonormal basis.
In both cases, $\delta$ interpolates between highly correlated ($\delta=0$) and uncorrelated states ($\delta\gg1$).
\BZ{In all these cases above, we take the reference and the system having the same size, $N_R = N_A$.}

Using Haar-random unitaries to model the chaotic evolution, we have the following theorem, where $d_A=2^{N_A}, d_B=2^{N_B}$ are dimensions of system and bath Hilbert spaces (see Appendix~\ref{app:theorem1} for a proof).
\begin{theorem}
\label{theorem_avgMI}
    \QZ{In quantum dynamics with mid-circuit measurements and reset}, the expected measurement-conditioned mutual information of an initial Bell state at time $t$ is asymptotically ($d_A\gg 1$) lower bounded by 
    \begin{align}
        &\E_{\rm Haar}\overline{I(R:A_t|\bfz)} \nonumber\\
        &\gtrsim 2(2t+1)N_A + 2 t N_B - 2\log_2\left[(d_A d_B + 1)^t (d_A + 1)^{t+1} \right] \nonumber\\
        &\quad - 2\log_2\left[\left(\frac{d_A d_B - 1}{d_A d_B + 1}\right)^t - \left(\frac{d_A - 1}{d_A + 1}\right)^{t+1}\right] + 2
        \label{eq:avgMI_lb}\\
        &\simeq 2N_A - 2\log_2\left[(1-1/d_B)t + 1\right]. \label{eq:avgMI_lb_simplify}
    \end{align}
\end{theorem}
In the large bath limit of $d_B \gg 1$, the dynamics of QMI converges to a universal form $\E_{\rm Haar} \overline{I(R:A_t|\bfz)} \gtrsim 2N_A - 2\log_2\left(t + 1\right)$. 
We can estimate the $\epsilon$-lifetime for measurement-conditioned QMI as
\be
    \tau_{\rm cond} \gtrsim \frac{d_B}{d_B - 1}\left(2^{(1-\epsilon) N_A} - 1\right) \sim \Omega(\exp(N_A)), \label{eq:avgMI_time}
\ee
which grows exponentially with the system size $N_A$ but is independent of bath size when $d_B \gg 1$.

\begin{figure}[t]
    \centering
    \includegraphics[width=0.45\textwidth]{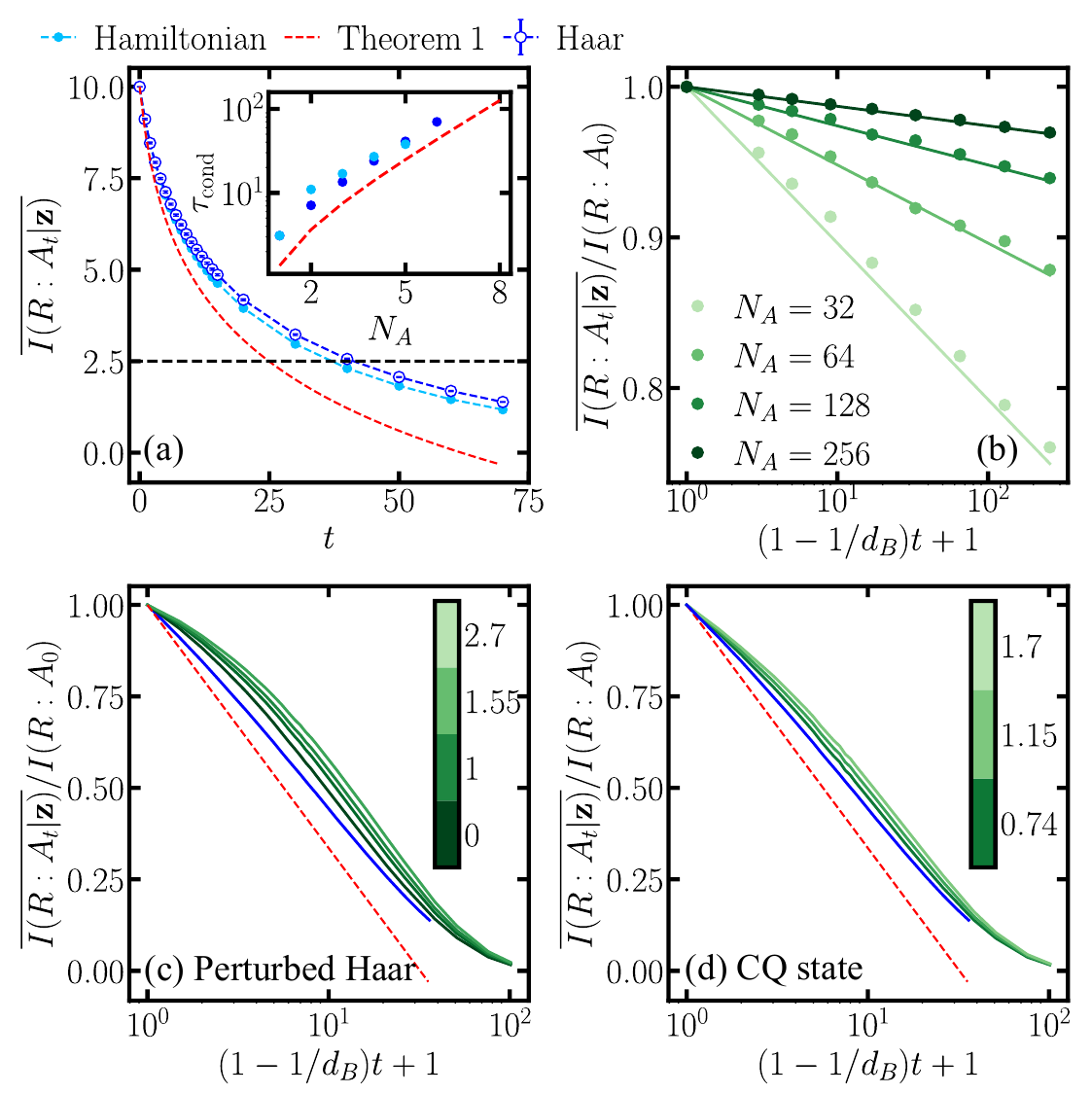}
    \caption{Measurement-conditioned QMI $\overline{I(R:A_t|\bfz)}$ in monitored dynamics with reset. In (a) we plot numerical simulation results of (von Neumann entropy version) $\overline{I(R:A_t|\bfz)}$ with Haar-random unitaries (dark blue dots) and fixed Ising Hamiltonian evolution (light blue dots) in a system of \BZ{$N_R =N_A=5, N_B=1$} qubits. Errorbars represent sample fluctuations of Haar unitary implementations. Red dashed line represents the theoretical asymptotic lower bound of Eq.~\eqref{eq:avgMI_lb_simplify} in Theorem~\ref{theorem_avgMI}. Horizontal black dashed line indicate the threshold of $\overline{I(R:A_t|\bfz)}=\overline{I(R:A_0|\bfz)}/4$. \BZ{The inset shows} the growth of lifetime versus system size $N_A$ of Haar unitary and Hamiltonian evolution. Dots are and red dashed line represent the numerical simulation results and theoretical result in Eq.~\eqref{eq:avgMI_time} with $\epsilon = 1/4$. \BZ{In (b), we plot QMI versus a rescaled time in the dynamics with random Clifford unitaries (dots) in a system of $N_R = N_A = 32, 64, 128, 256$ qubits (light to dark) and $N_B = 16$ qubits. The solid lines represent the theory of Eq.~\eqref{eq:avgMI_lb_simplify}.}
    In (c) and (d), we plot decay of normalized QMI in numerical simulations with initial states of perturbed Haar states $\phi_\delta$ and CQ states $\rho_\delta$ versus a shifted time separately. Blue lines are Bell states result with Haar unitary in (a) for reference. The color bar shows the value of $\delta$.}
    \label{fig:condQMI}
\end{figure}

Fig.~\ref{fig:condQMI}a shows numerical simulations of the measurement-conditioned QMI dynamics in the model where each step we apply a random unitary sampled from Haar ensemble (dark blue dots), and its decay behavior versus time steps is well characterized by our theoretical lower bound (red dashed) in Eq.~\eqref{eq:avgMI_lb_simplify} as long as the bound remains positive. The gap between theoretical lower bound and numerical simulation arises from the difference between R\'enyi-2 and von Neumann entropies, detailed in Appendix~\ref{app:numeric_detail}. The relatively small sample fluctuation shown by the error bars further indicates that the Haar average result can represent the dynamics with a typical implementation of $\bm U$. Moreover, the numerical results for dynamics of $\overline{I(R:A_t|\bfz)}$ utilizing a fixed Hamiltonian evolution in each step (light blue dots) also closely aligns with Haar results, thus indicating a wide applicability of our theory to different types of quantum dynamics. \BZ{In the inset of Fig.~\ref{fig:condQMI}a}, we interpret the QMI dynamics via the lifetime, and similar agreement between the Haar dynamics (blue dots), Hamiltonian dynamics (cyan dots) and the prediction (red dashed) from Eq.~\eqref{eq:avgMI_lb_simplify} can be found, confirming the exponential scaling of the lifetime with system size $N_A$.

\BZ{To verify the logarithmic decay indicated by Eq.~\eqref{eq:avgMI_lb_simplify} in the thermodynamic limit of system size, we also perform numerical simulations of the monitored dynamics using random Clifford unitaries up to $N_R = N_A = 256$ qubits. In Fig.~\ref{fig:condQMI}b, we plot the normalized QMI $I(R:A_t)/I(R:A_0)$ versus a rescaled time of $(1-1/d_B)t + 1$ in logarithmic scaling, as suggested by Eq.~\eqref{eq:avgMI_lb_simplify}. For stabilizer states, the von-Neumann and R\'enyi-2 entropy coincide, and we find exact agreement between our theoretical lower bound and numerical results. Moreover, the Clifford unitary suggests that Theorem~\ref{theorem_avgMI} requires at most a unitary $3$-design~\cite{webb2015clifford}. In fact, as shown in Appendix~\ref{app:theorem1}, we expect that a unitary $2$-design already suffices for our theoretical results.}

Besides the initial Bell state, we also extend numerical results to the QMI dynamics with various initial states, including the perturbed Haar-random states and CQ states, presented in Fig.~\ref{fig:condQMI}c and d. To compare with Theorem~\ref{theorem_avgMI}, we again plot with the \BZ{normalized} QMI versus a rescaled time $(1-1/d_B)t+1$. As shown in Fig.~\ref{fig:condQMI}c and d, the prediction of Theorem~\ref{theorem_avgMI} appears linear in such a rescaling (red dashed). Indeed, the results for various quantum states (green curves) align with the Bell state result (blue solid) and demonstrate agreement with the theory prediction (red dashed).
 
As the lower bound Eq.~\eqref{eq:avgMI_lb} (or the simplified version Eq.~\eqref{eq:avgMI_lb_simplify}) becomes negative at extremely late time, we expect deviations from the lower bound. Indeed, in Fig.~\ref{fig:condQMI}c and d the curves deviate from the red dashed lower bound curve. In Appendix~\ref{app:residual_dy}, we study this deviation numerically and identify an exponentially decaying tail of QMI. Taking late-time states as the initial states, we show that such a decay still yields a lifetime that is exponential in $N_A$ because of the exponentially suppressed exponent of the tail.

\subsection{\BZ{Generalization to $N_R<N_A$}}
\label{sec:exp_unequal}

\BZ{We further extend our study of measurement-conditioned QMI dynamics to the regime where the reference is smaller than the system, $N_R < N_A$. 
Without loss of generality and for analytical convenience, we choose the initial state to consist of $N_R$ Bell pairs shared between reference $R$ and $N_R$ qubits in system $A$, with the remaining $N_A-N_R$ qubits initialized in $\ket{0}$,
\be
    \ket{\Phi}_{RA_0} = \frac{1}{\sqrt{d_R}} \sum_{k=0}^{d_R - 1} \ket{k}_R \ket{k}_{A_0} \otimes \ket{0}^{\otimes(N_R - N_A)},
    \label{eq:init_state_lessR}
\ee
where $d_R = 2^{N_R}$ is the Hilbert space dimension of reference. We then have the following theorem for the QMI dynamics (see Appendix~\ref{app:theorem1} for a proof).
\begin{theorem}
\label{theorem_avgMI_lessR}
    In quantum dynamics with mid-circuit measurements and reset, the expected measurement-conditioned mutual information of an initial state of $N_R$ Bell pairs at time $t$ is asymptotically ($d_R, d_A\gg 1$) lower bounded by 
    \begin{align}
        &\E_{\rm Haar}\overline{I(R:A_t|\bfz)} \nonumber\\
        &\gtrsim 2N_R - 2\log_2\left[1+\frac{d_R t}{d_A}\left(1-\frac{d_R+1}{d_R d_B} + \frac{1}{2d_R}\right)\right]. \label{eq:avgMI_lb_simplify_lessR}
    \end{align}
\end{theorem}
The full version of the lower bound is given in Appendix~\ref{app:theorem1}.
In the limit of $N_R = N_A \gg 1$, it reduces to Eq.~\eqref{eq:avgMI_lb_simplify} in Theorem~\ref{theorem_avgMI}.
Moreover, in the large-bath limit of $d_B \gg 1$, the QMI dynamics approaches $\E_{\rm Haar}\overline{I(R:A_t|\bfz)} \gtrsim 2N_R - 2\log_2\left(1 + d_R t/d_A\right)$, which exhibits logarithmic decay with rescaled time $(d_R/d_A)t$. Consequently, the $\epsilon$-lifetime of the measurement-conditioned QMI is
\be
    \tau_{\rm cond} \gtrsim \frac{d_A}{d_R}\left(d_R^{1-\epsilon}-1\right) \sim 2^{N_A - \epsilon N_R},
\ee
which still grows exponentially with the system size $N_A$, and is enhanced by a factor $\sim (d_A/d_R)^\epsilon$ compared with Eq.~\eqref{eq:avgMI_time}. Another relative timescale is $t \lesssim d_A /d_R^{1-\epsilon}$, over which the QMI remains at least $2(1-\epsilon) N_R$, indicating near-perfect information protection. Physically, $N_A-N_R$ ``idler'' qubits act as an auxiliary memory: although they do not initially encode information, unitary scrambling transfers the encoded information into these degrees of freedom and helps protect it against decoherence from projective measurements.
In particular, for a constant encoded size $N_R \sim \mathcal{O}(1)$, this near-perfect information protection can persist up to $t\sim 2^{N_A}$.}

\begin{figure}[t]
    \centering
    \includegraphics[width=0.45\textwidth]{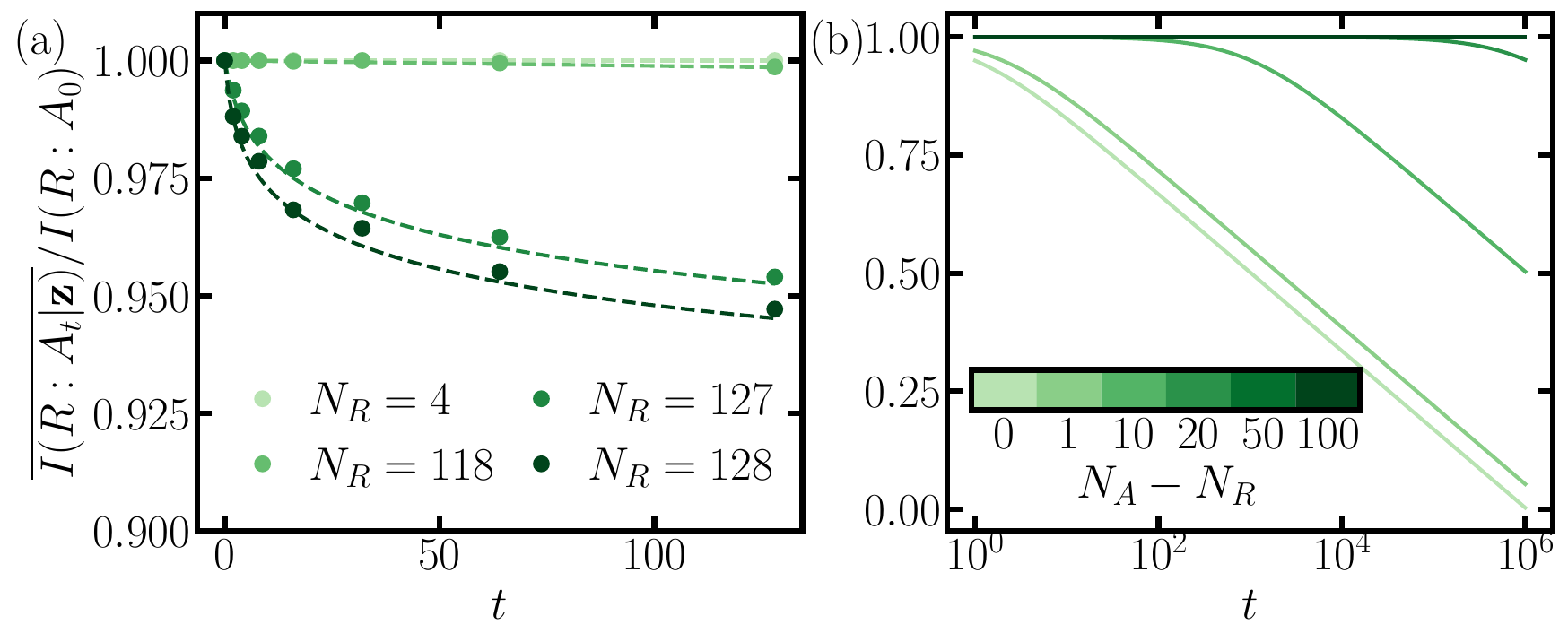}
    \caption{\BZ{Normalized measurement-conditioned QMI $\overline{I(R:A_t|\bfz)}$ for $N_R < N_A$ in monitored dynamics with reset. In (a) we plot numerical simulation results with random Clifford unitaries (dots) in a system of $N_A = 128, N_B=16$ qubits and different $N_R$ reference qubits. Dashed line represents the theoretical asymptotic lower bound of Eq.~\eqref{eq:avgMI_lb_simplify_lessR} in Theorem~\ref{theorem_avgMI_lessR} for the corresponding $N_R$. In (b), we plot the asymptotic dynamics of $\overline{I(R:A_t|\bfz)}/I(R:A_0) = 1 - \log_2(d_Rt/d_A + 1)/N_R$ with $N_R = 20$ and different $N_A$ (shown by the colorbar).}}
    \label{fig:condQMI_lessR}
\end{figure}

\BZ{In Fig.~\ref{fig:condQMI_lessR}a, we verify our theory (dashed lines) with numerical simulations of Clifford circuits (dots) for different reference sizes. For $d_R \ll d_A$ (e.g., $N_R = 4$ and $N_R =118$), we observe robust information protection over the plotted time window $t\le 256$. To highlight the dependence on ``idler'' size $N_A - N_R$, we plot the universal scaling of normalized conditional QMI in the regime of $d_R, d_B \gg 1$ as $\overline{I(R:A_t|\bfz)}/I(R:A_0) = 1-\log_2(1+d_Rt/d_A)/N_R$ in Fig.~\ref{fig:condQMI_lessR}b. 
For $N_R = 20$, when $N_A-N_R \sim \mathcal{O}(1)$, the QMI shows a logarithmic decay from the initial time $t=1$, consistent with Eq.~\eqref{eq:avgMI_lb_simplify}. As $N_A$ increases, the finite-time dynamics exhibits an increasingly long regime of near-perfect protection (light to dark curves).}

\section{QMI lifetime in unmonitored dynamics} 
\label{sec:uncond}

In this section, we study the unmonitored dynamics where the measurement trajectory of bath is not recorded during the dynamics described by Eq.~\eqref{unconditional_dynamics}. 

\subsection{Linear and constant lifetime for typical unitary \BZ{with $N_R = N_A$}}
\label{sec:linear_lifetime}

To characterize the decay of quantum information in the unmonitored scenario, we analyze the QMI of the unconditional state defined in Eq.~\eqref{eq:uncond_MI}. As in the monitored case, we begin with a maximally entangled Bell state \BZ{of $N_R = N_A$} to facilitate analytical calculations.
For computational convenience, we adopt a R\'enyi version of mutual information $I_2(R:A_t)$ defined by replacing the von Neumann entropy in Eq.~\eqref{eq:uncond_MI} with R\'enyi-2 entropies. 
Although this extension may not satisfy all property requirement of QMI, we numerically find that it fits well with the dynamics of the actual QMI. Indeed, for special classes of quantum states, R\'{e}nyi-2 based QMI well captures quantum correlation~\cite{adesso2012measuring} and agrees well with von Neumann based QMI~\cite{hamma2016mutual}, which is also utilized in the experiment of MIPT~\cite{google2023measurement}. With regard to the dynamics of R\'enyi extended mutual information, we thus have the following theorem  (see Appendix~\ref{app:thereom23} for a proof).


\begin{figure*}[t]
    \centering
    \includegraphics[width=0.65\textwidth]{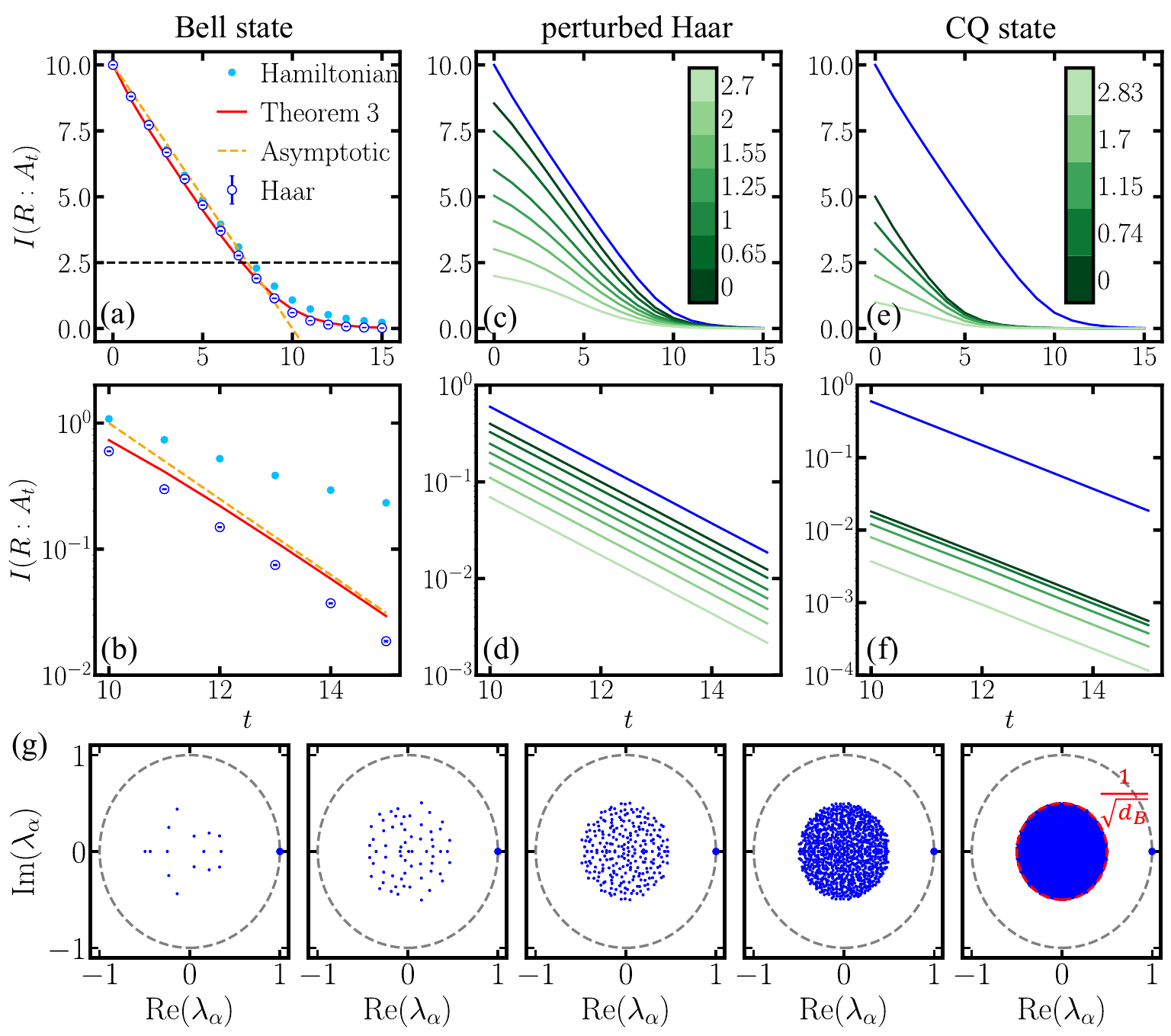}
    \caption{
    Measurement-unconditioned QMI $I(R:A_t)$ in unmonitored dynamics with reset. In (a) we plot numerical simulation results of $I(R:A_t)$ of a Bell initial state with random Haar unitary (dark blue dots) and fixed Ising Hamiltonian evolution (light blue dots) in a system of $N_A=5, N_B=1$ qubits. Errorbars represent sample fluctuations of Haar unitary implementations. Red and orange lines represent the exact and asymptotic theory of Eq.~\eqref{eq:MI_traceout} and Eq.~\eqref{eq:MI_traceout_asymp} in Theorem~\ref{MI_traceout_theorem}. Horizontal black dashed line indicate the threshold of $\epsilon=1/4$. Late-time dynamics is presented in (b) with orange line showing asymptotic theory of Eq.~\eqref{eq:MI_traceout_asymp_late} in Theorem~\ref{MI_traceout_theorem}.
    {In (c) and (e), we plot numerical simulation of $I(R:A_t)$ under Haar unitary dynamics with initial perturbed Haar states and CQ states separately. Dark to light green lines represent different perturbation $\delta$ indicated by the colorbar. Blue lines represent Haar result from (a) for reference. In both cases, we have 5 reference and data qubits. (d) and (f) are corresponding late-time dynamics, and blue lines is Haar result from (b). In (g), we show the eigen-spectrum of $\calP$ implemented by Haar unitary from $N_A=2$ to $6$ (from left to right) with $N_B=2$ bath qubits. The red dashed circle indicates the radius of $1/\sqrt{d_B}$.}
    }
    \label{fig:mi_traceout}
\end{figure*}

\begin{theorem}
\label{MI_traceout_theorem}
    The expected R\'enyi-2 extended measurement-unconditioned QMI of a Bell initial state in the quantum dynamics of $2$-design unitaries with mid-circuit measurements and reset at time $t$ is
    \begin{align}
        &\E_{\rm Haar} I_2(R:A_t)\nonumber\\
        &\simeq \log_2 \left(\frac{\left(d_A^2-1\right) d_B \left(\frac{\left(d_A^2-1\right) d_B}{d_A^2 d_B^2-1}\right)^t+d_B+1}{d_A^2 d_B+1}\right)\nonumber\\
        &\quad - \log_2 \left(\frac{d_A^2 (d_B+1)-\left(d_A^2-1\right) \left(\frac{\left(d_A^2-1\right) d_B}{d_A^2 d_B^2-1}\right)^t}{d_A^3 d_B+d_A}\right)+N_A.
        \label{eq:MI_traceout}
    \end{align}
In the asymptotic limit $d_A, d_B \gg 1$, when $t \ll 2N_A/N_B$, 
\begin{align}
    \E_{\rm Haar}I_2(R:A_t) \simeq 2N_A - t N_B,
    \label{eq:MI_traceout_asymp}
\end{align}
while for $t \gg 2N_A/N_B$, 
\be 
\E_{\rm Haar} I_2(R:A_t) \simeq  d_A^2 d_B^{-t}.
\label{eq:MI_traceout_asymp_late}
\ee 
\end{theorem}
At early time, Eq.~\eqref{eq:MI_traceout_asymp} indicates a linear decay of quantum information; while at late time, Eq.~\eqref{eq:MI_traceout_asymp_late} indicates an exponential decay of quantum information.

{
The late-time exponential decay of QMI in Eq.~\eqref{eq:MI_traceout_asymp_late} can be also explained from a quantum channel spectrum approach~\cite{hu2024overcoming}, which holds for general initial states.
Here, we focus on the case where the same unitary is applied repeatedly, and therefore $\calP_k=\calP$ for all time steps $1\le k\le t$. 
Consider the spectrum of the channel defined as the following: $\calP \sigma_\alpha=\lambda_\alpha \sigma_\alpha$, where we regard the quantum operators as vectors, e.g., expanded in the {Pauli} basis. Note that $\sigma_\alpha$'s are generally not quantum states. The largest eigenvalue corresponds to the fixed point state of the channel, $\lambda_0=1$ and $\sigma_0=\rho_{\rm fix}$ satisfying $\calP \rho_{\rm fix}=\rho_{\rm fix}$. The second largest eigenvalues (in terms of the norm) generally come in conjugate pairs $\{\lambda_1, \lambda_1^\ast\}$, with corresponding operator representation of eigenvector $\{\sigma_1, \sigma_1^\dagger\}$ given the real matrix representation of channel $\calP$. 
For the initial state $\rho_{R A_0}$, we can in general decompose it as 
\be 
\rho_{RA_0}=\sum_{i,j} c_{ij} \gamma_i \otimes \sigma_j=\sum_j o_j\otimes \sigma_j,
\ee 
where $\gamma_j$ forms a basis of operators on reference system $R$ and $o_j=\sum_i c_{ij}\gamma_i$ are operators. Note that $o_j$'s are generally not valid quantum states. Applying the channels $\calP$ for $t\gg1$ times following the unmonitored dynamics in Eq.~\eqref{unconditional_dynamics}, we have 
\begin{align}
\rho_{RA_t}&=\sum_j o_j \otimes \lambda_j^t\sigma_j
\\
&\simeq o_0 \otimes \rho_{\rm fix}+\lambda_1^t o_1\otimes \sigma_1 +\lambda_1^{\ast t} o_1^\dagger\otimes \sigma_1^\dagger.
\label{rho_expansion}
\end{align}
As QMI is continuous in the quantum states, from Taylor expansion the QMI
\begin{align} 
I(R:A_t)&\propto |\lambda_1|^{c t}
\label{IRA_late_time_general}
\end{align} 
where $c$ is a positive integer.} 

{
For Haar random unitary induced channel $\calP$, all eigenvalues filling a disk of radius $1/\sqrt{d_B}$ except for $\lambda_0$~\cite{kukulski2021generating} (see Fig.~\ref{fig:mi_traceout}g). Therefore, we have $I(R:A_t)\propto d_B^{-ct}$, which recovers the prediction from Theorem~\ref{MI_traceout_theorem}.}

\begin{figure}[t]
    \centering
    \includegraphics[width=0.45\textwidth]{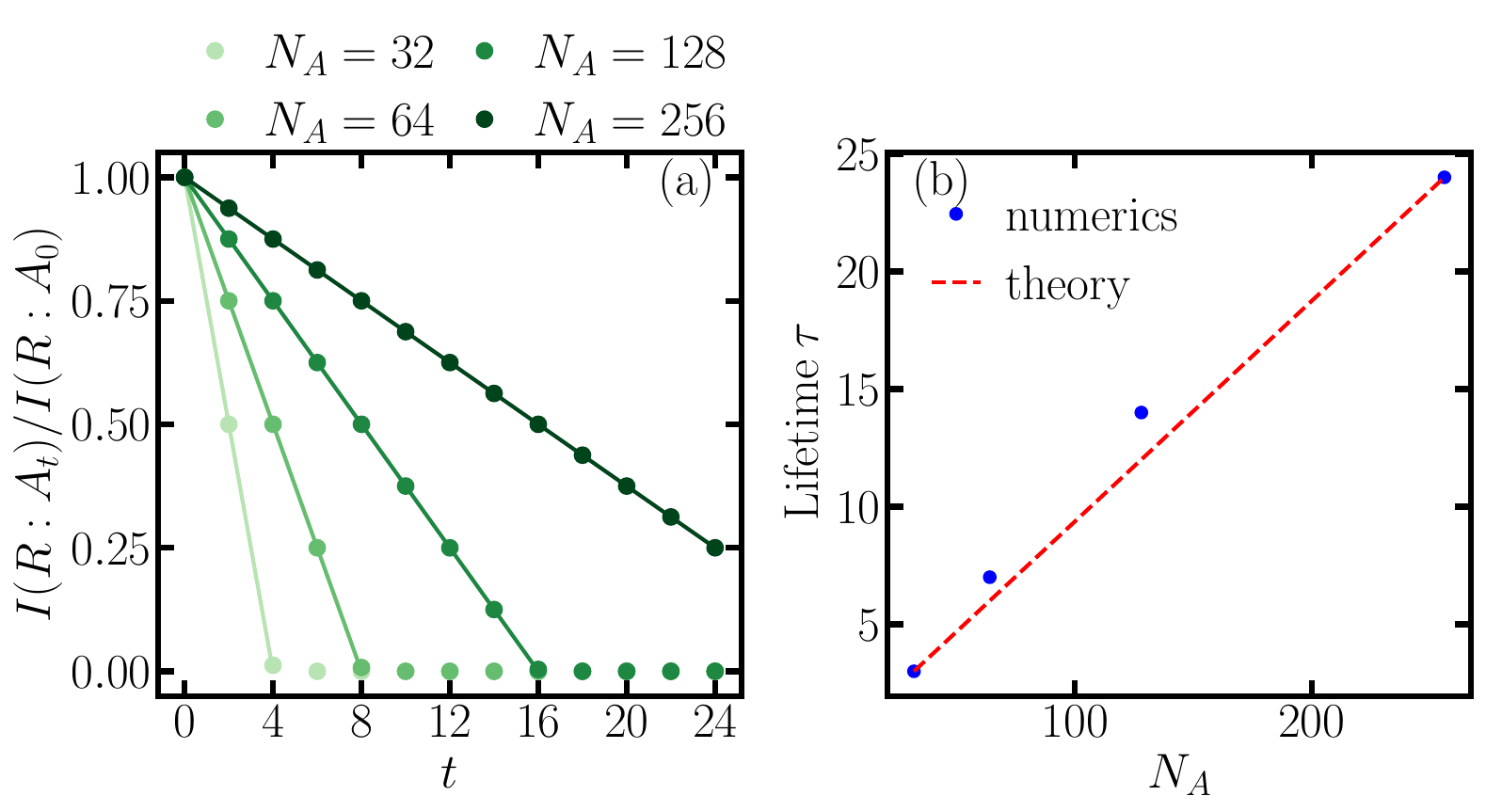}
    \caption{\BZ{Measurement-unconditioned QMI $I(R:A_t)$ in unmonitored Clifford dynamics with reset. The initial state is the Bell state with $N_R = N_A$. In subplot (a), we plot the numerical simulations of the normalized QMI $I(R:A_t)/I(R:A_0)$ (dots) in a system of various $N_A$ and $N_B = 16$. The solid lines represent the asymptotic result of Eq.~\eqref{eq:MI_traceout_asymp} in Theorem~\ref{MI_traceout_theorem}. In subplot (b), we show the QMI lifetime. Blue dots are numerical results obtained from (a) and the red dashed line is the theoretical prediction of Eq.~\eqref{eq:MI_traceout_time} with $\epsilon = 1/4$.}}
    \label{fig:uncondQMI_clifford}
\end{figure}

In Fig.~\ref{fig:mi_traceout}a, we verify the predictions in Theorem~\ref{MI_traceout_theorem} for Bell state, finding the close agreement between QMI of numerics (circles) and R\'enyi-2 extended QMI of theory (lines). The QMI indeed has a linear decay for both the Haar unitary dynamics and the Hamiltonian dynamics, when the environment is not monitored; in the late time region, we also observe an exponential decay (see Fig.~\ref{fig:mi_traceout}b). The Hamiltonian dynamics also shows similar behavior, except that the late time decay exponent is different from the theory prediction, which we will explore in detail in Section~\ref{sec:scar}. In Fig.~\ref{fig:mi_traceout}c-f, we further extend the numerical simulation to less correlated initial states, including various perturbed Haar random states and CQ states. Both the early time linear decay and late-time exponential decay are observed. \BZ{To validate our results for larger systems, we perform Clifford circuit simulations and show the results in Fig.~\ref{fig:uncondQMI_clifford}a. Up to $N_A = 256$ qubits, the numerics are in exact agreement with the early-time asymptotic prediction in Eq.~\eqref{eq:MI_traceout_asymp}. This agreement again follows from the coincidence of von-Neumann and R\'enyi-2 entropies for stabilizer states. 
Since stabilizer-state mutual information is quantized, the unconditioned QMI drops linearly to zero and does not exhibit the late-time exponential tail seen for Haar dynamics (Fig.~\ref{fig:mi_traceout}b.}

With Theorem~\ref{MI_traceout_theorem} and its generalizations established, we anticipate that the scaling behavior of QMI dynamics can vary depending on the choice of initial state, as explained below.

\begin{figure}[t]
    \centering
    \includegraphics[width=0.45\textwidth]{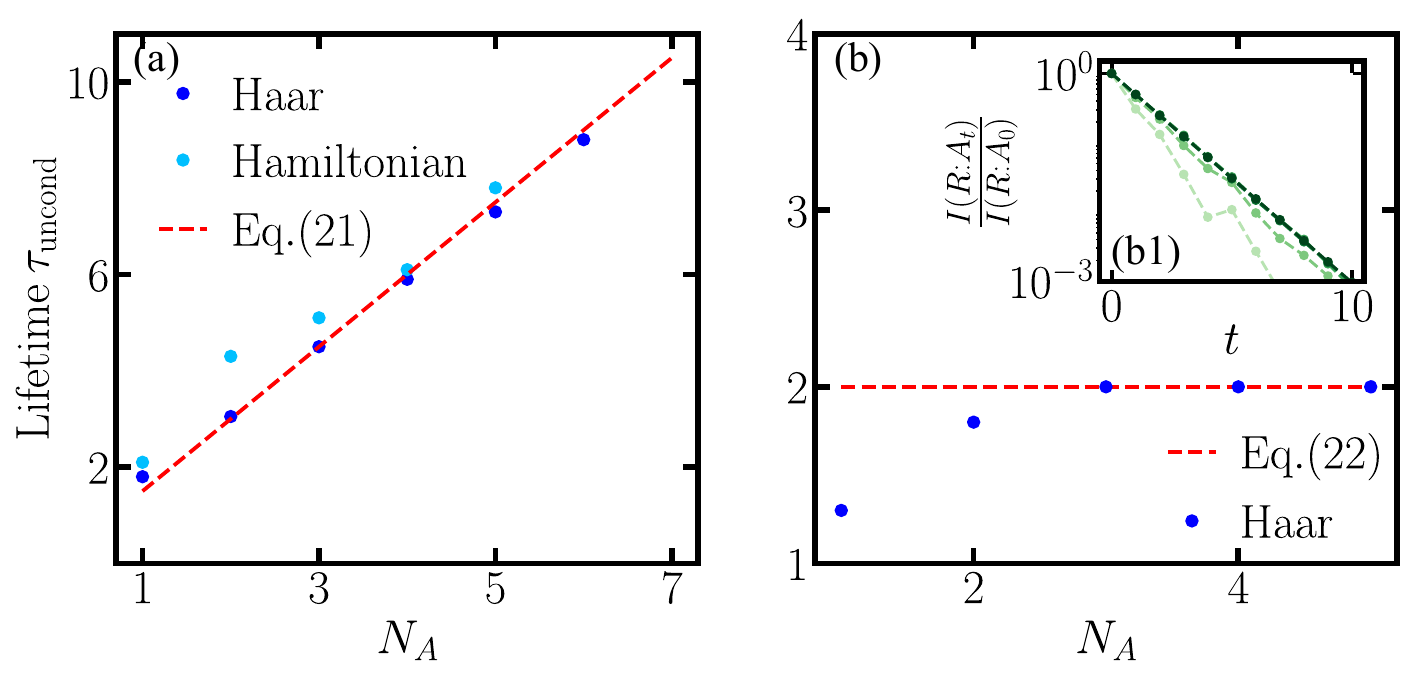}
    \caption{Lifetime of measurement-unconitioned QMI $I(R:A_t)$ in unmonitored dynamics with reset.
    In (a), we show QMI lifetime versus system size $N_A$ with Bell initial state for Haar (blue) and Hamiltonian (cyan) unitary implementation. Red dashed line represents the theoretical result in Eq.~\eqref{eq:MI_traceout_time}. In (b), we show QMI lifetime for Haar unitary with initial state $\rho_{RA_0}$ to be unconditional state in dynamics with Bell initial state at late-time. Red dashed line represents theory in Eq.~\eqref{tau_eig}. {Inset shows the decay of normalized QMI with $N_A=1$ to $5$ (light to dark). In both cases, we take $\epsilon=1/4$.}
    }
    \label{fig:lifetime_traceout}
\end{figure}

\subsubsection{Linear lifetime for typical cases}
For typical states of interest, the dynamics is dominated by the linear decay and we can obtain the unmonitored $\epsilon$-lifetime from Eq.~\eqref{eq:MI_traceout_asymp} as
\begin{align}
    \tau_{\rm uncond} \simeq 2(1-\epsilon)N_A/N_B.
    \label{eq:MI_traceout_time}
\end{align}
Interestingly, the lifetime without monitoring the bath agrees with the notion of ``thermalization time'' in {\em holographic deep thermalization}~\cite{zhang2025holographic} which scales $\sim N_A/N_B$ though with a factor of $2$ difference due to Bell initial state. On the contrary, the lifetime of Eq.~\eqref{eq:avgMI_time} with bath monitoring reveals a different time-scale of information lost toward equilibrium. In Fig.~\ref{fig:lifetime_traceout}a, we verify the linear QMI lifetime of Bell initial state for both Haar unitary and Hamiltonian evolution. \BZ{We further verify the scaling of QMI lifetime in random Clifford dynamics for larger systems in Fig.~\ref{fig:uncondQMI_clifford}b.} Moreover, through our observations in Fig.~\ref{fig:mi_traceout}c and e, we expect that the linear lifetime holds for other types of typical states as we considered there.

\subsubsection{Constant lifetime for residual dynamics}

{
For certain states with extremely low correlation, the late time dynamics of Eq.~\eqref{eq:MI_traceout_asymp_late} dominates. This happens for initial states whose expansion in Eq.~\eqref{rho_expansion} becomes precise even at an early time. Since Eq.~\eqref{eq:MI_traceout_asymp_late} indicates an exponential decay at the rate of $d_B^{-t}$, lifetime defined in Eq.~\eqref{epsilon_def} with time zero shifted to late time leads to 
\be 
\tau_{\rm uncond} \simeq \log_2(1/\epsilon)/N_B,
\label{tau_eig}
\ee 
a constant independent of data system size.
In the inset (Fig.~\ref{fig:lifetime_traceout}b1), we do see a collapse of exponential QMI decay with increasing system size $N_A$, when we adopt the initial state from the late-time ($t \gg 2N_A/N_B$) unconditional state (Eq.~\eqref{eq:state_uncond}) evolved from a Bell state $\ket{\Phi}_{RA_0}$. For such weakly correlated states, QMI lifetime does not depend on system size as shown in Fig.~\ref{fig:lifetime_traceout}b.}

\subsection{\BZ{Dynamical transition in the generalization to $N_R<N_A$}}
\label{sec:uncond_unequal}

\BZ{
In this section, we take the initial state to be Eq.~\eqref{eq:init_state_lessR} and study the measurement-unconditioned QMI dynamics. For analysis convenience, we focus on the R\'enyi-extended QMI, and obtain the following theorem (see Appendix~\ref{app:thereom23} for a proof) 
\begin{theorem}
\label{MI_traceout_lessR_theorem}
    In the quantum dynamics of $2$-design unitaries with mid-circuit measurements and reset, the expected R\'enyi-2 extended measurement-unconditioned QMI of an initial $N_R$ Bell pairs at time $t$ is asymptotically ($d_A, d_B \gg 1$)
    \begin{align}
        &\E_{\rm Haar} I_2(R:A_t)\nonumber\\
        &\simeq \log_2\left(1+d_R d_A d_B^{-t}\right) - \log_2\left[1 + \left(\frac{d_A}{d_R} - 1\right) d_B^{-t}\right].
        \label{eq:MI_traceout_lessR}
    \end{align}
\end{theorem}
The full version is given in Appendix~\ref{app:thereom23}. From the asymptotic dynamics in Eq.~\eqref{eq:MI_traceout_lessR}, we define two characteristic timescales 
\be
    \tau_{\mp} = (N_A \mp N_R)/N_B,
\ee
and the unconditional QMI $I_2(R:A_t)$ can be approximated as
\begin{align}
    \E_{\rm Haar} I_2(R:A_t) \simeq \begin{cases}
        2N_R &t \le \tau_{-},\\
        N_A + N_R-N_Bt & \tau_{-} \le t\le \tau_{+}, \\
        2^{N_R+N_A-N_Bt}/\log2\to 0 & t \ge \tau_{+}.
    \end{cases}
    \label{eq:MInorm_traceout_lessR}
\end{align}
This piecewise behavior indicates a dynamical transition of QMI. As a sanity check, setting $N_R = N_A$ in Eq.~\eqref{eq:MInorm_traceout_lessR} recovers the asymptotic expansion of Eq.~\eqref{eq:MI_traceout_asymp} and Eq.~\eqref{eq:MI_traceout_asymp_late} in Theorem~\ref{MI_traceout_theorem}.
For finite and non-vanishing timescales $\tau_{\mp}$, we consider the thermodynamic limit of system $N_A \to \infty$ but finite bath-to-system ratio $N_B/N_A \to b \in (0, 2)$. Under this assumption, the normalized QMI in Eq.~\eqref{eq:MInorm_traceout_lessR} exhibits a dynamical transition.
In particular, when the reference $N_R$ is asymptotically small, compared to the system, i.e., $N_R/N_A \to 0$, the two timescales coincide, $\tau_{-} = \tau_+ = N_A/N_B \eqqcolon \tau_c$, and the QMI at the critical time reduces to
$I_2(R:A_{\tau_c}) = N_R$. 
}


\BZ{
The nature of this dynamical transition can be understood from a QEC viewpoint. Via an equivalent expansion of the bath at different times, the final state can be viewed as a random QEC code that encodes $N_R$ logical qubits into $N_A + tN_B$ physical qubits (in spacetime). In the noise layer, $tN_B$ qubits undergo fully depolarizing channels. The {\em quantum hamming bound} then requires a functioning QEC code to satisfy
\be
    N_A + tN_B - N_R \ge \log_2 4^{tN_B} = 2tN_B,
\ee
which leads to the condition of $t\le (N_A-N_R)/N_B \equiv \tau_-$ in Eq.~\eqref{eq:MInorm_traceout_lessR}. While for $t > \tau_-$, with the increase of erasure errors, the dynamical random code can no longer correct all the errors. In particular, in the limit of $N_A \to \infty$ with $N_R/N_A \to 0$, the code undergoes a phase transition from a ``good'' code to a ``bad'' code at the critical code rate of $R_c = N_R/(2N_A)$. 
}

\begin{figure}[t]
    \centering
    \includegraphics[width=0.45\textwidth]{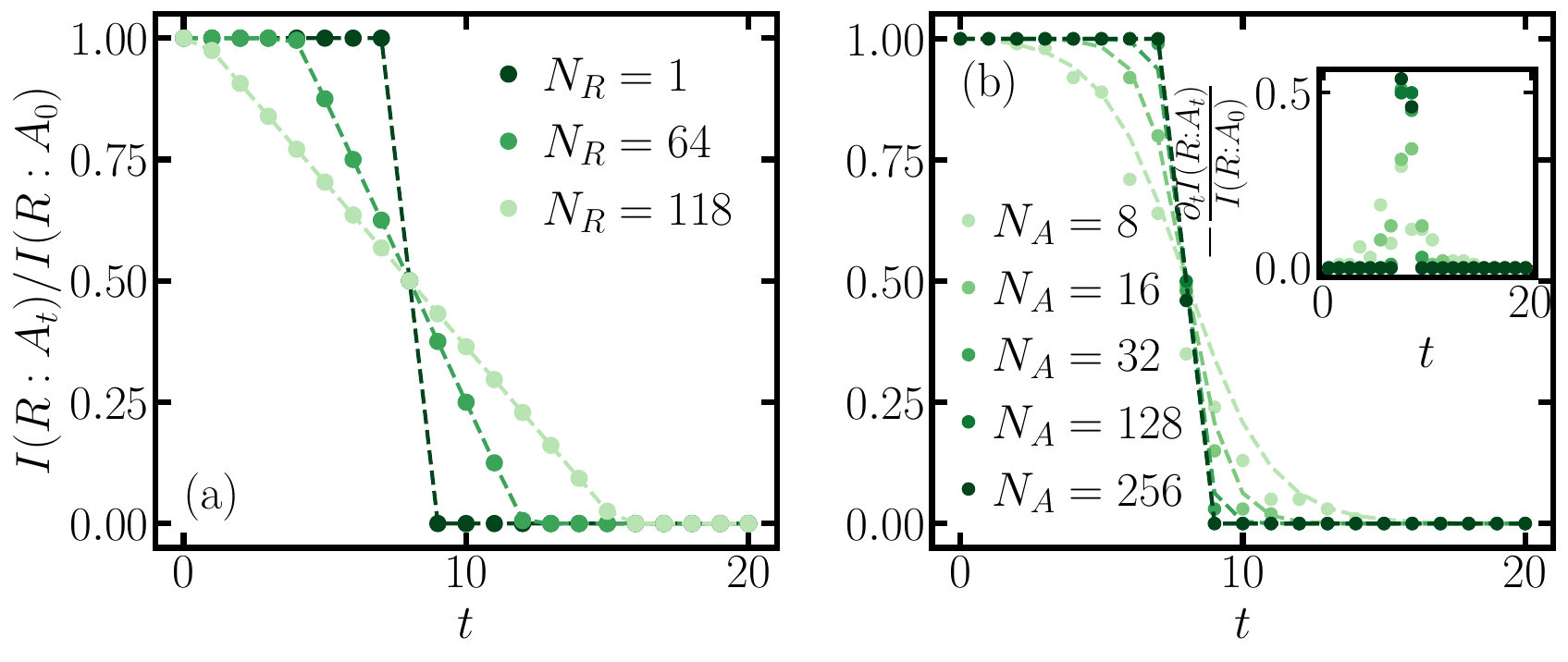}
    \caption{\BZ{Normalized measurement-unconditioned QMI $I(R:A_t)/I(R:A_0)$ in unmonitored dynamics with reset and $N_R<N_A$. In (a), we plot numerical simulations with $N_A = 128, N_B = 16$ qubits and various reference $N_R$ (dots). In (b), we set $N_R = 1$ and $N_B/N_A = 1/8$ for different system size $N_A$. The dashed lines in both subplots are theoretical results of Eq.~\eqref{eq:MI_traceout_lessR} in Theorem~\ref{MI_traceout_lessR_theorem}. 
    The inset shows the time derivative of the normalized QMI. The sharpness of transition is limited by the discrete $t$.}}
    \label{fig:uncondQMI_lessR}
\end{figure}

\BZ{
In Fig.~\ref{fig:uncondQMI_lessR}a, we numerically validate the dynamical transition of unconditioned QMI with $N_R<N_A$ qubits. Using Clifford circuits, the numerical results exactly agree with Eq.~\eqref{eq:MI_traceout_lessR} in Theorem~\ref{MI_traceout_lessR_theorem}, and for $t\ge \tau_+$, the QMI vanishes. For a fixed $N_A$, the time window $t \le \tau_-$ of near-perfect information protection and correctable errors shrinks as $N_R$ increases (light to dark). We next fix the encoded information to $N_R = 1$ (a single logical qubit) and set $N_B/N_A = 1/8$. As the system size $N_A$ increases (light to dark), the normalized unconditioned QMI dynamics collapse onto a universal scaling form (see Fig.~\ref{fig:uncondQMI_lessR}b), consistent with a dynamical phase transition. In the inset, we further plot temporal derivative of the normalized QMI, and observe a peak at the critical time $t = \tau_c$. Due to discrete temporal steps in the circuit, the peak only remains finite height rather than divering. 
}

\subsection{Anomaly in Hamiltonian dynamics: exponential lifetime from outliers}
\label{sec:scar}

Figure~\ref{fig:mi_traceout}b shows that, under the dynamics generated by the Hamiltonian in Eq.~\eqref{H_Ising}, the exponent governing the exponential decay of the QMI is significantly smaller than the theoretical prediction given in Theorem~\ref{MI_traceout_theorem}, indicating a notable deviation from Haar dynamics. In the following, we examine this anomaly in detail.

\begin{figure}[t]
    \centering
    \includegraphics[width=0.45\textwidth]{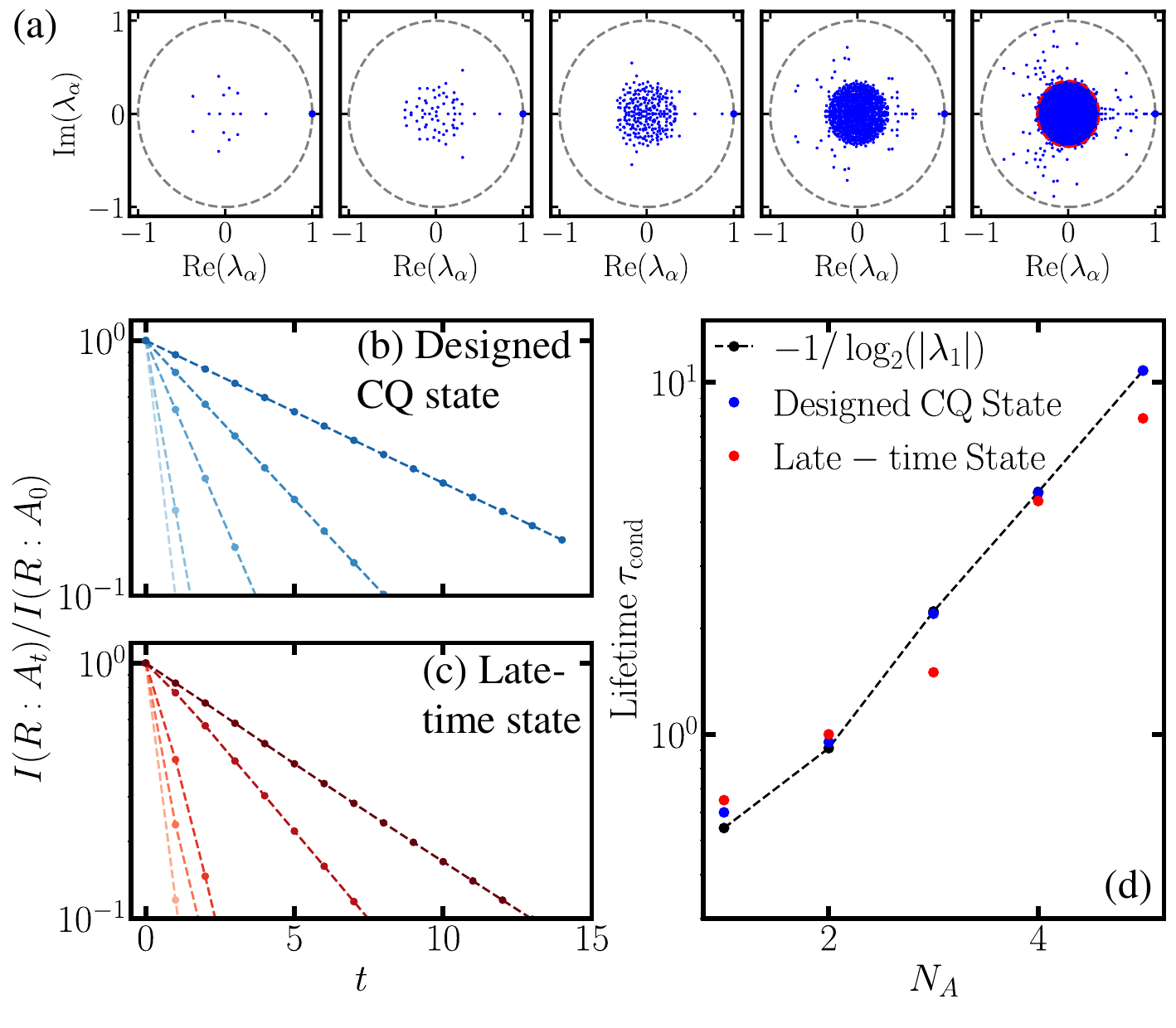}
    \caption{
    Measurement-unconditioned QMI in unmonitored Hamiltonian dynamics with reset. 
    In (a), we show the eigenspectrum of channel $\calP_0$ from $N_A=2$ to $6$ (left to right). {Red dashed circle indicates the radius of $1/\sqrt{d_B}$.
    In left bottom panel, we plot the decay of normalized QMI in different system size with initial state $\rho_{R A_0}$ to be (b) the designed CQ state in Eq.~\eqref{rho_RA0_classical} and (c)   
    unconditional state in dynamics with Bell initial state at late-time separately. 
    Colored lines from light to dark represent system size of $N_A=1$ to $5$.
    In (d), we show their corresponding lifetime and compare to the memory time $\tau_{\rm eig}$.  In all cases, we fix bath size $N_B=3$.}
    }
    \label{fig:spectrum}
\end{figure}

For the late-time dynamics, Eq.~\eqref{IRA_late_time_general}, derived using the quantum channel spectrum approach, remains generally applicable. Therefore, we start with an analysis based on the spectrum of the Hamiltonian-induced quantum channel. {In Fig.~\ref{fig:spectrum}a, we numerically find that for the Hamiltonian evolution considered, most eigenvalues concentrate within a constant-size disk of size $1/\sqrt{d_B}$ (red dashed circle) for increasing system size $N_A$, similar to the Haar case in Fig.~\ref{fig:mi_traceout}g. However, there exist isolated eigenvalues located between the bulk of the spectrum and the fixed point eigenvalue $\lambda_0 = 1$. These outliers contribute to the residual information retained in the system and serve as a phenomenological analog, in the channel spectrum, to the eigenstate spectrum observed in the ``many-body scar" Hamiltonian~\cite{serbyn2021quantum}. 
We note that the Hamiltonian in Eq.~\eqref{H_Ising} does not feature a low-entanglement energy tower, which is a one of the key hallmarks of many-body scarred systems. As a result, we find that the theoretical memory time, given by $\tau_{\rm eig} = -1/\log_2(|\lambda_1|)$, deviates from the Haar prediction and instead exhibits exponential growth with $N_A$, as indicated by the \BZ{black dashed line} in Fig.~\ref{fig:spectrum}d. }

As an example, we can construct a classically correlated state between the reference and memory system as
\be 
\rho_{RA_0}=\frac{1}{2}\left[\state{0}\otimes\rho_{\rm fix}+\state{1}\otimes \left(\rho_{\rm fix}+a (\sigma_1+\sigma_1^\dagger)\right)\right],
\label{rho_RA0_classical}
\ee  
for a properly chosen $a$ such that $\rho_{\rm fix}+a(\sigma_1 + \sigma_1^\dagger)$ is a valid quantum state. Here $\rho_{\rm fix}$ and $\sigma_1$ are the eigenvectors of the Hamiltonian induced quantum channel $\calP$. Iterating the dynamics with channel $\calP$ in Eq.~\eqref{unconditional_dynamics} to time step $t$, we have
\be 
\rho_{RA_t}=\frac{1}{2}\left[\state{0}\otimes\rho_{\rm fix}+\state{1}\otimes \left(\rho_{\rm fix}+a\lambda_1^t\sigma_1+a\lambda_1^{\ast t}\sigma_1^\dagger \right)\right].
\label{rho_RA_classical}
\ee 
which is a special case of Eq.~\eqref{rho_expansion}. Therefore, we can infer the lifetime as in Eq.~\eqref{IRA_late_time_general} with $|\lambda_1|$ following exponentially scaling as we have seen. {We see that the choice of initial state in Eq.~\eqref{rho_RA0_classical} guarantees the dominance of the first non-unity eigenvalue $\lambda_1$ in Eq.~\eqref{rho_expansion} even at early time. Therefore, the QMI decays exponentially at early time, as confirmed in Fig.~\ref{fig:spectrum}b.}

{
Besides the designed CQ state, we numerically evaluate the QMI with initial state from late-time ($t\gg 2N_A/N_B$) unconditional state evolved from a Bell state under the Hamiltonian dynamics in Fig.~\ref{fig:spectrum}c. In contrast to the collapsed behavior in the inset of Fig.~\ref{fig:lifetime_traceout}b1, QMI decays exponentially at different rates with increasing system size $N_A$. 
Figure~\ref{fig:spectrum}d clearly shows an exponential increase in the QMI lifetime, consistent with the scaling of the memory time $\tau_{\rm eig} = -1/\log_2(|\lambda_1|)$ discussed above.}

\section{Transition of QMI lifetime in partially monitored dynamics}
\label{sec:partial_monitor}

\begin{figure}
    \centering
    \includegraphics[width=0.45\textwidth]{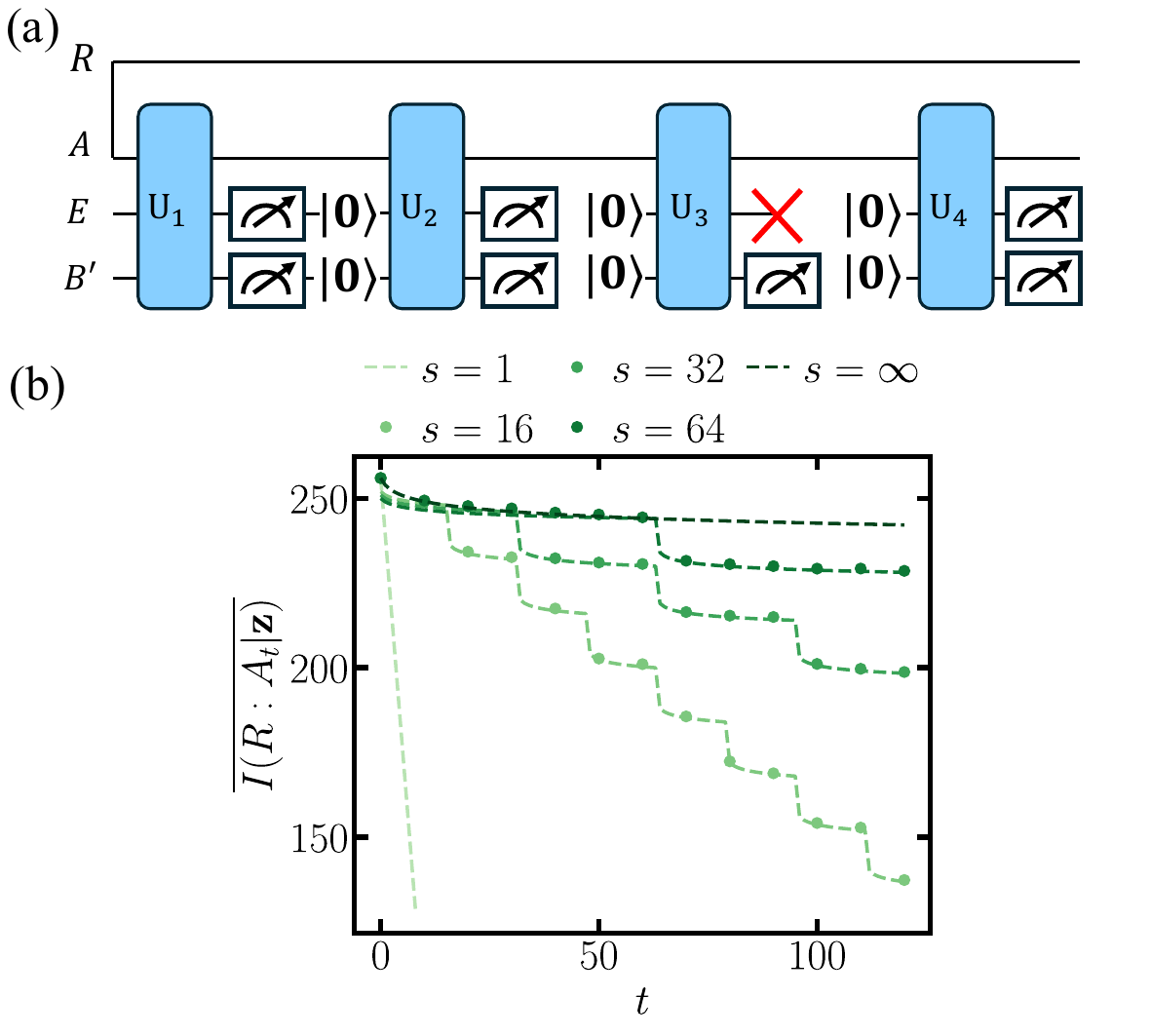}
    \caption{\BZ{Measurement-conditioned QMI $\overline{I(R:A_t|\bfz)}$ in partially monitored dynamics with reset. In (a), we show an example of the partially monitored circuit with $s = 3$. The red cross represents the erasure channel. In (b), we plot numerical simulations of Clifford circuits with $N_R = N_A = 128$, $N_{B'} = N_E = 16$ qubits (circles). Dashed lines represent the theoretical prediction of Eq.~\eqref{eq:avgQMI_partial_simplify} in Theorem~\ref{condMI_partial_theorem}.}}
    \label{fig:qmi_dope}
\end{figure}

\BZ{In this section, we study the dynamics with a tunable monitoring strength. We implement this by applying an erasure channel on $N_E\le N_B$ qubits in the bath every $s$ time steps (illustrated in Fig.~\ref{fig:qmi_dope}a). This is because when a bath qubit is erased (and to be reset in the next time step), it is equivalent to trace out---the quantum state becomes the average state even conditioned on $\bfz$. Equivalently, one can consider a fully depolarizing channel instead of erasure. The resulting measurement-conditioned dynamics therefore interpolates between fully monitored and unmonitored dynamics, as also discussed in Ref.~\cite{liu2024noise}.
In particular, $s = \infty$ reproduces the monitored dynamics of Section~\ref{sec:exp}; whereas $s = 1$ with $N_{E} = N_B$ yields the unmonitored dynamics of Section~\ref{sec:uncond}. 
For simplicity, we focus on $N_R = N_A$ (the extension to $N_R<N_A$ is analogous). Because the conditional state is also mixed, we work with the R\'enyi-2 extended conditional QMI. Assuming Haar random unitaries at each step, we obtain the following theorem where $d_E = 2^{N_E}$ is the Hilbert-space dimension of erased bath qubits (see Appendix~\ref{app:condMI_partial_proof} for a proof).
\begin{theorem}
\label{condMI_partial_theorem}
    In the quantum dynamics with mid-circuit measurements, reset and periodic erasure channel on $N_{E}$ qubits in every $s$ steps, the expected R\'enyi-2 extended measurement-conditioned QMI of a Bell initial state at time $t$ in the asymptotic limit of $d_A, d_B/d_E \gg 1$ is
    \begin{align}
        &\E_{\rm Haar} \overline{I_2(R:A_t|\bfz)} \nonumber\\
        &\simeq 2N_A - n_sN_E - \log_2\left(s\frac{1-d_E^{-n_s}}{d_E-1} + t-s n_s + 1\right) \nonumber\\
        &\quad- \log_2\left(\frac{s d_E\left(1-d_E^{-n_s}\right)}{d_E-1} + (t-s n_s+1)d_E^{-n_s}\right).
        \label{eq:avgQMI_partial}
    \end{align}
    Here $n_s = \lfloor t/s\rfloor$ is the number of periods for erasure channels. When $d_E \gg 1$, 
    \be
        \E_{\rm Haar} \overline{I_2(R:A_t|\bfz)} \simeq 2N_A - n_s N_E - \log_2(1+t-s n_s)-\log_2s.
        \label{eq:avgQMI_partial_simplify}
    \ee
\end{theorem}
}

\BZ{
Eq.~\eqref{eq:avgQMI_partial_simplify} shows that within each erasure period, the conditional QMI exhibits a logarithmic decay. At each erasure event, the QMI decreases by $N_E$, leading to a two-scale transition. More explicitly, for $t \in [n_s s, (n_s+1)s]$ with fixed $n_s$, $\E_{\rm Haar} \overline{I_2(R:A_t|\bfz)} \simeq \left(2N_A - n_s N_E\right) - \log_2 s -\log_2(1 + t - sn_s)$ presents a logarithmic decay of QMI with $t$. In contrast, if we probe times $t/s \in \mathbb{N}^+$ at the erasure events, the QMI $\E_{\rm Haar} \overline{I_2(R:A_t|\bfz)} \simeq  2N_A- \log_2 s - tN_E/s$ reduces to an effective linear decay on  macroscopic timescales. Overall, we observe this two-scale transition in the conditional QMI dynamics. This transition admits a statistical physics interpretation. The QMI can be mapped to the free energy of a permutation-spin model, where each spin takes values of identity or swap. 
An erasure event imposes an identity boundary condition at that time slice, biasing subsequent spins toward the identity. As a result, logarithmic decay survives only when erasures are rare $t/s \sim o(1)$, whereas for $t/s \sim \Omega(1)$, the effective behavior is linear. 
}

\BZ{For the $\epsilon$-QMI lifetime, we estimate the critical number of periods as $n_s^* = \left[(2-\epsilon)N_A - \log_2(s)\right]/N_E$, and so that $\tau_{\rm cond} = s n_s^* = s\left[(2-\epsilon)N_A - \log_2(s)\right]/N_E$. If $s \sim \mathcal{O}\left({\rm poly}(N_A)\right)$, then $\tau_{\rm cond} \sim \mathcal{O}\left({\rm poly}(N_A)\right)/N_E$. In contrast, for $s \sim {\rm exp}(N_A)$, the lifetime can grow exponentially, $\tau_{\rm cond} \sim {\rm exp}(N_A)$.}

\BZ{
In Fig.~\ref{fig:qmi_dope}b, we present Clifford circuit simulations of the conditional QMI in the partially monitored setting with different $s$. When $s \to \infty$, the conditional QMI decays logarithmically in time, consistent with Eq.~\eqref{eq:avgMI_lb_simplify} in Theorem~\ref{theorem_avgMI}. For a fixed time window (e.g., $t\le 128$ in Fig.~\ref{fig:qmi_dope}b), decreasing $s$ increases the frequency of erasure events, producing more step-like plateaus (``stairs'') and shortening the logarithmic-decay segments between successive drops. As a result, the dynamics crosses over continuously from a logarithmic to a linear scaling. In the extreme case of $s = 1$, we recover Eq.~\eqref{eq:MI_traceout_asymp} in Theorem~\ref{MI_traceout_theorem}.
}

\section{Interpretation in quantum circuit models}
\label{sec:interpretation}

Having completed the analysis of QMI lifetime, we now apply the framework introduced in Section~\ref{sec:preliminary} to interpret our results {within the context of quantum computing models, monitored quantum circuits and entanglement-assisted communication.}

\subsection{Quantum diffusion models}

\label{sec:qddpm}

In quantum diffusion models, i.e. QuDDPM~\cite{zhang2024generative}, we aims to learn the distribution of a target state ensemble, and generate new samples following this distribution, referred to as the quantum generative learning (see Fig.~\ref{fig:scheme}c). In the forward process of QuDDPM, the states samples from the target state ensemble are evolved with controllable scrambling unitaries to approach the Haar random states ensemble. The forward process stands as a reference dynamical process for the learning in backward process. In the backward learning process, the data and bath system are initially prepared in a random state and a trivial product state. Here we introduce the reference system to index the states on the data system for analysis convenience. In particular for the state ensemble $\{\ket{\phi_k}, P_{\phi_k}\}_k$, it can be equivalently represented as
\be
    \ket{\psi}_{RA} = \sum_k \sqrt{P_{\phi_k}}\ket{k}_R \ket{\phi_k}_A.
\ee
In the $t$-th step, the joint system of data and bath evolves through a parameterized unitary $U_t(\bm \theta_t)$ followed by mid-circuit measurements on the bath system. Ideally, we expect the output state ensemble in $t$-th step to match the corresponding state ensemble in the forward process, and as the backward denoising process proceeds, the final output state ensemble from backward circuits follows the same distribution of the target states ensemble of interest. As the output state depends on the whole measurement trajectory, the backward learning process refers to the monitored dynamics studied in Section~\ref{sec:exp}.

According to our theoretical results in Theorem~\ref{theorem_avgMI}, in the backward denoising process, if we simply apply unitary circuits with randomly initialized parameters {and focus on a single measurement trajectory}, it at least requires exponential number of steps to erase the correlation between the reference index system and the states in the data system, which makes the denoising process inefficient and hard for near-term devices and applications. This negative insight thus \BZ{indicates that one should not focus on single-trajectory learning}. \BZ{In addition, training may be able to resolve the issue for single-trajectory case.} {Indeed,} when we compare the output state ensemble of backward circuits and forward scrambling, minimizing a well-designed loss function can guide us to optimize the parameterized unitary $U_t (\bm \theta_t)$, which stores the information for mapping from input state ensemble from the prior step in backward process toward the output ensemble in corresponding forward step, therefore boosts the backward learning process in reducing the number of steps.


\subsection{Quantum reservoir computing}
\label{sec:qrc}

\begin{figure}[t]
    \centering
    \includegraphics[width=0.45\textwidth]{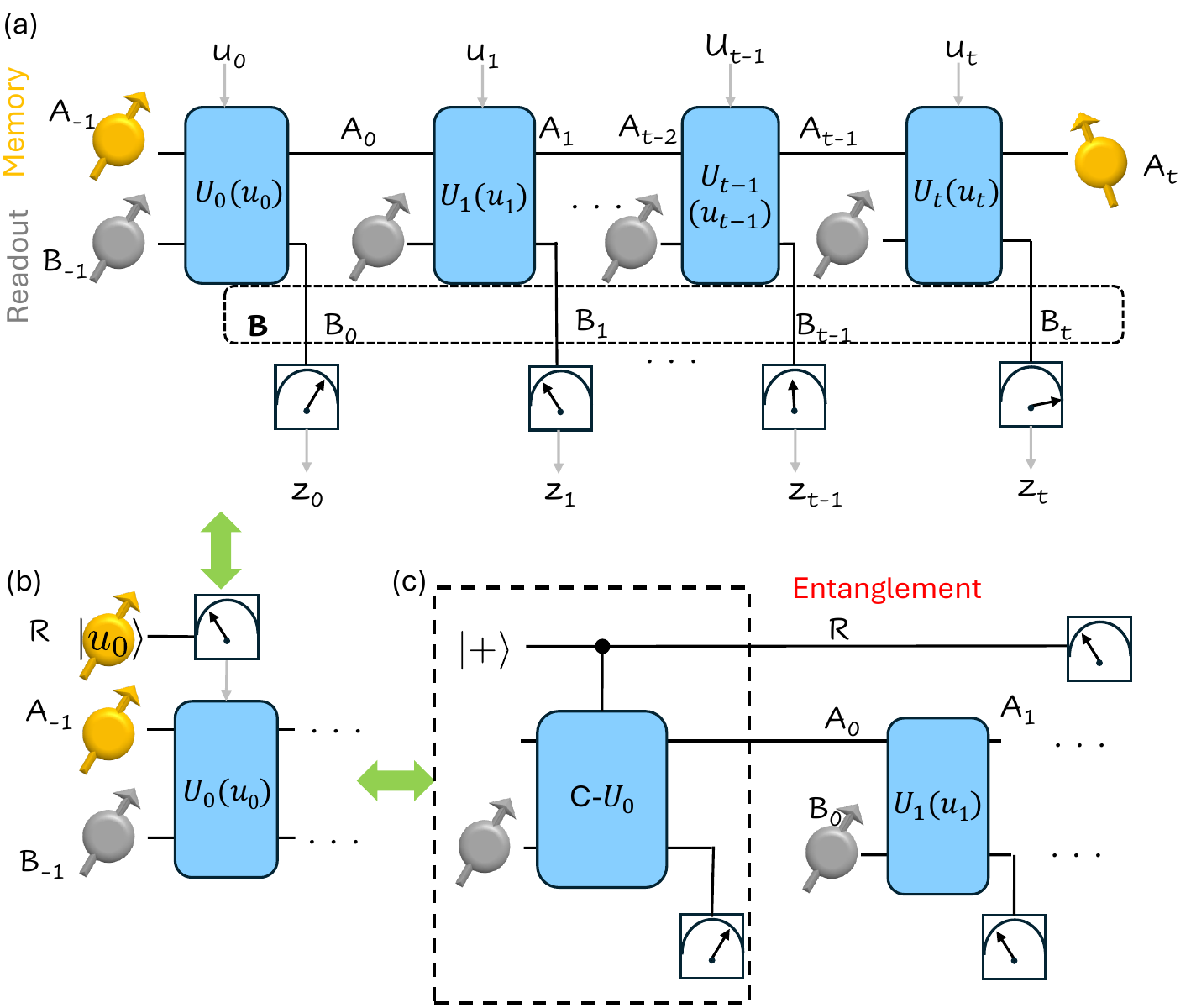}
    \caption{Reduction from the reservoir computing paradigm with classical inputs $u_0u_1\cdots u_t$ to the quantum circuit with entangled input $RA_0$.}
    \label{fig:scheme_RC}
\end{figure}

The objective of QRC is to learn a functional that maps a temporal sequence $\bm{u} = \{\cdots, u_0, u_1, \cdots, u_t, \cdots\}$ to an output sequence $\bm{y}$. In a mid-circuit-measurement based QRC setup~\cite{martinez-pena_quantum_2023, hu2024overcoming}, the system consists of a memory and readout register with $N_A$ and $N_B$ qubits, respectively. The system undergoes a sequence of unitary evolutions $U_t(u_t)$, conditioned on the classical input $u_t$ (see Fig.~~\ref{fig:scheme_RC}a). In typical implementations of $U_t(u_t)$, the unitary, for instance, can be generated by a parameterized Hamiltonian $H + u_t \sum_{i=1}^L \eta_i^z \sigma_i^z$ with $H$ defined in Eq.~\eqref{H_Ising}.

After implementing each unitary, one measures the readout system to obtain results $\{\cdots,z_0,z_1,\cdots, z_t,\cdots\}$, each $z_t\in\{0,1\}^{N_B}$ is a binary bitstring. 
Through repetitions of experiments, one can obtain the learning features $\bm x(t)$, a normalized $2^{N_B}$-dimensional vector of unconditional measurement statistics of $z_t \in \{0, 1\}^{N_B}$, with each component being the probability of the corresponding measurement outcome $z_t$. 
For example, with $N_B = 3$ readout qubits, $x_7(t) = \text{Prob}[z_t = 111]$. The final output $y_t$ is estimated as a weighted sum: $y_t = \bm{w} \cdot \bm{x}(t)$.

Consider the dynamics that starts with 
unitary $U_0(u_0)$ encoded with
the classical input $u_0$, as shown in Fig.~\ref{fig:scheme_RC}b, one can equivalently introduce a reference system $R$, whose quantum state $\ket{u_0}$ determines the dynamics of the memory-readout system $A_{-1}B_{-1}$ via a measurement: one measures $R$ in the encoding bases $\{\ketbra{u_0}\}$ and then performs the unitary $U_0(u_0)$ conditioned on the measurement result. Taking a step further in Fig.~\ref{fig:scheme_RC}c, one can consider $R$ in a superposition state $\ket{+}\propto\sum_{u_0}\ket{u_0}$ assuming equal prior probability of $u_0$ without losing generality and the measurement-conditioned unitary is realized as a control-unitary gate defined as ${\rm C}-U_0 = \sum_{u_0} \ketbra{u_0}{u_0}_R \otimes U_0(u_0)$. In this regard, assuming initial trivial product state, the $RA_{-1}B_{-1}$ system becomes entangled via the control unitary, and after the first measurement on the readout system, the state $RA_0$ remains entangled while $B_0$ is reset to $\ket{\bm 0}_{B_0}$, leading to a reduction towards the set-up in Fig.~\ref{fig:scheme}a. Since our theoretical analysis assumes typical and sufficiently complex unitary dynamics $U_1, \ldots, U_t$, we can neglect the subsequent inputs $u_1, \ldots, u_t$ when focusing on the memory of the initial input $u_0$, following the same reasoning as in Ref.~\cite{hu2024overcoming}.


In this context, the fading memory effect in reservoir computing can be quantified by the correlation between the initial input $u_0$ and the readout $z_t$, as a function of the time step $t$. Since QRC employs marginal measurement statistics to compute $y_t = \bm{w} \cdot \bm{x}(t)$, all relevant information is contained within the unconditional quantum state described in Eq.~\eqref{unconditional_dynamics}. Consequently, the correlation between $u_0$ and $z_t$ can be characterized by the unconditional classical mutual information $I^{\rm C}(u_0 : z_t)$, which is upper bounded by the quantum mutual information in Eq.~\eqref{eq:uncond_MI}, as dictated by the data processing inequality.

By initializing the system in a Bell state on $RA_0$, the {linear-in-system-size} lifetime of the unconditional QMI dynamics discussed in Section~\ref{sec:linear_lifetime} also implies a corresponding memory time {linear in the size of the memory subsystem} of QRC, assuming sufficient complex unitary dynamics. For Hamiltonian dynamics, on the other hand, the structure of Hamiltonian considered supports exponential lifetime {in system size}, as we recognize in Section~\ref{sec:scar} when tailoring the correlation to spectrum structure. 
However, such a specific weak initial correlation requires the input $u_0$ to affect the dynamics in a specific form and may not be utilized in a specific learning task. Overall, the general framework to interpret information loss in QRC is also applicable to recent advances in feedback-driven QRC with appropriate modifications~\cite{nakajima2019boosting}. 

{
To make use of more information from measurements, we propose to utilize a longer sequence of the memory outcome in multiple consecutive steps, instead of a single step. Let's begin with an extreme case, where at each step one adopts the full measurement outcome history $\bfz_t=(z_1,z_2,\cdots, z_t)$ in learning the approximation of the function. Considering the $N_Bt$ total number of qubits measured, the measurement statistics of $\bm z$ defines the feature $\bm x (t)=\{x_0(t),\cdots, x_{2^{N_Bt}-1}(t)\}$, i.e. the probability of the measurement outcomes, and output can be estimated by $y_t=\bm w \cdot \bm x (t)$. Note that such an extended scheme includes the case of adopting only the distribution of measurement result $z_t$ from the $t$-th step, via choosing a weight to reproduce the output only from marginal distribution. Consider the case of $t=2$ as an example, $\bm z\in\{00,01,10,11\}$ and $\bm x (2)=\{x_{00},x_{01},x_{10},x_{11}\}$, by limiting the weight to be $\bm w = (w_0, w_1, w_0, w_1)$, the output using the whole $\bm x(2)$ then becomes $y_2=w_0(x_{00}+x_{10})+w_1(x_{01}+x_{11})$, which matches the output from marginal distribution at $t=2$.}

{Apparently, now $y_t$ is derived from the statistics of the entire trajectory and the mutual information between $y_t$ and $u_0$ is upper bounded by
\be 
I^C(u_0:y_t)\le I^C(u_0:\bfz_t).
\label{I_extended}
\ee 
In this case, memory time is maximal as the early measurement result $z_1$ always appears in any $y_t$ as expected. However, we can ask how much additional information, $\Delta I^C_{t+1}$, does the additional measurement $z_{t+1}$ provides about $u_0$, to provide insight on whether increasing $t$ can assist in the learning of $u_0$. Taking the upper bound as an estimation, we have
\begin{align}
\Delta I^C_{t+1}
&\equiv I^C(u_0:\{z_1,\cdots,z_{t},z_{t+1}\})-I^C(u_0:\{z_1,\cdots,z_{t}\})
\\
&=I^C(u_0:z_{t+1}| \{z_1,\cdots,z_{t}\})\equiv I^C(u_0:z_{t+1}|\bfz_t).
\end{align}}
In this case, the {additional} correlation between $z_{t+1}$ and $u_0$ can be described by the classical conditional mutual information $I^C(u_0:z_{t+1}|\bfz)$, as previous measurement outcomes are taking into account in the learning. {Adopting data processing inequality in each trajectory of $\bfz_t$, we have $I^C(u_0:z_{t+1}|\bfz_t)\le I(R:A_t|\bfz_{t-1})$.} Accordingly, results in Section~\ref{sec:exp} indicates a slow logarithmic decay and therefore the additional information is appreciable for an exponentially long time.

\subsection{Monitored quantum circuits}
\label{sec:monitor_circuit}

\begin{figure}[t]
    \centering
    \includegraphics[width=0.45\textwidth]{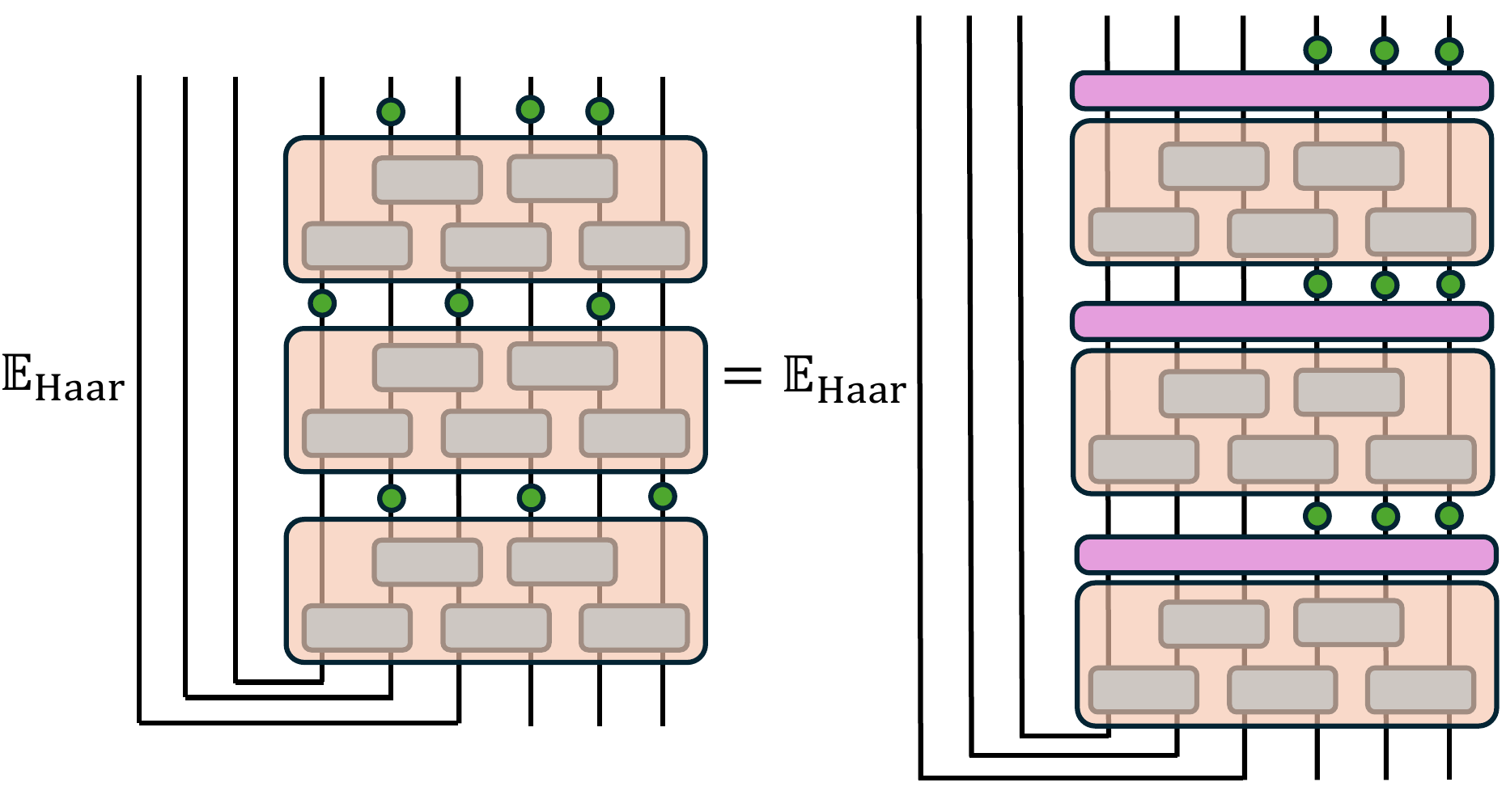}
    \caption{Equivalence of the monitored quantum circuit in the weak measurement limit (left) and the monitored circuit in the weak measurement limit with fixed qubits to be measured (right). Purple rectangles represent permutation operators.}
    \label{fig:scheme_MC}
\end{figure}

{In the monitored quantum circuit set-up, a brickwork-style unitary circuit is applied onto a trivial product initial state, with projective measurements randomly performed on qubits following every layers of unitary gates at a chosen probability $q$. Recent studies~\cite{skinner2019measurement, li2019measurement} have shown that with the increasing measurement probability $q$, the bipartite entanglement entropy of the resulting state transits from volume-law to area-law, known as the measurement-induced phase transition (MIPT). 
The model we considered in Fig.~\ref{fig:scheme}a can be regarded as a toy model of monitored circuits in the weak measurement limit,
where the data and bath system are always fully scrambled between each time of projective measurements. In addition, instead of measuring each qubit with a certain probability, we fix the number of qubits to be measured as $N_B$ among the $N_A+N_B$ qubits. 
For typical Haar random unitary with exponential complexity~\cite{hunter2019unitary}, the equivalent measurement probability $q \sim N_B/(N_A+N_B){\rm exp}(N_A + N_B)$ can be exponentially small; therefore, our model can be regarded as the monitored circuit in the weak-measurement limit and the output state follows volume-law entanglement for any cut within the data system. }

{
With the extension of the monitored circuit, initial information stored in a local system, i.e. the data system, propagates across the whole system via random unitary circuit, and gets lost due to the projective measurements. We focus on the information remained in the unmeasured subsystem through the dynamical process of monitored circuit in the weak measurement limit. To keep track of the initial information in the local subsystem, we introduce the same reference system that is maximally entangled with the data system. In the weak measurement limit, given that the unitary circuit between every two layers of measurements is sufficient complex (Fig.~\ref{fig:scheme_MC} left), we are allowed to insert an extra layer of permutation operation (purple) to rearrange the position of measured and unmeasured qubits in the system while without changing the ensemble average result (see Fig.~\ref{fig:scheme_MC} right).
With this equivalence transformation, the measurement-conditioned QMI $\overline{I(R:A_t|\bfz)}$ defined in Eq.~\eqref{QMI_cond} characterizes the typical amount of information remained in the unmeasured system, and according to our theoretical result of Theorem~\ref{theorem_avgMI} in Section~\ref{sec:exp}, the quantum information initially stored in a local subsystem can persist in the quantum system for exponentially long time through monitored circuits in the weak measurement limit. However, if we do not keep the record on the measurement outcome trajectory in the experiment, the quantum information can only persist in a linearly long time following Theorem~\ref{MI_traceout_theorem} in Section~\ref{sec:uncond}.}

\subsection{Entanglement-assisted communication}
\label{sec:capacity}

\begin{figure}[t]
    \centering
    \includegraphics[width=0.45\textwidth]{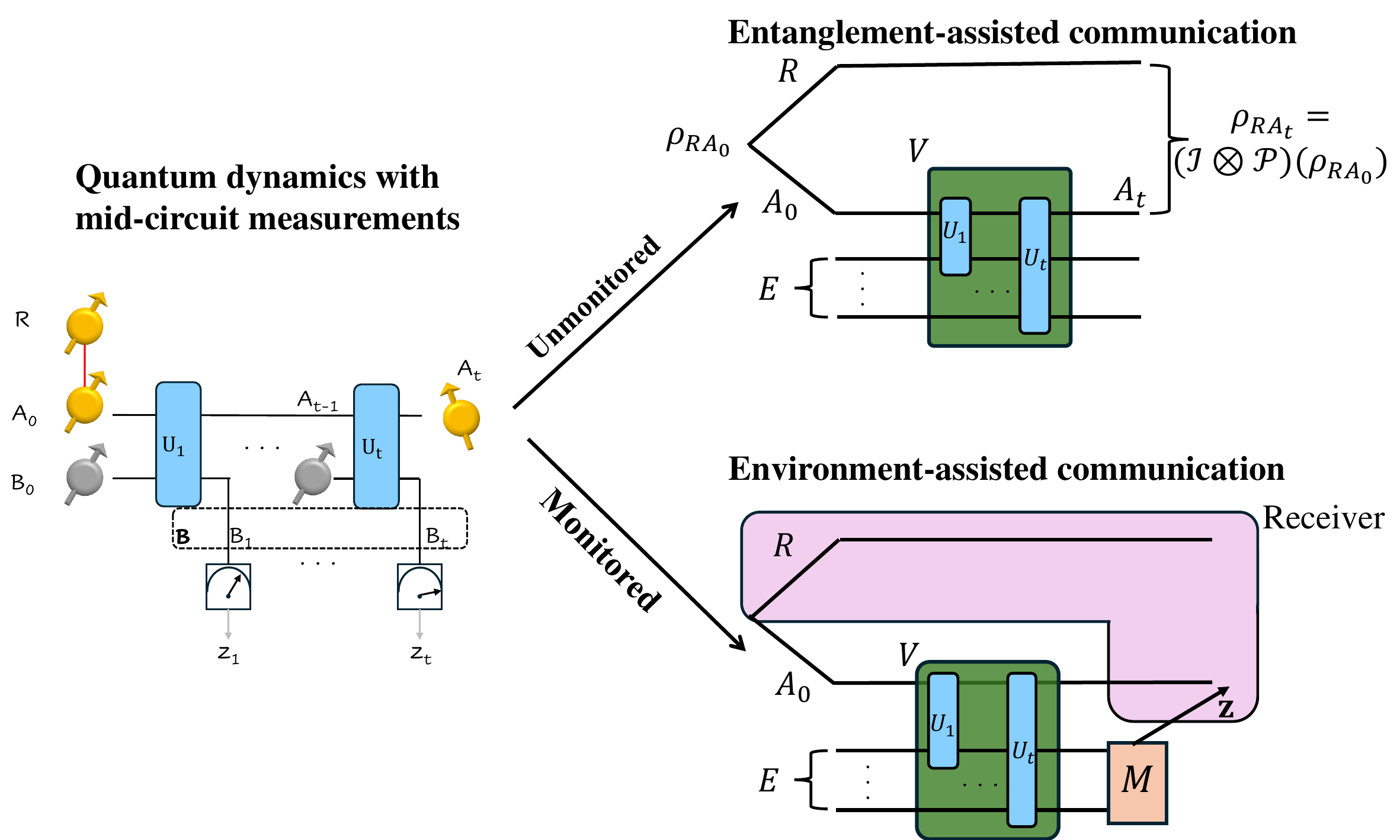}
    \caption{
    {Connection from quantum dynamics with mid-circuit measurements in unmonitored and monitored cases to entanglement-assisted communication without (top) and with (bottom) environment assistance.}}
    \label{fig:capacity}
\end{figure}

{To close this section, we interpret our results in the entanglement-assisted communication setting. While we focus on classical messages, quantum messages are equivalent at half of the communication rate due to quantum teleportation~\cite{bennett2002entanglement}.}

{
Let's begin with the measurement-unconditioned QMI in Eq.~\eqref{eq:uncond_MI}. As shown in Fig.~\ref{fig:capacity} top, the unmonitored dynamics naturally fits into the setup of entanglement-assisted communication~\cite{bennett2002entanglement}.
The sender aims to send a classical message via the input $A_0$ towards the quantum channel 
$\calI_R \otimes \bm \calP$ with $\bm \calP = \calP_t \circ \cdots \circ \calP_1$. Following the definition of channel $\calP$ of Eq.~\eqref{unconditional_dynamics}, we can represent the channel $\bm \calP$ via a unitary $V = \prod_{\ell=1}^t U_\ell$ applied on the sender $A_0$ and environment $E = B_0\cdots B_{t-1}$ which is a joint system of expanded baths initialized in $\ket{0^{\otimes N_B t}}_E$. A reference system $R$ to purify the input ${A_0}$ is sent through the identity channel as the entangled ancilla, which is required by definition of the communication setup and different from the cases discussed above. The entanglement-assisted classical capacity of the channel $\bm \calP$ quantifies the maximum bit transmission rate provided the shared entanglement, is known as~\cite{bennett2002entanglement}
\begin{align}
    C_E(\bm \calP) &= \max_{\rho_{A_0}} S(\rho_{A_0}) + S(\bm \calP(\rho_{A_0})) - S\left((\calI\otimes \bm \calP)\ketbra{\Phi}{\Phi}_{RA_0}\right)\\
    &=\max_{\rho_{A_0}} S(\rho_R) + S(\rho_{A_t}) - S\left(\rho_{RA_t}\right)\\
    &= \max_{\rho_{A_0}} I(R:A_t),
\end{align}
where in the second line we utilize the fact that $R$ purifies $A_0$. Therefore, our results on the unmonitored dynamics in Theorem~\ref{MI_traceout_theorem} provides an achievable lower bound on the entanglement-assisted capacity as Bell state input is one of the possible input states. 
Similarly, in {\em noisy super-dense coding}, the unconditional QMI $I(R:A_t)$ quantifies the amount of classical bits to be sent provided a shared entangled state $\rho_{RA_t}$~\cite{wilde2013quantum}.
}

{
Next, we connect the measurement-conditioned QMI defined by Eq.~\eqref{QMI_cond} in the scenario of environment-assisted classical communication. We still follow the setup of noisy super-dense coding, where Alice applied a quantum channel $\bm \calP$ to her side and send it to Bob, who owns the other side $R$ entangled with $A_0$ and $A_t$ at the end to recover the information sent from Alice, shown by the pink shaded area.  
However, unlike the earlier unconditional case, now at the output side of environment Charlie performs measurement $M$ and transmits the measurement outcome $\bfz$ to the receiver $A_t$ via classical communication. In each time conditioned on the measurement outcome $\bfz$, the maximum information Bob can learn about Alice's information is the QMI of state $\ket{\psi_\bfz}_{RA_t}$, therefore the measurement-conditioned QMI $\overline{I(R:A_t|\bfz)}$ represents an achievable rate of communication {\em on average case}, and the logarithmic decay of conditional QMI in Theorem~\ref{theorem_avgMI} indicates a large advantage in communication rate due to the assistance of classical information from environment.
}


{
Besides the entanglement-assisted communication scenario, we can further consider a connection to quantum communication. As $S(\rho_R)$ remains a time-dependent constant in our set-up (for instance $S(\rho_R) = N_A$ for Bell-state input), the decay of unconditional QMI in fact reveals the dynamics of coherent information 
\begin{align}
    I(R \rangle A_t) &\equiv S(\rho_{A_t}) - S(\rho_{RA_t})\\
    &=I(R:A_t) - S(\rho_R).
\end{align}
Following our definition of QMI lifetime in Eq.~\eqref{epsilon_def}, for a given initial input state $\rho_{RA_0}$, there always exists a threshold above which a positive quantum communication from $R$ to $A_t$ can be realized. The threshold corresponds to $I(R \rangle A_t) =  I(R:A_t) - S(\rho_R) = 0$, and since 
from Eq.~\eqref{epsilon_def} we define $I(R:A_t)=\epsilon I(R:A_0)$, we have the threshold value $\epsilon = {S(\rho_R)}/{I(R:A_0)}$.
Specifically, for a pure entangled state $\ket{\Phi}_{RA_0}$, we have $\epsilon = 1/2$.}

\section{QMI dynamics with and without reset}
\label{sec:reset}

Up to this point, our analysis has focused on the quantum dynamics with mid-circuit measurements followed by a  reset operation, which erases any prior information in the bath and reset it to a trivial \BZ{pure} state -- e.g., $\ket{0}^{\otimes N_B}$ -- for the next time step. \BZ{In this section, we compare with the QMI dynamics without the reset operation, where the measurement outcome is retained in the bath qubit and carried into the next time step.}

For the measurement-conditioned QMI $\overline{I(R:A_t|\bfz)}$, we prove that \BZ{the dynamics is unchanged by whether the reset operation is applied (see proof in Appendix~\ref{app:reset})}. \BZ{When we do not perform any reset operation to erase the measurement outcome $z$ in the preceding step, the initial state of bath in the following step is $\ket{z}$. The physical intuition behind this insensitivity is that we can insert an adaptive layer of local Pauli-X gates that transforms the bath state $\ket{z}$ to the trivial reset state $\ket{0}$. Since local gates do {\em not} change entanglement, they leave the conditional QMI unchanged. Equivalently, these $X$ gates can be absorbed into the subsequent global unitary layer.} We verify the conclusion through numerical simulations for the logarithmic decay in Fig.~\ref{fig:reset}a (see verification on the residual dynamics in Appendix~\ref{app:residual_dy}).

\BZ{For measurement-unconditioned QMI $I(R:A_t)$, if no reset is applied after each bath measurement, the joint reference-system-bath state becomes a classical mixture over measurement-conditioned states. At step $t$, the resulting unconditioned dynamics can be written as the channel
\be
    \mathcal{Q}_t(\cdot) = \sum_{z_t} \ketbra{z_t}{z_t}_{B_t} \left( U_t (\cdot) U_t^\dagger \right) \ketbra{z_t}{z_t}_{B_t},
    \label{eq:channel_noreset}
\ee 
which corresponds to a fully dephasing channel on the bath and also considered in Ref.~\cite{hu2024overcoming}.
The corresponding output joint state at step $t$ can be expressed analogously to Eq.~\eqref{unconditional_dynamics} as
\be
    \rho_{RA_tB_t} = \mathcal{I} \otimes \left(\mathcal{Q}_t \circ \mathcal{Q}_{t-1} \circ \cdots \mathcal{Q}_1\right) (\ketbra{\Phi}{\Phi}_{RA_0}\ketbra{0}{0}_{B_0}). 
\ee
For analytical convenience, we consider the R\'enyi-2 extended unconditional QMI and obtain the following theorem (see Appendix~\ref{app:thereom23} for a proof).
\begin{theorem}
\label{MI_traceout_woReset_theorem}
In the quantum dynamics of $2$-design unitaries with mid-circuit measurements and without reset, the expected R\'enyi-2 extended measurement-unconditioned QMI of a Bell initial state at time $t\ge 1$ is
\begin{align}
    &\E_{\rm Haar}I_2(R:A_t)\nonumber\\
    &\simeq - \log_2\left(\frac{(d_A^2-1)(d_B-1)}{d_A^2 d_B-1}\left(\frac{d_A^2 d_B-1}{d_A^2 d_B^2-1}\right)^t + 1\right)\nonumber\\
    &\quad + \log_2\left((d_A^2-1) \left(\frac{d_A^2 d_B-1}{d_A^2 d_B^2-1}\right)^t + 1\right)
    \label{eq:MI_traceout_wo}
\end{align}
In the asymptotic limit $d_A, d_B \gg 1$, when $t\ll 2N_A/N_B$, $\E_{\rm Haar}I_2(R:A_t) \simeq 2N_A - t N_B$, while for $t \gg 2N_A/N_B$, $\E_{\rm Haar}I_2(R:A_t) \simeq d_A^2 d_B^{-t}$. 
\end{theorem}
Theorem~\ref{MI_traceout_woReset_theorem} shows that, the unconditioned QMI dynamics without reset operation coincides with the reset case of Theorem~\ref{MI_traceout_theorem} in the asymptotic limit.
Intuitively, because the unitaries are random and strongly scrambling, the measurement-conditioned states obtained without reset are effectively typical, so the bath's initial state has little influence on the resulting QMI. 
We verify this numerically by comparing the unconditioned dynamics without reset (orange circles) with the reset case (blue circles) in Fig.~\ref{fig:reset}b1 and b2, finding a strong overlap. Our theoretical prediction of Eq.~\eqref{eq:MI_traceout_wo} in Theorem~\ref{MI_traceout_woReset_theorem} also agrees with the numerical simulations, reproducing the early-time linear decay and the late-time exponential tail.
On the other hand, if the bath is reset to a fully mixed state, the unconditioned QMI decays twice faster than pure state reset case (see Appendix.~\ref{app:reset} for details).}

\BZ{We also analyze the channel spectrum of Eq.~\eqref{eq:channel_noreset} for interpreting the late-time behavior as illustrated in Sec~\ref{sec:uncond} (see Appendix~\ref{app:MI_traceout_woReset_theorem} for some details). 
In Fig.~\ref{fig:reset}c, we plot the eigenvalue distribution of the quantum channel $\mathcal{Q}$ for a Haar random unitrary $U$, and also observe that the spectrum is bounded within a disk of radius $1/\sqrt{d_B}$ (red circle) in large $N_A$ limit. This behavior is consistent with the phenomenon observed in the unmonitored dynamics with reset (see Fig.~\ref{fig:mi_traceout}g).
Therefore, we conclude that whether adopting the reset or not does not affect QMI in both monitored and unmonitored dynamics.
Finally, we emphasize that even though the QMI memory time of the unconditioned states is identical in both the reset-present and reset-free cases, this quantity only characterizes memory during the transient regime of QRC.
In the more general time-asymptotic regime $t \to \infty$, the reset-free QRC exhibits no response to any new input $u_t$, whereas the reset-present QRC retains a nontrivial persistent memory to new inputs.
This is a fundamentally crucial difference between reset-present and reset-free settings when applying the unconditioned dynamics model to QRC.}

\begin{figure}[t]
    \centering
    \includegraphics[width=0.45\textwidth]{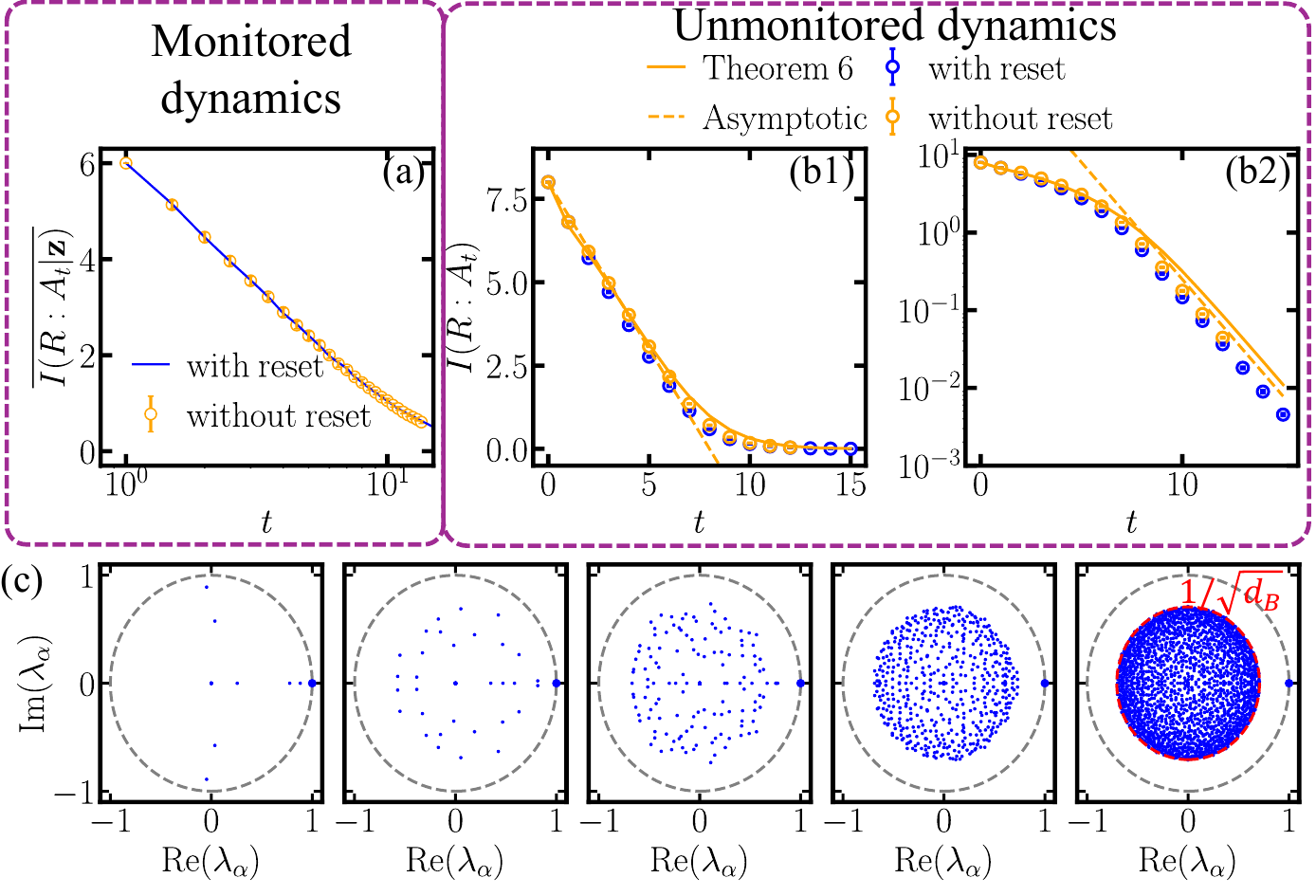}
    \caption{
    \BZ{QMI dynamics in monitored / unmonitored dynamics with and without reset. In (a), we plot decay of measurement-conditioned QMI with reset (blue) and without reset (orange). In (b1) and (b2), we plot numerical simulation of measurement-unconditioned QMI with pure state reset (blue circles), without pure state reset (orange circles) in early and late time separately. Orange solid and dashed lines show the theory of Eq.~\eqref{eq:MI_traceout_wo} in Theorem~\ref{MI_traceout_woReset_theorem} and asymptotic theory, respectively. 
    Errorbars represent sample fluctuations of Haar unitary implementations. 
    The system consists of (a) $N_A=3$ and (b1-b2) $N_A=4$ data qubits, and $N_B=1$ bath qubits.
    In (c), we show the spectrum of Haar channel without reset of system size from $N_A=1$ to $5$ (left to right) with $N_B = 1$ bath qubits. The red dashed circle indicates the radius of $1/\sqrt{d_B}$.}}
    \label{fig:reset}
\end{figure}

Due to the intrinsic chaotic property of Hamiltonian dynamics considered here, we expect that the reset strategy would lead to similar scaling laws. Meanwhile, the anomaly case of QMI lifetime in Hamiltonian dynamics without the reset strategy remains unexplored, and we leave it as an interesting open question. 

\section{Experiment of QMI dynamics on near-term hardware}
\label{sec:experiment}

In this section, we validate the exponential separation in measurement-conditioned and unconditioned QMI dynamics through near-term hardware compatible realizations. With a focus on the superconducting platform of IBM Quantum devices, we implement 
\BZ{the brickwork circuit} to mimic the implementation of Haar random unitary in theoretical modeling of Fig.~\ref{fig:scheme}a with appropriate modifications to reduce circuit compilation cost (see Appendix~\ref{app:experiment_detail}). We perform both ideal and noisy simulations of IBM quantum devices via Qiskit~\cite{qiskit2024}. \BZ{For the IBM Quantum Torino we targeted at, it has median $T_1$ as $185.04 \mu s$, median control-Z (CZ) gate error $2.566\times 10^{-3}$, median SX error \BZ{$3.262 \times10^{-4}$}, median readout (measurement) error $3.052 \times 10^{-2}$.} \BZ{We also perform the noisy simulation of IBM Quantum Sherbrooke and present the results in Appendix~\ref{app:experiment_detail}.}

In Fig.~\ref{fig:qmi_noisy}a, we evaluate the QMI dynamics implemented with \BZ{brickwork circuit}. For the ideal simulation, both measurement-conditioned (light blue dots) and unconditioned (light green dots) QMI align with the simulation results with Haar unitaries (solid lines), demonstrating its scrambling ability to mimic the dynamics with random unitaries. 

The above simulation results on the dynamics of QMI replies on a complete tomography of quantum states. In practice, one could consider the alternatives based on R\'enyi entropy~\cite{google2023measurement}. However, for R\'enyi entropy estimation via the randomized measurement, there still exists a statistical error factor of $1/\sqrt{M}$ for applying $M$ sets of measurement unitaries due to central limit theorem, and a large number of shots are still required for each choice of measurement unitaries~\cite{elben2019statistical}. To mitigate the resource cost for practical applications, inspired by the cross-entropy benchmark for verifying MIPT~\cite{li2023cross}, we propose a quantum-to-classical (Q2C) mutual information protocol. Through the same dynamics as in Fig.~\ref{fig:scheme}a, 
we define the measurement-conditioned Q2C mutual information as
\begin{align}
    &\overline{I^{\rm Q2C}(R:A_t|\bfz)} \nonumber\\
    &\equiv \E_{\bfz} D_{\rm KL}(P_{\bm U|\bfz}(z_R, z_{A_t})||P_{\bm U|\bfz}(z_R)\otimes P_{\bm U|\bfz}(z_{A_t})),
    \label{eq:q2cMI_cond}
\end{align}
where $P_{\bm U|\bfz}(z_R) = |{}_R\braket{z_R|\psi_\bfz}_{RA_t}|^2$ is the marginal distribution of measurement result on subsystem $R$ and similar for $P_{\bm U|\bfz}(z_{A_t})$, while $P_{\bm U}(z_R,z_{A_t}) = |\braket{z_R,z_{A_t}|\psi_\bfz}_{RA_t}|^2$ is the corresponding joint distribution of $(z_R, z_{A_t})$ on the conditional state $\ket{\psi_\bfz}_{RA_t}$ (see definition in Eq.~\eqref{state_RAt}), and $D_{\rm KL}$ is the KL divergence.
In the same spirit, we can also define the measurement-unconditioned Q2C mutual information as
\be
    I^{\rm Q2C}(R:A_t) \equiv D_{\rm KL}(P_{\bm U}(z_R, z_{A_t})||P_{\bm U}(z_R)\otimes P_{\bm U}(z_{A_t})),
    \label{eq:q2cMI_uncond}
\ee
with marginal and joint distributions measured on the unconditional state $\rho_{RA_t}$ defined in Eq.~\eqref{eq:state_uncond}, i.e. $P_{\bm U}(z_R) = \tr(\rho_{RA_t}\ketbra{z_R}{z_R}_R)$ and $P_{\bm U}(z_R,z_{A_t}) = \tr(\rho_{RA_t}\ketbra{z_R,z_{A_t}}{z_R,z_{A_t}}_{RA_t})$.

\begin{figure*}[t]
    \centering
    \includegraphics[width=0.65\textwidth]{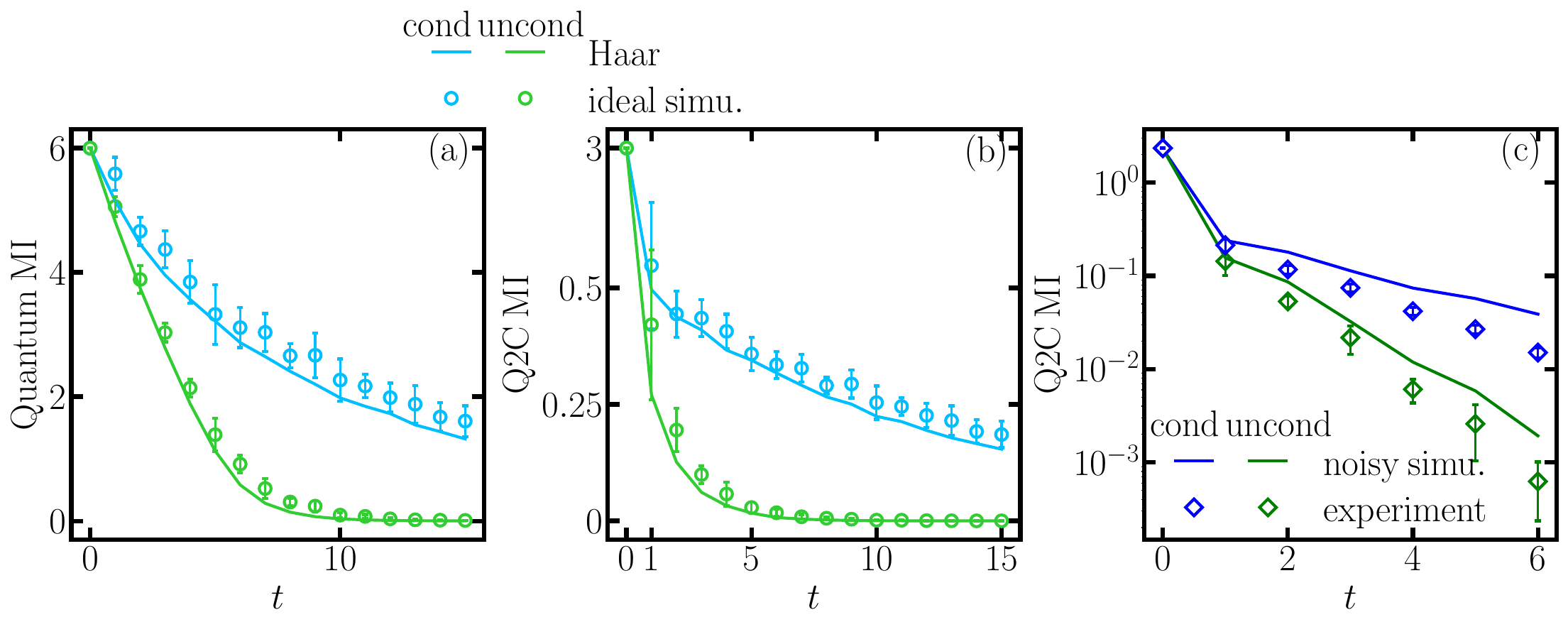}
    \caption{\BZ{Dynamics of measurement-conditioned and unconditioned (a) QMI and (b)-(c) Q2C MI. 
    Solid lines in (a) and (b) represent Haar simulation results. Light colored dots with errorbars represent ideal simulation results using \BZ{brickwork circuit}. In (c), solid lines and diamonds represent noisy simulation and experimental results on IBM Quantum, Torino.  
    Here we perform simulation on a system of $N_A=3, N_B=1$ qubits with circuit consisting of $L=4$ layers \BZ{in each step}. Each data point is averaged over $10$ randomized circuits.}}
    \label{fig:qmi_noisy}
\end{figure*}

Through the conversion of quantum correlation to state measurement outcome statistics, we only need to perform computational basis measurement in the Q2C protocol without any extra quantum operation which significantly reduces the sample complexity. The measurement results $\{(z_R, z_{A_t}, \bfz)\}$ with measurement trajectory of bath $\bfz$ can be further utilized to estimate all statistics involved in Eqs.~\eqref{eq:q2cMI_cond} and~\eqref{eq:q2cMI_uncond}. This protocol can save more resources in running quantum devices following the spirit of {\em ``measure first, ask questions later''}~\cite{elben2023randomized} though at the cost of losing characterization of quantum information partially through the quantum to classical conversion. We leave an in-depth analysis of complexity and statistical error as future studies.

In Fig.~\ref{fig:qmi_noisy}b, we show the dynamics of Q2C mutual information in the ideal setup. Similar to the behavior of the QMI dynamics discussed above, the measurement-conditioned Q2C mutual information (light blue circles) also decays logarithmically with steps, while the unconditioned one (light green circles) simply decays linearly, and both of them agree with the corresponding Haar simulation results (solid lines), \BZ{presenting a large gap there}. 
\BZ{We further perform experiments on IBM Quantum devices, shown in Fig.~\ref{fig:qmi_noisy}c. Despite the effect of the device and shot noise, we still observe a significant gap between the measurement-conditioned Q2C mutual information (blue diamonds) and unconditioned one (green diamonds), which enlarges with time and approaches an order of magnitude at $t=6$. The noisy simulations (lines) including gate noise and idling decoherence also align with the experimental results, though small deviations remain due to unmodeled noise resources such as crosstalk.
Overall, the Q2C protocol provides a practical diagnostic of the exponential separation in mutual information lifetime on near-term devices with reduced experimental overhead.}

\section{Discussion}

In this work, we demonstrate that quantum information can exhibit a lifetime that scales exponentially with system size, even in the presence of interactions with an arbitrary environment, provided the environment (bath) is monitored at each time step. In contrast, when the environment is left unmonitored, we show that the lifetime of quantum information -- such as that stored in a Bell state -- scales only linearly with system size. \BZ{In the partially monitored dynamics, we uncover a two-scale transition: the QMI decays logarithmically on microscopic time scales and linearly on macroscopic scales.}

In the case of Hamiltonian dynamics, we observe an anomalously long exponential lifetime of classical correlations. This behavior is attributed to Hamiltonian-dependent outliers in the spectrum of the corresponding single-step quantum channel -- a phenomenon whose precise origin remains an intriguing open question from the perspective of quantum many-body physics. \BZ{While we have focused on all-to-all coupled Ising Hamiltonian as an illustrative example in the main text, we also observe same behavior in the mixing-field Ising model with only nearest-neighbor interaction in Appendix.~\ref{app:MFIM}.}

\BZ{To reduce experimental resources, we propose a hardware-efficient Q2C mutual information protocol to diagnose the separation of QMI in monitored and unmonitored dynamics, and demonstrate the exponential separation in both ideal simulations and experiments on IBM Quantum devices.}

Our QMI framework provides a useful perspective for understanding memory dynamics in mid-circuit-measurement-based QRC algorithms. We propose extending memory time by leveraging statistics from sequences of measurement outcomes. This approach holds potential for improved performance, warranting further investigation through detailed numerical and experimental studies.

Our results also offer valuable insights for the field of quantum machine learning, including applications such as quantum diffusion models and quantum reservoir computing. For generative quantum machine learning models such as QuDDPM that incorporate recorded mid-circuit measurements, our theoretical result on the exponentially long QMI lifetime reveals a limitation: {for single-trajectory learning,} efficient learning of the transformation from random initial states to a target ensemble is challenging for chaotic circuits with high representation power. Instead, one should go beyond the states generated under a single measurement trajectory.
Additionally, the linear lifetime of mutual information in unmonitored dynamics offers guidance for setting the appropriate temporal window in QRC when learning individual signals. Incorporating statistics from extended sequences of measurement outcomes may further enhance the ability to learn non-Markovian sequences. Finally, our findings provide theoretical support for the security of holographic deep thermalization against entanglement attacks, as identified in Ref.~\cite{zhang2025holographic}.


We point out some future directions. The robustness of the exponential separation in quantum information lifetime against various types of noise is interesting to explore for both theoretical understanding and practical applications on near-term devices. 
\BZ{Extending our analysis to different Hamiltonian classes (e.g., chaotic, integrable, many-body-localized, and symmetry-constrained) is an interesting direction to understand how Hamiltonian structure protects information against measurement-induced loss~\cite{feng2025}.} For practical experimental verification, a statistical analysis of the proposed Q2C mutual information is useful for uncovering both its strengths and limitations. Another experimental direction for demonstrating QMI dynamics is to design a reliable witness of the exponential separation in QMI behavior that avoids post-selection, thereby further reducing the required resources. An intriguing question is whether the decay of QMI observed in this work can be linked to local system thermalization and the emergent state design in holographic deep thermalization~\cite{zhang2025holographic}.

\begin{acknowledgements}
QZ, BZ and RM acknowledge support from NSF (CCF-2240641, OMA-2326746, 2350153), ONR N00014-23-1-2296, AFOSR MURI FA9550-24-1-0349 and DARPA (HR0011-24-9-0362, HR00112490453, D24AC00153-02). This work was partially funded by an unrestricted gift from Google.
HET, FH and TC acknowledge support from the DARPA contract HR00112190072, AFOSR award FA9550-20-1-0177, and AFOSR MURI award FA9550-22-1-0203. The views, opinions, and findings expressed are solely the authors' and not the U.S. government's. The writing of the manuscript is completed in part at Aspen Center for Physics, which is supported by National Science Foundation grant PHY-2210452.
\end{acknowledgements}

\section*{Data Availability}
    The data and code of numerical simulation and experiments used in this study are publicly available in GitHub~\cite{github}.


\begin{thebibliography}{54}%
	\makeatletter
	\providecommand \@ifxundefined [1]{%
		\@ifx{#1\undefined}
	}%
	\providecommand \@ifnum [1]{%
		\ifnum #1\expandafter \@firstoftwo
		\else \expandafter \@secondoftwo
		\fi
	}%
	\providecommand \@ifx [1]{%
		\ifx #1\expandafter \@firstoftwo
		\else \expandafter \@secondoftwo
		\fi
	}%
	\providecommand \natexlab [1]{#1}%
	\providecommand \enquote  [1]{``#1''}%
	\providecommand \bibnamefont  [1]{#1}%
	\providecommand \bibfnamefont [1]{#1}%
	\providecommand \citenamefont [1]{#1}%
	\providecommand \href@noop [0]{\@secondoftwo}%
	\providecommand \href [0]{\begingroup \@sanitize@url \@href}%
	\providecommand \@href[1]{\@@startlink{#1}\@@href}%
	\providecommand \@@href[1]{\endgroup#1\@@endlink}%
	\providecommand \@sanitize@url [0]{\catcode `\\12\catcode `\$12\catcode
		`\&12\catcode `\#12\catcode `\^12\catcode `\_12\catcode `\%12\relax}%
	\providecommand \@@startlink[1]{}%
	\providecommand \@@endlink[0]{}%
	\providecommand \url  [0]{\begingroup\@sanitize@url \@url }%
	\providecommand \@url [1]{\endgroup\@href {#1}{\urlprefix }}%
	\providecommand \urlprefix  [0]{URL }%
	\providecommand \Eprint [0]{\href }%
	\providecommand \doibase [0]{https://doi.org/}%
	\providecommand \selectlanguage [0]{\@gobble}%
	\providecommand \bibinfo  [0]{\@secondoftwo}%
	\providecommand \bibfield  [0]{\@secondoftwo}%
	\providecommand \translation [1]{[#1]}%
	\providecommand \BibitemOpen [0]{}%
	\providecommand \bibitemStop [0]{}%
	\providecommand \bibitemNoStop [0]{.\EOS\space}%
	\providecommand \EOS [0]{\spacefactor3000\relax}%
	\providecommand \BibitemShut  [1]{\csname bibitem#1\endcsname}%
	\let\auto@bib@innerbib\@empty
	\bibitem [{\citenamefont {Ho}\ and\ \citenamefont {Choi}(2022)}]{ho2022exact}%
	\BibitemOpen
	\bibfield  {author} {\bibinfo {author} {\bibfnamefont {W.~W.}\ \bibnamefont
			{Ho}}\ and\ \bibinfo {author} {\bibfnamefont {S.}~\bibnamefont {Choi}},\
	}\bibfield  {title} {\bibinfo {title} {Exact emergent quantum state designs
			from quantum chaotic dynamics},\ }\href@noop {} {\bibfield  {journal}
		{\bibinfo  {journal} {Phys. Rev. Lett.}\ }\textbf {\bibinfo {volume} {128}},\
		\bibinfo {pages} {060601} (\bibinfo {year} {2022})}\BibitemShut {NoStop}%
	\bibitem [{\citenamefont {Cotler}\ \emph {et~al.}(2023)\citenamefont {Cotler},
		\citenamefont {Mark}, \citenamefont {Huang}, \citenamefont {Hernandez},
		\citenamefont {Choi}, \citenamefont {Shaw}, \citenamefont {Endres},\ and\
		\citenamefont {Choi}}]{cotler2023emergent}%
	\BibitemOpen
	\bibfield  {author} {\bibinfo {author} {\bibfnamefont {J.~S.}\ \bibnamefont
			{Cotler}}, \bibinfo {author} {\bibfnamefont {D.~K.}\ \bibnamefont {Mark}},
		\bibinfo {author} {\bibfnamefont {H.-Y.}\ \bibnamefont {Huang}}, \bibinfo
		{author} {\bibfnamefont {F.}~\bibnamefont {Hernandez}}, \bibinfo {author}
		{\bibfnamefont {J.}~\bibnamefont {Choi}}, \bibinfo {author} {\bibfnamefont
			{A.~L.}\ \bibnamefont {Shaw}}, \bibinfo {author} {\bibfnamefont
			{M.}~\bibnamefont {Endres}},\ and\ \bibinfo {author} {\bibfnamefont
			{S.}~\bibnamefont {Choi}},\ }\bibfield  {title} {\bibinfo {title} {Emergent
			quantum state designs from individual many-body wave functions},\ }\href@noop
	{} {\bibfield  {journal} {\bibinfo  {journal} {PRX quantum}\ }\textbf
		{\bibinfo {volume} {4}},\ \bibinfo {pages} {010311} (\bibinfo {year}
		{2023})}\BibitemShut {NoStop}%
	\bibitem [{\citenamefont {Li}\ \emph {et~al.}(2019)\citenamefont {Li},
		\citenamefont {Chen},\ and\ \citenamefont {Fisher}}]{li2019measurement}%
	\BibitemOpen
	\bibfield  {author} {\bibinfo {author} {\bibfnamefont {Y.}~\bibnamefont
			{Li}}, \bibinfo {author} {\bibfnamefont {X.}~\bibnamefont {Chen}},\ and\
		\bibinfo {author} {\bibfnamefont {M.~P.}\ \bibnamefont {Fisher}},\ }\bibfield
	{title} {\bibinfo {title} {Measurement-driven entanglement transition in
			hybrid quantum circuits},\ }\href@noop {} {\bibfield  {journal} {\bibinfo
			{journal} {Phys. Rev. B}\ }\textbf {\bibinfo {volume} {100}},\ \bibinfo
		{pages} {134306} (\bibinfo {year} {2019})}\BibitemShut {NoStop}%
	\bibitem [{\citenamefont {Skinner}\ \emph {et~al.}(2019)\citenamefont
		{Skinner}, \citenamefont {Ruhman},\ and\ \citenamefont
		{Nahum}}]{skinner2019measurement}%
	\BibitemOpen
	\bibfield  {author} {\bibinfo {author} {\bibfnamefont {B.}~\bibnamefont
			{Skinner}}, \bibinfo {author} {\bibfnamefont {J.}~\bibnamefont {Ruhman}},\
		and\ \bibinfo {author} {\bibfnamefont {A.}~\bibnamefont {Nahum}},\ }\bibfield
	{title} {\bibinfo {title} {Measurement-induced phase transitions in the
			dynamics of entanglement},\ }\href@noop {} {\bibfield  {journal} {\bibinfo
			{journal} {Phys. Rev. X}\ }\textbf {\bibinfo {volume} {9}},\ \bibinfo {pages}
		{031009} (\bibinfo {year} {2019})}\BibitemShut {NoStop}%
	\bibitem [{\citenamefont {Calderbank}\ and\ \citenamefont
		{Shor}(1996)}]{calderbank1996good}%
	\BibitemOpen
	\bibfield  {author} {\bibinfo {author} {\bibfnamefont {A.~R.}\ \bibnamefont
			{Calderbank}}\ and\ \bibinfo {author} {\bibfnamefont {P.~W.}\ \bibnamefont
			{Shor}},\ }\bibfield  {title} {\bibinfo {title} {Good quantum
			error-correcting codes exist},\ }\href@noop {} {\bibfield  {journal}
		{\bibinfo  {journal} {Phys. Rev. A}\ }\textbf {\bibinfo {volume} {54}},\
		\bibinfo {pages} {1098} (\bibinfo {year} {1996})}\BibitemShut {NoStop}%
	\bibitem [{\citenamefont {{Google Quantum AI and
				Collaborators}}(2025)}]{acharya_quantum_2025}%
	\BibitemOpen
	\bibfield  {author} {\bibinfo {author} {\bibnamefont {{Google Quantum AI and
					Collaborators}}},\ }\bibfield  {title} {\bibinfo {title} {Quantum error
			correction below the surface code threshold},\ }\href
	{https://doi.org/10.1038/s41586-024-08449-y} {\bibfield  {journal} {\bibinfo
			{journal} {Nature}\ }\textbf {\bibinfo {volume} {638}},\ \bibinfo {pages}
		{920} (\bibinfo {year} {2025})}\BibitemShut {NoStop}%
	\bibitem [{\citenamefont {Zhang}\ \emph
		{et~al.}(2025{\natexlab{a}})\citenamefont {Zhang}, \citenamefont {Xu},
		\citenamefont {Chen},\ and\ \citenamefont {Zhuang}}]{zhang2025holographic}%
	\BibitemOpen
	\bibfield  {author} {\bibinfo {author} {\bibfnamefont {B.}~\bibnamefont
			{Zhang}}, \bibinfo {author} {\bibfnamefont {P.}~\bibnamefont {Xu}}, \bibinfo
		{author} {\bibfnamefont {X.}~\bibnamefont {Chen}},\ and\ \bibinfo {author}
		{\bibfnamefont {Q.}~\bibnamefont {Zhuang}},\ }\bibfield  {title} {\bibinfo
		{title} {Holographic deep thermalization for secure and efficient quantum
			random state generation},\ }\href@noop {} {\bibfield  {journal} {\bibinfo
			{journal} {Nature Communications}\ }\textbf {\bibinfo {volume} {16}},\
		\bibinfo {pages} {6341} (\bibinfo {year} {2025}{\natexlab{a}})}\BibitemShut
	{NoStop}%
	\bibitem [{\citenamefont {Zhang}\ and\ \citenamefont
		{Zhuang}(2025)}]{zhang2025quantum}%
	\BibitemOpen
	\bibfield  {author} {\bibinfo {author} {\bibfnamefont {B.}~\bibnamefont
			{Zhang}}\ and\ \bibinfo {author} {\bibfnamefont {Q.}~\bibnamefont {Zhuang}},\
	}\bibfield  {title} {\bibinfo {title} {Quantum advantage from effective $200
			$-qubit holographic random circuit sampling},\ }\href@noop {} {\bibfield
		{journal} {\bibinfo  {journal} {arXiv preprint arXiv:2511.05433}\ } (\bibinfo
		{year} {2025})}\BibitemShut {NoStop}%
	\bibitem [{\citenamefont {C{\'o}rcoles}\ \emph {et~al.}(2021)\citenamefont
		{C{\'o}rcoles}, \citenamefont {Takita}, \citenamefont {Inoue}, \citenamefont
		{Lekuch}, \citenamefont {Minev}, \citenamefont {Chow},\ and\ \citenamefont
		{Gambetta}}]{corcoles2021exploiting}%
	\BibitemOpen
	\bibfield  {author} {\bibinfo {author} {\bibfnamefont {A.~D.}\ \bibnamefont
			{C{\'o}rcoles}}, \bibinfo {author} {\bibfnamefont {M.}~\bibnamefont
			{Takita}}, \bibinfo {author} {\bibfnamefont {K.}~\bibnamefont {Inoue}},
		\bibinfo {author} {\bibfnamefont {S.}~\bibnamefont {Lekuch}}, \bibinfo
		{author} {\bibfnamefont {Z.~K.}\ \bibnamefont {Minev}}, \bibinfo {author}
		{\bibfnamefont {J.~M.}\ \bibnamefont {Chow}},\ and\ \bibinfo {author}
		{\bibfnamefont {J.~M.}\ \bibnamefont {Gambetta}},\ }\bibfield  {title}
	{\bibinfo {title} {Exploiting dynamic quantum circuits in a quantum algorithm
			with superconducting qubits},\ }\href@noop {} {\bibfield  {journal} {\bibinfo
			{journal} {Phys. Rev. Lett.}\ }\textbf {\bibinfo {volume} {127}},\ \bibinfo
		{pages} {100501} (\bibinfo {year} {2021})}\BibitemShut {NoStop}%
	\bibitem [{\citenamefont {DeCross}\ \emph {et~al.}(2023)\citenamefont
		{DeCross}, \citenamefont {Chertkov}, \citenamefont {Kohagen},\ and\
		\citenamefont {Foss-Feig}}]{decross2023qubit}%
	\BibitemOpen
	\bibfield  {author} {\bibinfo {author} {\bibfnamefont {M.}~\bibnamefont
			{DeCross}}, \bibinfo {author} {\bibfnamefont {E.}~\bibnamefont {Chertkov}},
		\bibinfo {author} {\bibfnamefont {M.}~\bibnamefont {Kohagen}},\ and\ \bibinfo
		{author} {\bibfnamefont {M.}~\bibnamefont {Foss-Feig}},\ }\bibfield  {title}
	{\bibinfo {title} {Qubit-reuse compilation with mid-circuit measurement and
			reset},\ }\href@noop {} {\bibfield  {journal} {\bibinfo  {journal} {Phys.
				Rev. X}\ }\textbf {\bibinfo {volume} {13}},\ \bibinfo {pages} {041057}
		(\bibinfo {year} {2023})}\BibitemShut {NoStop}%
	\bibitem [{\citenamefont {B{\"a}umer}\ \emph {et~al.}(2024)\citenamefont
		{B{\"a}umer}, \citenamefont {Tripathi}, \citenamefont {Wang}, \citenamefont
		{Rall}, \citenamefont {Chen}, \citenamefont {Majumder}, \citenamefont
		{Seif},\ and\ \citenamefont {Minev}}]{baumer2024efficient}%
	\BibitemOpen
	\bibfield  {author} {\bibinfo {author} {\bibfnamefont {E.}~\bibnamefont
			{B{\"a}umer}}, \bibinfo {author} {\bibfnamefont {V.}~\bibnamefont
			{Tripathi}}, \bibinfo {author} {\bibfnamefont {D.~S.}\ \bibnamefont {Wang}},
		\bibinfo {author} {\bibfnamefont {P.}~\bibnamefont {Rall}}, \bibinfo {author}
		{\bibfnamefont {E.~H.}\ \bibnamefont {Chen}}, \bibinfo {author}
		{\bibfnamefont {S.}~\bibnamefont {Majumder}}, \bibinfo {author}
		{\bibfnamefont {A.}~\bibnamefont {Seif}},\ and\ \bibinfo {author}
		{\bibfnamefont {Z.~K.}\ \bibnamefont {Minev}},\ }\bibfield  {title} {\bibinfo
		{title} {Efficient long-range entanglement using dynamic circuits},\
	}\href@noop {} {\bibfield  {journal} {\bibinfo  {journal} {PRX Quantum}\
		}\textbf {\bibinfo {volume} {5}},\ \bibinfo {pages} {030339} (\bibinfo {year}
		{2024})}\BibitemShut {NoStop}%
	\bibitem [{\citenamefont {Piroli}\ \emph {et~al.}(2024)\citenamefont {Piroli},
		\citenamefont {Styliaris},\ and\ \citenamefont
		{Cirac}}]{piroli2024approximating}%
	\BibitemOpen
	\bibfield  {author} {\bibinfo {author} {\bibfnamefont {L.}~\bibnamefont
			{Piroli}}, \bibinfo {author} {\bibfnamefont {G.}~\bibnamefont {Styliaris}},\
		and\ \bibinfo {author} {\bibfnamefont {J.~I.}\ \bibnamefont {Cirac}},\
	}\bibfield  {title} {\bibinfo {title} {Approximating many-body quantum states
			with quantum circuits and measurements},\ }\href@noop {} {\bibfield
		{journal} {\bibinfo  {journal} {Phys. Rev. Lett.}\ }\textbf {\bibinfo
			{volume} {133}},\ \bibinfo {pages} {230401} (\bibinfo {year}
		{2024})}\BibitemShut {NoStop}%
	\bibitem [{\citenamefont {Buhrman}\ \emph {et~al.}(2024)\citenamefont
		{Buhrman}, \citenamefont {Folkertsma}, \citenamefont {Loff},\ and\
		\citenamefont {Neumann}}]{buhrman2024state}%
	\BibitemOpen
	\bibfield  {author} {\bibinfo {author} {\bibfnamefont {H.}~\bibnamefont
			{Buhrman}}, \bibinfo {author} {\bibfnamefont {M.}~\bibnamefont {Folkertsma}},
		\bibinfo {author} {\bibfnamefont {B.}~\bibnamefont {Loff}},\ and\ \bibinfo
		{author} {\bibfnamefont {N.~M.}\ \bibnamefont {Neumann}},\ }\bibfield
	{title} {\bibinfo {title} {State preparation by shallow circuits using feed
			forward},\ }\href@noop {} {\bibfield  {journal} {\bibinfo  {journal}
			{Quantum}\ }\textbf {\bibinfo {volume} {8}},\ \bibinfo {pages} {1552}
		(\bibinfo {year} {2024})}\BibitemShut {NoStop}%
	\bibitem [{\citenamefont {Smith}\ \emph {et~al.}(2024)\citenamefont {Smith},
		\citenamefont {Khan}, \citenamefont {Clark}, \citenamefont {Girvin},\ and\
		\citenamefont {Wei}}]{smith2024constant}%
	\BibitemOpen
	\bibfield  {author} {\bibinfo {author} {\bibfnamefont {K.~C.}\ \bibnamefont
			{Smith}}, \bibinfo {author} {\bibfnamefont {A.}~\bibnamefont {Khan}},
		\bibinfo {author} {\bibfnamefont {B.~K.}\ \bibnamefont {Clark}}, \bibinfo
		{author} {\bibfnamefont {S.}~\bibnamefont {Girvin}},\ and\ \bibinfo {author}
		{\bibfnamefont {T.-C.}\ \bibnamefont {Wei}},\ }\bibfield  {title} {\bibinfo
		{title} {Constant-depth preparation of matrix product states with adaptive
			quantum circuits},\ }\href {https://doi.org/10.1103/PRXQuantum.5.030344}
	{\bibfield  {journal} {\bibinfo  {journal} {PRX Quantum}\ }\textbf {\bibinfo
			{volume} {5}},\ \bibinfo {pages} {030344} (\bibinfo {year}
		{2024})}\BibitemShut {NoStop}%
	\bibitem [{\citenamefont {Iqbal}\ \emph {et~al.}(2024)\citenamefont {Iqbal},
		\citenamefont {Tantivasadakarn}, \citenamefont {Verresen}, \citenamefont
		{Campbell}, \citenamefont {Dreiling}, \citenamefont {Figgatt}, \citenamefont
		{Gaebler}, \citenamefont {Johansen}, \citenamefont {Mills}, \citenamefont
		{Moses} \emph {et~al.}}]{iqbal2024non}%
	\BibitemOpen
	\bibfield  {author} {\bibinfo {author} {\bibfnamefont {M.}~\bibnamefont
			{Iqbal}}, \bibinfo {author} {\bibfnamefont {N.}~\bibnamefont
			{Tantivasadakarn}}, \bibinfo {author} {\bibfnamefont {R.}~\bibnamefont
			{Verresen}}, \bibinfo {author} {\bibfnamefont {S.~L.}\ \bibnamefont
			{Campbell}}, \bibinfo {author} {\bibfnamefont {J.~M.}\ \bibnamefont
			{Dreiling}}, \bibinfo {author} {\bibfnamefont {C.}~\bibnamefont {Figgatt}},
		\bibinfo {author} {\bibfnamefont {J.~P.}\ \bibnamefont {Gaebler}}, \bibinfo
		{author} {\bibfnamefont {J.}~\bibnamefont {Johansen}}, \bibinfo {author}
		{\bibfnamefont {M.}~\bibnamefont {Mills}}, \bibinfo {author} {\bibfnamefont
			{S.~A.}\ \bibnamefont {Moses}}, \emph {et~al.},\ }\bibfield  {title}
	{\bibinfo {title} {Non-abelian topological order and anyons on a trapped-ion
			processor},\ }\href@noop {} {\bibfield  {journal} {\bibinfo  {journal}
			{Nature}\ }\textbf {\bibinfo {volume} {626}},\ \bibinfo {pages} {505}
		(\bibinfo {year} {2024})}\BibitemShut {NoStop}%
	\bibitem [{\citenamefont {Cao}\ and\ \citenamefont
		{Eisert}(2025)}]{cao2025measurement}%
	\BibitemOpen
	\bibfield  {author} {\bibinfo {author} {\bibfnamefont {C.}~\bibnamefont
			{Cao}}\ and\ \bibinfo {author} {\bibfnamefont {J.}~\bibnamefont {Eisert}},\
	}\bibfield  {title} {\bibinfo {title} {Measurement-driven quantum advantages
			in shallow circuits},\ }\href@noop {} {\bibfield  {journal} {\bibinfo
			{journal} {arXiv preprint arXiv:2505.04705}\ } (\bibinfo {year}
		{2025})}\BibitemShut {NoStop}%
	\bibitem [{\citenamefont {Zhang}\ \emph {et~al.}(2024)\citenamefont {Zhang},
		\citenamefont {Xu}, \citenamefont {Chen},\ and\ \citenamefont
		{Zhuang}}]{zhang2024generative}%
	\BibitemOpen
	\bibfield  {author} {\bibinfo {author} {\bibfnamefont {B.}~\bibnamefont
			{Zhang}}, \bibinfo {author} {\bibfnamefont {P.}~\bibnamefont {Xu}}, \bibinfo
		{author} {\bibfnamefont {X.}~\bibnamefont {Chen}},\ and\ \bibinfo {author}
		{\bibfnamefont {Q.}~\bibnamefont {Zhuang}},\ }\bibfield  {title} {\bibinfo
		{title} {Generative quantum machine learning via denoising diffusion
			probabilistic models},\ }\href@noop {} {\bibfield  {journal} {\bibinfo
			{journal} {Phys. Rev. Lett.}\ }\textbf {\bibinfo {volume} {132}},\ \bibinfo
		{pages} {100602} (\bibinfo {year} {2024})}\BibitemShut {NoStop}%
	\bibitem [{\citenamefont {Chen}\ \emph {et~al.}(2020)\citenamefont {Chen},
		\citenamefont {Nurdin},\ and\ \citenamefont {Yamamoto}}]{chen_temporal_2020}%
	\BibitemOpen
	\bibfield  {author} {\bibinfo {author} {\bibfnamefont {J.}~\bibnamefont
			{Chen}}, \bibinfo {author} {\bibfnamefont {H.~I.}\ \bibnamefont {Nurdin}},\
		and\ \bibinfo {author} {\bibfnamefont {N.}~\bibnamefont {Yamamoto}},\
	}\bibfield  {title} {\bibinfo {title} {Temporal {Information} {Processing} on
			{Noisy} {Quantum} {Computers}},\ }\href
	{https://doi.org/10.1103/PhysRevApplied.14.024065} {\bibfield  {journal}
		{\bibinfo  {journal} {Physical Review Applied}\ }\textbf {\bibinfo {volume}
			{14}},\ \bibinfo {pages} {024065} (\bibinfo {year} {2020})}\BibitemShut
	{NoStop}%
	\bibitem [{\citenamefont {Hu}\ \emph {et~al.}(2024)\citenamefont {Hu},
		\citenamefont {Khan}, \citenamefont {Bronn}, \citenamefont {Angelatos},
		\citenamefont {Rowlands}, \citenamefont {Ribeill},\ and\ \citenamefont
		{T{\"u}reci}}]{hu2024overcoming}%
	\BibitemOpen
	\bibfield  {author} {\bibinfo {author} {\bibfnamefont {F.}~\bibnamefont
			{Hu}}, \bibinfo {author} {\bibfnamefont {S.~A.}\ \bibnamefont {Khan}},
		\bibinfo {author} {\bibfnamefont {N.~T.}\ \bibnamefont {Bronn}}, \bibinfo
		{author} {\bibfnamefont {G.}~\bibnamefont {Angelatos}}, \bibinfo {author}
		{\bibfnamefont {G.~E.}\ \bibnamefont {Rowlands}}, \bibinfo {author}
		{\bibfnamefont {G.~J.}\ \bibnamefont {Ribeill}},\ and\ \bibinfo {author}
		{\bibfnamefont {H.~E.}\ \bibnamefont {T{\"u}reci}},\ }\bibfield  {title}
	{\bibinfo {title} {Overcoming the coherence time barrier in quantum machine
			learning on temporal data},\ }\href@noop {} {\bibfield  {journal} {\bibinfo
			{journal} {Nat. Commun.}\ }\textbf {\bibinfo {volume} {15}},\ \bibinfo
		{pages} {7491} (\bibinfo {year} {2024})}\BibitemShut {NoStop}%
	\bibitem [{\citenamefont {Bennett}\ \emph {et~al.}(2002)\citenamefont
		{Bennett}, \citenamefont {Shor}, \citenamefont {Smolin},\ and\ \citenamefont
		{Thapliyal}}]{bennett2002entanglement}%
	\BibitemOpen
	\bibfield  {author} {\bibinfo {author} {\bibfnamefont {C.~H.}\ \bibnamefont
			{Bennett}}, \bibinfo {author} {\bibfnamefont {P.~W.}\ \bibnamefont {Shor}},
		\bibinfo {author} {\bibfnamefont {J.~A.}\ \bibnamefont {Smolin}},\ and\
		\bibinfo {author} {\bibfnamefont {A.~V.}\ \bibnamefont {Thapliyal}},\
	}\bibfield  {title} {\bibinfo {title} {Entanglement-assisted capacity of a
			quantum channel and the reverse shannon theorem},\ }\href@noop {} {\bibfield
		{journal} {\bibinfo  {journal} {IEEE Trans. Inf. Theo.}\ }\textbf {\bibinfo
			{volume} {48}},\ \bibinfo {pages} {2637} (\bibinfo {year}
		{2002})}\BibitemShut {NoStop}%
	\bibitem [{\citenamefont {Nakajima}\ \emph {et~al.}(2019)\citenamefont
		{Nakajima}, \citenamefont {Fujii}, \citenamefont {Negoro}, \citenamefont
		{Mitarai},\ and\ \citenamefont {Kitagawa}}]{nakajima2019boosting}%
	\BibitemOpen
	\bibfield  {author} {\bibinfo {author} {\bibfnamefont {K.}~\bibnamefont
			{Nakajima}}, \bibinfo {author} {\bibfnamefont {K.}~\bibnamefont {Fujii}},
		\bibinfo {author} {\bibfnamefont {M.}~\bibnamefont {Negoro}}, \bibinfo
		{author} {\bibfnamefont {K.}~\bibnamefont {Mitarai}},\ and\ \bibinfo {author}
		{\bibfnamefont {M.}~\bibnamefont {Kitagawa}},\ }\bibfield  {title} {\bibinfo
		{title} {Boosting computational power through spatial multiplexing in quantum
			reservoir computing},\ }\href@noop {} {\bibfield  {journal} {\bibinfo
			{journal} {Phys. Rev. Applied}\ }\textbf {\bibinfo {volume} {11}},\ \bibinfo
		{pages} {034021} (\bibinfo {year} {2019})}\BibitemShut {NoStop}%
	\bibitem [{\citenamefont {Fujii}\ and\ \citenamefont
		{Nakajima}(2017)}]{fujii2017harnessing}%
	\BibitemOpen
	\bibfield  {author} {\bibinfo {author} {\bibfnamefont {K.}~\bibnamefont
			{Fujii}}\ and\ \bibinfo {author} {\bibfnamefont {K.}~\bibnamefont
			{Nakajima}},\ }\bibfield  {title} {\bibinfo {title} {Harnessing
			disordered-ensemble quantum dynamics for machine learning},\ }\href@noop {}
	{\bibfield  {journal} {\bibinfo  {journal} {Phys. Rev. Applied}\ }\textbf
		{\bibinfo {volume} {8}},\ \bibinfo {pages} {024030} (\bibinfo {year}
		{2017})}\BibitemShut {NoStop}%
	\bibitem [{\citenamefont {Mujal}\ \emph {et~al.}(2023)\citenamefont {Mujal},
		\citenamefont {Mart{\'{\i}}nez-Pe{\~{n}}a}, \citenamefont {Giorgi},
		\citenamefont {Soriano},\ and\ \citenamefont
		{Zambrini}}]{mujal_time-series_2023}%
	\BibitemOpen
	\bibfield  {author} {\bibinfo {author} {\bibfnamefont {P.}~\bibnamefont
			{Mujal}}, \bibinfo {author} {\bibfnamefont {R.}~\bibnamefont
			{Mart{\'{\i}}nez-Pe{\~{n}}a}}, \bibinfo {author} {\bibfnamefont {G.~L.}\
			\bibnamefont {Giorgi}}, \bibinfo {author} {\bibfnamefont {M.~C.}\
			\bibnamefont {Soriano}},\ and\ \bibinfo {author} {\bibfnamefont
			{R.}~\bibnamefont {Zambrini}},\ }\bibfield  {title} {\bibinfo {title}
		{Time-series quantum reservoir computing with weak and projective
			measurements},\ }\href {https://doi.org/10.1038/s41534-023-00682-z}
	{\bibfield  {journal} {\bibinfo  {journal} {npj Quantum Information}\
		}\textbf {\bibinfo {volume} {9}},\ \bibinfo {pages} {16} (\bibinfo {year}
		{2023})}\BibitemShut {NoStop}%
	\bibitem [{\citenamefont {Yasuda}\ \emph {et~al.}(2023)\citenamefont {Yasuda},
		\citenamefont {Suzuki}, \citenamefont {Kubota}, \citenamefont {Nakajima},
		\citenamefont {Gao}, \citenamefont {Zhang}, \citenamefont {Shimono},
		\citenamefont {Nurdin},\ and\ \citenamefont
		{Yamamoto}}]{yasuda_quantum_2023}%
	\BibitemOpen
	\bibfield  {author} {\bibinfo {author} {\bibfnamefont {T.}~\bibnamefont
			{Yasuda}}, \bibinfo {author} {\bibfnamefont {Y.}~\bibnamefont {Suzuki}},
		\bibinfo {author} {\bibfnamefont {T.}~\bibnamefont {Kubota}}, \bibinfo
		{author} {\bibfnamefont {K.}~\bibnamefont {Nakajima}}, \bibinfo {author}
		{\bibfnamefont {Q.}~\bibnamefont {Gao}}, \bibinfo {author} {\bibfnamefont
			{W.}~\bibnamefont {Zhang}}, \bibinfo {author} {\bibfnamefont
			{S.}~\bibnamefont {Shimono}}, \bibinfo {author} {\bibfnamefont {H.~I.}\
			\bibnamefont {Nurdin}},\ and\ \bibinfo {author} {\bibfnamefont
			{N.}~\bibnamefont {Yamamoto}},\ }\bibfield  {title} {\bibinfo {title}
		{Quantum reservoir computing with repeated measurements on superconducting
			devices},\ }\href {https://arxiv.org/abs/2310.06706} {\bibfield  {journal}
		{\bibinfo  {journal} {arXiv:2310.06706 [quant-ph]}\ } (\bibinfo {year}
		{2023})}\BibitemShut {NoStop}%
	\bibitem [{\citenamefont {Martínez-Peña}\ and\ \citenamefont
		{Ortega}(2023)}]{martinez-pena_quantum_2023}%
	\BibitemOpen
	\bibfield  {author} {\bibinfo {author} {\bibfnamefont {R.}~\bibnamefont
			{Martínez-Peña}}\ and\ \bibinfo {author} {\bibfnamefont {J.-P.}\
			\bibnamefont {Ortega}},\ }\bibfield  {title} {\bibinfo {title} {Quantum
			reservoir computing in finite dimensions},\ }\href
	{http://dx.doi.org/10.1103/PhysRevE.107.035306} {\bibfield  {journal}
		{\bibinfo  {journal} {Physical Review E}\ }\textbf {\bibinfo {volume}
			{107}},\ \bibinfo {pages} {035306} (\bibinfo {year} {2023})}\BibitemShut
	{NoStop}%
	\bibitem [{\citenamefont {Jaeger}(2002)}]{jaeger2002short}%
	\BibitemOpen
	\bibfield  {author} {\bibinfo {author} {\bibfnamefont {H.}~\bibnamefont
			{Jaeger}},\ }\bibfield  {title} {\bibinfo {title} {Short term memory in echo
			state networks. gmd-report 152},\ }\href
	{https://www.ai.rug.nl/minds/uploads/STMEchoStatesTechRep.pdf} {\bibfield
		{journal} {\bibinfo  {journal} {GMD-German National Research Institute for
				Computer Science (2002)}\ } (\bibinfo {year} {2002})}\BibitemShut {NoStop}%
	\bibitem [{\citenamefont {Ganguli}\ \emph {et~al.}(2008)\citenamefont
		{Ganguli}, \citenamefont {Huh},\ and\ \citenamefont
		{Sompolinsky}}]{ganguli_memory_2008}%
	\BibitemOpen
	\bibfield  {author} {\bibinfo {author} {\bibfnamefont {S.}~\bibnamefont
			{Ganguli}}, \bibinfo {author} {\bibfnamefont {D.}~\bibnamefont {Huh}},\ and\
		\bibinfo {author} {\bibfnamefont {H.}~\bibnamefont {Sompolinsky}},\
	}\bibfield  {title} {\bibinfo {title} {Memory traces in dynamical systems},\
	}\href {https://doi.org/10.1073/pnas.0804451105} {\bibfield  {journal}
		{\bibinfo  {journal} {Proceedings of the National Academy of Sciences}\
		}\textbf {\bibinfo {volume} {105}},\ \bibinfo {pages} {18970–18975}
		(\bibinfo {year} {2008})}\BibitemShut {NoStop}%
	\bibitem [{\citenamefont {Dambre}\ \emph {et~al.}(2012)\citenamefont {Dambre},
		\citenamefont {Verstraeten}, \citenamefont {Schrauwen},\ and\ \citenamefont
		{Massar}}]{dambre_information_2012}%
	\BibitemOpen
	\bibfield  {author} {\bibinfo {author} {\bibfnamefont {J.}~\bibnamefont
			{Dambre}}, \bibinfo {author} {\bibfnamefont {D.}~\bibnamefont {Verstraeten}},
		\bibinfo {author} {\bibfnamefont {B.}~\bibnamefont {Schrauwen}},\ and\
		\bibinfo {author} {\bibfnamefont {S.}~\bibnamefont {Massar}},\ }\bibfield
	{title} {\bibinfo {title} {Information {Processing} {Capacity} of {Dynamical}
			{Systems}},\ }\href {https://doi.org/10.1038/srep00514} {\bibfield  {journal}
		{\bibinfo  {journal} {Scientific Reports}\ }\textbf {\bibinfo {volume} {2}},\
		\bibinfo {pages} {514} (\bibinfo {year} {2012})}\BibitemShut {NoStop}%
	\bibitem [{\citenamefont {Verstraeten}\ \emph {et~al.}(2010)\citenamefont
		{Verstraeten}, \citenamefont {Dambre}, \citenamefont {Dutoit},\ and\
		\citenamefont {Schrauwen}}]{verstraeten_memory_2010}%
	\BibitemOpen
	\bibfield  {author} {\bibinfo {author} {\bibfnamefont {D.}~\bibnamefont
			{Verstraeten}}, \bibinfo {author} {\bibfnamefont {J.}~\bibnamefont {Dambre}},
		\bibinfo {author} {\bibfnamefont {X.}~\bibnamefont {Dutoit}},\ and\ \bibinfo
		{author} {\bibfnamefont {B.}~\bibnamefont {Schrauwen}},\ }\bibfield  {title}
	{\bibinfo {title} {Memory versus non-linearity in reservoirs},\ }in\ \href
	{https://doi.org/10.1109/ijcnn.2010.5596492} {\emph {\bibinfo {booktitle}
			{The 2010 International Joint Conference on Neural Networks (IJCNN)}}}\
	(\bibinfo  {publisher} {IEEE},\ \bibinfo {year} {2010})\ p.\ \bibinfo {pages}
	{1–8}\BibitemShut {NoStop}%
	\bibitem [{\citenamefont {Inubushi}\ and\ \citenamefont
		{Yoshimura}(2017)}]{inubushi_reservoir_2017}%
	\BibitemOpen
	\bibfield  {author} {\bibinfo {author} {\bibfnamefont {M.}~\bibnamefont
			{Inubushi}}\ and\ \bibinfo {author} {\bibfnamefont {K.}~\bibnamefont
			{Yoshimura}},\ }\bibfield  {title} {\bibinfo {title} {Reservoir computing
			beyond memory-nonlinearity trade-off},\ }\bibfield  {journal} {\bibinfo
		{journal} {Scientific Reports}\ }\textbf {\bibinfo {volume} {7}},\ \href
	{https://doi.org/10.1038/s41598-017-10257-6} {10.1038/s41598-017-10257-6}
	(\bibinfo {year} {2017})\BibitemShut {NoStop}%
	\bibitem [{\citenamefont {Wigner}(1995)}]{wigner1995remarks}%
	\BibitemOpen
	\bibfield  {author} {\bibinfo {author} {\bibfnamefont {E.~P.}\ \bibnamefont
			{Wigner}},\ }\bibfield  {title} {\bibinfo {title} {Remarks on the mind-body
			question},\ }in\ \href@noop {} {\emph {\bibinfo {booktitle} {Philosophical
				reflections and syntheses}}}\ (\bibinfo  {publisher} {Springer},\ \bibinfo
	{year} {1995})\ pp.\ \bibinfo {pages} {247--260}\BibitemShut {NoStop}%
	\bibitem [{\citenamefont {Kobayashi}\ \emph {et~al.}(2024)\citenamefont
		{Kobayashi}, \citenamefont {Fujii},\ and\ \citenamefont
		{Yamamoto}}]{kobayashi2024feedback}%
	\BibitemOpen
	\bibfield  {author} {\bibinfo {author} {\bibfnamefont {K.}~\bibnamefont
			{Kobayashi}}, \bibinfo {author} {\bibfnamefont {K.}~\bibnamefont {Fujii}},\
		and\ \bibinfo {author} {\bibfnamefont {N.}~\bibnamefont {Yamamoto}},\
	}\bibfield  {title} {\bibinfo {title} {Feedback-driven quantum reservoir
			computing for time-series analysis},\ }\href@noop {} {\bibfield  {journal}
		{\bibinfo  {journal} {PRX Quantum}\ }\textbf {\bibinfo {volume} {5}},\
		\bibinfo {pages} {040325} (\bibinfo {year} {2024})}\BibitemShut {NoStop}%
	\bibitem [{Note1()}]{Note1}%
	\BibitemOpen
	\bibinfo {note} {In the context of the NISQRC algorithm of Ref.~\cite
		{hu2024overcoming}, these are referred to as the Memory (M) and Readout (R)
		subsystems.}\BibitemShut {Stop}%
	\bibitem [{\citenamefont {Webb}(2015)}]{webb2015clifford}%
	\BibitemOpen
	\bibfield  {author} {\bibinfo {author} {\bibfnamefont {Z.}~\bibnamefont
			{Webb}},\ }\bibfield  {title} {\bibinfo {title} {The clifford group forms a
			unitary 3-design},\ }\href@noop {} {\bibfield  {journal} {\bibinfo  {journal}
			{arXiv:1510.02769}\ } (\bibinfo {year} {2015})}\BibitemShut {NoStop}%
	\bibitem [{\citenamefont {Adesso}\ \emph {et~al.}(2012)\citenamefont {Adesso},
		\citenamefont {Girolami},\ and\ \citenamefont
		{Serafini}}]{adesso2012measuring}%
	\BibitemOpen
	\bibfield  {author} {\bibinfo {author} {\bibfnamefont {G.}~\bibnamefont
			{Adesso}}, \bibinfo {author} {\bibfnamefont {D.}~\bibnamefont {Girolami}},\
		and\ \bibinfo {author} {\bibfnamefont {A.}~\bibnamefont {Serafini}},\
	}\bibfield  {title} {\bibinfo {title} {Measuring gaussian quantum information
			and correlations using the r{\'e}nyi entropy of order 2},\ }\href@noop {}
	{\bibfield  {journal} {\bibinfo  {journal} {Phys. Rev. Lett.}\ }\textbf
		{\bibinfo {volume} {109}},\ \bibinfo {pages} {190502} (\bibinfo {year}
		{2012})}\BibitemShut {NoStop}%
	\bibitem [{\citenamefont {Hamma}\ \emph {et~al.}(2016)\citenamefont {Hamma},
		\citenamefont {Giampaolo},\ and\ \citenamefont
		{Illuminati}}]{hamma2016mutual}%
	\BibitemOpen
	\bibfield  {author} {\bibinfo {author} {\bibfnamefont {A.}~\bibnamefont
			{Hamma}}, \bibinfo {author} {\bibfnamefont {S.~M.}\ \bibnamefont
			{Giampaolo}},\ and\ \bibinfo {author} {\bibfnamefont {F.}~\bibnamefont
			{Illuminati}},\ }\bibfield  {title} {\bibinfo {title} {Mutual information and
			spontaneous symmetry breaking},\ }\href@noop {} {\bibfield  {journal}
		{\bibinfo  {journal} {Phys. Rev. A}\ }\textbf {\bibinfo {volume} {93}},\
		\bibinfo {pages} {012303} (\bibinfo {year} {2016})}\BibitemShut {NoStop}%
	\bibitem [{goo(2023)}]{google2023measurement}%
	\BibitemOpen
	\bibfield  {title} {\bibinfo {title} {Measurement-induced entanglement and
			teleportation on a noisy quantum processor},\ }\href@noop {} {\bibfield
		{journal} {\bibinfo  {journal} {Nature}\ }\textbf {\bibinfo {volume} {622}},\
		\bibinfo {pages} {481} (\bibinfo {year} {2023})}\BibitemShut {NoStop}%
	\bibitem [{\citenamefont {Kukulski}\ \emph {et~al.}(2021)\citenamefont
		{Kukulski}, \citenamefont {Nechita}, \citenamefont {Pawela}, \citenamefont
		{Pucha{\l}a},\ and\ \citenamefont {{\.Z}yczkowski}}]{kukulski2021generating}%
	\BibitemOpen
	\bibfield  {author} {\bibinfo {author} {\bibfnamefont {R.}~\bibnamefont
			{Kukulski}}, \bibinfo {author} {\bibfnamefont {I.}~\bibnamefont {Nechita}},
		\bibinfo {author} {\bibfnamefont {{\L}.}~\bibnamefont {Pawela}}, \bibinfo
		{author} {\bibfnamefont {Z.}~\bibnamefont {Pucha{\l}a}},\ and\ \bibinfo
		{author} {\bibfnamefont {K.}~\bibnamefont {{\.Z}yczkowski}},\ }\bibfield
	{title} {\bibinfo {title} {Generating random quantum channels},\ }\href@noop
	{} {\bibfield  {journal} {\bibinfo  {journal} {J. Math. Phys.}\ }\textbf
		{\bibinfo {volume} {62}} (\bibinfo {year} {2021})}\BibitemShut {NoStop}%
	\bibitem [{\citenamefont {Serbyn}\ \emph {et~al.}(2021)\citenamefont {Serbyn},
		\citenamefont {Abanin},\ and\ \citenamefont {Papi{\'c}}}]{serbyn2021quantum}%
	\BibitemOpen
	\bibfield  {author} {\bibinfo {author} {\bibfnamefont {M.}~\bibnamefont
			{Serbyn}}, \bibinfo {author} {\bibfnamefont {D.~A.}\ \bibnamefont {Abanin}},\
		and\ \bibinfo {author} {\bibfnamefont {Z.}~\bibnamefont {Papi{\'c}}},\
	}\bibfield  {title} {\bibinfo {title} {Quantum many-body scars and weak
			breaking of ergodicity},\ }\href@noop {} {\bibfield  {journal} {\bibinfo
			{journal} {Nat. Phys.}\ }\textbf {\bibinfo {volume} {17}},\ \bibinfo {pages}
		{675} (\bibinfo {year} {2021})}\BibitemShut {NoStop}%
	\bibitem [{\citenamefont {Liu}\ \emph {et~al.}(2024)\citenamefont {Liu},
		\citenamefont {Li}, \citenamefont {Zhang}, \citenamefont {Jian},\ and\
		\citenamefont {Yao}}]{liu2024noise}%
	\BibitemOpen
	\bibfield  {author} {\bibinfo {author} {\bibfnamefont {S.}~\bibnamefont
			{Liu}}, \bibinfo {author} {\bibfnamefont {M.-R.}\ \bibnamefont {Li}},
		\bibinfo {author} {\bibfnamefont {S.-X.}\ \bibnamefont {Zhang}}, \bibinfo
		{author} {\bibfnamefont {S.-K.}\ \bibnamefont {Jian}},\ and\ \bibinfo
		{author} {\bibfnamefont {H.}~\bibnamefont {Yao}},\ }\bibfield  {title}
	{\bibinfo {title} {Noise-induced phase transitions in hybrid quantum
			circuits},\ }\href@noop {} {\bibfield  {journal} {\bibinfo  {journal}
			{Physical Review B}\ }\textbf {\bibinfo {volume} {110}},\ \bibinfo {pages}
		{064323} (\bibinfo {year} {2024})}\BibitemShut {NoStop}%
	\bibitem [{\citenamefont {Hunter-Jones}(2019)}]{hunter2019unitary}%
	\BibitemOpen
	\bibfield  {author} {\bibinfo {author} {\bibfnamefont {N.}~\bibnamefont
			{Hunter-Jones}},\ }\bibfield  {title} {\bibinfo {title} {Unitary designs from
			statistical mechanics in random quantum circuits},\ }\href@noop {} {\bibfield
		{journal} {\bibinfo  {journal} {arXiv:1905.12053}\ } (\bibinfo {year}
		{2019})}\BibitemShut {NoStop}%
	\bibitem [{\citenamefont {Wilde}(2013)}]{wilde2013quantum}%
	\BibitemOpen
	\bibfield  {author} {\bibinfo {author} {\bibfnamefont {M.~M.}\ \bibnamefont
			{Wilde}},\ }\href@noop {} {\emph {\bibinfo {title} {Quantum information
				theory}}}\ (\bibinfo  {publisher} {Cambridge university press},\ \bibinfo
	{year} {2013})\BibitemShut {NoStop}%
	\bibitem [{\citenamefont {Javadi-Abhari}\ \emph {et~al.}(2024)\citenamefont
		{Javadi-Abhari}, \citenamefont {Treinish}, \citenamefont {Krsulich},
		\citenamefont {Wood}, \citenamefont {Lishman}, \citenamefont {Gacon},
		\citenamefont {Martiel}, \citenamefont {Nation}, \citenamefont {Bishop},
		\citenamefont {Cross}, \citenamefont {Johnson},\ and\ \citenamefont
		{Gambetta}}]{qiskit2024}%
	\BibitemOpen
	\bibfield  {author} {\bibinfo {author} {\bibfnamefont {A.}~\bibnamefont
			{Javadi-Abhari}}, \bibinfo {author} {\bibfnamefont {M.}~\bibnamefont
			{Treinish}}, \bibinfo {author} {\bibfnamefont {K.}~\bibnamefont {Krsulich}},
		\bibinfo {author} {\bibfnamefont {C.~J.}\ \bibnamefont {Wood}}, \bibinfo
		{author} {\bibfnamefont {J.}~\bibnamefont {Lishman}}, \bibinfo {author}
		{\bibfnamefont {J.}~\bibnamefont {Gacon}}, \bibinfo {author} {\bibfnamefont
			{S.}~\bibnamefont {Martiel}}, \bibinfo {author} {\bibfnamefont {P.~D.}\
			\bibnamefont {Nation}}, \bibinfo {author} {\bibfnamefont {L.~S.}\
			\bibnamefont {Bishop}}, \bibinfo {author} {\bibfnamefont {A.~W.}\
			\bibnamefont {Cross}}, \bibinfo {author} {\bibfnamefont {B.~R.}\ \bibnamefont
			{Johnson}},\ and\ \bibinfo {author} {\bibfnamefont {J.~M.}\ \bibnamefont
			{Gambetta}},\ }\href {https://doi.org/10.48550/arXiv.2405.08810} {\bibinfo
		{title} {Quantum computing with {Q}iskit}} (\bibinfo {year} {2024}),\ \Eprint
	{https://arxiv.org/abs/2405.08810} {arXiv:2405.08810 [quant-ph]} \BibitemShut
	{NoStop}%
	\bibitem [{\citenamefont {Elben}\ \emph {et~al.}(2019)\citenamefont {Elben},
		\citenamefont {Vermersch}, \citenamefont {Roos},\ and\ \citenamefont
		{Zoller}}]{elben2019statistical}%
	\BibitemOpen
	\bibfield  {author} {\bibinfo {author} {\bibfnamefont {A.}~\bibnamefont
			{Elben}}, \bibinfo {author} {\bibfnamefont {B.}~\bibnamefont {Vermersch}},
		\bibinfo {author} {\bibfnamefont {C.~F.}\ \bibnamefont {Roos}},\ and\
		\bibinfo {author} {\bibfnamefont {P.}~\bibnamefont {Zoller}},\ }\bibfield
	{title} {\bibinfo {title} {Statistical correlations between locally
			randomized measurements: A toolbox for probing entanglement in many-body
			quantum states},\ }\href@noop {} {\bibfield  {journal} {\bibinfo  {journal}
			{Phys. Rev. A}\ }\textbf {\bibinfo {volume} {99}},\ \bibinfo {pages} {052323}
		(\bibinfo {year} {2019})}\BibitemShut {NoStop}%
	\bibitem [{\citenamefont {Li}\ \emph {et~al.}(2023)\citenamefont {Li},
		\citenamefont {Zou}, \citenamefont {Glorioso}, \citenamefont {Altman},\ and\
		\citenamefont {Fisher}}]{li2023cross}%
	\BibitemOpen
	\bibfield  {author} {\bibinfo {author} {\bibfnamefont {Y.}~\bibnamefont
			{Li}}, \bibinfo {author} {\bibfnamefont {Y.}~\bibnamefont {Zou}}, \bibinfo
		{author} {\bibfnamefont {P.}~\bibnamefont {Glorioso}}, \bibinfo {author}
		{\bibfnamefont {E.}~\bibnamefont {Altman}},\ and\ \bibinfo {author}
		{\bibfnamefont {M.~P.}\ \bibnamefont {Fisher}},\ }\bibfield  {title}
	{\bibinfo {title} {Cross entropy benchmark for measurement-induced phase
			transitions},\ }\href@noop {} {\bibfield  {journal} {\bibinfo  {journal}
			{Phys. Rev. Lett.}\ }\textbf {\bibinfo {volume} {130}},\ \bibinfo {pages}
		{220404} (\bibinfo {year} {2023})}\BibitemShut {NoStop}%
	\bibitem [{\citenamefont {Elben}\ \emph {et~al.}(2023)\citenamefont {Elben},
		\citenamefont {Flammia}, \citenamefont {Huang}, \citenamefont {Kueng},
		\citenamefont {Preskill}, \citenamefont {Vermersch},\ and\ \citenamefont
		{Zoller}}]{elben2023randomized}%
	\BibitemOpen
	\bibfield  {author} {\bibinfo {author} {\bibfnamefont {A.}~\bibnamefont
			{Elben}}, \bibinfo {author} {\bibfnamefont {S.~T.}\ \bibnamefont {Flammia}},
		\bibinfo {author} {\bibfnamefont {H.-Y.}\ \bibnamefont {Huang}}, \bibinfo
		{author} {\bibfnamefont {R.}~\bibnamefont {Kueng}}, \bibinfo {author}
		{\bibfnamefont {J.}~\bibnamefont {Preskill}}, \bibinfo {author}
		{\bibfnamefont {B.}~\bibnamefont {Vermersch}},\ and\ \bibinfo {author}
		{\bibfnamefont {P.}~\bibnamefont {Zoller}},\ }\bibfield  {title} {\bibinfo
		{title} {The randomized measurement toolbox},\ }\href@noop {} {\bibfield
		{journal} {\bibinfo  {journal} {Nat. Rev. Phys.}\ }\textbf {\bibinfo {volume}
			{5}},\ \bibinfo {pages} {9} (\bibinfo {year} {2023})}\BibitemShut {NoStop}%
	\bibitem [{\citenamefont {Feng}\ and\ \citenamefont {Zhuang}(2025)}]{feng2025}%
	\BibitemOpen
	\bibfield  {author} {\bibinfo {author} {\bibfnamefont {J.}~\bibnamefont
			{Feng}}\ and\ \bibinfo {author} {\bibfnamefont {Q.}~\bibnamefont {Zhuang}},\
	}\bibfield  {title} {\bibinfo {title} {Spectrum and quantum information in
			reset-driven floquet hamiltonian dynamics},\ }\href@noop {} {\bibfield
		{journal} {\bibinfo  {journal} {in preparation}\ } (\bibinfo {year}
		{2025})}\BibitemShut {NoStop}%
	\bibitem [{\citenamefont {Zhang}\ \emph
		{et~al.}(2025{\natexlab{b}})\citenamefont {Zhang}, \citenamefont {Hu},
		\citenamefont {Mo}, \citenamefont {Chen}, \citenamefont {Türeci},\ and\
		\citenamefont {Zhuang}}]{github}%
	\BibitemOpen
	\bibfield  {author} {\bibinfo {author} {\bibfnamefont {B.}~\bibnamefont
			{Zhang}}, \bibinfo {author} {\bibfnamefont {F.}~\bibnamefont {Hu}}, \bibinfo
		{author} {\bibfnamefont {R.}~\bibnamefont {Mo}}, \bibinfo {author}
		{\bibfnamefont {T.}~\bibnamefont {Chen}}, \bibinfo {author} {\bibfnamefont
			{H.~E.}\ \bibnamefont {Türeci}},\ and\ \bibinfo {author} {\bibfnamefont
			{Q.}~\bibnamefont {Zhuang}},\ }\href@noop {} {\bibinfo {title}
		{{QMI\_monitored\_dynamics}}},\ \bibinfo {howpublished}
	{\url{https://github.com/bzGit06/QMI_monitored_dynamics}} (\bibinfo {year}
	{2025}{\natexlab{b}})\BibitemShut {NoStop}%
	\bibitem [{\citenamefont {Nahum}\ \emph {et~al.}(2017)\citenamefont {Nahum},
		\citenamefont {Ruhman}, \citenamefont {Vijay},\ and\ \citenamefont
		{Haah}}]{nahum2017quantum}%
	\BibitemOpen
	\bibfield  {author} {\bibinfo {author} {\bibfnamefont {A.}~\bibnamefont
			{Nahum}}, \bibinfo {author} {\bibfnamefont {J.}~\bibnamefont {Ruhman}},
		\bibinfo {author} {\bibfnamefont {S.}~\bibnamefont {Vijay}},\ and\ \bibinfo
		{author} {\bibfnamefont {J.}~\bibnamefont {Haah}},\ }\bibfield  {title}
	{\bibinfo {title} {Quantum entanglement growth under random unitary
			dynamics},\ }\href@noop {} {\bibfield  {journal} {\bibinfo  {journal}
			{Physical Review X}\ }\textbf {\bibinfo {volume} {7}},\ \bibinfo {pages}
		{031016} (\bibinfo {year} {2017})}\BibitemShut {NoStop}%
	\bibitem [{\citenamefont {QuantumSavory}()}]{quantumclifford}%
	\BibitemOpen
	\bibfield  {author} {\bibinfo {author} {\bibnamefont {QuantumSavory}},\
	}\href@noop {} {\bibinfo {title} {Quantumclifford.jl}},\ \bibinfo
	{howpublished} {\url{https://github.com/QuantumSavory/QuantumClifford.jl}},\
	\bibinfo {note} {accessed: 2025-11-19}\BibitemShut {NoStop}%
	\bibitem [{\citenamefont {Hayden}\ and\ \citenamefont
		{Preskill}(2007)}]{hayden2007black}%
	\BibitemOpen
	\bibfield  {author} {\bibinfo {author} {\bibfnamefont {P.}~\bibnamefont
			{Hayden}}\ and\ \bibinfo {author} {\bibfnamefont {J.}~\bibnamefont
			{Preskill}},\ }\bibfield  {title} {\bibinfo {title} {Black holes as mirrors:
			quantum information in random subsystems},\ }\href@noop {} {\bibfield
		{journal} {\bibinfo  {journal} {J. High Energy Phys.}\ }\textbf {\bibinfo
			{volume} {2007}}\bibinfo  {number} { (09)},\ \bibinfo {pages}
		{120}}\BibitemShut {NoStop}%
	\bibitem [{\citenamefont {Ippoliti}\ and\ \citenamefont
		{Ho}(2022)}]{ippoliti2022solvable}%
	\BibitemOpen
	\bibfield  {number} {  }\bibfield  {author} {\bibinfo {author} {\bibfnamefont
			{M.}~\bibnamefont {Ippoliti}}\ and\ \bibinfo {author} {\bibfnamefont {W.~W.}\
			\bibnamefont {Ho}},\ }\bibfield  {title} {\bibinfo {title} {Solvable model of
			deep thermalization with distinct design times},\ }\href@noop {} {\bibfield
		{journal} {\bibinfo  {journal} {Quantum}\ }\textbf {\bibinfo {volume} {6}},\
		\bibinfo {pages} {886} (\bibinfo {year} {2022})}\BibitemShut {NoStop}%
	\bibitem [{\citenamefont {Zhang}\ \emph {et~al.}(2023)\citenamefont {Zhang},
		\citenamefont {Allcock}, \citenamefont {Wan}, \citenamefont {Liu},
		\citenamefont {Sun}, \citenamefont {Yu}, \citenamefont {Yang}, \citenamefont
		{Qiu}, \citenamefont {Ye}, \citenamefont {Chen} \emph
		{et~al.}}]{zhang2023tensorcircuit}%
	\BibitemOpen
	\bibfield  {author} {\bibinfo {author} {\bibfnamefont {S.-X.}\ \bibnamefont
			{Zhang}}, \bibinfo {author} {\bibfnamefont {J.}~\bibnamefont {Allcock}},
		\bibinfo {author} {\bibfnamefont {Z.-Q.}\ \bibnamefont {Wan}}, \bibinfo
		{author} {\bibfnamefont {S.}~\bibnamefont {Liu}}, \bibinfo {author}
		{\bibfnamefont {J.}~\bibnamefont {Sun}}, \bibinfo {author} {\bibfnamefont
			{H.}~\bibnamefont {Yu}}, \bibinfo {author} {\bibfnamefont {X.-H.}\
			\bibnamefont {Yang}}, \bibinfo {author} {\bibfnamefont {J.}~\bibnamefont
			{Qiu}}, \bibinfo {author} {\bibfnamefont {Z.}~\bibnamefont {Ye}}, \bibinfo
		{author} {\bibfnamefont {Y.-Q.}\ \bibnamefont {Chen}}, \emph {et~al.},\
	}\bibfield  {title} {\bibinfo {title} {Tensorcircuit: a quantum software
			framework for the nisq era},\ }\href@noop {} {\bibfield  {journal} {\bibinfo
			{journal} {Quantum}\ }\textbf {\bibinfo {volume} {7}},\ \bibinfo {pages}
		{912} (\bibinfo {year} {2023})}\BibitemShut {NoStop}%
	\bibitem [{\citenamefont {Belyansky}\ \emph {et~al.}(2020)\citenamefont
		{Belyansky}, \citenamefont {Bienias}, \citenamefont {Kharkov}, \citenamefont
		{Gorshkov},\ and\ \citenamefont {Swingle}}]{belyansky2020minimal}%
	\BibitemOpen
	\bibfield  {author} {\bibinfo {author} {\bibfnamefont {R.}~\bibnamefont
			{Belyansky}}, \bibinfo {author} {\bibfnamefont {P.}~\bibnamefont {Bienias}},
		\bibinfo {author} {\bibfnamefont {Y.~A.}\ \bibnamefont {Kharkov}}, \bibinfo
		{author} {\bibfnamefont {A.~V.}\ \bibnamefont {Gorshkov}},\ and\ \bibinfo
		{author} {\bibfnamefont {B.}~\bibnamefont {Swingle}},\ }\bibfield  {title}
	{\bibinfo {title} {Minimal model for fast scrambling},\ }\href@noop {}
	{\bibfield  {journal} {\bibinfo  {journal} {Phys. Rev. Lett.}\ }\textbf
		{\bibinfo {volume} {125}},\ \bibinfo {pages} {130601} (\bibinfo {year}
		{2020})}\BibitemShut {NoStop}%
\end{thebibliography}

%

\appendix

\section{Clifford circuit simulation with stabilizer formalism}
\label{app:clifford}

\BZ{
In this section, we discuss the stabilizer formalism for Clifford circuit simulation, with further details presented in Refs.~\cite{nahum2017quantum, li2019measurement}.
}

\BZ{
A (mixed) stabilizer state $\rho$ of $N$ qubits can be fully described as
\be
    \rho = \prod_{i=1}^{|G|} \frac{\bI + g_i}{2},
\ee
where mutually commuting Pauli operators $G = \{g_1, \dots, g_{|G|}\}$ formulates a set of generators for the stabilizer group of the state. The size of generator set satisfies $|G|\le N$ where the equality holds for pure states. For analysis and calculation convenience, we can write the generators $g\in G$ in the {\it clipped gauge}, such that there are exactly two generators with left or right end on any single qubit. Furthermore, if the two generators have only left or right ends on a given qubit, then the Pauli operator of this qubit must be different in the two generators. The advantage of {\it clipped gauge} is that the entanglement in a subregion $A$ can be directly read out from the generators as
\be
    S_A = |A| - |\{g|\:{\rm supp}(g) \subseteq A\}|,
\ee
where ${\rm supp}(g)$ is the support of operator $g$. 
The mutual information between two neighboring region $A$ and $B$ can be written out in a similar way. Our numerical simulation of the Clifford circuit is performed with \texttt{QuantumClifford.jl}~\cite{quantumclifford}.
}

\section{QMI dynamics with and without reset}
\label{app:reset}

In this section, we prove that the dynamics of measurement-conditioned QMI remains the same regardless of reset or not. Similar to Eq.~\eqref{state_RAt}, the conditional state at step $t$ without reset strategy is
\begin{align}
&\ket{\psi^\prime_{\bfz}(\bm U)}_{RA_t} 
\\
&\propto {}_{\bm B}\bra{\bm z} U_t\cdots U_2U_1 \left[\ket{\Phi}_{RA_0}\otimes \left(\ket{0}_{B_0}\otimes_{k=1}^{t-1} \ket{z_k}_{B_k}\right)\right]
\\
&\propto {}_{\bm B}\bra{\bm z} U_t\cdots U_2U_1 \left[\ket{\Phi}_{RA_0}\otimes \left(\otimes_{k=1}^{t-1} X^{z_k}\ket{\bm 0}_{\bm B}\right)\right]
\\
&=\ket{\psi_{\bf z}(\bm U^{\bfz})}_{RA_t},
\label{state_RAt_wo_reset}
\end{align}
where $\bm B = B_t \cdots B_2 B_1$ is the joint system of bath system in each step and we have defined $\bm U^{\bfz}=U_t X^{z_{t-1}}U_3X^{z_2} U_2X^{z_1} U_1$ with $\{X^{z_k}\}_{k=1}^{t-1}$ applied on the bath system only.

Then from the definition of measurement-conditioned QMI in Eq.~\eqref{QMI_cond},
\begin{align}
&\E_{\rm Haar} \overline{I(R:A_t|\bfz)}^\prime    
\\
&=\E_{\bm U\sim {\rm Haar}} 2\E_{\bfz}   S\left(\tr_{R}\left(\ketbra{\psi^\prime_\bfz(\bm U)}{\psi^\prime_\bfz(\bm U)}_{RA_t}\right)\right)
\\
&=2\E_{\bm U\sim {\rm Haar}} \sum_{\bm z} P_{\bm U^{\bm z}}(\bm z) S\left(\tr_{R}\left(\ketbra{\psi_\bfz(\bm U^{\bfz})}{\psi_{\bfz}(\bm U^{\bfz})}_{RA_t}\right)\right)
\\
&=2\sum_{\bm z} \E_{\bm U\sim {\rm Haar}} P_{\bm U^{\bm z}}(\bm z) S\left(\tr_{R}\left(\ketbra{\psi_\bfz(\bm U^{\bfz})}{\psi_{\bfz}(\bm U^{\bfz})}_{RA_t}\right)\right)
\\
&=\E_{\rm Haar} \overline{I(R:A_t|\bfz)}
\end{align}
equals the original measurement-conditioned QMI with reset strategy.

\BZ{We further consider the unmonitored dynamics but reset the bath state to a fully-mixed state in each step instead.
We expect that it can provide a faster decay as it is equivalent to a fully depolarizing channel. For the R\'enyi-2 extended unconditioned QMI, we have the following theorem (see Appendix~\ref{app:thereom23} for a proof).}
\begin{theorem}
\label{MI_traceout_id_theorem}
    The expected R\'enyi-2 extended measurement-unconditioned QMI of a Bell initial state in the quantum dynamics of $2$-design unitaries with mid-circuit measurements and fully-mixed state reset at time $t$ is
    \begin{align}
        &\E_{\rm Haar} I_2(R:A_t) \nonumber\\
        &\simeq \log_2\left(\left(d_B - \frac{1}{d_A^2}\right) \left(\frac{d_A^2-1}{d_A^2 d_B^2 - 1}\right)^t + \frac{1}{d_A^2}\right) \nonumber \\
        &\quad - \log_2\left(\frac{d_B-1}{d_A}\left(\frac{d_A^2-1}{d_A^2 d_B^2 - 1}\right)^t + \frac{1}{d_A}\right) + N_A.
        \label{eq:MI_traceout_id}
    \end{align}
In the asymptotic limit $d_A, d_B \gg 1$, when $t \ll N_A/N_B + 1/2$, 
\begin{align}
    \E_{\rm Haar} I_2(R:A_t) \simeq 2N_A + N_B - 2N_Bt,
    \label{eq:MI_traceout_id_asymp}
\end{align}
while for $t \gg N_A/N_B + 1/2$,
\begin{align}
    \E_{\rm Haar} I_2(R:A_t) \simeq d_A^2 d_B^{1-2t}.
    \label{eq:MI_traceout_id_asymp_late}
\end{align}
\end{theorem}
Compared to Theorem~\ref{MI_traceout_theorem}, the information dynamics in both early and late time still exhibits a corresponding linear / exponential decay, but at a decay rate twice larger. \BZ{Here, we provide the physical intuition for the different decay rates. In each step of reset, we can purify the fully mixed state by a Bell pair of dimension of $d_B^2$, which can be regarded as an expanded bath system with twice the number of qubits. Furthermore, we can also regard the whole unmonitored dynamics with Bell purification as the Hayden-Preskill protocol~\cite{hayden2007black}, although the scrambling unitary over $N_A + N_B t$ qubits is slightly different. According to the decoupling theorem~\cite{hayden2007black}, when $N_A < 1/2(N_A + N_Bt)$ thus $t> N_A/N_B$, the output side of system is nearly fully decoupled from the reference, which also indicates the decay rate of unconditioned QMI to be $2N_B$.}

Accordingly, the $\epsilon$-lifetime \BZ{with fully-mixed state reset} is estimated as
\begin{align}
    \tau_{\rm uncond} \simeq 
    \begin{cases}
        (1-\epsilon)\left(N_A/N_B + 1/2\right), &t \ll N_A/N_B + 1/2,\\
        \log_2(1/\epsilon)/2N_B, & t \gg N_A/N_B + 1/2.
    \end{cases}
\label{eq:MI_traceout_id_time}
\end{align}
which is only approximately half of the corresponding lifetime compared to the one with zero state reset strategy in Eq.~\eqref{eq:MI_traceout_time} and Eq.~\eqref{tau_eig}.

\begin{figure}[t]
    \centering
    \includegraphics[width=0.45\textwidth]{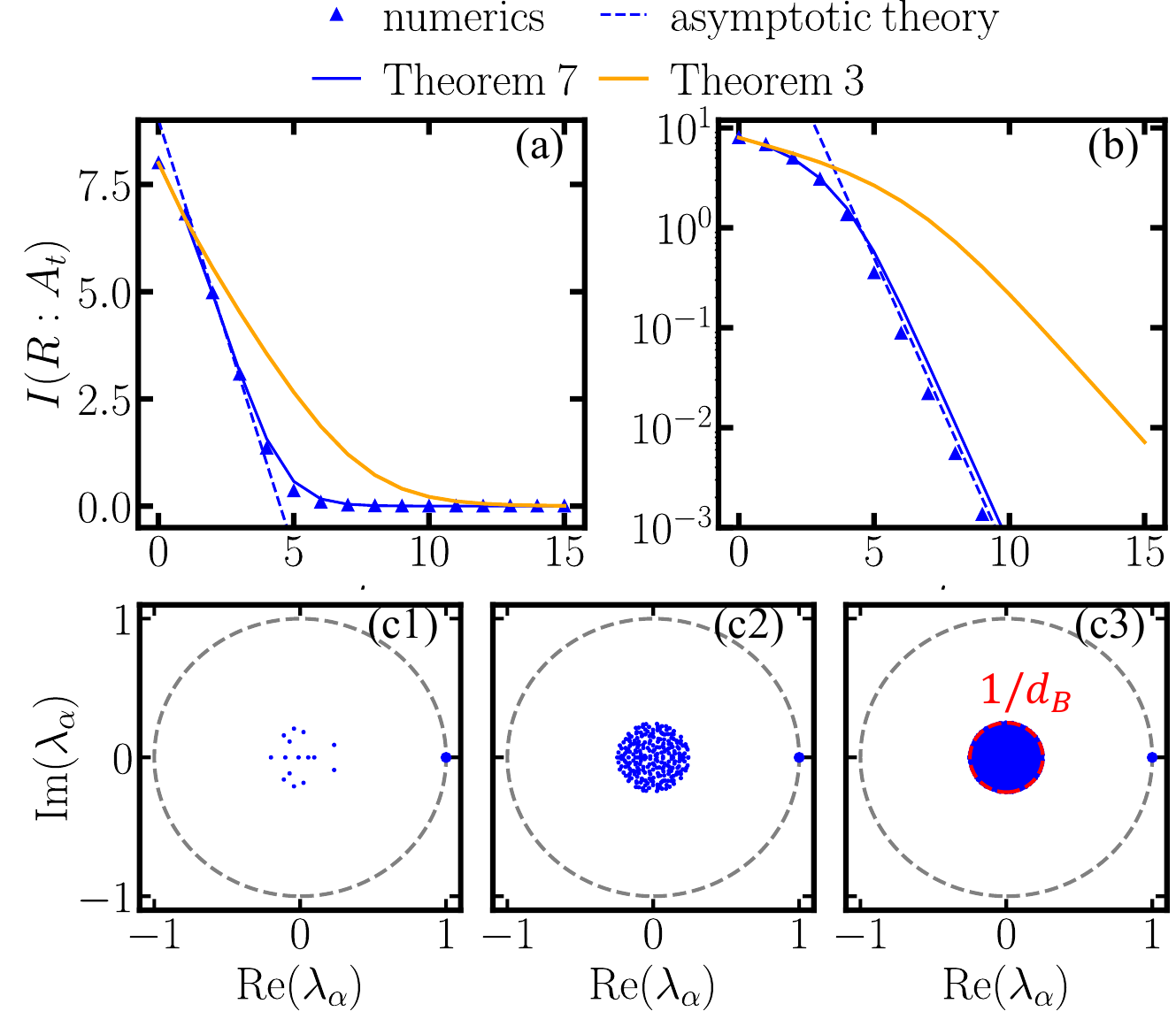}
    \caption{
    \BZ{Measurement-unconditioned QMI in unmonitored dynamics with full-mixed state reset. In (a) and (b), we plot numerical simulation of measurement-unconditioned QMI with fully mixed state reset (blue triangles) in early and late time separately. Blue solid lines show the analytical theory of Eq.~\eqref{eq:MI_traceout_id} in Theorem~\ref{MI_traceout_id_theorem}.
    Blue dashed lines in (a) and (b) show the corresponding asymptotic theory. Orange solid line represent the theory for with pure state reset in Eq.~\eqref{eq:MI_traceout}.   
    The system consists of $N_A=4$ data qubits, and $N_B=1$ bath qubits in (a-b).
    In (c), we show the spectrum of Haar channel with fully mixed state reset of system size $N_A=2, 4, 6$ (left to right) with $N_B = 2$ bath qubits. The red dashed circle indicates the radius of $1/d_B$.}}
    \label{fig:reset_mm}
\end{figure}

\BZ{Through numerical simulation with Haar unitaries (see Fig.~\ref{fig:reset_mm}a and b), the measurement-unconditioned QMI dynamics with fully-mixed state reset (blue triangles) overlaps with our theory prediction in Theorem~\ref{MI_traceout_id_theorem}, thus demonstrates the double-speed linear and exponential decay compared to zero state reset strategy (orange line) in early and late stage.
Moreover, the eigenspectrum of channel with Haar unitary and fully mixed state reset strategy concentrates in a disk of radius $1/d_B$~\cite{kukulski2021generating}, shown in Fig.~\ref{fig:reset_mm}c, providing supports to the intuition and theorem proposed above.}

\section{Identical unitary or random unitary }
\label{app:indentical_or_not}

\begin{figure}[t]
    \centering
    \includegraphics[width=0.45\textwidth]{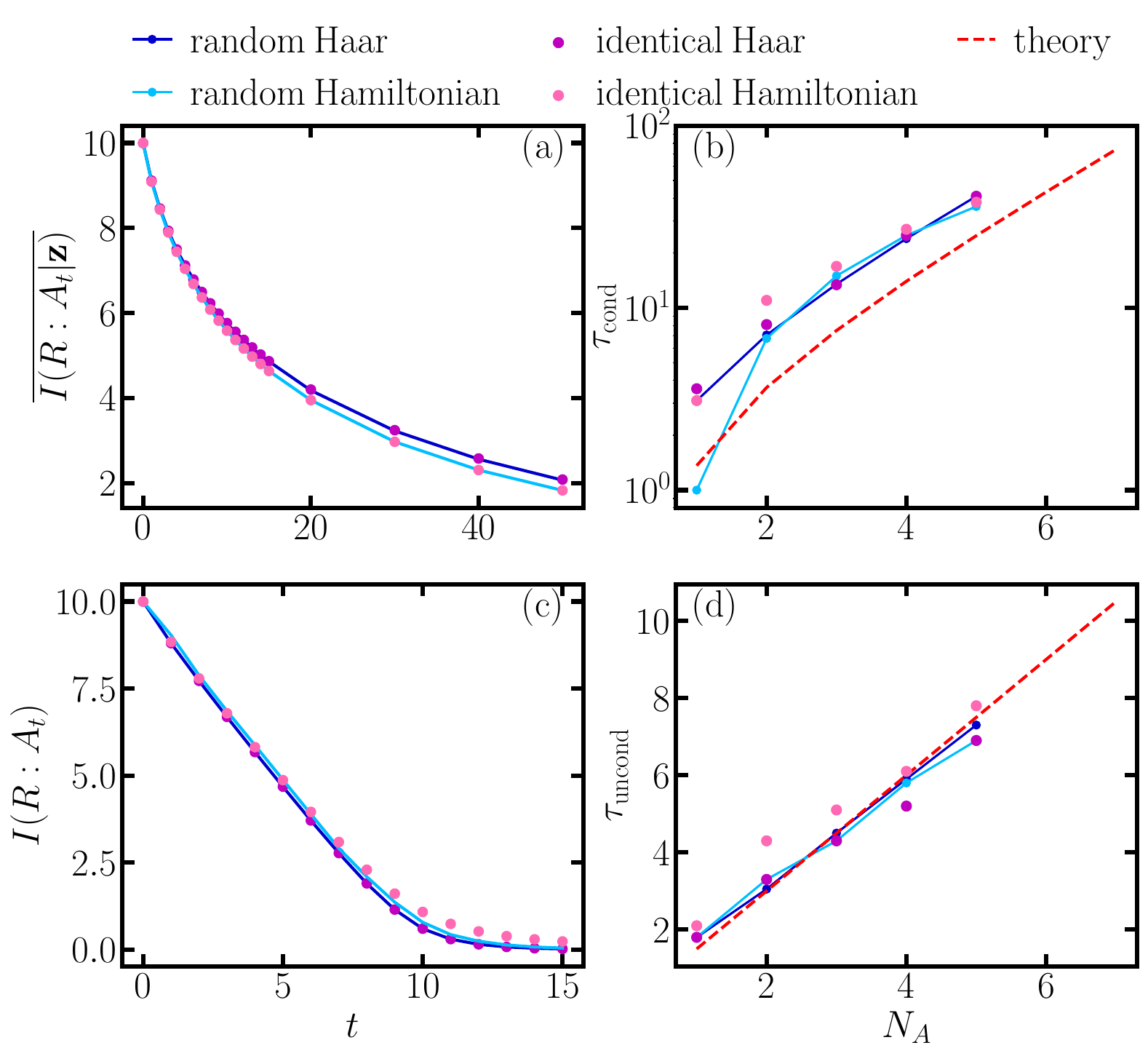}
    \caption{Comparisons on dynamics of QMI and the corresponding lifetime in quantum dynamics with random unitary and identical unitary. We present measurement-conditioned QMI and measurement-unconditioned QMI in top and bottom panels. Dark and light blue lines represent numerical simulations with randomly sampled Haar unitary and random Hamiltonian evolution separately. Purple and pink dots show results with identical fixed Haar unitary and fixed Hamiltonian evolution separately. Red dashed line in (b) and (d) represent theory from Eq.~\eqref{eq:avgMI_time} and Eq.~\eqref{eq:MI_traceout_time} with $\epsilon = 1/4$. In (a) and (c), we have system of $N_A=5$ data qubits. In all cases, the bath consists of $N_B=1$ qubits.}
    \label{fig:unitary_compare}
\end{figure}

In this section, we provide additional numerical simulations to compare the QMI dynamics and its corresponding lifetime in quantum dynamics with randomly-sampled unitary and identical unitaries. To be precise, in the main text, for Haar quantum dynamics, we randomly sample $t$ unitaries $\{U_k\}_{k=1}^t$ from Haar ensemble. Therefore, we regard this as random Haar dynamics; meanwhile for Hamiltonian-based quantum dynamics, we randomly choose a group of parameters $\{\eta_i^x, \eta_i^z, J_{ij}\}$ and keep it fixed for all $\{U_k\}_{k=1}^t$, which we regard as identical Hamiltonian dynamics. We also consider two additional cases here. For identical Haar quantum dynamics, we randomly sample a single unitary $U$ from Haar ensemble and utilize it across the dynamics for $t$ steps; for random Hamiltonian-based quantum dynamics, we randomly choose $t$ groups of parameters for the Hamiltonian in Eq.~\eqref{H_Ising}, and take their evolution for unitaries in each step. 

In Fig.~\ref{fig:unitary_compare}, we numerically simulate the measurement-conditioned and unconditioned QMI dynamics for the four cases above and estimate the corresponding QMI lifetime. The QMI dynamics for identical Haar and Hamiltonian evolution (purple and pink) overlap with the results for random Haar and Hamiltonian (dark and light blue) for a system of $N_A=5$ data qubits. The QMI lifetime among the four cases also approach each other in the same asymptotic scaling though with deviations at small $N_A$. To conclude, given a large system size of $N_A$, the QMI dynamics with random unitaries or identical unitaries become indistinguishable, making our theories of Theorem~\ref{theorem_avgMI} and Theorem~\ref{MI_traceout_theorem} applicable to wide scenarios of quantum dynamics from holographic thermalization to reservoir computing.

\section{QMI dynamics with nearest-neighbor Hamiltonian}
\label{app:MFIM}

\begin{figure}[t]
    \centering
    \includegraphics[width=0.45\textwidth]{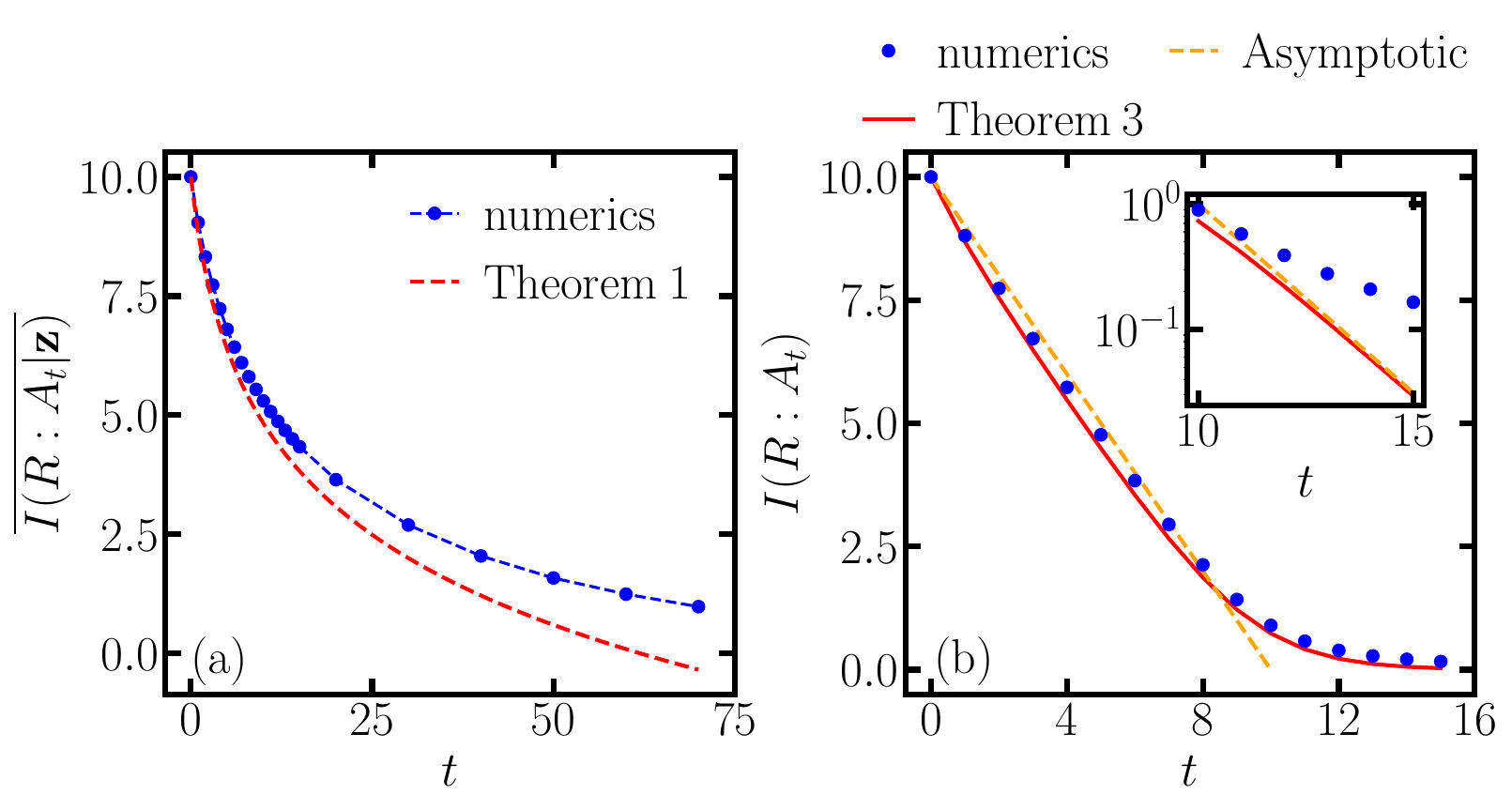}
    \caption{(a) Measurement-conditioned QMI and (b) measurement-unconditioned QMI dynamics with MFIM Hamiltonian and reset. Blue circles represent numerical simulations of $N_R = N_A = 5$ and $N_B = 1$ qubits. Red dashed line in (a) shows Eq.~\eqref{eq:avgMI_lb_simplify} in Theorem~\ref{theorem_avgMI}. Red solid line and orange dashed line in (b) represent Eq.~\eqref{eq:MI_traceout} and Eq.~\eqref{eq:MI_traceout_asymp} in Theorem~\ref{MI_traceout_theorem}. The inset shows the late-time dynamics in logarithmic scaling and the orange dashed line is Eq.~\eqref{eq:MI_traceout_asymp_late}.}
    \label{fig:MI_MFIM}
\end{figure}

In the main text, we consider the Hamiltonian to be an all-to-all coupled Ising Hamiltonian with random parameters. In this section, we turn to another type of Hamiltonian, mixing-field Ising model (MFIM), which only involves nearest-neighbor couplings,
\be
    H_{\rm MFIM} = h_x\sum_i \sigma_i^x + h_y \sum_i \sigma_i^y + \sum_{i=1}^{N-1} \sigma_i^x \sigma_{i+1}^x.
\ee
Under the choice of $h_x, h_y = 0.8090, 0.9045$, this Hamiltonian dynamics is consistent with the eigenstate thermalization hypothesis and is also adopted in the studies of deep thermalization~\cite{cotler2023emergent}.Here we adopt this parameter option to numerically validate the corresponding QMI dynamics with and without bath monitoring in Fig.~\ref{fig:MI_MFIM}. In both monitored and unmonitored dynamics, we find that the QMI dynamics behaves similarly to the all-to-all Ising Hamiltonian considered in Fig.~\ref{fig:condQMI}a and~\ref{fig:mi_traceout}a, and it is well described by our theory in Theorem~\ref{theorem_avgMI} and~\ref{MI_traceout_theorem}. Even for the late-time unmonitored dynamics, the unconditioned QMI still decays exponentially though deviating from the theory (inset).

\section{Residual monitored dynamics}
\label{app:residual_dy}

\begin{figure*}[t]
    \centering
    \includegraphics[width=0.65\textwidth]{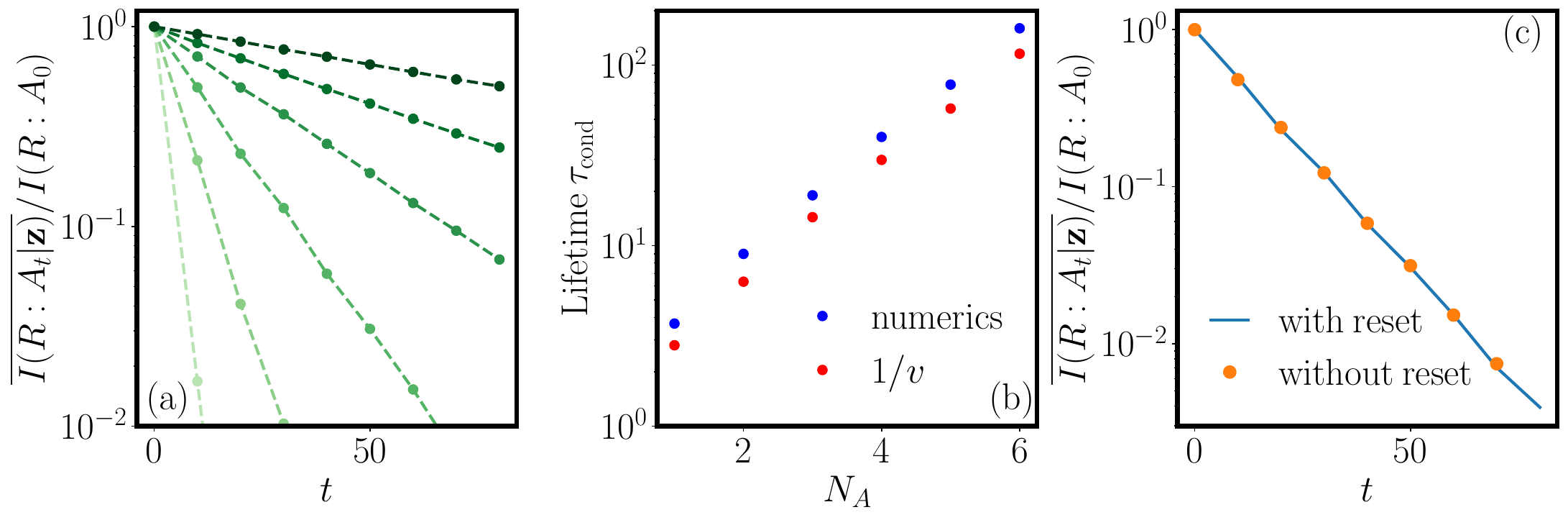}
    \caption{{Measurement-conditioned QMI $\overline{I(R:A_t|\bfz)}$ in monitored residual dynamics. The initial state $\ket{\psi}_{RA_0}$ is chosen to be the late-time conditional state in dynamics with Bell initial state. In (a), we plot the normalized QMI dynamics for $N_A=1$ to $6$ (light to dark). In (b), blue dots correspond to the estimated lifetime from numerical simulation in (a) for $\epsilon=1/4$ and red dots represent the scaling of $1/v$ where $v$ is the rate from linear fitting of $\overline{I(R:A_t|\bfz)} \sim e^{-v t}$. In (c), we compare the numerical simulation of QMI dynamics with (blue) and without reset (orange) in a system of $N_A=3$ qubits. In all cases, we take $N_B = 1$ bath qubits.}}
    \label{fig:condQMI_late}
\end{figure*}

{In this section, we provide numerical evidence on the decay of QMI in time beyond the lifetime $\tau_{\rm cond}$ in Eq.~\eqref{eq:avgMI_time}.
As the lower bound Eq.~\eqref{eq:avgMI_lb} (or the simplified version Eq.~\eqref{eq:avgMI_lb_simplify}) becomes negative at extremely late time, we expect deviations from the lower bound. For the quantum states under consideration though, these residual dynamics correspond to $t\gg \tau_{\rm cond}$, which is beyond exponential in system size. Moreover, at this point the QMI is already infinitesimal and less interesting. Nevertheless, we present some numerical results demonstrating the exponential decay for $t\gg \tau_{\rm cond}$ in Fig.~\ref{fig:condQMI_late}a. Here we choose the initial state for different $N_A$ to be the conditional state in Eq.~\eqref{state_RAt} with Bell state $\ket{\Phi}$ and $\bf z = (0,\cdots 0)$ at $t \gg \tau_{\rm cond}$. 
The exponential decay of QMI further indicates that the logarithmic decay of QMI identified in Theorem~\ref{theorem_avgMI} is {\it not} universal for {\it all} states, despite proven for Bell state and generic for various states considered above.
Counterexample can be constructed by initializing the quantum state in the state at $t\gg \tau_{\rm cond}$ in those cases, as demonstrated by the exponential decay starting from time zero in our numerical examples. Despite the exponential decay with time, we find that the lifetime versus the system size $N_A$ is still exponential, as shown in Fig.~\ref{fig:condQMI_late}b (blue dots). This is because the exponent $v$ in the exponential decay $\overline{I(R:A_t|\bfz)} \sim e^{-v t}$ is vanishing as the system size increases (red dots). Furthermore, we would like to point out that our conclusion on the equivalence between cases with and without reset in measurement-conditioned QMI still holds, as shown via the comparison in Fig.~\ref{fig:condQMI_late}c.
}

\begin{widetext}

\section{Derivation for measurement-conditioned QMI dynamics (Theorem~\ref{theorem_avgMI} and Theorem~\ref{theorem_avgMI_lessR})}
\label{app:theorem1}

In this section, we derive the dynamics of measurement-conditioned QMI. 
We begin with the lower bound to connect the measurement-conditioned QMI to purity of reduced state.
\begin{align}
    \overline{I(R:A_t|\bfz)} &\equiv \E_{\bfz} \left[S(R|{\bfz}) + S(A_t|{\bfz}) - S(RA_t|{\bfz})\right] \label{eq:avgMI_def}\\
    &= \E_{\bfz} 2S(A_t|{\bfz})  \\
    & \ge \E_{\bfz} 2S_2(A_t|{\bfz}) = -2\E_{\bfz} \log_2\left(\tr(\rho_{A_t|\bfz}^2)\right) \label{eq:third_line}\\
    & \ge -2\log_2\left(\E_{\bfz} \tr(\rho_{A_t|\bfz}^2)\right),
    \label{eq:avgMI_r2}
\end{align}
where $S(A_t|{\bfz})$ is the von Neumann entanglement entropy of the reduced state $\rho_{A_t|\bfz} = \tr_{R}(\state{\psi_{\bfz}}_{RA_t})$ and so as others. The second line comes from the fact that the conditional state $\ket{\psi_{\bfz}}_{RA_t}$ is a pure state. We utilize the monotonicity in R\'enyi entropy to obtain Ineq.~\eqref{eq:third_line}, and obtain the last inequality utilizing the concavity of logarithmic function. Utilizing the concavity property again, we can obtain the lower bound for the Haar-averaged measurement-conditioned QMI as
\be
    \E_{\rm Haar} \overline{I(R:A_t|\bfz)} \ge -2\log_2\left(\E_{\rm Haar}\E_{\bfz} \tr(\rho_{A_t|\bfz}^2)\right).
    \label{eq:avgMI_lb_sm}
\ee

\begin{figure}[t]
    \centering
    \includegraphics[width=0.5\textwidth]{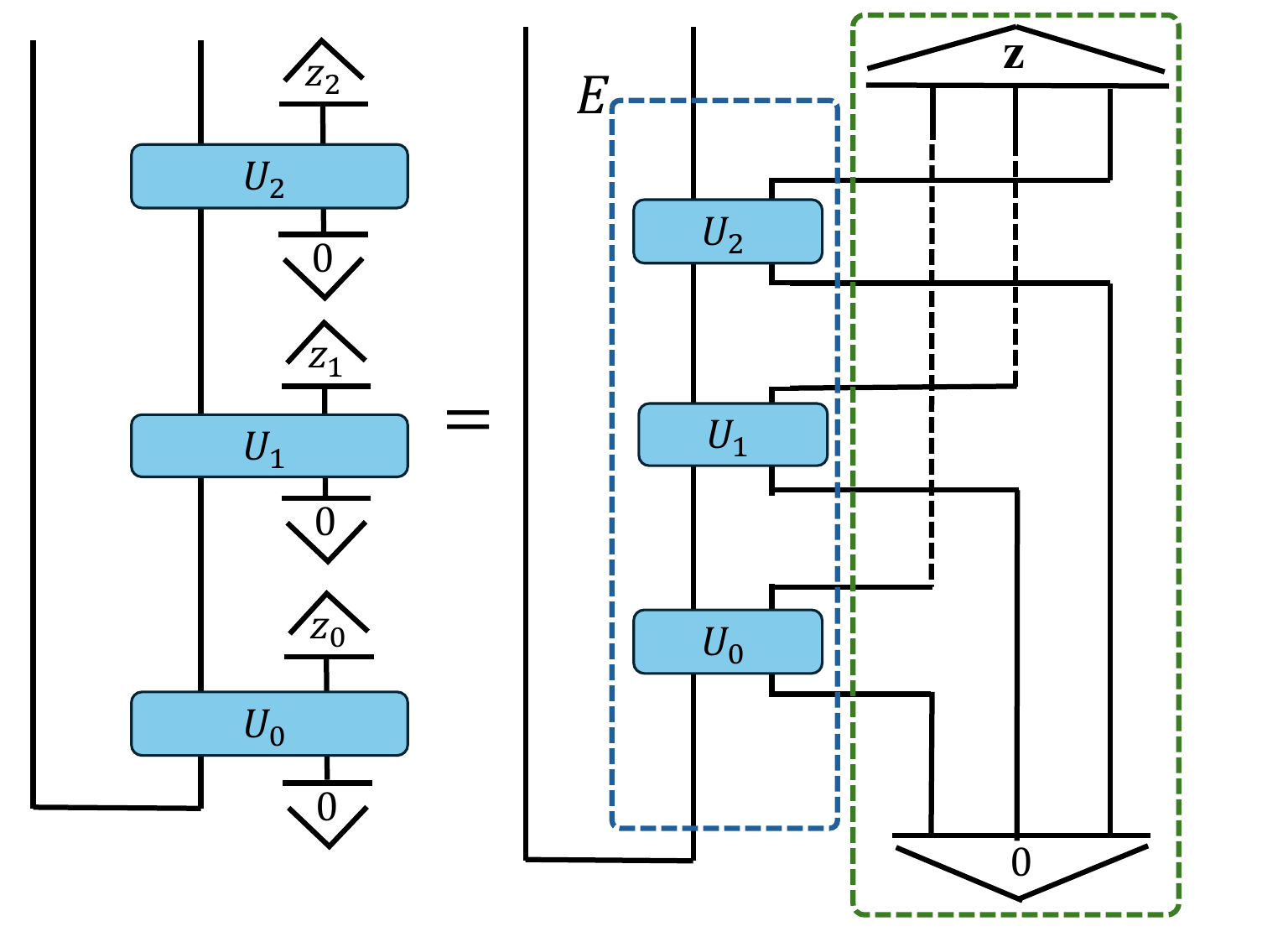}
    \caption{Equivalent relation on the quantum dynamics with mid-circuit measurements and zero state reset via expanding the bath. The linear map $E$ (blue box) is applied on the initial maximally entangled state. The dashed lines indicate that those outputs are directly connected to the measurement results $\ket{\bm z}$. Blue box shows the linear map $E$, green dashed box shows the 
    temporal input $\ket{\phi_\bfz}$. Here we show an example for $t=3$.}
    \label{fig:cond_diagram}
\end{figure}

\subsection{Proof of Theorem~\ref{theorem_avgMI} for $N_R = N_A$}

\BZ{In this subsection, we begin with the proof for Theorem~\ref{theorem_avgMI} where the initial state is a maximally entangled state thus $N_R = N_A$.} Via the bath equivalent expansion in Fig.~\ref{fig:cond_diagram}, we can first write out the linear map $E$ (blue box) as
\be
    E = \sum_{\bm a, \bm b} \prod_{k=0}^{t-1} \braket{a_{k+1} b_{2k+1}|U_k|a_k b_{2k}} \ket{a_t}_{A_t}\bra{a_0}_{A_0} \otimes_{k=0}^{t-1} \bra{b_{2k+1}}\bra{b_{2k}},
    \label{eq:E_cond_def}
\ee
where $\bm a = (a_0, \cdots, a_t)$ and $\bm b = (b_0, b_1, \cdots, b_{2t-2}, b_{2t-1})$ are vector representations of indices.
By applying the linear map $E$ on the maximally entangled state $\ket{\Phi}_{RA_0}$, we have 
\begin{align}
    E' \equiv (\mathbb{I}_R \otimes E) \ket{\Phi}_{RA_0} = \frac{1}{\sqrt{d_A}} \sum_{\bm a, \bm b} \prod_{k=0}^{t-1} \braket{a_{k+1} b_{2k+1}|U_k|a_k b_{2k}} \ket{a_0}_{R} \ket{a_t}_{A_t} \otimes_{k=0}^{t-1} \bra{b_{2k+1}}\bra{b_{2k}},
\end{align}
which is a linear map from the expanded bath including input and measurement to the output conditional state in system $R A_t$. The same formalism is also adopted in Ref.~\cite{ippoliti2022solvable, zhang2025holographic}. From Fig.~\ref{fig:cond_diagram}, the temporal input generated from the green dashed box is $\ket{\phi_\bfz} = \otimes_{k=0}^{t-1} \ket{z_t} \ket{0}$, and therefore we can obtain the output conditional state and measurement probability as
\begin{align}
    \begin{cases}
        \ket{\psi_\bfz}_{RA_t} &= E'\ket{\phi_\bfz}/\sqrt{P(\bfz)}\\
        P(\bfz) &= \braket{\phi_\bfz|E'^\dagger E' |\phi_\bfz}.
    \end{cases}
\end{align}
The corresponding density operator is $\state{\psi_\bfz}_{R A_t} = P(\bfz)^{-1} E'\state{\phi_\bfz}E'^\dagger$, and the reduced state $\rho_{A_t|\bfz}$ is
\be
    \rho_{A_t|\bfz} = P(\bfz)^{-1} \tr_R(E'\state{\phi_\bfz}E'^\dagger).
\ee
The measured-conditioned purity of $\rho_{A_t|\bfz}$ is thus
\begin{align}
    \overline{\gamma} &\equiv \E_\bfz \tr(\rho_{A_t|\bfz}^2) = \sum_{\bfz} P(\bfz) \tr(\rho_{A_t|\bfz}^2)\\
    &= \sum_{\bfz} P(\bfz) P(\bfz)^{-2} \tr(\tr_R(E'\state{\phi_\bfz}E'^\dagger)^2)\\
    &= \sum_{\bfz} P(\bfz)^{-1} \tr(E'^{\otimes 2}\state{\phi_\bfz}^{\otimes 2} E'^\dagger{}^{\otimes 2} (\mathbb{I}_R \otimes \mathbb{S}_{A_t}))\\
    &= \sum_{\bfz} \braket{\phi_\bfz|E'^\dagger E'|\phi_\bfz}^{-1} \tr(E'^{\otimes 2}\state{\phi_\bfz}^{\otimes 2} E'^\dagger{}^{\otimes 2} (\mathbb{I}_R \otimes \mathbb{S}_{A_t})),
\end{align}
where $\mathbb{S}_{A_t}$ is the swap operator over two replicas in system $A_t$. To evaluate it analytically, we introduce the pseudo measurement-conditioned purity from replica trick as
\begin{align}
    \overline{\gamma}^{(m)} &= \sum_{\bfz} \braket{\phi_\bfz|E'^\dagger E'|\phi_\bfz}^{m} \tr(E'^{\otimes 2}\state{\phi_\bfz}^{\otimes 2} E'^\dagger{}^{\otimes 2} (\mathbb{I}_R \otimes \mathbb{S}_{A_t}))\\
    &= \sum_{\bfz} \tr(E'\state{\phi_\bfz}E'^\dagger)^m \tr(E'^{\otimes 2}\state{\phi_\bfz}^{\otimes 2} E'^\dagger{}^{\otimes 2} (\mathbb{I}_R \otimes \mathbb{S}_{A_t}))\\
    &= \sum_{\bfz} \tr(E'^{\otimes m}\state{\phi_\bfz}^{\otimes m} E'^\dagger{}^{\otimes m}) \tr(E'^{\otimes 2}\state{\phi_\bfz}^{\otimes 2} E'^\dagger{}^{\otimes 2} (\mathbb{I}_R \otimes \mathbb{S}_{A_t}))\\
    &= \sum_{\bfz} \tr(E'^{\otimes (m+2)}\state{\phi_\bfz}^{\otimes (m+2)} E'^\dagger{}^{\otimes (m+2)} (\mathbb{I}_R \otimes (\mathbb{I}_m \otimes \mathbb{S}_{A_t})))\\
    &= \tr\left(E'^{\otimes (m+2)} \left(\sum_{\bfz}\state{\phi_\bfz}^{\otimes (m+2)}\right) E'^\dagger{}^{\otimes (m+2)} (\mathbb{I}_R \otimes (\mathbb{I}_m \otimes \mathbb{S}_{A_t}))\right),
    \label{eq:pseudo_purity1}
\end{align}
where in the last step $\mathbb{I}_R$ is changed to the identity operator on $m+2$ replicas and $\mathbb{I}_m \otimes \mathbb{S}_{A_t}$ is the swap operator on $A_t$ with trivial identity on first $m$ replicas. Note that
\begin{align}
    \sum_\bfz \state{\phi_\bfz}^{\otimes (m+2)} &= \sum_\bfz \otimes_{k=0}^{t-1} \state{z_k}^{\otimes (m+2)} \otimes \state{0}^{\otimes (m+2)}\\
    &= \otimes_{k=0}^{t-1} \left(\sum_{z_k} \state{z_k}^{\otimes (m+2)} \otimes \state{0}^{\otimes (m+2)}\right)\\
    &\equiv \otimes_{k=0}^{t-1} \left(D_{m+2} \otimes \state{0}^{\otimes (m+2)}\right),
\end{align}
where we define $D_{m+2} = \sum_{z} \state{z}^{\otimes (m+2)}$. Then we can evaluate $E'^{\otimes (m+2)} \left(\sum_{\bfz}\state{\phi_\bfz}^{\otimes (m+2)}\right) E'^\dagger{}^{\otimes (m+2)}$ as
\begin{align}
    &E'^{\otimes (m+2)} \left(\sum_{\bfz}\state{\phi_\bfz}^{\otimes (m+2)}\right) E'^\dagger{}^{\otimes (m+2)} \nonumber\\
    &= d_A^{-(m+2)} \sum_{\substack{\bm \alpha, \bm \beta,\\ \bm \alpha', \bm \beta'}} \prod_{k=0}^{t-1} \braket{\alpha_{k+1} \beta_{2k+1}|U_k^{\otimes (m+2)}|\alpha_k \beta_{2k}} \braket{\alpha_k' \beta_{2k}'|U_k^\dagger{}^{\otimes (m+2)}|\alpha_{k+1}' \beta_{2k+1}'} \ketbra{\alpha_0}{\alpha_0'}_R \ketbra{\alpha_t}{\alpha_t'}_{A_t} \nonumber\\
    &\qquad \qquad \qquad \times \otimes_{k=0}^{t-1} \braket{\beta_{2k+1}|D_{m+2}|\beta_{2k+1}'} \braket{\beta_{2k}|0^{m+2}}\braket{0^{m+2}|\beta_{2k}'},
\end{align}
where $\ket{\alpha_k} = \otimes_{j=1}^{m+2}\ket{a_k^{(j)}}, \ket{\beta_k} = \otimes_{j=1}^{m+2}\ket{b_k^{(j)}}, \ket{\alpha_k'} = \otimes_{j=1}^{m+2}\ket{a_k'^{(j)}}, \ket{\beta_k'} = \otimes_{j=1}^{m+2}\ket{b_k'^{(j)}}$ are the simplified notation for $m+2$ replicas of basis. $\bm \alpha = (\alpha_0, \cdots, \alpha_{t-1})$, $\bm \beta = (\beta_0, \beta_1, \cdots, \beta_{2k-2}, \beta_{2k-1})$ are the vector representations and so as the others $\bm \alpha', \bm \beta'$. Utilizing the facts 
\begin{align}
    &\braket{\beta_{2k+1}|D_{m+2}|\beta_{2k+1}'} = \sum_{z_k} \braket{\beta_{2k+1}|z_k^{\otimes (m+2)}}\braket{z_k^{\otimes (m+2)}|\beta_{2k+1}'} = \sum_{z_k} \delta_{\beta_{2k+1}, z_k} \delta_{\beta_{2k+1}', z_k},\\
    &\braket{\beta_{2k}|0^{m+2}}\braket{0^{m+2}|\beta_{2k}'}  = \delta_{\beta_{2k}, 0} \delta_{\beta_{2k}', 0},
\end{align}
where $\delta_{\beta, z} = \prod_{j=1}^{m+2} \delta_{b^{(j)}, z}$ is a simplified notation of Kronecker deltas on each replicas. Then we have
\begin{align}
    &E'^{\otimes (m+2)} \left(\sum_{\bfz}\state{\phi_\bfz}^{\otimes (m+2)}\right) E'^\dagger{}^{\otimes (m+2)} \nonumber\\
    &= d_A^{-(m+2)} \sum_{\bm \alpha, \bm \alpha'} \sum_{\bfz} \prod_{k=0}^{t-1} \braket{\alpha_{k+1} z_k^{\otimes (m+2)}|U_k^{\otimes (m+2)}|\alpha_k 0^{\otimes (m+2)}} \braket{\alpha_k' 0^{\otimes (m+2)}|U_k^\dagger{}^{\otimes (m+2)}|\alpha_{k+1}' z_k^{\otimes (m+2)}} \ketbra{\alpha_0}{\alpha_0'}_R \ketbra{\alpha_t}{\alpha_t'}_{A_t} \\
    &= d_A^{-(m+2)} \sum_{\bm \alpha, \bm \alpha'} \sum_{\bfz} \prod_{k=0}^{t-1} \tr\left(U_k^{\otimes (m+2)}\left(\ketbra{\alpha_k}{\alpha_k'}\otimes \state{0}^{\otimes (m+2)}\right)U_k^\dagger{}^{\otimes (m+2)} \left(\ketbra{\alpha_{k+1}'}{\alpha_{k+1}}\otimes \state{z_k}^{\otimes (m+2)}\right)\right) \ketbra{\alpha_0}{\alpha_0'}_R \ketbra{\alpha_t}{\alpha_t'}_{A_t}.
\end{align}
The pseudo measurement-conditioned purity (Eq.~\eqref{eq:pseudo_purity1}) then becomes
\begin{align}
    &\overline{\gamma}^{(m)} = \tr\left(E'^{\otimes (m+2)} \left(\sum_{\bfz}\state{\phi_\bfz}^{\otimes (m+2)}\right) E'^\dagger{}^{\otimes (m+2)} (\mathbb{I}_R \otimes (\mathbb{I}_m \otimes \mathbb{S}_{A_t}))\right) \\
    &= d_A^{-(m+2)} \sum_{\bm \alpha, \bm \alpha'} \sum_{\bfz} \prod_{k=0}^{t-1} \tr\left(U_k^{\otimes (m+2)}\left(\ketbra{\alpha_k}{\alpha_k'}\otimes \state{0}^{\otimes (m+2)}\right)U_k^\dagger{}^{\otimes (m+2)} \left(\ketbra{\alpha_{k+1}'}{\alpha_{k+1}}\otimes \state{z_k}^{\otimes (m+2)}\right)\right) \nonumber\\
    &\qquad \qquad \qquad \times \braket{\alpha_0'|\alpha_0}_R \braket{\alpha_t'|\mathbb{I}_m \otimes \mathbb{S}_{A_t}|\alpha_t}_{A_t}\\
    &= d_A^{-(m+2)} \sum_{\bm \alpha, \bm \alpha'} \prod_{k=0}^{t-1} \tr\left(U_k^{\otimes (m+2)}\left(\ketbra{\alpha_k}{\alpha_k'}\otimes \state{0}^{\otimes (m+2)}\right)U_k^\dagger{}^{\otimes (m+2)} \left(\ketbra{\alpha_{k+1}'}{\alpha_{k+1}}\otimes D_{m+2}\right)\right) \nonumber\\
    & \qquad \qquad \qquad \times \prod_{j=1}^{m+2}\delta_{a_0^{(j)}, a_0'^{(j)}} \prod_{j=1}^{m}\delta_{a_t^{(j)}, a_t'^{(j)}} 
    \delta_{a_t'^{(m+1)}, a_t^{(m+2)}} \delta_{a_t'^{(m+2)}, a_t^{(m+1)}}.
    \label{eq:pseudo_purity2}
\end{align}

Next, we evaluate Haar ensemble average on the pseudo measurement-conditioned purity $\overline{\gamma}^{(m)}$. For Haar unitary twirling, it is known that
\begin{align}
    &\E_{\rm Haar}\left[U^{\otimes (m+2)} \otimes U^*{}^{\otimes (m+2)}\right] = \sum_{\sigma, \pi \in S_{m+2}} {\rm Wg}(\sigma^{-1}\pi, m+2) |\hat{\sigma})(\hat{\pi}| \\
    &= {\rm Wg}(e, m+2)\sum_{\pi} |\hat{\pi})(\hat{\pi}| + \sum_{\sigma \neq \pi} {\rm Wg}(\sigma^{-1}\pi, m+2) |\hat{\sigma})(\hat{\pi}|,
\end{align}
where $S_{m+2}$ is the permutation group of $m+2$ replicas, and $\hat{\pi}$ is the operator representation of permutation $\pi$. ${\rm Wg}(\sigma^{-1}\pi, m+2) = (d^{|\sigma^{-1}\pi|})^{-1}$ is the Weingarten coefficient with $|\sigma^{-1}\pi|$ to be the number of cycles of permutation $\sigma^{-1}\pi$. Here $e$ is the identity.

\begin{figure}[t]
    \centering
    \includegraphics[width=0.2\textwidth]{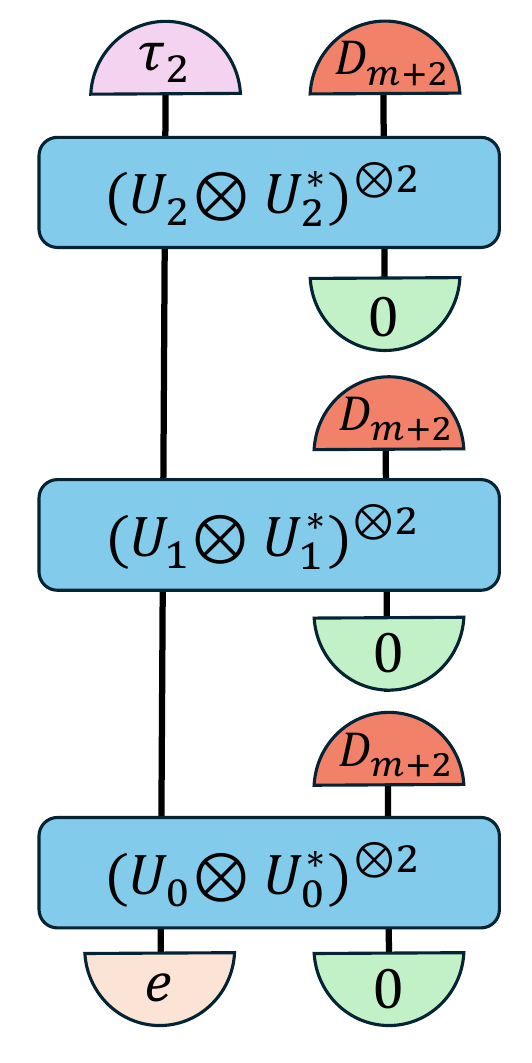}
    \caption{Tensor network representation of Haar-averaged  pseudo measurement-conditioned purity of $\rho_{A_t|\bfz}$ with time steps $t=3$. Here $\tau_2 = \mathbb{I}_m \otimes \mathbb{S}$ is the swap operator on the last two replicas.}
    \label{fig:cond_purity_diagram}
\end{figure}

We start from the unitary twirling on the last step boundary condition,
\begin{align}
    & \E_{\rm Haar}\left[U_{t-1}^{\otimes (m+2)} \otimes U_{t-1}^*{}^{\otimes (m+2)}\right] |\tau_2)_A |D_{m+2})_B \nonumber\\
    &= {\rm Wg}(e, m+2)\sum_{\pi} |\hat{\pi})_{AB}(\hat{\pi}|\tau_2)_A (\hat{\pi}|D_{m+2})_B + \sum_{\sigma \neq \pi} {\rm Wg}(\sigma^{-1}\pi, m+2) |\hat{\sigma})_{AB} (\hat{\pi}|\tau_2)_A (\hat{\pi}|D_{m+2})_B.
    \label{eq:twirling_expansion}
\end{align}
In the following, we evaluate the two terms above separately. For the first term, note that $(\hat{\pi}|D_{m+2})_B = \tr(\hat{\pi}^\dagger \sum_z \state{z}^{\otimes (m+2)}) = \sum_z \braket{z^{\otimes (m+2)}|\hat{\pi}^\dagger|z^{\otimes (m+2)}} = d_B$, we then have
\be
    {\rm Wg}(e, m+2)\sum_{\pi} |\hat{\pi})_{AB}(\hat{\pi}|\tau_2)_A (\hat{\pi}|D_{m+2})_B =  {\rm Wg}(e, m+2) d_B \sum_{\pi} d_A^{|\pi^\dagger \tau_2|} |\hat{\pi})_{AB},
\ee
and further connecting $|D_{m+2})(0|_B$ on the bath system, we have
\begin{align}
    & |D_{m+2})(0|_B {\rm Wg}(e, m+2) \sum_{\pi} |\hat{\pi})_{AB} (\hat{\pi}|\tau_2)_A (\hat{\pi}|D_{m+2})_B \nonumber\\
    & = {\rm Wg}(e, m+2) d_B \sum_{\pi} d_A^{|\pi^\dagger \tau_2|} |\hat{\pi})_A |D_{m+2})_B \\
    & \simeq {\rm Wg}(e, m+2) d_B \left( d_A^{m+2}|\tau_2)_A + d_A^{m+1} |e)_A \right) |D_{m+2})_B,
\end{align}
where we keep the leading order terms $\pi = \tau_2, \mathbb{I}$ in the last step given large system size $d_A \gg 1$. Similarly, for the second term in Eq.~\eqref{eq:twirling_expansion}, we have
\begin{align}
    & |D_{m+2})(0|_B  \sum_{\sigma \neq \pi} {\rm Wg}(\sigma^{-1}\pi, m+2) |\hat{\sigma})_{AB} (\hat{\pi}|\tau_2)_A (\hat{\pi}|D_{m+2})_B \nonumber\\
    &\simeq d_B \left[{\rm Wg}(e\tau_2, m+2) d_A^{m+2} |e)_A |D_{m+2})_B + {\rm Wg}(\tau_2^{-1}e, m+2) d_A^{m+1} |\tau_2)_A |D_{m+2})_B \right]\\
    &= {\rm Wg}(e\tau_2, m+2) d_A^{m+1} d_B \left(d_A |e)_A + |\tau_2)_A\right) |D_{m+2})_B,
\end{align}
where in the second line we still focus on the leading order $(\sigma, \pi) = (\tau_2, e), (e, \tau_2)$.
Combining these two terms together, we have
\begin{align}
    &|D_{m+2})(0|_B \E_{\rm Haar}\left[U_{t-1}^{\otimes (m+2)} \otimes U_{t-1}^*{}^{\otimes (m+2)}\right] |\tau_2)_A |D_{m+2})_B \nonumber\\
    &\simeq {\rm Wg}(e, m+2) d_A^{m+1} d_B \left( d_A|\tau_2)_A + |e)_A \right) |D_{m+2})_B + {\rm Wg}(e\tau_2, m+2) d_A^{m+1} d_B \left(d_A |e)_A + |\tau_2)_A\right) |D_{m+2})_B\\
    &\simeq \frac{d_A^{m+1} d_B}{d_A^{m+2} d_B^{m+2}} \left( d_A|\tau_2)_A + |e)_A \right) |D_{m+2})_B - \frac{d_A^{m+1} d_B}{d_A^{m+3} d_B^{m+3}} \left(d_A |e)_A + |\tau_2)_A\right) |D_{m+2})_B\\
    &= d_B^{-(m+1)}\left(1 - d_A^{-2}d_B^{-1}\right)|\tau_2)_A |D_{m+2})_B + d_A^{-1}d_B^{-(m+1)}\left(1 - d_B^{-1}\right)|e)_A |D_{m+2})_B,
    \label{eq:twirling_taud}
\end{align}
where in the third line we approximate it by ${\rm Wg}(e, m+2) \simeq d_A^{-(m+2)}d_B^{-(m+2)}$ and ${\rm Wg}(e\tau_2, m+2) \simeq -d_A^{-(m+3)}d_B^{-(m+3)}$ given $d_A \gg 1$. Let's move to the next twirling of $U_{t-2}$ and bath operation $|D_{m+2})(0|_B$. For the boundary condition $|e)_A |D_{m+2})_B$, we have
\begin{align}
    & |D_{m+2})(0|_B \E_{\rm Haar}\left[U_{t-2}^{\otimes (m+2)} \otimes U_{t-2}^*{}^{\otimes (m+2)}\right] |e)_A |D_{m+2})_B \nonumber \\
    &= |D_{m+2})(0|_B \sum_{\sigma, \pi} {\rm Wg}(\sigma^{-1}\pi, m+2) |\hat{\sigma})_{AB} (\hat{\pi}|e)_A (\hat{\pi}|D_{m+2})_B\\
    &= d_B \sum_{\sigma, \pi} {\rm Wg}(\sigma^{-1}\pi, m+2) d_A^{|\pi^\dagger|} |\hat{\sigma})_{A} |D_{m+2})_{B}\\
    &= d_B \left[{\rm Wg}(e, m+2)\sum_{\pi} d_A^{|\pi^\dagger|} |\hat{\pi})_A + \sum_{\sigma \neq \pi} {\rm Wg}(\sigma^{-1}\pi, m+2) d_A^{|\pi^\dagger|} |\hat{\sigma})_{A} \right] |D_{m+2})_{B}\\
    &\simeq d_B \left[{\rm Wg}(e, m+2)\left(d_A^{m+2} |e)_A + d_A^{m+1} |\tau_2)_A\right) + {\rm Wg}(\tau_2^{-1}e, m+2) d_A^{m+2} |\tau_2)_{A} + {\rm Wg}(e\tau_2, m+2) d_A^{m+1} |e)_{A}\right] |D_{m+2})_{B} \\
    &= d_A^{m+1} d_B \left[\left({\rm Wg}(e, m+2) + {\rm Wg}(\tau_2^{-1}e, m+2) d_A\right)|\tau_2)_A +\left({\rm Wg}(e, m+2) d_A + {\rm Wg}(e\tau_2, m+2)\right)|e)_{A}\right] |D_{m+2})_{B} \\
    &\simeq d_A^{m+1} d_B \left[\left(d_A^{-(m+2)}d_B^{-(m+2)} - d_A^{-(m+3)}d_B^{-(m+3)} d_A\right)|\tau_2)_A \left(d_A^{-(m+2)}d_B^{-(m+2)} d_A - d_A^{-(m+3)}d_B^{-(m+3)}\right)|e)_{A}\right] |D_{m+2})_{B} \\
    &= d_A^{-1} d_B^{-(m+1)}\left(1 - d_B^{-1}\right)|\tau_2)_A |D_{m+2})_{B} + d_B^{-(m+1)} \left(1 - d_A^{-2}d_B^{-1}\right)|e)_{A}|D_{m+2})_{B},
    \label{eq:twirling_ed}
\end{align}
where we make the same approximations as we utilized above. To summarize, we have the following linear transformation on boundary conditions in every step from Eqs.~\eqref{eq:twirling_taue},~\eqref{eq:twirling_ed}
\begin{align}
    \begin{cases}
        |\tau_2)_A |D_{n+2})_B \to q_\tau |\tau_2)_A |D_{m+2})_B + q_e |e)_A |D_{m+2})_B \\
        |e)_A |D_{n+2})_B \to q_e |\tau_2)_A |D_{m+2})_B + q_\tau |e)_A |D_{m+2})_B
    \end{cases},
\end{align}
where $q_\tau = d_B^{-(m+1)}\left(1 - d_A^{-2}d_B^{-1}\right)$ and $q_e = d_A^{-1}d_B^{-(m+1)}\left(1 - d_B^{-1}\right)$ are two coefficients. We can thus write the linear transformation in the form of a two-by-two matrix
\be
    Q = 
    \begin{pmatrix}
        q_\tau & q_e \\
        q_e & q_\tau
    \end{pmatrix},
    \label{eq:Q_def}
\ee
with bases $|\tau_2)_A |D_{m+2})_B \to (1, 0)^T, |e)_A |D_{m+2})_B \to (0, 1)^T$. Via diagonalization, we can write out $Q^t$ explicitly as
\be
    Q^t = \frac{1}{2}
    \begin{pmatrix}
        \nu_+^t + \nu_-^t & \nu_+^t - \nu_-^t \\
        \nu_+^t - \nu_-^t & \nu_+^t + \nu_-^t
    \end{pmatrix},
\ee
where $\nu_{\pm} = q_{\tau} \pm q_e$ are eigenvalues of $Q$. Considering the initial boundary condition at step $t$ in Fig.~\ref{fig:cond_purity_diagram}, we obtain the boundary condition at step $0$ as
\begin{align}
    Q^t (1, 0)^T \to \frac{1}{2}\left[(q_\tau + q_e)^t + (q_\tau - q_e)^t\right] |\tau_2)_A |D_{m+2})_B + \frac{1}{2}\left[(q_\tau + q_e)^t - (q_\tau - q_e)^t\right] |e)_A |D_{m+2})_B,
    \label{eq:cond_boundaryA_evo}
\end{align}
and thus we have the dynamical solution for pseudo measurement-conditioned purity
\begin{align}
    \E_{\rm Haar}\overline{\gamma}^{(m)} &\simeq d_A^{-(m+2)} (e|_A (0|_B \left(\frac{1}{2}\left[(q_\tau + q_e)^t + (q_\tau - q_e)^t\right] |\tau_2)_A |D_{m+2})_B + \frac{1}{2}\left[(q_\tau + q_e)^t - (q_\tau - q_e)^t\right] |e)_A |D_{m+2})_B\right)\\
    &= \frac{1}{2} d_A^{-(m+2)}  \left(\left[(q_\tau + q_e)^t + (q_\tau - q_e)^t\right] d_A^{m+1}  + \left[(q_\tau + q_e)^t - (q_\tau - q_e)^t\right] d_A^{m+2} \right)\\
    &= \frac{1}{2d_A} \left[(1+d_A) (q_\tau + q_e)^t + (1-d_A) (q_\tau - q_e)^t\right]\\
    &= \frac{1}{2d_A} \left[(1+d_A) \left(d_A^{-2}d_B^{-(m+2)}(d_A + 1)(d_A d_B - 1)\right)^t + (1-d_A) \left(d_A^{-2}d_B^{-(m+2)}(d_A - 1)(d_A d_B + 1)\right)^t\right] \\
    &= \frac{1}{2d_A} d_A^{-2t}d_B^{-(m+2) t} \left[(d_A + 1)^{t+1}(d_A d_B - 1)^t - (d_A-1)^{t+1} (d_A d_B + 1)^t\right],
\end{align}
where we utilize the definition of $q_\tau, q_e$ defined above in the second to last line.
Via taking the limit $m \to -1$, we solve the Haar-averaged dynamical solution for the measurement-conditioned purity of $\rho_{A_t|\bfz}$ as
\begin{align}
    \E_{\rm Haar}\E_{\bfz} \tr(\rho_{A_t|\bfz}^2) = \E_{\rm Haar} \overline{\gamma} &= \lim_{m \to -1} \E_{\rm Haar}\overline{\gamma}^{(m)} \\
    &\simeq \frac{1}{2} d_A^{-(2t+1)}d_B^{-t} \left[(d_A + 1)^{t+1}(d_A d_B - 1)^t - (d_A-1)^{t+1} (d_A d_B + 1)^t\right].
    \label{eq:cond_purity_sol}
\end{align}
Finally, we have the asymptotic lower bound for Haar-averaged measurement-conditioned QMI from Eq.~\eqref{eq:avgMI_lb}
\begin{align}
    &\E_{\rm Haar} \overline{I(R:A_t|\bfz)} \ge -2\log_2\left(\E_{\rm Haar}\E_{\bfz} \tr(\rho_{A_t|\bfz}^2)\right) \\
    &\gtrsim  -2\log_2\left( \frac{1}{2} d_A^{-(2t+1)}d_B^{-t} \left[(d_A + 1)^{t+1}(d_A d_B - 1)^t - (d_A-1)^{t+1} (d_A d_B + 1)^t\right]\right)\\
    &= 2 + 2(2t+1)N_A + 2 t N_B - 2\log_2\left[(d_A + 1)^{t+1}(d_A d_B - 1)^t - (d_A-1)^{t+1} (d_A d_B + 1)^t\right]\\
    &= 2 + 2(2t+1)N_A + 2 t N_B - 2\log_2\left[(d_A d_B + 1)^t (d_A + 1)^{t+1} \right] - 2\log_2\left[\left(\frac{d_A d_B - 1}{d_A d_B + 1}\right)^t - \left(\frac{d_A - 1}{d_A + 1}\right)^{t+1}\right],
    \label{eq:avgMI_sol_sm}
\end{align}
which is the solution of Eq.~\eqref{eq:avgMI_lb} of Theorem.~\ref{theorem_avgMI} in the main text. 

At the end of this section, we perform an asymptotic analysis on the scaling of above result in the limit of large system size $d_A \gg 1$.
\begin{align}
    &\E_{\rm Haar} \overline{I(R:A_t|\bfz)} \nonumber\\
    &= 2 + 2(2t+1)N_A + 2 t N_B - 2t \log_2(d_A d_B + 1) - 2(t+1)\log_2(d_A + 1) - 2\log_2\left[\left(1-\frac{2}{d_A d_B +1}\right)^t - \left(1-\frac{2}{d_A + 1}\right)^{t+1}\right] \\
    &\simeq 2 + 2(2t+1)N_A + 2 t N_B - 2t (N_A+N_B) - 2(t+1)N_A - 2\log_2\left[1-\frac{2t}{d_A d_B +1} - \left(1-\frac{2(t+1)}{d_A + 1}\right)\right]\\
    &= 2 - 2\log_2\left(\frac{2(t+1)}{d_A + 1} -\frac{2t}{d_A d_B +1}\right)\\
    &\simeq 2 - 2\log_2\left(\frac{2(t+1)}{d_A} -\frac{2t}{d_A d_B}\right)\\
    & = 2 - 2\log_2\left(\frac{2}{d_A}(t+1 -t/d_B)\right)\\
    &= 2N_A - \log_2\left((1-1/d_B)t + 1\right).
    \label{eq:cond_lb_asymp_sm}
\end{align}

\subsection{Proof of Theorem~\ref{theorem_avgMI_lessR} for $N_R<N_A$}

In this subsection, we provide the proof of Theorem~\ref{theorem_avgMI_lessR} where $N_R < N_A$ and thus there are only $N_R$ Bell pairs shared between reference and the system. The derivation of the measurement-conditioned QMI is similar to the Theorem~\ref{theorem_avgMI} presented above.

The initial state on reference $R$ and system $A$ is
\be
    \ket{\Phi}_{RA_0} = \frac{1}{\sqrt{d_R}}\sum_{i=0}^{d_R-1} \ket{i}_R \ket{i}_{A_1} \ket{0}_{A_2},
\ee
where $|A_1| = N_R$ for qubits in the system entangled with the reference, while $A_2$ stands for the idling qubits. The measurement-conditioned state is
\be
    \rho_{A|\bfz} = \tr_{RB}\left(U \ketbra{\Phi}{\Phi}_{RA_0}\ketbra{0}{0}_B U^\dagger \ketbra{\bfz}{\bfz}_B\right)/p_\bfz \equiv \tilde{\rho}_{RA|\bfz}/p_\bfz,
\ee
where $p_\bfz = \tr(\rho_{A|\bfz})$ is the corresponding probability. Here $U$ represents the whole circuit.
The measurement-conditioned purity is 
\be
    \overline{\gamma} = \sum_\bfz p_\bfz \tr(\rho_{A|\bfz}^2) =\sum_\bfz p_\bfz^{-1} \tr(\tilde{\rho}_{A|\bfz}^2).
\ee
We again introduce the pseudo purity as
\begin{align}
    \overline{\gamma}^{(m)} &\equiv \sum_\bfz p_\bfz^{m} \tr(\tilde{\rho}_{A|\bfz}^2) \\
    &= \sum_\bfz \tr(U^{\otimes m} \ketbra{\Phi}{\Phi}_{RA}^{\otimes m} \ketbra{0}{0}_B^{\otimes m} U^\dagger{}^{\otimes m} \bI_R^{\otimes m} \bI_A^{\otimes m} \ketbra{\bfz}{\bfz}_B^{\otimes m}) \nonumber\\
    &\qquad \quad \times \tr(U^{\otimes 2} \ketbra{\Phi}{\Phi}_{RA}^{\otimes 2} \ketbra{0}{0}_B^{\otimes 2} U^\dagger{}^{\otimes 2} \bI_R^{\otimes 2} \tau_A \ketbra{\bfz}{\bfz}_B^{\otimes 2})\\
    &= \sum_\bfz \tr(U^{\otimes (m+2)} \ketbra{\Phi}{\Phi}_{RA}^{\otimes (m+2)} \ketbra{0}{0}_B^{\otimes (m+2)} U^\dagger{}^{\otimes (m+2)} \bI_R^{\otimes (m+2)} \bI_A^{\otimes m}\tau_A \ketbra{\bfz}{\bfz}_B^{\otimes (m+2)})\\
    &= d_R^{-(m+2)} \tr(U^{\otimes (m+2)} \bI_{A_1}^{\otimes (m+2)} \ketbra{0}{0}_{A_2}^{\otimes (m+2)} \ketbra{0}{0}_B^{\otimes (m+2)} U^\dagger{}^{\otimes (m+2)} \bI_A^{\otimes m}\tau_A \sum_\bfz \ketbra{\bfz}{\bfz}_B^{\otimes (m+2)}).
\end{align}
Compared to the expression in Eq.~\eqref{eq:pseudo_purity2}, we notice that the only difference is the initial boundary condition as $\bI_A^{\otimes (m+2)} \to \bI_{A_1}^{\otimes (m+2)} \otimes \ketbra{0}{0}_{A_2}^{\otimes m+2}$, while the twirling evolution and final boundary condition remain the same as expected. Therefore, we can still utilize the result in Eq.~\eqref{eq:cond_boundaryA_evo}, and the dynamical solution for the pseudo measurement-conditioned purity becomes
\begin{align}
    \E_{\rm Haar}\overline{\gamma}^{(m)} &\simeq d_R^{-(m+2)} (e|_{A_1}(0|_{A_2}(0|_B \left(\frac{1}{2}\left[(q_\tau+q_e)^t + (q_\tau-q_e)^t\right] |\tau_2)_{A}|D_{m+2})_B + \frac{1}{2}\left[(q_\tau+q_e)^t - (q_\tau-q_e)^t\right] |e)_A|D_{m+2})_B\right) \\
    &= \frac{1}{2}d_R^{-(m+2)} \left(\left[(q_\tau+q_e)^t + (q_\tau-q_e)^t\right] d_R^{m+1} + \left[(q_\tau+q_e)^t - (q_\tau-q_e)^t\right] d_R^{m+2}\right)\\
    &= \frac{1}{2d_R} \left[(q_\tau+q_e)^t (1+d_R) + (q_\tau-q_e)^t (1-d_R)\right] \\
    &= \frac{1}{2d_R} \left[(1+d_R) \left(d_A^{-2}d_B^{-(m+2)}(d_A + 1)(d_A d_B - 1)\right)^t + (1-d_R) \left(d_A^{-2}d_B^{-(m+2)}(d_A - 1)(d_A d_B + 1)\right)^t\right] \\
    &= \frac{1}{2d_R} d_A^{-2t}d_B^{-(m+2) t} \left[(d_A + 1)^t(d_A d_B - 1)^t(1+d_R) - (d_A-1)^{t} (d_A d_B + 1)^t (d_R-1)\right].  
\end{align}
Take the replica limit of $m \to -1$, we have
\be
    \E_{\rm Haar} \overline{\gamma} = \frac{1}{2d_R} d_A^{-2t}d_B^{-t} \left[(d_A + 1)^t(d_A d_B - 1)^t(1+d_R) - (d_A-1)^{t} (d_A d_B + 1)^t (d_R-1)\right],
\ee
and the asymptotic lower bound for the Haar-averaged measurement-conditioned QMI from Eq.~\eqref{eq:avgMI_lb}
\begin{align}
    &\E_{\rm Haar} \overline{I(R:A_t|\bfz)} \ge -2\log_2\left(\E_{\rm Haar}\overline{\gamma}\right) \\
    &\gtrsim  -2\log_2\left( \frac{1}{2d_R} d_A^{-2t}d_B^{-t} \left[(d_A + 1)^t(d_A d_B - 1)^t(1+d_R) - (d_A-1)^{t} (d_A d_B + 1)^t (d_R-1)\right]\right)\\
    &= 2 + 2N_R + 4t N_A + 2 t N_B - 2\log_2\left[(d_A + 1)^{t}(d_A d_B - 1)^t (1+d_R) - (d_A-1)^{t} (d_A d_B + 1)^t (d_R-1)\right]\\
    &= 2 + 2N_R + 4tN_A + 2 t N_B - 2\log_2\left[(d_A+1)^t (d_A d_B+1)^t (d_R+1)\right] - 2\log_2\left[\left(\frac{d_A d_B - 1}{d_A d_B + 1}\right)^t - \left(\frac{d_A - 1}{d_A + 1}\right)^t\left(\frac{d_R-1}{d_R + 1}\right)\right].
    \label{eq:avgMI_sol_lessR_sm}
\end{align}

To simplify the results, we perform an asymptotic anslysis in the limit of $d_A, d_R \gg 1$, then we have
\begin{align}
    &\E_{\rm Haar} \overline{I(R:A_t|\bfz)} \nonumber\\
    &= 2 + 2N_R + 4tN_A + 2 t N_B - 2\log_2\left[(d_A+1)^t (d_A d_B+1)^t (d_R+1)\right] \nonumber\\
    &\quad - 2\log_2\left[\left(1-\frac{2}{d_A d_B+1}\right)^t - \left(1-\frac{2}{d_A + 1}\right)^t\left(1-\frac{2}{d_R + 1}\right)\right]\\
    &\simeq 2 + 2N_R + 4tN_A + 2tN_B-2tN_A - 2t(N_A + N_B) - 2N_R - 2\log_2\left[\left(1-\frac{2t}{d_A d_B+1}\right) - \left(1-\frac{2t}{d_A + 1}\right)\left(1-\frac{2}{d_R + 1}\right)\right]\\
    &= 2 - 2\log_2\left[\frac{2}{d_R+1} + \frac{2t}{d_A}\left(1-\frac{2}{d_R+1} - \frac{1}{d_B}\right)\right]\\
    &= 2\log_2(d_R + 1) - 2\log_2\left[1+\frac{d_R t}{d_A}\left(1- \frac{d_R + 1}{d_R d_B} + \frac{1}{2d_R}\right)\right]\\
    &\simeq 2N_R - 2\log_2\left[1+\frac{d_R t}{d_A}\left(1- \frac{d_R + 1}{d_R d_B} + \frac{1}{2d_R}\right)\right].
\end{align}

\section{Derivation for measurement-unconditioned QMI (Theorems~\ref{MI_traceout_theorem},~\ref{MI_traceout_lessR_theorem},~\ref{MI_traceout_woReset_theorem} and \ref{MI_traceout_id_theorem})}
\label{app:thereom23}

In this section, we provide the proof on the dynamics for measurement-unconditioned QMI with pure and mixed state reset strategy. As we point out in the main text, for computation convenience, we focus on the R\'enyi-extended QMI defined by
\be
    I_2(R:A_t) = S_2(\rho_R) + S_2(\rho_{A_t}) - S_2(\rho_{R A_t}),
    \label{eq:renyi_qmi}
\ee
where $S_2(\cdot)$ is the R\'enyi-2 entropy.

In the following, Theorem~\ref{MI_traceout_theorem} is proven in Section~\ref{app:pure_reset}, Theorem~\ref{MI_traceout_lessR_theorem} is proven in Section~\ref{app:pure_reset_lessR},
Theorem~\ref{MI_traceout_woReset_theorem} is proven in Section~\ref{app:MI_traceout_woReset_theorem}
and Theorem~\ref{MI_traceout_id_theorem} is proven in Section~\ref{app:mixed_reset}.

\subsection{Pure-state reset with $N_R = N_A$ (Theorem~\ref{MI_traceout_theorem})}
\label{app:pure_reset}
\begin{figure}[t]
    \centering
    \includegraphics[width=0.5\textwidth]{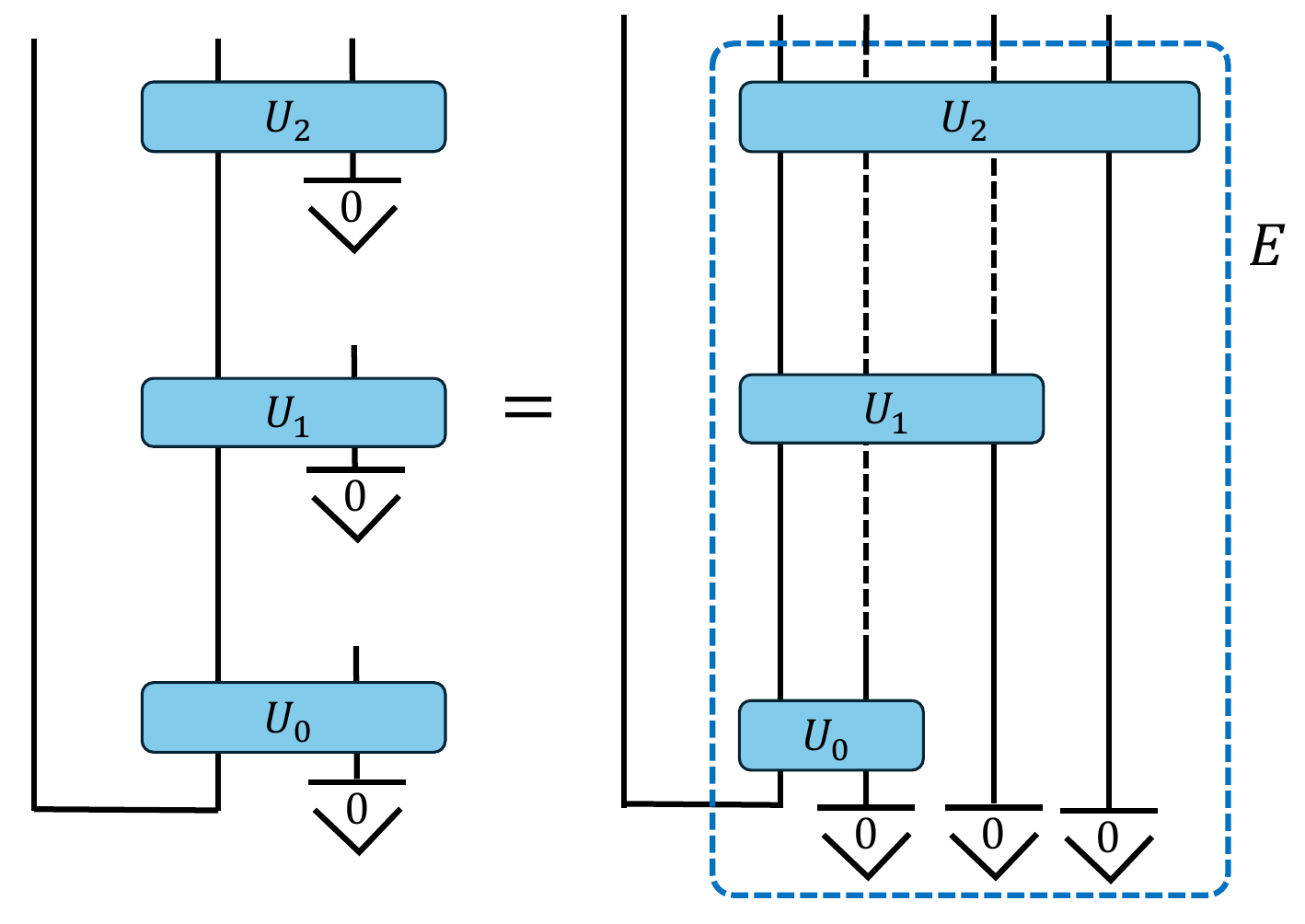}
    \caption{Equivalent relation on the quantum dynamics with mid-circuit measurements and $\ket
    {0}_B$ reset via expanding the bath. The linear map $E$ (blue box) is applied on the initial maximally entangled state. The dashed lines indicate that those outputs do not interact with the unitaries to be passed through. Here we show an example for $t=3$.}
    \label{fig:uncond_diagram}
\end{figure}

We begin with the setup of $N_R = N_A$. For pure-state reset strategy, without loosing generality, we simply assume in every step, the bath system is reset to the trivial product state $\ket{0}$.
From the equivalence relation in Fig.~\ref{fig:uncond_diagram}, we can write out the linear map $E$ as
\be
    E = \sum_{\bm a, \bm b} \prod_{k=0}^{t-1} \braket{a_{k+1} b_{k+1}|U_k|a_{k} 0} \ket{a_t}_{A_t} \otimes_{k=0}^{t-1} \ket{b_{k+1}}_{B_{k+1}} \bra{a_0}_{A_0},
    \label{eq:E_def_uncond}
\ee
where $\bm a = (a_0, \cdots, a_t)$ and $\bm b = (b_1, \cdots, b_t)$ are vector representation of the corresponding indices on data and bath system. Given the initial maximally-entangled state $\ket{\Phi}_{RA_0} = \frac{1}{\sqrt{d_A}} \sum_j \ket{j}_R \ket{j}_{A_0}$, then we have the output state on $R A_t \bm B$ to be
\begin{align}
    &\ket{\psi}_{R A_t \bm B} = (\mathbb{I}_R \otimes E) \ket{\Phi}_{RA_0}\\
    &= \sum_{\bm a, \bm b} \prod_{k=0}^{t-1} \braket{a_{k+1} b_{k+1}|U_k|a_{k} 0} \ket{a_t}_{A_t} \otimes_{k=0}^{t-1} \ket{b_{k+1}}_{B_{k+1}} \left(\bI_R \otimes \bra{a_0}_{A_0}\right) \ket{\Phi}_{RA_0} \\
    &= \frac{1}{\sqrt{d_A}} \sum_{\bm a, \bm b} \prod_{k=0}^{t-1} \braket{a_{k+1} b_{k+1}|U_k|a_{k} 0} \ket{a_0}_R \ket{a_t}_{A_t} \otimes_{k=0}^{t-1} \ket{b_{k+1}}_{B_{k+1}}. 
\end{align}
With one more step, we can write out the corresponding density operator of $\ket{\psi}_{R A_t \bm B}$ as
\begin{align}
    \rho_{R A_t \bm B} = \frac{1}{d_A} \sum_{\substack{\bm a, \bm a'\\ \bm b, \bm b'}}  \prod_{k=0}^{t-1} \braket{a_{k+1} b_{k+1}|U_k|a_{k} 0} \braket{a_k' 0|U_k^\dagger|a_{k+1}' b_{k+1}'} \ketbra{a_0}{a_0'}_R \ketbra{a_t}{a_t'}_{A_t} \otimes_{k=0}^{t-1} \ketbra{b_{k+1}}{b_{k+1}'}_{B_{k+1}}.
    \label{eq:rho_uncond}
\end{align}

Recall the definition of R\'enyi-extended QMI in Eq.~\eqref{eq:renyi_qmi}, now we evluate the corresponding reduced states and their R\'enyi-2 entropy. We begin with $\rho_R$.
\begin{align}
    &\rho_R = \tr_{A_t \bm B}(\rho_{R A_t \bm B})\\
    &= \frac{1}{d_A} \sum_{\substack{\bm a, \bm a'\\ \bm b, \bm b'}}  \prod_{k=0}^{t-1} \braket{a_{k+1} b_{k+1}|U_k|a_{k} 0} \braket{a_k' 0|U_k^\dagger|a_{k+1}' b_{k+1}'} \ketbra{a_0}{a_0'}_R \delta_{a_t, a_t'} \prod_{k=0}^{t-1} \delta_{b_{k+1}, b_{k+1}'}\\
    &= \frac{1}{d_A} \sum_{\substack{\bm a, \bm a'\\ \bm b, \bm b'}}  \prod_{k=0}^{t-1} \tr\left(U_k \left(\ketbra{a_k}{a_k'}\otimes\ketbra{0}{0}\right) U_k^\dagger \left(\ketbra{a_{k+1}'}{a_{k+1}} \otimes \ketbra{b_{k+1}'}{b_{k+1}}\right) \right) \ketbra{a_0}{a_0'}_R \delta_{a_t, a_t'} \prod_{k=0}^{t-1} \delta_{b_{k+1}, b_{k+1}'}.
\end{align}
The purity of $\rho_R$ then becomes
\begin{align}
    &\tr(\rho_R^2) \nonumber\\
    &= \frac{1}{d_A^2} \sum_{\substack{\bm a, \bm a'\\ \bm b, \bm b'}}  \prod_{k=0}^{t-1} \tr\left(U_k \left(\ketbra{a_k}{a_k'}\otimes\ketbra{0}{0}\right) U_k^\dagger \left(\ketbra{a_{k+1}'}{a_{k+1}} \otimes \ketbra{b_{k+1}'}{b_{k+1}}\right) \right) \delta_{a_t, a_t'} \prod_{k=0}^{t-1} \delta_{b_{k+1}, b_{k+1}'} \nonumber\\
    &\qquad \quad \times \sum_{\substack{\bm \alpha, \bm \alpha'\\ \bm \beta, \bm \beta'}}  \prod_{k=0}^{t-1} \tr\left(U_k \left(\ketbra{\alpha_k}{\alpha_k'}\otimes\ketbra{0}{0}\right) U_k^\dagger \left(\ketbra{\alpha_{k+1}'}{\alpha_{k+1}} \otimes \ketbra{\beta_{k+1}'}{\beta_{k+1}}\right) \right)  \delta_{\alpha_t, \alpha_t'} \prod_{k=0}^{t-1} \delta_{\beta_{k+1}, \beta_{k+1}'} \delta_{a_0, \alpha_0'} \delta_{a_0', \alpha_0}\\
    &= \frac{1}{d_A^2} \sum_{\substack{\bm a, \bm a'\\ \bm b, \bm b'}} \sum_{\substack{\bm \alpha, \bm \alpha'\\ \bm \beta, \bm \beta'}} \prod_{k=0}^{t-1} \tr\left(U_k^{\otimes 2} \left(\ketbra{a_k \alpha_k}{a_k' \alpha_k'}\otimes\ketbra{0}{0}^{\otimes 2}\right) U_k^\dagger{}^{\otimes 2} \left(\ketbra{a_{k+1}' \alpha_{k+1}'}{a_{k+1} \alpha_{k+1}} \otimes \ketbra{b_{k+1}' \beta_{k+1}'}{b_{k+1} \beta_{k+1}}\right)\right)\nonumber\\
    &\qquad \qquad \qquad \times \delta_{a_0, \alpha_0'} \delta_{a_0', \alpha_0} \delta_{a_t, a_t'} \delta_{\alpha_t, \alpha_t'} \prod_{k=0}^{t-1} \delta_{b_{k+1}, b_{k+1}'} \delta_{\beta_{k+1}, \beta_{k+1}'},
    \label{eq:rhoR_purity}
\end{align}
which can be equivalently expressed in the diagram shown in Fig.~\ref{fig:uncond_purity_diagram}a with a choice of $t=3$.

\begin{figure}[t]
    \centering
    \includegraphics[width=0.5\textwidth]{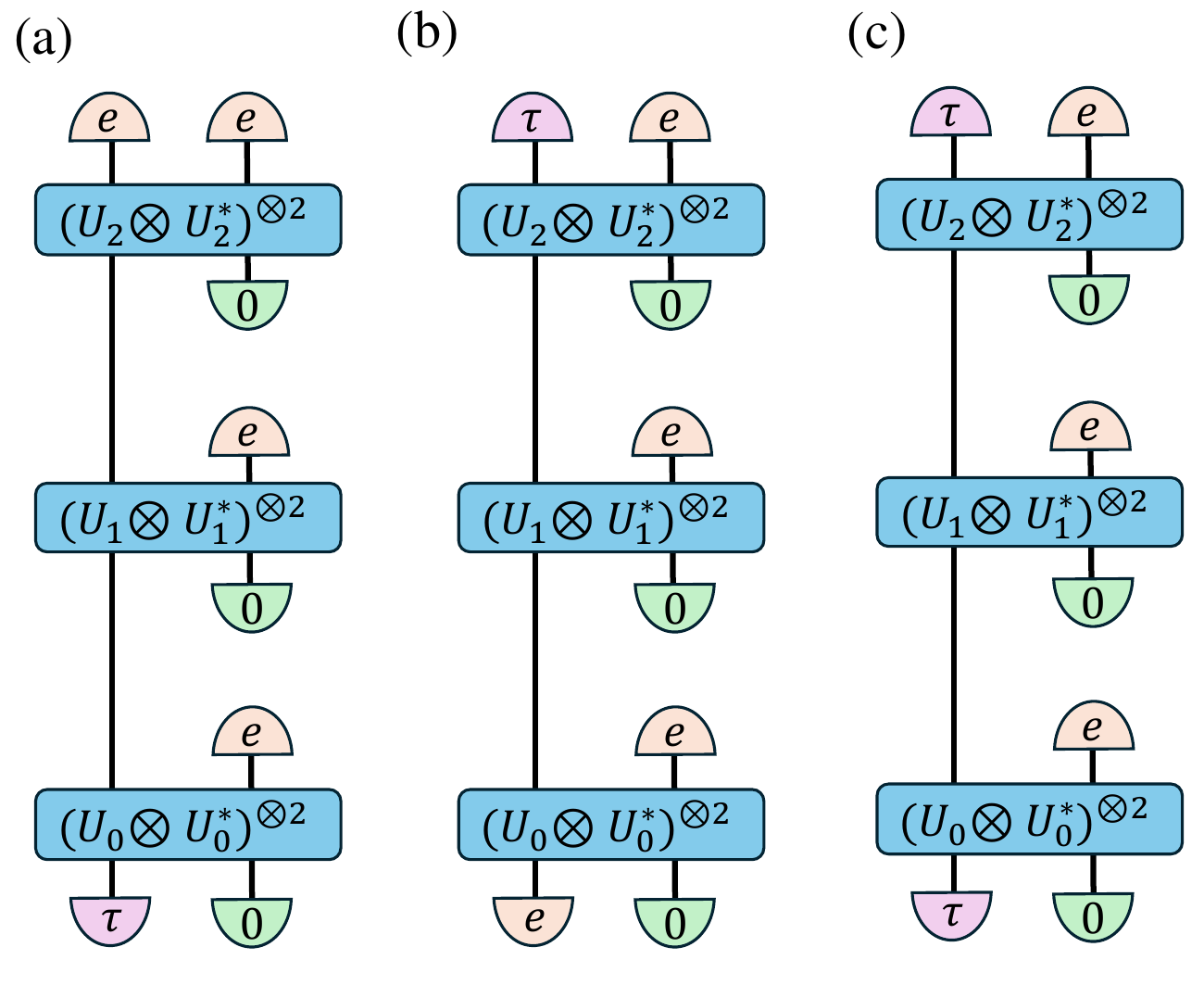}
    \caption{Tensor network representation of Haar-averaged purity for reduced states (a) $\rho_R$, (b) $\rho_{A_t}$ and (c) $\rho_{R A_t}$ with zero-state reset and time steps $t=3$.}
    \label{fig:uncond_purity_diagram}
\end{figure}

For Haar ensemble average of unitary twirling over two replicas, we have
\be
    \E_{\rm Haar}\left[U^{\otimes 2} \otimes U^*{}^{\otimes 2}\right] = \frac{1}{d^2-1} \left[|e)(e| - \frac{1}{d}|e)(\tau| + |\tau)(\tau| - \frac{1}{d}|\tau)(e|\right],
    \label{eq:haar_2design}
\ee
where $|e)$ and $|\tau)$ are the vectorized operator for identity and transposition between two replicas, following the convention of Ref.~\cite{ippoliti2022solvable, zhang2025holographic}. The inner product of any two operator thus can be conveniently represented as $\tr(A^\dagger B) = (A|B)$. The unitary twirling on the step $t$ becomes
\begin{align}
    &\E_{\rm Haar}\left[U_{t-1}^{\otimes 2} \otimes U_{t-1}^*{}^{\otimes 2}\right]|e)_{A}|e)_{B} \nonumber\\
    &=\frac{1}{d_A^2d_B^2-1} \left[d_A^2d_B^2|e)_{AB} - |e)_{AB} + d_A d_B|\tau)_{AB} - d_A d_B|\tau)_{AB}\right]\\
    &= |e)_{AB},
    \label{eq:twirling_ee}
\end{align}
where we utilize $(e|e)_A = d_A^2, (\tau|e)_A = d_A$ in the second line.
Via the reseting with $(0|_B$ and replacing it to $|e)_B$ in the $t-1$ step, we reobtain the initial boundary condition $|e)_A |e)_B$. Therefore, the Haar-averaged purity becomes
\be
\E_{\rm Haar} \tr(\rho_R^2) = \frac{1}{d_A^2} (\tau|_A (0|_B |e)_{A} |e)_B = \frac{1}{d_A}.
\label{eq:rhoR_purity_sol}
\ee

Next, we move to the reduced state $\rho_{A_t}$. Following Eq.~\eqref{eq:rho_uncond}, we have
\begin{align}
    &\rho_{A_t} = \tr_{R \bm B}(\rho_{R A_t \bm B})\\
    &= \frac{1}{d_A} \sum_{\substack{\bm a, \bm a'\\ \bm b, \bm b'}}  \prod_{k=0}^{t-1} \tr\left(U_k \left(\ketbra{a_k}{a_k'}\otimes\ketbra{0}{0}\right) U_k^\dagger \left(\ketbra{a_{k+1}'}{a_{k+1}} \otimes \ketbra{b_{k+1}'}{b_{k+1}}\right) \right) \delta_{a_0, a_0'} \ketbra{a_t}{a_t'}_{A_t} \prod_{k=0}^{t-1} \delta_{b_{k+1},b_{k+1}'},
\end{align}
and the purity becomes
\begin{align}
    &\tr(\rho_{A_t}^2) \nonumber\\
    &= \frac{1}{d_A^2} \sum_{\substack{\bm a, \bm a'\\ \bm b, \bm b'}} \sum_{\substack{\bm \alpha, \bm \alpha'\\ \bm \beta, \bm \beta'}} \prod_{k=0}^{t-1} \tr\left(U_k^{\otimes 2} \left(\ketbra{a_k \alpha_k}{a_k' \alpha_k'}\otimes\ketbra{0}{0}^{\otimes 2}\right) U_k^\dagger{}^{\otimes 2} \left(\ketbra{a_{k+1}' \alpha_{k+1}'}{a_{k+1} \alpha_{k+1}} \otimes \ketbra{b_{k+1}' \beta_{k+1}'}{b_{k+1} \beta_{k+1}}\right)\right) \nonumber\\
    &\qquad \qquad \qquad \times \delta_{a_0, a_0'} \delta_{\alpha_0, \alpha_0'} \delta_{a_t, \alpha_t'}\delta_{a_t', \alpha_t} \prod_{k=0}^{t-1} \delta_{b_{k+1},b_{k+1}'}\delta_{\beta_{k+1},\beta_{k+1}'},
\end{align}
which is equivalently expressed in the diagram shown in Fig.~\ref{fig:uncond_purity_diagram}b.
The Haar ensemble average from step $t$ is 
\begin{align}
    &\E_{\rm Haar}\left[U_{t-1}^{\otimes 2} \otimes U_{t-1}^*{}^{\otimes 2}\right]|\tau)_{A}|e)_{B}  \nonumber \\
    &= \frac{1}{d_A^2d_B^2-1} \left[d_A d_B^2|e) - \frac{d_A^2 d_B}{d_A d_B}|e) + d_A^2 d_B|\tau) - \frac{d_A d_B^2}{d_A d_B}|\tau)\right]\\
    &= \frac{d_A(d_B^2-1)}{d_A^2d_B^2-1} |e)_{AB} + \frac{d_B(d_A^2 - 1)}{d_A^2d_B^2-1} |\tau)_{AB},
    \label{eq:twirling_taue}
\end{align}
where in the second line we utilize the fact $(\tau|\tau)_A = d_A^2$.
Catenating with $|e)(0|_B$ results in
\be
    |e)(0|_B \E_{\rm Haar}\left[U_{t-1}^{\otimes 2} \otimes U_{t-1}^*{}^{\otimes 2}\right]|\tau)_{A}|e)_{B} = \frac{d_A(d_B^2-1)}{d_A^2d_B^2-1} |e)_A|e)_B + \frac{d_B(d_A^2 - 1)}{d_A^2 d_B^2-1} |\tau)_A|e)_B.
\ee
We can find that the second term shares the same boundary condition $|\tau)_A|e)_B$ as the step $t$ we have evaluated, and from the calculation in Eq.~\eqref{eq:twirling_ee} we know that the boundary $|e)_A |e)_B$ remains the same through unitary twirling. Therefore, we can separately evaluate the contribution from $|\tau)_A|e)_B$ and $|e)_A |e)_B$ as follows.
\begin{itemize}
    \item $|\tau)_A|e)_B$: 
    \be
        \frac{1}{d_A^2}\left(\frac{d_B(d_A^2 - 1)}{d_A^2 d_B^2-1}\right)^t (e|\tau)_A (0|e)_B = \frac{1}{d_A}\left(\frac{d_B(d_A^2 - 1)}{d_A^2 d_B^2-1}\right)^t.
    \ee
    \item $|e)_A|e)_B$: 
    \be
        \frac{1}{d_A^2} \frac{d_A(d_B^2-1)}{d_A^2 d_B^2-1} \sum_{k=0}^{t-1} \left(\frac{d_B(d_A^2 - 1)}{d_A^2 d_B^2-1}\right)^k (e|e)_A (0|e)_B = \frac{d_A(d_B+1)}{d_A^2 d_B + 1} \left(1 - \left(\frac{d_B(d_A^2 - 1)}{d_A^2 d_B^2-1}\right)^t\right).
    \ee
\end{itemize}
Combining the above two terms together, we obtain the dynamical solution of Haar-averaged purity for $\rho_{A_t}$
\begin{align}
    \E_{\rm Haar} \tr(\rho_{A_t}^2) &= \frac{1}{d_A}\left(\frac{d_B(d_A^2 - 1)}{d_A^2 d_B^2-1}\right)^t + \frac{d_A(d_B+1)}{d_A^2 d_B + 1} \left(1 - \left(\frac{d_B(d_A^2 - 1)}{d_A^2 d_B^2-1}\right)^t\right)\\
    &= \frac{d_A (1+d_B)}{1 + d_A^2 d_B} + \frac{1-d_A^2}{d_A(1 + d_A^2 d_B)} \left(\frac{d_B(d_A^2 - 1)}{d_A^2 d_B^2-1}\right)^t.
    \label{eq:rhoA_purity_sol}
\end{align}

Finally, we move to the reduced state $\rho_{R A_t}$.
\begin{align}
    &\rho_{R A_t} = \tr_{\bm B}(\rho_{R A_t \bm B})\\
    &= \frac{1}{d_A} \sum_{\substack{\bm a, \bm a'\\ \bm b, \bm b'}}  \prod_{k=0}^{t-1} \tr\left(U_k \left(\ketbra{a_k}{a_k'}\otimes\ketbra{0}{0}\right) U_k^\dagger \left(\ketbra{a_{k+1}'}{a_{k+1}} \otimes \ketbra{b_{k+1}'}{b_{k+1}}\right) \right) \left(\ketbra{a_0}{a_0'}_R \otimes \ketbra{a_t}{a_t'}_{A_t}\right) \prod_{k=0}^{t-1} \delta_{b_{k+1},b_{k+1}'},
\end{align}
and the purity becomes
\begin{align}
    &\tr(\rho_{R A_t}^2) \nonumber\\
    &= \frac{1}{d_A^2} \sum_{\substack{\bm a, \bm a'\\ \bm b, \bm b'}} \sum_{\substack{\bm \alpha, \bm \alpha'\\ \bm \beta, \bm \beta'}} \prod_{k=0}^{t-1} \tr\left(U_k^{\otimes 2} \left(\ketbra{a_k \alpha_k}{a_k' \alpha_k'}\otimes\ketbra{0}{0}^{\otimes 2}\right) U_k^\dagger{}^{\otimes 2} \left(\ketbra{a_{k+1}' \alpha_{k+1}'}{a_{k+1} \alpha_{k+1}} \otimes \ketbra{b_{k+1}' \beta_{k+1}'}{b_{k+1} \beta_{k+1}}\right)\right) \nonumber\\
    &\qquad \qquad \qquad \times \delta_{a_0, \alpha_0'} \delta_{a_0', \alpha_0} \delta_{a_t, \alpha_t'}\delta_{a_t', \alpha_t} \prod_{k=0}^{t-1} \delta_{b_{k+1},b_{k+1}'}\delta_{\beta_{k+1},\beta_{k+1}'},
\end{align}
which is equivalently expressed in the diagram shown in Fig.~\ref{fig:uncond_purity_diagram}c.
Since it shares the same boundary condition as in Fig.~\ref{fig:uncond_purity_diagram}b, we thus can directly write out the contributions from $|\tau)_A|e)_B$ and $|e)_A |e)_B$ as
\begin{itemize}
    \item $|\tau)_A|e)_B$: 
    \be
        \frac{1}{d_A^2}\left(\frac{d_B(d_A^2 - 1)}{d_A^2 d_B^2-1}\right)^t (\tau|\tau)_A (0|e)_B = \left(\frac{d_B(d_A^2 - 1)}{d_A^2 d_B^2-1}\right)^t.
    \ee
    \item $|e)_A|e)_B$: 
    \be
        \frac{1}{d_A^2} \frac{d_A(d_B^2-1)}{d_A^2 d_B^2-1} \sum_{k=0}^{t-1} \left(\frac{d_B(d_A^2 - 1)}{d_A^2 d_B^2-1}\right)^k (\tau|e)_A (0|e)_B = \frac{d_B+1}{d_A^2 d_B + 1} \left(1 - \left(\frac{d_B(d_A^2 - 1)}{d_A^2 d_B^2-1}\right)^t\right).
    \ee
\end{itemize}
The dynamical solution of Haar-averaged purity for $\rho_{R A_t}$ becomes
\begin{align}
    \E_{\rm Haar} \tr(\rho_{R A_t}^2) &= \left(\frac{d_B(d_A^2 - 1)}{d_A^2 d_B^2-1}\right)^t + \frac{d_B+1}{d_A^2 d_B + 1} \left(1 - \left(\frac{d_B(d_A^2 - 1)}{d_A^2 d_B^2-1}\right)^t\right)\\
    &= \frac{1+d_B}{1 + d_A^2 d_B} + \frac{d_B(d_A^2-1)}{1 + d_A^2 d_B} \left(\frac{d_B(d_A^2 - 1)}{d_A^2 d_B^2-1}\right)^t.
    \label{eq:rhoRA_purity_sol}
\end{align}

To evaluate the Haar-averaged of measurement-uncondioned R\'enyi-extended QMI in Eq.~\eqref{eq:uncond_MI}, we switch the order of ensemble average and logarithmic, which does not change the final result as long as the fluctuation of entropy is small. Combining the results in Eqs.~\eqref{eq:rhoR_purity_sol},~\eqref{eq:rhoA_purity_sol} and~\eqref{eq:rhoRA_purity_sol}, we have
\begin{align}
    &\E_{\rm Haar} I_2(R:A_t) \simeq -\log_2 \E_{\rm Haar}\tr(\rho_R^2) - \log_2 \E_{\rm Haar}\tr(\rho_{A_t}^2) + \log_2 \E_{\rm Haar}\tr(\rho_{R A_t}^2) \\
    &= N_A - \log_2 \left[\frac{d_A (1+d_B)}{1 + d_A^2 d_B} + \frac{1-d_A^2}{d_A(1 + d_A^2 d_B)} \left(\frac{d_B(d_A^2 - 1)}{d_A^2 d_B^2-1}\right)^t\right] + \log_2\left[\frac{1+d_B}{1 + d_A^2 d_B} + \frac{d_B(d_A^2-1)}{1 + d_A^2 d_B} \left(\frac{d_B(d_A^2 - 1)}{d_A^2 d_B^2-1}\right)^t\right],
\end{align}
which is the result of Eq.~\eqref{eq:MI_traceout} in the main text. Assuming large system size $d_A, d_B \gg 1$, we can perform asymptotic analysis, 
\begin{align}
    &\E_{\rm Haar} I_2(R:A_t) = N_A - \log_2 \left[\frac{d_A^2 (1+d_B) + (1-d_A^2)\left(\frac{d_B(d_A^2 - 1)}{d_A^2 d_B^2-1}\right)^t}{d_A(1 + d_A^2 d_B)} \right] + \log_2\left[\frac{1+d_B + d_B(d_A^2-1)\left(\frac{d_B(d_A^2 - 1)}{d_A^2 d_B^2-1}\right)^t}{1 + d_A^2 d_B}\right] \nonumber\\
    &\simeq N_A - \log_2\left(\frac{1}{d_A} - \frac{1}{d_A d_B} \frac{1}{d_B^t}\right) + \log_2\left(\frac{1}{d_A^2} + \frac{1}{d_B^t}\right)\\
    &= N_A - \log_2 \frac{1}{d_A}\left( 1 - \frac{1}{d_B^{t+1}}\right) + \log_2\left(\frac{1}{d_A^2} + \frac{1}{d_B^t}\right)\\
    &\simeq 2N_A - d_B^{-(t+1)} + \log_2\left(d_A^{-2} + d_B^{-t}\right).
\end{align}
For early time $t \ll 2N_A/N_B$, we can further approximate the QMI as
\begin{align}
    \E_{\rm Haar} I_2(R:A_t) &\simeq 2N_A - d_B^{-(t+1)} + \log_2\left(d_A^{-2} + d_B^{-t}\right)\nonumber\\
    &= 2N_A - d_B^{-(t+1)} + \log_2 d_B^{-t}\left(1 + d_A^{-2} d_B^t \right)\\
    &\simeq 2N_A - d_B^{-(t+1)} -t N_B + d_A^{-2} d_B^t\\
    &\simeq 2N_A - tN_B + \mathcal{O}(d_B^{t}/d_A^2),
\end{align}
which recovers the linear scaling in the main text.
On the other hand for late time $t \gg 2N_A/N_B$, with similar approximation method, we have
\begin{align}
    \E_{\rm Haar} I_2(R:A_t) &\simeq 2N_A - d_B^{-(t+1)} + \log_2\left(d_A^{-2} + d_B^{-t}\right)\nonumber\\
    &= 2N_A - d_B^{-(t+1)} + \log_2 d_A^{-2}\left(1 + d_A^{2} d_B^{-t} \right)\\
    &\simeq  2N_A - d_B^{-(t+1)} -2 N_A + d_A^{2} d_B^{-t}\\
    &= d_A^2 d_B^{-t} + \mathcal{O}(d_B^{-(t+1)}).
\end{align}

\subsection{Pure-state reset with $N_R < N_A$ (Theorem~\ref{MI_traceout_lessR_theorem})}
\label{app:pure_reset_lessR}

In this part, we consider a smaller reference of $N_R < N_A$ and still adopt pure reset strategy. The proof is similar to the Appendix~\ref{app:pure_reset} shown above.

The initial state on reference $R$ and system $A$ is still written as Eq.~\eqref{eq:init_state_lessR}, and the measurement-unconditioned state is
\be
    \rho_{RA} = \tr_{B}\left(U \ketbra{\Phi}{\Phi}_{RA_0}\ketbra{0}{0}_B U^\dagger\right),
\ee
where $U$ is still the whole circuit unitary. Similarly, we can also write out the unconditioned state only for reference and system separately. The purity for the reduced state on reference only is
\begin{align}
    \tr(\rho_R^2) &= \tr\left(U^{\otimes 2} \ketbra{\Phi}{\Phi}_{RA_0}^{\otimes 2} \ketbra{0}{0}_B^{\otimes 2} U^\dagger{}^{\otimes 2} \tau_R \bI_A^{\otimes 2} \bI_B^{\otimes 2}\right)\\
    &=  d_R^{-2}\tr\left(U^2 \tau_{A_1} \ketbra{0}{0}^{\otimes 2}_{A_2} \ketbra{0}{0}_B^{\otimes 2} U^\dagger{}^{\otimes 2} \bI_A^{\otimes 2} \bI_{B}^{\otimes 2}\right),
\end{align}
which is same as Eq.~\eqref{eq:rhoR_purity} except for the initial boundary condition. Therefore, the Haar-averaged purity becomes
\be
    \E_{\rm Haar} \tr(\rho_R^2) = \frac{1}{d_R^2} (\tau|_{A_1} (0|_{A_2} (0|_B |e)_{A} |e)_B = \frac{1}{d_R}.
    \label{eq:purityR_lessR}
\ee

For the purity of reduced state on system, we have
\begin{align}
    \tr(\rho_A^2) &= \tr\left(U^{\otimes 2} \ketbra{\Phi}{\Phi}_{RA_0}^{\otimes 2} \ketbra{0}{0}_B^{\otimes 2} U^\dagger{}^{\otimes 2} \bI_R^{\otimes 2} \tau_A \bI_B^{\otimes 2}\right)\\
    &=  d_R^{-2}\tr\left(U^2 \bI_{A_1}^{\otimes 2} \ketbra{0}{0}^{\otimes 2}_{A_2} \ketbra{0}{0}_B^{\otimes 2} U^\dagger{}^{\otimes 2} \tau_A \bI_{B}^{\otimes 2}\right),
\end{align}
and utilizing the twirling dynamics illustrated in Eq.~\eqref{eq:twirling_taue} and Eq.~\eqref{eq:twirling_ee}, the Haar-averaged purity becomes
\begin{align}
    \E_{\rm Haar}  \tr(\rho_A^2) &= \frac{1}{d_R^2} \left[\left(\frac{d_B(d_A^2-1)}{d_A^2 d_B^2 - 1}\right)^t (e|\tau)_{A_1} + \frac{d_A(d_B + 1)}{d_A^2 d_B + 1}\left(1 - \left(\frac{d_B(d_A^2-1)}{d_A^2 d_B^2 - 1}\right)^t\right)(e|e)_{A_1}\right]\\
    &= \left(\frac{d_B(d_A^2-1)}{d_A^2 d_B^2 - 1}\right)^t \frac{1}{d_R} + \frac{d_A(d_B + 1)}{d_A^2 d_B + 1}\left(1 - \left(\frac{d_B(d_A^2-1)}{d_A^2 d_B^2 - 1}\right)^t\right)\\
    &= \frac{d_A(d_B + 1)}{d_A^2 d_B + 1} + \left(\frac{1}{d_R} - \frac{d_A(d_B + 1)}{d_A^2 d_B + 1}\right) \left(\frac{d_B(d_A^2-1)}{d_A^2 d_B^2 - 1}\right)^t.
    \label{eq:purityA_lessR}
\end{align}

Finally, we consider the purity of $\rho_{RA}$, which can be written as
\begin{align}
    \tr(\rho_{RA}^2) &= \tr\left(U^{\otimes 2} \ketbra{\Phi}{\Phi}_{RA_0}^{\otimes 2} \ketbra{0}{0}_B^{\otimes 2} U^\dagger{}^{\otimes 2} \tau_R \tau_A \bI_B^{\otimes 2}\right)\\
    &=  d_R^{-2}\tr\left(U^2 \tau_{A_1} \ketbra{0}{0}^{\otimes 2}_{A_2} \ketbra{0}{0}_B^{\otimes 2} U^\dagger{}^{\otimes 2} \tau_A \bI_{B}^{\otimes 2}\right),
\end{align}
and we can utilize the result in Eq.~\eqref{eq:purityA_lessR} above to derive
\begin{align}
    \E_{\rm Haar}  \tr(\rho_{RA}^2) &= \frac{1}{d_R^2} \left[\left(\frac{d_B(d_A^2-1)}{d_A^2 d_B^2 - 1}\right)^t (\tau|\tau)_{A_1} + \frac{d_A(d_B + 1)}{d_A^2 d_B + 1}\left(1 - \left(\frac{d_B(d_A^2-1)}{d_A^2 d_B^2 - 1}\right)^t\right)(\tau|e)_{A_1}\right]\\
    &= \left(\frac{d_B(d_A^2-1)}{d_A^2 d_B^2 - 1}\right)^t  + \frac{d_A(d_B + 1)}{d_A^2 d_B + 1}\left(1 - \left(\frac{d_B(d_A^2-1)}{d_A^2 d_B^2 - 1}\right)^t\right) \frac{1}{d_R}\\
    &= \frac{d_A(d_B + 1)}{d_R(d_A^2 d_B + 1)} + \left(1 - \frac{d_A(d_B + 1)}{d_R(d_A^2 d_B + 1)}\right) \left(\frac{d_B(d_A^2-1)}{d_A^2 d_B^2 - 1}\right)^t.
    \label{eq:purityRA_lessR}
\end{align}
Therefore, combing Eqs.~\eqref{eq:purityR_lessR},~\eqref{eq:purityA_lessR} and~\eqref{eq:purityRA_lessR}, we have the Haar-averaged R\'enyi-2 extended mutual information as
\begin{align}
    \E_{\rm Haar}I_2(R:A_t) &\simeq -\log_2 \E_{\rm Haar}\tr(\rho_R^2) - \log_2 \E_{\rm Haar}\tr(\rho_{A}^2) + \log_2 \E_{\rm Haar}\tr(\rho_{R A}^2) \\
    &= N_R - \log_2 \left[\frac{d_A(d_B + 1)}{d_A^2 d_B + 1} + \left(\frac{1}{d_R} - \frac{d_A(d_B + 1)}{d_A^2 d_B + 1}\right) \left(\frac{d_B(d_A^2-1)}{d_A^2 d_B^2 - 1}\right)^t \right] \nonumber\\
    &\quad + \log_2\left[\frac{d_A(d_B + 1)}{d_R(d_A^2 d_B + 1)} + \left(1 - \frac{d_A(d_B + 1)}{d_R(d_A^2 d_B + 1)}\right) \left(\frac{d_B(d_A^2-1)}{d_A^2 d_B^2 - 1}\right)^t\right]. 
    \label{eq:uncondQMI_lessR_full_sol}
\end{align}

In the following, we simplify the expression in the asymptotic limit of $d_A, d_B \gg 1$, and Eq.~\eqref{eq:uncondQMI_lessR_full_sol} can be reduced to
\begin{align}
    \E_{\rm Haar}I_2(R:A_t) &= N_R - \log_2 \left[\frac{d_A(d_B + 1)}{d_A^2 d_B + 1} + \left(\frac{1}{d_R} - \frac{d_A(d_B + 1)}{d_A^2 d_B + 1}\right) \left(\frac{d_B(d_A^2-1)}{d_A^2 d_B^2 - 1}\right)^t \right] \nonumber\\
    &\quad + \log_2\left[\frac{d_A(d_B + 1)}{d_R(d_A^2 d_B + 1)} + \left(1 - \frac{d_A(d_B + 1)}{d_R(d_A^2 d_B + 1)}\right) \left(\frac{d_B(d_A^2-1)}{d_A^2 d_B^2 - 1}\right)^t\right]\\
    &\simeq N_R - \log_2 \left[\frac{1}{d_A}+ \left(\frac{1}{d_R} - \frac{1}{d_A}\right) \frac{1}{d_B^t} \right] + \log_2\left[\frac{1}{d_R d_A} + \left(1-\frac{1}{d_R d_A}\right)\frac{1}{d_B^t}\right]\\
    &= N_R + N_A - \log_2\left[1 + \left(\frac{d_A}{d_R}-1\right)d_B^{-t}\right] - (N_R + N_A) + \log_2\left[1 + (d_R d_A - 1)d_B^{-t}\right] \\
    &= \log_2\left(1 + d_R d_A d_B^{-t}\right) - \log_2\left[1 + \left(\frac{d_A}{d_R}-1\right)d_B^{-t}\right],
    \label{eq:MI_traceout_lessR_sm}
\end{align}
which is the theoretical result of Eq.~\eqref{eq:MI_traceout_lessR} in Theorem~\ref{MI_traceout_lessR_theorem} in the main text.

Therefore, through the two logarithmic terms above, we can identify two characteristic time scales $\tau_{\mp} = (N_A\mp N_R)/N_B$. When $t \le \tau_-$, Eq.~\eqref{eq:MI_traceout_lessR_sm} becomes
\be
    \E_{\rm Haar}I_2(R:A_t) \simeq \log_2\left(d_R d_Ad_B^{-t}\right) - \log_2\left(d_A/d_R d_B^{-t}\right) = 2N_R,
\ee
meanwhile for $\tau_- \le t \le \tau_+$, we have
\be
    \E_{\rm Haar}I_2(R:A_t) \simeq \log_2\left(d_R d_A d_B^{-t}\right) - \frac{(d_A/d_R-1) d_B^{-t}}{\log 2} \simeq N_R + N_A - N_B t,
\ee
and finally for $t \ge \tau_+$, we have
\be
    \E_{\rm Haar}I_2(R:A_t) \simeq \frac{1}{\log 2}\left(d_R d_A d_B^{-t} - (d_A/d_R-1) d_B^{-t}\right)\simeq \frac{d_R d_A d_B^{-t}}{\log 2} \to 0.
\ee
Therefore, we complete the dynamical transition in Eq.~\eqref{eq:MInorm_traceout_lessR} in the main text up to a normalization constant.

\subsection{Dynamics without reset (Theorem~\ref{MI_traceout_woReset_theorem})}
\label{app:MI_traceout_woReset_theorem}

In this part, we prove the dynamics of measurement-unconditioned QMI with bath measured but without reset. The conditioned state $\ket{\psi_\bfz}$ in the dynamics without reset can be written out as
\begin{align}
    \ket{\psi_\bfz} &= {}_B\bra{\bfz}\left(U\ket{\Phi}_{RA} X^{\tilde{\bfz}}\ket{0}_B\right)/\sqrt{p_\bfz}\\
    &= {}_B\bra{\bfz}\left(U\ket{\Phi}_{RA} \ket{\tilde{\bfz}}_B\right)/\sqrt{p_\bfz} \equiv \ket{\tilde{\psi}_\bfz}/\sqrt{p_\bfz},
\end{align}
where $p_\bfz = \braket{\tilde{\psi}_\bfz | \tilde{\psi}_\bfz}$ and $\tilde{\bfz} = (0,z_1,\dots, z_{t-1})$. The mixed state of interest in the unconditioned dynamics is 
\be
    \rho_{RA_t} = \sum_\bfz p_\bfz \ketbra{\psi_\bfz}{\psi_\bfz} = \sum_\bfz \ketbra{\tilde{\psi}_\bfz}{\tilde{\psi}_\bfz}. 
    \label{eq:rho_RA_uncond_wo}
\ee

In the following, we derive the R\'enyi-2 extended mutual information of $\rho_{RA_t}$. We begin with the purity of reduced state on reference.
\begin{align}
    \tr(\rho_R^2) &= \tr(\rho_R^{\otimes 2} \tau_R)\\
    &= \sum_{\bfz, \bfz'} \tr(\ketbra{\tilde{\psi}_\bfz}{\tilde{\psi}_\bfz} \otimes \ketbra{\tilde{\psi}_{\bfz'}}{\tilde{\psi}_{\bfz'}} \tau_R \bI_A^{\otimes 2})\\
    &= \sum_{\bfz, \bfz'} \tr(U^{\otimes 2} \ketbra{\Phi}{\Phi}_{RA}^{\otimes 2} \ketbra{\tilde{\bfz}}{\tilde{\bfz}}_B\otimes \ketbra{\tilde{\bfz}'}{\tilde{\bfz}'}_B U^\dagger{}^{\otimes 2} \tau_R \bI^{\otimes 2}_A \ketbra{\bfz}{\bfz}_B\otimes \ketbra{\bfz'}{\bfz'}_B)\\
    &= d_A^{-2} \sum_{\bfz, \bfz'} \tr(U^{\otimes 2} \tau_A \ketbra{\tilde{\bfz}}{\tilde{\bfz}}_B\otimes \ketbra{\tilde{\bfz}'}{\tilde{\bfz}'}_B U^\dagger{}^{\otimes 2} \bI^{\otimes 2}_A \ketbra{\bfz}{\bfz}_B\otimes \ketbra{\bfz'}{\bfz'}_B).
\end{align}
We still start the Haar unitary twirling from $U_t$ as
\begin{align}
    &\sum_{z_{t-1},z_{t-1}'}|z_{t-1},z_{t-1}')(z_{t-1},z_{t-1}'|_B\E_{\rm Haar}\left[U_t^{\otimes 2} \otimes U_t^*{}^{\otimes 2}\right] |e)_A |D_1^{\otimes 2})_B\\
    &= \frac{1}{d^2-1} \sum_{z_{t-1},z_{t-1}'}|z_{t-1},z_{t-1}')(z_{t-1},z_{t-1}'|_B \left[|e)(e| + |\tau)(\tau|- \frac{1}{d}|e)(\tau| - \frac{1}{d}|\tau)(e|\right] |e)_A |D_1^{\otimes 2})_B\\
    &= \frac{1}{d^2-1} \sum_{z_{t-1},z_{t-1}'}|z_{t-1},z_{t-1}')(z_{t-1},z_{t-1}'|_B \left[|e)d_A^2 d_B^2 + |\tau)d_A d_B- \frac{d_A d_B}{d}|e) - \frac{d_A^2 d_B^2}{d}|\tau)\right] \\
    &= |e)_A \sum_{z_{t-1},z_{t-1}'}|z_{t-1},z_{t-1}')_B (z_{t-1},z_{t-1}'|e)_B\\
    &= |e)_A |D_1^{\otimes 2})_B,
\end{align}
where we denote $D_1 = \sum_z \ketbra{z}{z}$ following the convention, and we can immediately see that $(e|D_1^{\otimes 2})_B = d_B^2$ and $(\tau|D_1^{\otimes 2})_B = d_B$. Therefore, we directly have the purity of reference reduced state as
\begin{align}
    \E_{\rm Haar} \tr(\rho_R^2) &= d_A^{-2} \E_{\rm Haar}\sum_{\bfz, \bfz'} \tr(U^{\otimes 2} \tau_A \ketbra{\tilde{\bfz}}{\tilde{\bfz}}_B\otimes \ketbra{\tilde{\bfz}'}{\tilde{\bfz}'}_B U^\dagger{}^{\otimes 2} \bI^{\otimes 2}_A \ketbra{\bfz}{\bfz}_B\otimes \ketbra{\bfz'}{\bfz'}_B) \nonumber\\
    &= d_A^{-2} (\tau|e)_A (0|D_1^{\otimes 2})_B = \frac{1}{d_A}.
    \label{eq:purity_R_uncond_wo}
\end{align}

For the purity of reduced state on system $A$, we have
\begin{align}
    \tr(\rho_A^2) &= \tr(\rho_A^{\otimes 2} \tau_A)\\
    &= \sum_{\bfz, \bfz'} \tr(\ketbra{\tilde{\psi}_\bfz}{\tilde{\psi}_\bfz} \otimes \ketbra{\tilde{\psi}_{\bfz'}}{\tilde{\psi}_{\bfz'}} \bI_R^{\otimes 2} \tau_A)\\
    &= \sum_{\bfz, \bfz'} \tr(U^{\otimes 2} \ketbra{\Phi}{\Phi}_{RA}^{\otimes 2} \ketbra{\tilde{\bfz}}{\tilde{\bfz}}_B\otimes \ketbra{\tilde{\bfz}'}{\tilde{\bfz}'}_B U^\dagger{}^{\otimes 2} \bI^{\otimes 2}_R \tau_A \ketbra{\bfz}{\bfz}_B\otimes \ketbra{\bfz'}{\bfz'}_B)\\
    &= d_A^{-2} \sum_{\bfz, \bfz'} \tr(U^{\otimes 2} \bI^{\otimes 2}_A \ketbra{\tilde{\bfz}}{\tilde{\bfz}}_B\otimes \ketbra{\tilde{\bfz}'}{\tilde{\bfz}'}_B U^\dagger{}^{\otimes 2} \tau_A  \ketbra{\bfz}{\bfz}_B\otimes \ketbra{\bfz'}{\bfz'}_B).
\end{align}
For Haar twirling of $U_t$, we first have
\begin{align}
    &\sum_{z_{t-1},z_{t-1}'}|z_{t-1},z_{t-1}')(z_{t-1},z_{t-1}'|_B\E_{\rm Haar}\left[U_t^{\otimes 2} \otimes U_t^*{}^{\otimes 2}\right] |\tau)_A|D_1^{\otimes 2})_B \nonumber\\
    &= \frac{1}{d^2-1} \sum_{z_{t-1},z_{t-1}'}|z_{t-1},z_{t-1}')(z_{t-1},z_{t-1}'|_B \left[|e)(e| + |\tau)(\tau|- \frac{1}{d}|e)(\tau| - \frac{1}{d}|\tau)\right] |\tau)_A |D_1^{\otimes 2})_B \\
    &= \frac{1}{d^2-1} \sum_{z_{t-1},z_{t-1}'}|z_{t-1},z_{t-1}')(z_{t-1},z_{t-1}'|_B \left[|e)d_A d_B^2 + |\tau)d_A^2 d_B- \frac{d_A^2 d_B}{d}|e) - \frac{d_A d_B^2}{d}|\tau)(e|\right]\\
    &= \sum_{z_{t-1},z_{t-1}'}|z_{t-1},z_{t-1}')(z_{t-1},z_{t-1}'|_B \left[\frac{d_A}{d^2-1}(d_B^2 - 1)|e) + \frac{d_B}{d^2-1} (d_A^2 -1 )|\tau)\right]\\
    &=\frac{d_A}{d^2-1}(d_B^2 - 1)|e)_A |D_1^{\otimes 2})_B + \frac{d_B}{d^2-1} (d_A^2 -1 )|\tau)_A |D_2)_B,
\end{align}
where $D_2 = \sum_z \ketbra{z}{z}^{\otimes 2}$. It leads to two boundary conditions, where the first one of $|e)_A |D_1^{\otimes 2})_B$ remains invariant under twirling while the second one $|\tau)_A |D_2)_B$ is a new boundary condition to evaluate.
\begin{align}
    &\sum_{z_{t-2},z_{t-2}'}|z_{t-2},z_{t-2}')(z_{t-2},z_{t-2}'|_B\E_{\rm Haar}\left[U_{t-1}^{\otimes 2} \otimes U_{t-1}^*{}^{\otimes 2}\right] |\tau)_A|D_2)_B \nonumber\\
    &= \frac{1}{d^2-1} \sum_{z_{t-2},z_{t-2}'}|z_{t-2},z_{t-2}')(z_{t-2},z_{t-2}'|_B \left[|e)(e| + |\tau)(\tau|- \frac{1}{d}|e)(\tau| - \frac{1}{d}|\tau)(e|\right] |\tau)_A |D_2)_B \\
    &= \frac{d_B}{d^2-1} \sum_{z_{t-2},z_{t-2}'}|z_{t-2},z_{t-2}')(z_{t-2},z_{t-2}'|_B \left[|e)d_A  + |\tau)d_A^2- \frac{d_A^2}{d}|e) - \frac{d_A}{d}|\tau)\right]\\
    &= \frac{1}{d^2-1}\sum_{z_{t-1},z_{t-1}'}|z_{t-1},z_{t-1}')(z_{t-1},z_{t-1}'|_B \left[d_A(d_B-1)|e) + (d_A^2 d_B-1)|\tau)\right]\\
    &=\frac{d_A}{d^2-1}(d_B - 1)|e)_A |D_1^{\otimes 2})_B + \frac{d_A^2 d_B-1}{d^2-1} |\tau)_A |D_2)_B,
\end{align}
where we utilize $(\pi|D_2)_B = d_B$ in the third line for any permutation $\pi$. We can now still utilize a $2$-dimensional linear system to model the twirling dynamics with $|\tau)_A |D_2)_B \to (1,0)^T$ and $|e)_A |D_1^{\otimes 2})_B \to (0, 1)^T$. The transfer matrix is
\begin{align}
    M = \frac{1}{d^2-1}\begin{pmatrix}
        d_A^2 d_B-1 & 0\\
        d_A(d_B-1) & d^2-1
    \end{pmatrix},
\end{align}
and the purity for $t\ge 1$ then becomes
\begin{align}
    \E_{\rm Haar}\tr(\rho_A^2) &= d_A^{-2} (e|_A (0|_B M^{t-1} \left(\frac{d_B(d_A^2-1)}{d^2-1}, \frac{d_A(d_B^2-1)}{d^2-1}\right)^T \\
    &=d_A^{-2}\left[\frac{\left(d_A^2-1\right) d_B \left(\frac{d_A^2 d_B-1}{d_A^2 d_B^2-1}\right)^t}{d_A^2 d_B-1} d_A + \frac{1-\frac{\left(d_A^2-1\right) \left(\frac{d_A^2 d_B-1}{d_A^2 d_B^2-1}\right)^t}{d_A^2 d_B-1}}{d_A} d_A^2\right]\\
    &= \frac{(d_A^2-1)(d_B-1)}{d_A(d_A^2 d_B-1)}\left(\frac{d_A^2 d_B-1}{d_A^2 d_B^2-1}\right)^t + \frac{1}{d_A}.
    \label{eq:purity_A_uncond_wo}
\end{align}
Similarly, the purity of the joint state of reference and system is
\begin{align}
    \tr(\rho_{RA}^2) &= \tr(\rho_{RA}^{\otimes 2} \tau_R\tau_A)\\
    &= \sum_{\bfz, \bfz'} \tr(\ketbra{\tilde{\psi}_\bfz}{\tilde{\psi}_\bfz} \otimes \ketbra{\tilde{\psi}_{\bfz'}}{\tilde{\psi}_{\bfz'}} \tau_R \tau_A)\\
    &= \sum_{\bfz, \bfz'} \tr(U^{\otimes 2} \ketbra{\Phi}{\Phi}_{RA}^{\otimes 2} \ketbra{\tilde{\bfz}}{\tilde{\bfz}}_B\otimes \ketbra{\tilde{\bfz}'}{\tilde{\bfz}'}_B U^\dagger{}^{\otimes 2} \tau_R \tau_A \ketbra{\bfz}{\bfz}_B\otimes \ketbra{\bfz'}{\bfz'}_B)\\
    &= d_A^{-2} \sum_{\bfz, \bfz'} \tr(U^{\otimes 2} \tau_A \ketbra{\tilde{\bfz}}{\tilde{\bfz}}_B\otimes \ketbra{\tilde{\bfz}'}{\tilde{\bfz}'}_B U^\dagger{}^{\otimes 2} \tau_A  \ketbra{\bfz}{\bfz}_B\otimes \ketbra{\bfz'}{\bfz'}_B).
\end{align}
The Haar-averaged purity is thus
\begin{align}
    \E_{\rm Haar}\tr(\rho_{RA}^2) &= d_A^{-2} (\tau|_A (0|_B M^{t-1} \left(\frac{d_B(d_A^2-1)}{d^2-1}, \frac{d_A(d_B^2-1)}{d^2-1}\right)^T \\
    &= d_A^{-2}\left[\frac{\left(d_A^2-1\right) d_B \left(\frac{d_A^2 d_B-1}{d_A^2 d_B^2-1}\right)^t}{d_A^2 d_B-1} d_A^2 + \frac{1-\frac{\left(d_A^2-1\right) \left(\frac{d_A^2 d_B-1}{d_A^2 d_B^2-1}\right)^t}{d_A^2 d_B-1}}{d_A} d_A\right]\\
    &= \frac{d_A^2-1}{d_A^2} \left(\frac{d_A^2 d_B-1}{d_A^2 d_B^2-1}\right)^t + \frac{1}{d_A^2}.
    \label{eq:purity_RA_uncond_wo}
\end{align}
Combining Eqs.~\eqref{eq:purity_R_uncond_wo},~\eqref{eq:purity_A_uncond_wo} and~\eqref{eq:purity_RA_uncond_wo}, we have the Haar-averaged R\'enyi-2 extended QMI as
\begin{align}
    &\E_{\rm Haar}I_2(R:A_t) \simeq -\log_2\E_{\rm Haar}\tr(\rho_R^2) - \log_2\E_{\rm Haar}\tr(\rho_A^2) + \log_2\E_{\rm Haar}\tr(\rho_{RA}^2)\\
    &= -\log_2\left(\frac{1}{d_A}\right) - \log_2\left(\frac{(d_A^2-1)(d_B-1)}{d_A(d_A^2 d_B-1)}\left(\frac{d_A^2 d_B-1}{d_A^2 d_B^2-1}\right)^t + \frac{1}{d_A}\right) + \log_2\left(\frac{d_A^2-1}{d_A^2} \left(\frac{d_A^2 d_B-1}{d_A^2 d_B^2-1}\right)^t + \frac{1}{d_A^2}\right)\\
    &= \log_2\left((d_A^2-1) \left(\frac{d_A^2 d_B-1}{d_A^2 d_B^2-1}\right)^t + 1\right) - \log_2\left(\frac{(d_A^2-1)(d_B-1)}{d_A^2 d_B-1}\left(\frac{d_A^2 d_B-1}{d_A^2 d_B^2-1}\right)^t + 1\right). 
\end{align}
In the asymptotic limit of $d_A, d_B \gg 1$, we can further reduce it to
\begin{align}
    \E_{\rm Haar}I_2(R:A_t) &= \log_2\left(d_A^2 d_B^{-t} + 1\right) - \log_2\left(d_B^{-t} + 1\right)\\
    &\simeq 2N_A - d_B^{-t} + \log_2\left(d_B^{-t} + d_A^{-2}\right).
\end{align}
For early time $t \ll 2N_A /N_B$, the QMI becomes
\begin{align}
    \E_{\rm Haar}I_2(R:A_t) &\simeq 2N_A - d_B^{-t} + \log_2d_B^{-t}\left(1 + d_A^{-2}d_B^t\right)\\
    &= 2N_A - d_B^{-t} -tN_B + d_A^{-2}d_B^t\\
    &= 2N_A - tN_B + \mathcal{O}(d_B^t/d_A^2).
\end{align}
On the other hand, for late time of $t \gg 2N_A/N_B$, we have
\begin{align}
    \E_{\rm Haar}I_2(R:A_t) &\simeq 2N_A - d_B^{-t} + \log_2d_A^{-2}\left(d_A^2 d_B^{-t} + 1\right)\\
    &= d_A^2 d_B^{-t} + \mathcal{O}(d_B^{-t}).
\end{align}
We complete the proof of Theorem~\ref{MI_traceout_woReset_theorem}.

We end this section by providing an alternative way of describing the channel of Eq.\,(\ref{eq:channel_noreset}) adopted in the QRC literature \cite{hu2024overcoming}. As detailed in Eq.\,(10) of Supplementary Note 2 in Ref.\,\cite{hu2024overcoming}, the quantum channel can be described by the superoperator
\be
    \mathcal{Q}_t(\cdot) = \mathcal{M} (U_t (\cdot) U_t^\dagger).
\ee
Here, we introduce the \textit{measurement-induced decoherence superoperator} $\mathcal{M}(\cdot) = (\mathbf{1}_A \otimes \bm{I}_B ) \odot (\cdot)$ to represent the full decoherence of the subsystem $B$ \cite{hu2024overcoming}, where $\mathbf{1}_A \in \mathbb{R}^{d_A \times d_A}$ is an all-one square matrix on subsystem $A$, $\bm{I}_B \in \mathbb{R}^{d_B \times d_B}$ is the identity matrix on subsystem $B$, and the Hadamard product $\odot$ is the entry-wise product between two matrices.

\subsection{Fully-mixed state reset (Theorem~\ref{MI_traceout_id_theorem})}
\label{app:mixed_reset}

In this part, we focus on a different reset strategy, the bath system is initialized with a fully-mixed state $\omega = \bI_B/d_B$ starting from the second step.

\begin{figure}[t]
    \centering
    \includegraphics[width=0.5\textwidth]{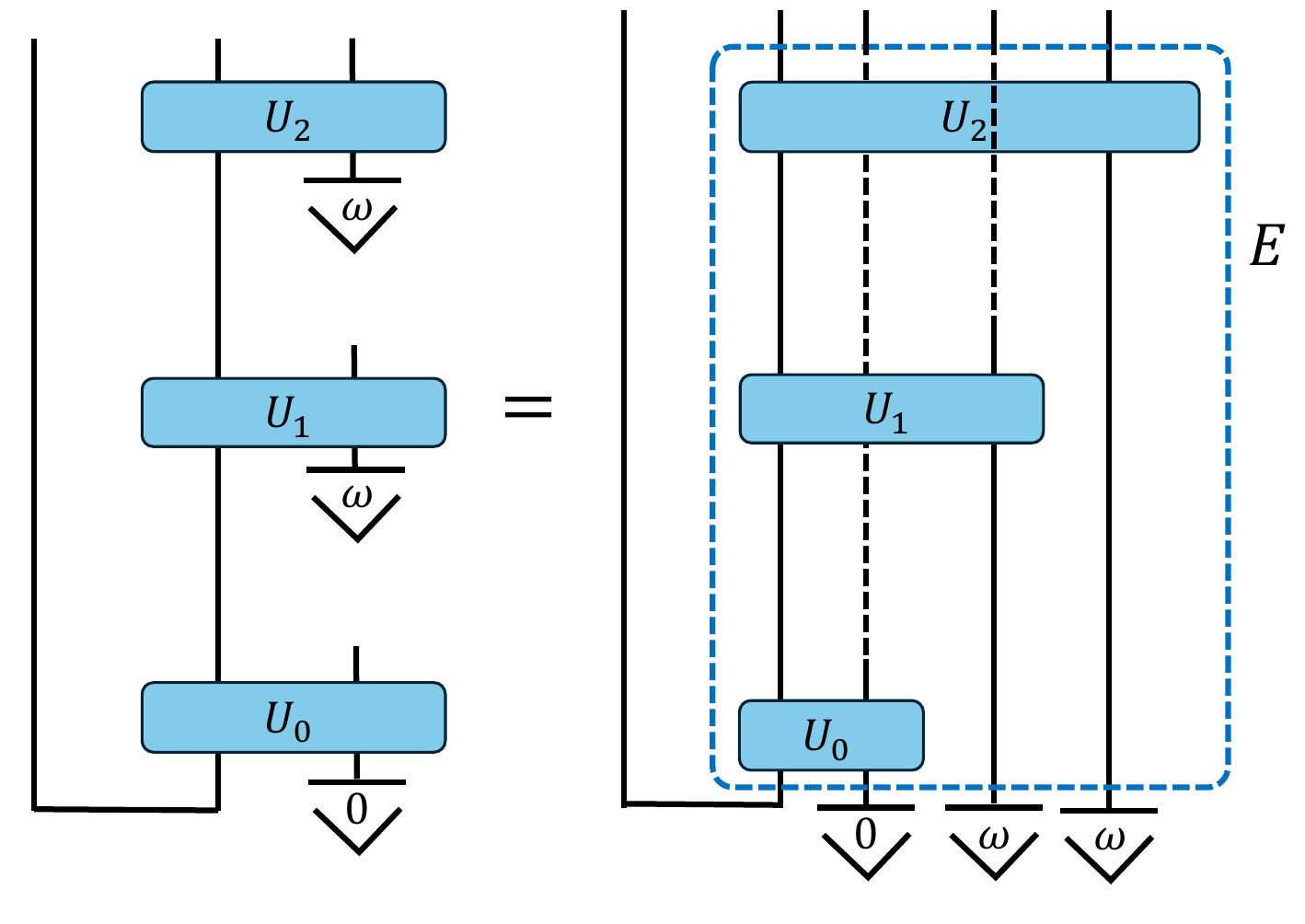}
    \caption{Equivalent relation on the quantum dynamics with mid-circuit measurements and diagonal-mixed-state reset via expanding the bath. Here $\omega = \bI_B/d_B$ represents the fully-mixed state. The linear map $E$ (blue box) is applied on the initial maximally entangled state. The dashed lines indicate that those outputs do not interact with the unitaries to be passed through. Here we show an example for $t=3$.}
    \label{fig:uncond_mix_diagram}
\end{figure}

Similar to the derivation in pure-state reset strategy, we first write out the linear map $E$ (shown by blue box in Fig.~\ref{fig:uncond_mix_diagram}) as
\be
    E = \sum_{\bm a, \bm b} \prod_{k=0}^{t-1} \braket{a_{k+1} b_{2k+1}|U_k|a_k b_{2k}} \ket{a_t}_{A_t}\bra{a_0}_{A_0} \otimes_{k=0}^{t-1} \ketbra{b_{2k+1}}{b_{2k}},
    \label{eq:E_def_uncond_mix}
\ee
where different from above we define the vector representation $\bm b = (b_0, b_1, \cdots, b_{2t-2}, b_{2t-1})$. Via applying the linear map $E$ on maximally entangled state $\ket{\Phi}_{RA_0}$ and the expanded bath system (shown in Fig.~\ref{fig:uncond_mix_diagram}), we have the output state over the whole system as
\begin{align}
    &\rho_{R A_t \bm B} = (\mathbb{I}_R \otimes E) \left(\ketbra{\Phi}{\Phi}_{RA_0} \left(\otimes_{k=0}^{t-1} \omega\right) \otimes \ketbra{0}{0}\right) (\mathbb{I}_R \otimes E)^\dagger\\
    &= \frac{1}{d_A} \sum_{\substack{\bm a, \bm b,\\ \bm a', \bm b'}}\prod_{k=0}^{t-1} \braket{a_{k+1} b_{2k+1}|U_k|a_k b_{2k}} \braket{a_k' b_{2k}'|U_k^\dagger|a_{k+1}' b_{2k+1}'} \ketbra{a_0}{a_0'}_R \ketbra{a_t}{a_t'}_{A_t} \left(\otimes_{k=0}^{t-1} \ketbra{b_{2k+1}}{b_{2k+1}'}\right) \braket{b_0|0}\braket{0|b_0'}\otimes_{k=1}^{t-1} \braket{b_{2k}|\omega|b_{2k}'}\\
    &= \frac{1}{d_A} \frac{1}{d_B^{t-1}}\sum_{\substack{\bm a, \bm b,\\ \bm a', \bm b'}}\prod_{k=0}^{t-1} \braket{a_{k+1} b_{2k+1}|U_k|a_k b_{2k}} \braket{a_k' b_{2k}'|U_k^\dagger|a_{k+1}' b_{2k+1}'} \ketbra{a_0}{a_0'}_R \ketbra{a_t}{a_t'}_{A_t} \left(\otimes_{k=0}^{t-1} \ketbra{b_{2k+1}}{b_{2k+1}'}\right) \delta_{b_0, 0}\delta_{b_0',0} \prod_{k=1}^{t-1} \delta_{b_{2k},b_{2k}'}. 
\end{align}

For the reduced state $\rho_R$, we have
\begin{align}
    &\rho_R = \tr_{A_t \bm B}(\rho_{R A_t \bm B})\\
    &= \frac{1}{d_A} \frac{1}{d_B^{t-1}}\sum_{\substack{\bm a, \bm b,\\ \bm a', \bm b'}}\prod_{k=0}^{t-1} \braket{a_{k+1} b_{2k+1}|U_k|a_k b_{2k}} \braket{a_k' b_{2k}'|U_k^\dagger|a_{k+1}' b_{2k+1}'} \ketbra{a_0}{a_0'}_R \delta_{a_t, a_t'} \prod_{k=0}^{t-1} \delta_{b_{2k+1},b_{2k+1}'} \delta_{b_0, 0}\delta_{b_0',0} \prod_{k=1}^{t-1} \delta_{b_{2k},b_{2k}'}\\
    &= \frac{1}{d_A} \frac{1}{d_B^{t-1}} \sum_{\bm a, \bm a'}  \prod_{k=1}^{t-1} \tr\left(U_k \left(\ketbra{a_k}{a_k'}\otimes \bI\right) U_k^\dagger \left(\ketbra{a_{k+1}'}{a_{k+1}}\otimes \bI\right)\right) \tr\left(U_0 \left(\ketbra{a_0}{a_0'}\otimes \ketbra{0}{0}\right) U_0^\dagger \left(\ketbra{a_1'}{a_1}\otimes \bI\right)\right) \ketbra{a_0}{a_0'}_R \delta_{a_t, a_t'}.
\end{align}
The purity of $\rho_R$ then can be evaluated as
\begin{align}
    &\tr(\rho_R^2)\nonumber\\
    &= \frac{1}{d_A^2} \frac{1}{d_B^{2t-2}} \sum_{\bm a, \bm a'}  \prod_{k=1}^{t-1} \tr\left(U_k \left(\ketbra{a_k}{a_k'}\otimes \bI\right) U_k^\dagger \left(\ketbra{a_{k+1}'}{a_{k+1}}\otimes \bI\right)\right) \tr\left(U_0 \left(\ketbra{a_0}{a_0'}\otimes \ketbra{0}{0}\right) U_0^\dagger \left(\ketbra{a_1'}{a_1}\otimes \bI\right)\right) \delta_{a_t, a_t'} \nonumber\\
    & \qquad \quad \times\sum_{\bm \alpha, \bm \alpha'}  \prod_{k=1}^{t-1} \tr\left(U_k \left(\ketbra{\alpha_k}{\alpha_k'}\otimes \bI \right) U_k^\dagger \left(\ketbra{\alpha_{k+1}'}{\alpha_{k+1}}\otimes \bI\right)\right) \tr\left(U_0 \left(\ketbra{\alpha_0}{\alpha_0'}\otimes \ketbra{0}{0}\right) U_0^\dagger \left(\ketbra{\alpha_1'}{\alpha_1}\otimes \bI\right)\right) \delta_{\alpha_t, \alpha_t'} \delta_{a_0, \alpha_0'}\delta_{a_0',\alpha_0} \\
    &= \frac{1}{d_A^2} \frac{1}{d_B^{2t-2}} \sum_{\substack{\bm a, \bm a', \\ \bm \alpha, \bm \alpha'}} \prod_{k=1}^{t-1} \tr\left(U_k^{\otimes 2} \left(\ketbra{a_k \alpha_k}{a_k'\alpha_k'}\otimes \bI^{\otimes 2}\right) U_k^\dagger{}^{\otimes 2} \left(\ketbra{a_{k+1}' \alpha_{k+1}'}{a_{k+1} \alpha_{k+1}}\otimes \bI^{\otimes 2}\right)\right) \nonumber\\
    &\qquad \qquad \qquad \times \tr\left(U_0^{\otimes 2} \left(\ketbra{a_0 \alpha_0}{a_0' \alpha_0'}\otimes \ketbra{0}{0}^{\otimes 2}\right) U_0^\dagger{}^{\otimes 2} \left(\ketbra{a_1' \alpha_1'}{a_1 \alpha_1}\otimes \bI^{\otimes 2}\right)\right) \delta_{a_t, a_t'} \delta_{\alpha_t, \alpha_t'} \delta_{a_0, \alpha_0'}\delta_{a_0',\alpha_0},
\end{align}
which is represented in tensor network in Fig.~\ref{fig:uncond_purity_mix_diagram}a.

Recall that $\E_{\rm Haar}\left[U_{t-1}^{\otimes 2} \otimes U_{t-1}^*{}^{\otimes 2}\right]|e)_A |e)_B = |e)_{AB}$ from Eq.~\eqref{eq:twirling_ee}, the following concatenation with $|e)(e|_B$ results in
\be
    |e)(e|_B\E_{\rm Haar}\left[U_{t-1}^{\otimes 2} \otimes U_{t-1}^*{}^{\otimes 2}\right]|e)_A |e)_B = d_B^2 |e)_A |e)_B,
\ee
As the boundary condition is exactly reproduced in the following step with an extra coefficient $d_B^2$, then we can directly obtain the solution of purity as
\be
    \E_{\rm Haar} \tr(\rho_R^2) = \frac{1}{d_A^2} \frac{1}{d_B^{2t-2}} d_B^{2(t-1)} (\tau|_A (0|_B |e)_{A} |e)_B = \frac{1}{d_A}.
    \label{eq:rhoR_purity_mix_sol}
\ee

\begin{figure}[t]
    \centering
    \includegraphics[width=0.5\textwidth]{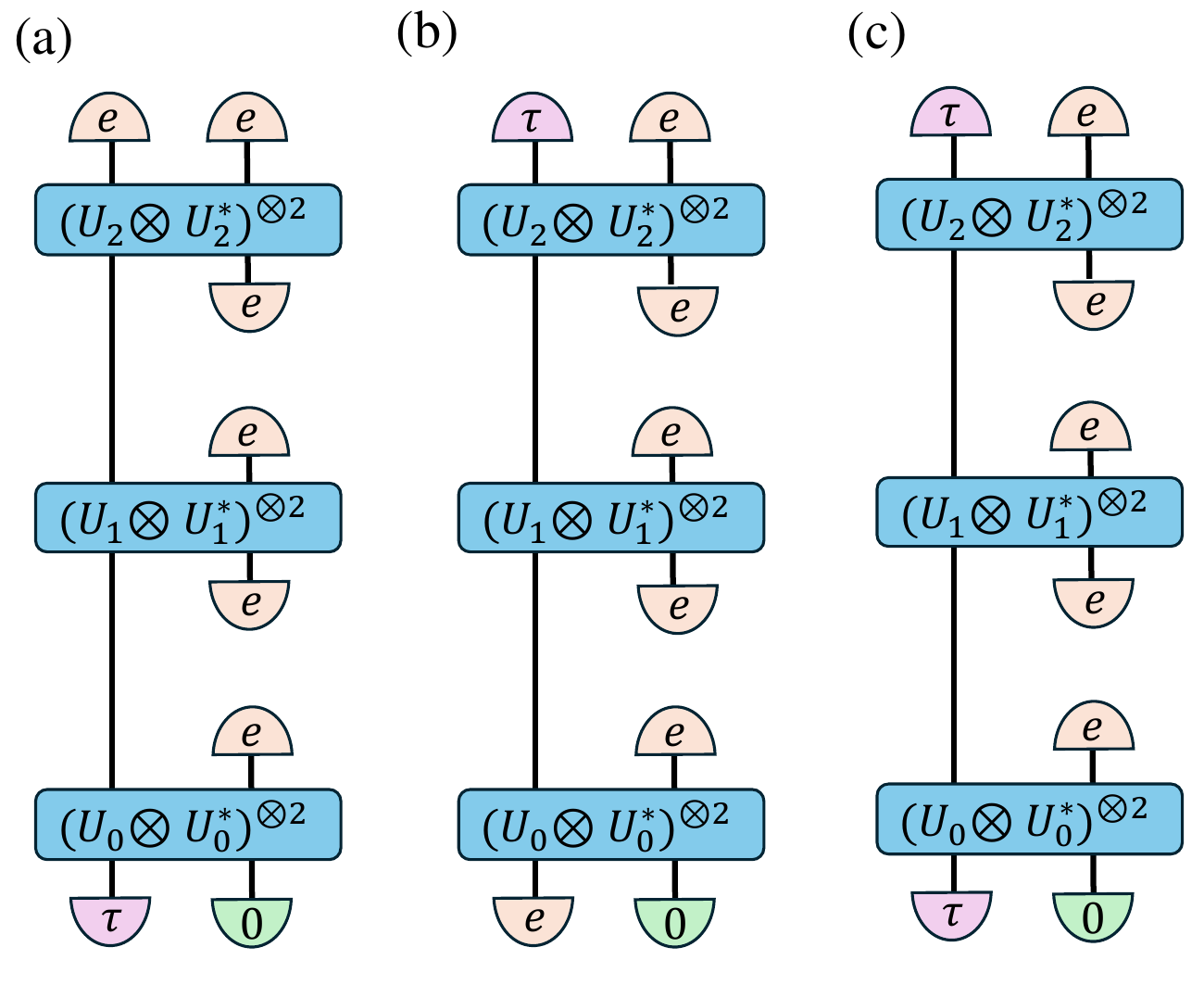}
    \caption{Tensor network representation of Haar-averaged purity for reduced states (a) $\rho_R$, (b) $\rho_{A_t}$ and (c) $\rho_{R A_t}$ with fully-mixed state reset and time steps $t=3$.}
    \label{fig:uncond_purity_mix_diagram}
\end{figure}

Next, we move to the reduced state $\rho_{A_t}$.
\begin{align}
    &\rho_{A_t} = \tr_{R \bm B}(\rho_{R A_t \bm B})\\
    &= \frac{1}{d_A}\frac{1}{d_B^{t-1}} \sum_{\substack{\bm a, \bm b,\\ \bm a', \bm b'}}\prod_{k=0}^{t-1} \braket{a_{k+1} b_{2k+1}|U_k|a_k b_{2k}} \braket{a_k' b_{2k}'|U_k^\dagger|a_{k+1}' b_{2k+1}'} \delta_{a_0, a_0'} \ketbra{a_t}{a_t'}_{A_t} \prod_{k=0}^{t-1} \delta_{b_{2k+1},b_{2k+1}'} \delta_{b_0, 0}\delta_{b_0',0} \prod_{k=1}^{t-1} \delta_{b_{2k},b_{2k}'}\\
    &= \frac{1}{d_A} \frac{1}{d_B^{t-1}} \sum_{\bm a, \bm a'}  \prod_{k=1}^{t-1} \tr\left(U_k \left(\ketbra{a_k}{a_k'}\otimes \bI\right) U_k^\dagger \left(\ketbra{a_{k+1}'}{a_{k+1}}\otimes \bI\right)\right) \tr\left(U_0 \left(\ketbra{a_0}{a_0'}\otimes \ketbra{0}{0}\right) U_0^\dagger \left(\ketbra{a_1'}{a_1}\otimes \bI\right)\right) \delta_{a_0, a_0'} \ketbra{a_t}{a_t'}_{A_t},
\end{align}
and the purity is
\begin{align}
    &\tr(\rho_{A_t}^2)\nonumber\\
    &= \frac{1}{d_A^2} \frac{1}{d_B^{2t-2}} \sum_{\bm a, \bm a'}  \prod_{k=1}^{t-1} \tr\left(U_k \left(\ketbra{a_k}{a_k'}\otimes \bI\right) U_k^\dagger \left(\ketbra{a_{k+1}'}{a_{k+1}}\otimes \bI\right)\right) \tr\left(U_0 \left(\ketbra{a_0}{a_0'}\otimes \ketbra{0}{0}\right) U_0^\dagger \left(\ketbra{a_1'}{a_1}\otimes \bI\right)\right) \delta_{a_0, a_0'} \nonumber\\
    & \qquad \quad \times \sum_{\bm \alpha, \bm \alpha'}  \prod_{k=1}^{t-1} \tr\left(U_k \left(\ketbra{\alpha_k}{\alpha_k'}\otimes \bI\right) U_k^\dagger \left(\ketbra{\alpha_{k+1}'}{\alpha_{k+1}}\otimes \bI\right)\right) \tr\left(U_0 \left(\ketbra{\alpha_0}{\alpha_0'}\otimes \ketbra{0}{0}\right) U_0^\dagger \left(\ketbra{\alpha_1'}{\alpha_1}\otimes \bI\right)\right) \delta_{\alpha_0, \alpha_0'} \delta_{a_t, \alpha_t'}\delta_{a_t',\alpha_t}\\
    &= \frac{1}{d_A^2}\frac{1}{d_B^{2t-2}} \sum_{\substack{\bm a, \bm a', \\ \bm \alpha, \bm \alpha'}} \prod_{k=1}^{t-1} \tr\left(U_k^{\otimes 2} \left(\ketbra{a_k \alpha_k}{a_k'\alpha_k'}\otimes \bI^{\otimes 2}\right) U_k^\dagger{}^{\otimes 2} \left(\ketbra{a_{k+1}' \alpha_{k+1}'}{a_{k+1} \alpha_{k+1}}\otimes \bI^{\otimes 2}\right)\right) \nonumber\\
    &\qquad \qquad \qquad \times \tr\left(U_0^{\otimes 2} \left(\ketbra{a_0 \alpha_0}{a_0' \alpha_0'}\otimes \ketbra{0}{0}^{\otimes 2}\right) U_0^\dagger{}^{\otimes 2} \left(\ketbra{a_1' \alpha_1'}{a_1 \alpha_1}\otimes \bI^{\otimes 2}\right)\right) \delta_{a_0, a_0'} \delta_{\alpha_0, \alpha_0'} \delta_{a_t, \alpha_t'}\delta_{a_t',\alpha_t},
\end{align}
which is shown in Fig.~\ref{fig:uncond_purity_mix_diagram}b.

Recall from Eq.~\eqref{eq:twirling_taue} $\E_{\rm Haar}\left[U_{t-1}^{\otimes 2} \otimes U_{t-1}^*{}^{\otimes 2}\right]|\tau)_{A}|e)_{B} = \frac{d_A(d_B^2-1)}{d_A^2d_B^2-1} |e)_{AB} + \frac{d_B(d_A^2 - 1)}{d_A^2d_B^2-1} |\tau)_{AB}$, we next apply $|e)(e|_B$ and have
\be
    |e)(e|_B \E_{\rm Haar}\left[U_{t-1}^{\otimes 2} \otimes U_{t-1}^*{}^{\otimes 2}\right]|\tau)_{A}|e)_{B} = \frac{d_A d_B^2(d_B^2-1)}{d_A^2 d_B^2-1} |e)_A |e)_B + \frac{d_B^2(d_A^2 - 1)}{d_A^2d_B^2-1} |\tau)_A |e)_B,
\ee
Moving downward in Fig.~\ref{fig:uncond_purity_mix_diagram}b up to $U_1$, we have the contribution from the two boundary conditions above as
\begin{itemize}
    \item $|\tau)_A|e)_B$: $\left(\frac{d_B^2(d_A^2 - 1)}{d_A^2 d_B^2-1}\right)^{t-1} |\tau)_A |e)_B$.
    
    \item $|e)_A|e)_B$: $\frac{d_A d_B^2(d_B^2-1)}{d_A^2 d_B^2-1} \sum_{k=0}^{t-2} \left(\frac{d_B^2 (d_A^2 - 1)}{d_A^2 d_B^2-1}\right)^k d_B^{2(t-2-k)} |e)_A |e)_B$.
\end{itemize}
After finishing the twirling over $U_0$, we have the dynamical solution for purity of $\rho_{A_t}$
\begin{align}
    &\E_{\rm Haar} \tr(\rho_{A_t}^2)\nonumber\\
    &= \frac{1}{d_A^2} \frac{1}{d_B^{2(t-1)}}\left[\left(\frac{d_B^2(d_A^2 - 1)}{d_A^2 d_B^2-1}\right)^{t-1} \frac{d_B(d_A^2 - 1)}{d_A^2d_B^2-1} d_A + \left(\frac{d_B^2(d_A^2 - 1)}{d_A^2 d_B^2-1}\right)^{t-1} \frac{d_A(d_B^2-1)}{d_A^2d_B^2-1} d_A^2 \nonumber \right.\\
    &\left. \qquad \qquad \qquad + \frac{d_A d_B^2(d_B^2-1)}{d_A^2 d_B^2-1} \sum_{k=0}^{t-2} \left(\frac{d_B^2 (d_A^2 - 1)}{d_A^2 d_B^2-1}\right)^k d_B^{2(t-2-k)} d_A^2\right]\\
    &= \frac{1}{d_A^2} \frac{1}{d_B^{2(t-1)}} \left[\frac{d_A}{d_B}\left(\frac{d_B^2(d_A^2 - 1)}{d_A^2 d_B^2-1}\right)^{t} + \frac{d_A^3 (d_B^2-1)}{(d_A^2-1)d_B^2}\left(\frac{d_B^2(d_A^2 - 1)}{d_A^2 d_B^2-1}\right)^{t} + \frac{d_A d_B^{2(t-1)}}{d_A^2 - 1}\left(d_A^2-1-(d_A^2 d_B^2-1) \left(\frac{d_A^2 - 1}{d_A^2 d_B^2 - 1}\right)^t \right)   \right] \\
    &= \frac{d_B}{d_A}\left(\frac{d_A^2 - 1}{d_A^2 d_B^2-1}\right)^{t} + \frac{d_A (d_B^2-1)}{d_A^2-1}\left(\frac{d_A^2 - 1}{d_A^2 d_B^2-1}\right)^{t} + \frac{1}{d_A(d_A^2 - 1)}\left(d_A^2-1-(d_A^2 d_B^2-1) \left(\frac{d_A^2 - 1}{d_A^2 d_B^2 - 1}\right)^t \right)\\
    &= \frac{d_B-1}{d_A} \left(\frac{d_A^2 - 1}{d_A^2 d_B^2-1}\right)^{t} + \frac{1}{d_A}.
    \label{eq:rhoA_purity_mix_sol}
\end{align}

Last, the reduce state $\rho_{R A_t}$ is
\begin{align}
    &\rho_{R A_t} = \tr_{\bm B}(\rho_{R A_t \bm B})\\
    &= \frac{1}{d_A} \frac{1}{d_B^{t-1}}\sum_{\substack{\bm a, \bm b,\\ \bm a', \bm b'}}\prod_{k=0}^{t-1} \braket{a_{k+1} b_{2k+1}|U_k|a_k b_{2k}} \braket{a_k' b_{2k}'|U_k^\dagger|a_{k+1}' b_{2k+1}'} \ketbra{a_0}{a_0'}_R \ketbra{a_t}{ a_t'}_{A_t} \prod_{k=0}^{t-1} \delta_{b_{2k+1},b_{2k+1}'} \delta_{b_0, 0} \delta_{b_0',0} \prod_{k=1}^{t-1} \delta_{b_{2k},b_{2k}'}\\
    &= \frac{1}{d_A} \frac{1}{d_B^{t-1}} \sum_{\bm a, \bm a'}  \prod_{k=1}^{t-1} \tr\left(U_k \left(\ketbra{a_k}{a_k'}\otimes \bI \right) U_k^\dagger \left(\ketbra{a_{k+1}'}{a_{k+1}}\otimes \bI\right)\right) \tr\left(U_0 \left(\ketbra{a_0}{a_0'}\otimes \ketbra{0}{0}\right) U_0^\dagger \left(\ketbra{a_1'}{a_1}\otimes \bI\right)\right) \ketbra{a_0}{a_0'}_R \ketbra{a_t}{ a_t'}_{A_t},
\end{align}
and the purity is
\begin{align}
    &\tr(\rho_{R A_t}^2)\nonumber\\
    &= \frac{1}{d_A^2} \frac{1}{d_B^{2t-2}} \sum_{\bm a, \bm a'}  \prod_{k=1}^{t-1} \tr\left(U_k \left(\ketbra{a_k}{a_k'}\otimes \bI \right) U_k^\dagger \left(\ketbra{a_{k+1}'}{a_{k+1}}\otimes \bI\right)\right) \tr\left(U_0 \left(\ketbra{a_0}{a_0'}\otimes \ketbra{0}{0}\right) U_0^\dagger \left(\ketbra{a_1'}{a_1}\otimes \bI\right)\right) \delta_{a_0, \alpha_0'} \delta_{a_0', \alpha_0} \nonumber\\
    & \qquad \quad \times \sum_{\bm \alpha, \bm \alpha'}  \prod_{k=1}^{t-1} \tr\left(U_k \left(\ketbra{\alpha_k}{\alpha_k'}\otimes \bI \right) U_k^\dagger \left(\ketbra{\alpha_{k+1}'}{\alpha_{k+1}}\otimes \bI\right)\right) \tr\left(U_0 \left(\ketbra{\alpha_0}{\alpha_0'}\otimes \ketbra{0}{0}\right) U_0^\dagger \left(\ketbra{\alpha_1'}{\alpha_1}\otimes \bI\right)\right) \delta_{a_t, \alpha_t'} \delta_{a_t', \alpha_t}\\
    &= \frac{1}{d_A^2} \frac{1}{d_B^{2t-2}} \sum_{\substack{\bm a, \bm a', \\ \bm \alpha, \bm \alpha'}} \prod_{k=1}^{t-1} \tr\left(U_k^{\otimes 2} \left(\ketbra{a_k \alpha_k}{a_k'\alpha_k'}\otimes \bI^{\otimes 2}\right) U_k^\dagger{}^{\otimes 2} \left(\ketbra{a_{k+1}' \alpha_{k+1}'}{a_{k+1} \alpha_{k+1}}\otimes \bI^{\otimes 2}\right)\right) \nonumber\\
    &\qquad \qquad \qquad \times \tr\left(U_0^{\otimes 2} \left(\ketbra{a_0 \alpha_0}{a_0' \alpha_0'}\otimes \ketbra{0}{0}^{\otimes 2}\right) U_0^\dagger{}^{\otimes 2} \left(\ketbra{a_1' \alpha_1'}{a_1 \alpha_1}\otimes \bI^{\otimes 2}\right)\right) \delta_{a_0, \alpha_0'} \delta_{a_0', \alpha_0} \delta_{a_t, \alpha_t'} \delta_{a_t', \alpha_t},
\end{align}
which is represented as tensor network as in Fig.~\ref{fig:uncond_purity_mix_diagram}c. Since it shares the same boundary condition as in Fig.~\ref{fig:uncond_purity_mix_diagram}b, the contribution toward $U_1$ remains the same as in $\tr(\rho_{A_t}^2)$.
The dynamical solution for the purity of $\rho_{RA_t}$ is
\begin{align}
    &\E_{\rm Haar} \tr(\rho_{R A_t}^2) \nonumber\\
    &= \frac{1}{d_A^2} \frac{1}{d_B^{2(t-1)}}\left[\left(\frac{d_B^2(d_A^2 - 1)}{d_A^2 d_B^2-1}\right)^{t-1} \frac{d_B(d_A^2 - 1)}{d_A^2d_B^2-1} d_A^2 + \left(\frac{d_B^2(d_A^2 - 1)}{d_A^2 d_B^2-1}\right)^{t-1} \frac{d_A(d_B^2-1)}{d_A^2d_B^2-1} d_A \nonumber \right.\\
    &\left. \qquad \qquad \qquad + \frac{d_A (d_B^2-1)}{d_A^2 d_B^2-1} d_B^{2(t-1)}\sum_{k=0}^{t-2} \left(\frac{d_B^2 (d_A^2 - 1)}{d_A^2 d_B^2-1}\right)^k d_B^{-2k} d_A\right]\\
    &= \frac{1}{d_A^2} \frac{1}{d_B^{2(t-1)}} \left[\frac{d_A^2}{d_B}\left(\frac{d_B^2(d_A^2 - 1)}{d_A^2 d_B^2-1}\right)^{t} + \frac{d_A^2 (d_B^2-1)}{(d_A^2-1)d_B^2}\left(\frac{d_B^2(d_A^2 - 1)}{d_A^2 d_B^2-1}\right)^{t} + \frac{d_B^{2(t-1)}}{d_A^2 - 1}\left(d_A^2-1-(d_A^2 d_B^2-1) \left(\frac{d_A^2 - 1}{d_A^2 d_B^2 - 1}\right)^t \right)   \right] \\
    &= \left(d_B - \frac{1}{d_A^2}\right) \left(\frac{d_A^2 - 1}{d_A^2 d_B^2-1}\right)^{t} + \frac{1}{d_A^2}.
    \label{eq:rhoRA_purity_mix_sol}
\end{align}

We again switch the order of ensemble average and logarithmic, and combine the results in Eqs.~\eqref{eq:rhoR_purity_mix_sol},~\eqref{eq:rhoA_purity_mix_sol} and~\eqref{eq:rhoRA_purity_mix_sol}.
\begin{align}
    &\E_{\rm Haar} I_2(R:A_t) \simeq -\log_2 \E_{\rm Haar}\tr(\rho_R^2) - \log_2 \E_{\rm Haar}\tr(\rho_{A_t}^2) + \log_2 \E_{\rm Haar}\tr(\rho_{R A_t}^2) \\
    &= N_A - \log_2 \left[\frac{d_B-1}{d_A} \left(\frac{d_A^2 - 1}{d_A^2 d_B^2-1}\right)^{t} + \frac{1}{d_A}\right] + \log_2\left[\left(d_B - \frac{1}{d_A^2}\right) \left(\frac{d_A^2 - 1}{d_A^2 d_B^2-1}\right)^{t} + \frac{1}{d_A^2}\right],
\end{align}
which is the result of Eq.~\eqref{eq:MI_traceout_id} in the main text. In the large limit of system $d_A, d_B \gg 1$, we perform the asymptotic analysis as
\begin{align}
    \E_{\rm Haar} I_2(R:A_t) &\simeq N_A - \log_2 \left(\frac{1}{d_A} d_B^{1-2t} + \frac{1}{d_A}\right) + \log_2\left(d_B^{1-2t} + \frac{1}{d_A^2}\right) \\
    &\simeq 2N_A - d_B^{1-2t} + \log_2 \left(d_A^{-2} + d_B^{1-2t}\right).
\end{align}
In the early time $t \ll N_A/N_B + 1/2$, we can further approximate it as
\begin{align}
    \E_{\rm Haar} I_2(R:A_t) &\simeq 2N_A - d_B^{1-2t} + \log_2 d_B^{1-2t} \left(d_A^{-2}d_B^{2t-1} + 1\right)\\
    &\simeq 2N_A - d_B^{1-2t} + (1-2t)N_B + d_A^{-2}d_B^{2t-1}\\
    &\simeq 2N_A + N_B - 2t N_B + \mathcal{O}(d_B^{1-2t}). 
\end{align}
While for the late time $t \gg N_A/N_B + 1/2$, we can further approximate it as
\begin{align}
    \E_{\rm Haar} I_2(R:A_t) &\simeq 2N_A - d_B^{1-2t} + \log_2 d_A^{-2} \left(1 + d_A^{2}d_B^{1-2t}\right)\\
    &\simeq 2N_A - d_B^{1-2t} -2 N_A + d_A^2 d_B^{1-2t}\\
    & \simeq d_A^2 d_B^{1-2t}.
\end{align}

\section{Derivation of measurement-conditioned QMI in partially monitored dynamics (Theorem~\ref{condMI_partial_theorem})}
\label{app:condMI_partial_proof}

In this section, we derive the dynamics of the measurement-conditioned QMI in a monitored dynamics with periodically doped erasure channels.
Due to the existence of erasure channels, the conditioned state is a mixed state in general, and thus we consider the R\'enyi-2 extended measurement-conditioned QMI for analysis convenience as
\begin{align}
    \overline{I_2(R:A_t|\bfz)} &= \E_\bfz I_2(R:A_t|\bfz) = \E_\bfz \left[S_2(R|\bfz) + S_2(A_t|\bfz) - S_2(RA_t|\bfz)\right]\\
    &\simeq -\log_2\left(\E_\bfz \tr\left(\rho_{R|\bfz}^2\right)\right) - \log_2\left(\E_\bfz \tr\left(\rho_{A_t|\bfz}^2\right)\right) + \log_2\left(\E_\bfz \tr\left(\rho_{RA_t|\bfz}^2\right)\right),
    \label{eq:condQMI_partial_def}
\end{align}
where the approximation can be regarded as annealed approximation to the quenched quantity. Here we still assume that the initial state of reference and system is a maximally entangled state with $N_R = N_A$.

For notation convenience, we denote $B'$ to represent the bath qubits that are not erasured in arbitrary step. The conditioned state on the reference and system at step $t$ is
\be
    \rho_{RA_t|\bfz} = \tr_{B}\left(U \ketbra{\Phi}{\Phi}_{RA} \ketbra{0}{0}_{\bm B} U^\dagger \bI_{\bm E} \ketbra{\bfz}{\bfz}_{\bm B'}\right)/p_\bfz \equiv \tilde{\rho}_{RA_t|\bfz} / p_\bfz,
\ee
where $p_\bfz = \tr(\tilde{\rho}_{RA_t|\bfz})$ is the probability of measurement trajectory $\bfz$. Here $\bm B, \bm E, \bm B'$ represent the expanded bath system over all past temporal steps.
The reduced state of $\rho_{R|\bfz}$ and $\rho_{A_t|\bfz}$ can be written out similarly.
Now we evaluate the conditioned-purity of $\rho_{R|\bfz}$, which is
\begin{align}
    \overline{\gamma}_R &= \sum_\bfz p_\bfz \tr \rho_{R|\bfz}^2 \\
    &= \sum_\bfz p_\bfz^{-1} \tr(\tilde{\rho}_{R|\bfz}^2),
\end{align}
and thus we utilize the replica trick to defined the pseudo conditioned-purity as
\begin{align}
    \overline{\gamma}_R^{(m)} &\equiv  \sum_\bfz p_\bfz^{m} \tr(\tilde{\rho}_{R|\bfz}^2) \\
    &= \sum_\bfz \tr\left(U^{\otimes (m+2)} \ketbra{\Phi}{\Phi}^{\otimes (m+2)}_{RA} \ketbra{0}{0}^{\otimes (m+2)}_{\bm B} U^\dagger{}^{\otimes (m+2)} \bI^{\otimes m}_R\tau_R \bI^{\otimes (m+2)}_A \bI^{\otimes (m+2)}_{\bm E} \ketbra{\bfz}{\bfz}_{\bm B'}^{\otimes (m+2)}\right)\\
    &= d_A^{-(m+2)}\tr\left(U^{\otimes (m+2)} \bI_A^{\otimes m} \tau_A \ketbra{0}{0}^{\otimes (m+2)}_{\bm B} U^\dagger{}^{\otimes (m+2)} \bI^{\otimes (m+2)}_A \bI^{\otimes (m+2)}_{\bm E} \sum_\bfz \ketbra{\bfz}{\bfz}_{\bm B'}^{\otimes (m+2)}\right).
\end{align}
Therefore, there exist two types of boundary condition for bath in the unitary twirling, $|D_{m+2})_{B}$ and $|e)_E |D_{m+2})_{B'}$, which correspond to the case without and with erasure channels, respectively.

For $|D_{m+2})_{B}$, we have derived the twirling results in Appendix~\ref{app:theorem1}, and the transition can be efficiently represented by the linear transformation matrix $Q$ (see also in Eq.~\eqref{eq:Q_def}) as
\be
    Q = \begin{pmatrix}
        q_\tau, &q_e\\
        q_e, & q_\tau
    \end{pmatrix}
    \simeq d_B^{-(m+1)}\begin{pmatrix}
        1 & d_A^{-1} \\
        d_A^{-1} & 1
    \end{pmatrix},
    \label{eq:Q_def_asymp}
\ee
where the approximation holds when $d_A, d_B \gg 1$. We remind the readers that here we take the mapping of $|\tau_2)_A \to (1,0)^T$ and $|e)_A \to (0, 1)^T$ omitting the operator on bath.

On the other hand, for $|e)_E |D_{m+2})_{B'}$, when the system boundary condition is $|e)_A$, we have
\begin{align}
    &(0|_{EB'}\E_{\rm Haar}\left[U^{\otimes (m+2)} \otimes U^*{}^{\otimes (m+2)}\right] |e)_A |e)_E |D_{m+2})_{B'} = \sum_{\sigma, \pi \in S_{m+2}} {\rm Wg}(\sigma^{-1}\pi,m+2) |\sigma)_A (\pi|e)_A (\pi|e)_E (\pi|D_{\rm m+2})_{B'}\\
    &= d_{B'} \sum_{\sigma, \pi} {\rm Wg}(\sigma^{-1}\pi,m+2) |\sigma)_A d_{AE}^{|\pi|}\\
    &\simeq d_{B'} \left[{\rm Wg}(e,m+2)\left(|e)_A d_{AE}^{m+2} + |\tau_2)_A d_{AE}^{m+1} \right) + {\rm Wg}(\tau_2,m+2)\left(|e)_A d_{AE}^{m+1} + |\tau_2)_A d_{AE}^{m+2} \right) \right]\\
    &\simeq d_{B'} \left[d_{AB}^{-(m+2)}\left(|e)_A d_{AE}^{m+2} + |\tau_2)_A d_{AE}^{m+1} \right) - d_{AB}^{-(m+3)}\left(|e)_A d_{AE}^{m+1} + |\tau_2)_A d_{AE}^{m+2} \right) \right]\\
    &= d_{B'}^{-(m+1)}\left(1-d_{AE}^{-2}d_{B'}^{-1}\right) |e)_A + d_{B'}^{-(m+1)}d_{AE}^{-1}\left(1-d_{B'}^{-1}\right)|\tau_2)_A.
\end{align}
Meanwhile, for the system boundary condition to be $|\tau_2)_A$, we have
\begin{align}
    &(0|_{EB'}\E_{\rm Haar}\left[U^{\otimes (m+2)} \otimes U^*{}^{\otimes (m+2)}\right] |\tau_2)_A |e)_E |D_{m+2})_{B'} = \sum_{\sigma, \pi \in S_{m+2}} {\rm Wg}(\sigma^{-1}\pi,m+2) |\sigma)_A (\pi|\tau_2)_A (\pi|e)_E (\pi|D_{\rm m+2})_{B'}\\
    &= d_{B'}\sum_{\sigma, \pi} {\rm Wg}(\sigma^{-1}\pi,m+2) |\sigma)_A d_{A}^{|\pi^{-1}\tau_2|}d_E^{|\pi|}\\
    &\simeq d_{B'} \left[{\rm Wg}(e,m+2)\left(|e)_A d_{A}^{m+1} d_E^{m+2} + |\tau_2)_A d_{A}^{m+2} d_E^{m+1} \right) + {\rm Wg}(\tau_2,m+2)\left(|e)_A d_{A}^{m+2} d_E^{m+1} + |\tau_2)_A d_{A}^{m+1} d_E^{m+2} \right) \right]\\
    &\simeq d_{B'} \left[d_{AB}^{-(m+2)}\left(|e)_A d_{A}^{m+1} d_E^{m+2} + |\tau_2)_A d_{A}^{m+2} d_E^{m+1} \right) -d_{AB}^{-(m+3)}\left(|e)_A d_{A}^{m+2} d_E^{m+1} + |\tau_2)_A d_{A}^{m+1} d_E^{m+2} \right) \right]\\
    &= d_{B'}^{-(m+1)} d_A^{-1}\left(1-d_{B'}^{-1}d_E^{-2}\right) |e)_A + d_{B'}^{-(m+1)}d_E^{-1}\left(1-d_A^{-2} d_{B'}^{-1}\right) |\tau_2)_A.
\end{align}
Now we can summarize the transformation by
\be
    Q' = d_{B'}^{-(m+1)}\begin{pmatrix}
        d_E^{-1}(1-d_{A}^{-2} d_{B'}^{-1}) & d_{AE}^{-1}(1-d_{B'}^{-1})\\
        d_A^{-1}(1-d_{E}^{-2} d_{B'}^{-1}) & 1-d_{AE}^{-2}d_{B'}^{-1}
    \end{pmatrix}
    \simeq d_{B'}^{-(m+1)} \begin{pmatrix}
        d_E^{-1} & d_{A}^{-1} d_E^{-1} \\
        d_A^{-1} & 1
    \end{pmatrix},
    \label{eq:Qprime_def_asymp}
\ee
where the approximation holds when $d_A, d_{B'} \gg 1$.

With the representation of $Q$ and $Q'$, we can in general write out the whole dynamics in permutation basis as
\begin{align}
    (0|_B \E_{\rm Haar}\left[U^{\otimes (m+2)} \otimes U^*{}^{\otimes (m+2)}\right] |e)_{\bm E}|D_{m+2})_{\bm B'} \to Q^{t-s n_s}\left(Q' Q^{s-1}\right)^{n_s},
    \label{eq:twirling_partial_full}
\end{align}
where $n_s = \lfloor t/s\rfloor$ counts the number of erasure channel periods. Overall, one can apply this linear transformation to different boundary conditions and obtain the corresponding purities of interest, which lead to the R\'enyi-2 extended measurement-conditioned QMI in Eq.~\eqref{eq:condQMI_partial_def}.

The exact analytical formula for the QMI is challenging due to the fact that the matrix $Q'$ is not symmetric. Therefore, to obtain analytical insight, we consider the asymptotic limit of $d_A, d_{B'} \gg 1$, and thus only focus on the asymptotic form of $Q$ and $Q'$ in Eqs.~\eqref{eq:Q_def_asymp} and~\eqref{eq:Qprime_def_asymp}. We further notice that the asymptotic form of $Q'$ is directly connected to $Q$ by another diagonal matrix multiplication as
\begin{align}
    Q' = {\rm diag}(d_E^{-1}, 1) Q \equiv \Lambda Q,
\end{align}
and $Q = d_B^{-(m+1)} (\bI + d_A^{-1}\sigma_x)$ where $\sigma_x$ is the Pauli-X matrix.
The above transformation in Eq.~\eqref{eq:twirling_partial_full} can be reduced to 
\begin{align}
    Q^{t-s n_s}\left(Q' Q^{s-1}\right)^{n_s}  &= Q^{t-s n_s}\left(\Lambda Q^{s}\right)^{n_s}\\
    &= d_B^{-(m+1)(t-s n_s)}\left(\bI+\frac{\sigma_x}{d_A}\right)^{t-s n_s}\left[ \Lambda \left(d_B^{-(m+1)s}\left(\bI+\frac{\sigma_x}{d_A}\right)^{s}\right)\right]^{n_s}\\
    &= d_B^{-(m+1)(t-s n_s)}\left(\bI+(t-sn_s)\frac{\sigma_x}{d_A}\right) d_B^{-(m+1)s n_s} \left(\Lambda+\frac{s}{d_A}\Lambda\sigma_x\right)^{n_s} + \mathcal{O}\left(\frac{1}{d_A^2}\right)\\
    &= d_B^{-(m+1)t} \left(\bI+(t-sn_s)\frac{\sigma_x}{d_A}\right) \left(\Lambda+\frac{s}{d_A}\Lambda\sigma_x\right)^{n_s}.
\end{align}
For $\left(\Lambda+\frac{s}{d_A}\Lambda\sigma_x\right)^{n_s}$, we have
\begin{align}
    \left(\Lambda+\frac{s}{d_A}\Lambda\sigma_x\right)^{n_s} &= \Lambda^{n_s} + \frac{s}{d_A} \sum_{i=0}^{n_s-1} \Lambda^{n_s-1-i}\Lambda \sigma_x \Lambda^{i} + \mathcal{O}\left(\frac{1}{d_A^2}\right) \\
    &= \Lambda^{n_s} + \frac{s}{d_A} \sum_{i=0}^{n_s-1} \Lambda^{n_s-i}\sigma_x \Lambda^{i}\\
    &= \Lambda^{n_s} + \frac{s}{d_A} \sum_{i=0}^{n_s-1} \begin{pmatrix}
        0 & d_E^{-(n_s-i)}\\
        d_E^{-i} & 0
    \end{pmatrix}\\
    &= \Lambda^{n_s} + \frac{s}{d_A} \begin{pmatrix}
        0 & \left(1-d_E^{-n_s}\right)/(d_E-1)\\
        d_E\left(1-d_E^{-n_s}\right)/(d_E-1) & 0
    \end{pmatrix}\equiv \Lambda^{n_s} + \frac{s}{d_A}C.
\end{align}
Therefore, the original $Q^{t-s n_s}\left(Q' Q^{s-1}\right)^{n_s}$ above becomes
\begin{align}
    Q^{t-s n_s}\left(Q' Q^{s-1}\right)^{n_s} &= d_B^{-(m+1)t} \left(\bI+(t-sn_s)\frac{\sigma_x}{d_A}\right) \left(\Lambda+\frac{s}{d_A}\Lambda\sigma_x\right)^{n_s}\\
    &= d_B^{-(m+1)t} \left(\bI+(t-sn_s)\frac{\sigma_x}{d_A}\right) \left(\Lambda^{n_s} + \frac{s}{d_A} C\right)\\
    &= d_B^{-(m+1)t} \left[\Lambda^{n_s} + \frac{s}{d_A} C + \frac{t-sn_s}{d_A} \sigma_x \Lambda^{n_s} + \mathcal{O}
    \left(\frac{1}{d_A^2}\right)\right]\\
    &= d_B^{-(m+1)t} \begin{pmatrix}
        d_E^{-n_s} & \frac{1}{d_A}\left(s\frac{1-d_E^{-n_s}}{d_E-1} + t-s n_s\right)\\
        \frac{1}{d_A}\left(\frac{s d_E\left(1-d_E^{-n_s}\right)}{d_E-1} + (t-s n_s)d_E^{-n_s}\right) & 1
    \end{pmatrix} \equiv \calT.
\end{align}

With the global transformation matrix $\calT$, we can turn to evaluate all the conditioned pseudopurities of interest. For pseudopurity of $\rho_{R|\bfz}$, we have
\begin{align}
    \E_{\rm Haar}\overline{\gamma}_R^{(m)} &= d_A^{-(m+2)} (\tau_2| \calT (0,1)^T \\
    &= d_A^{-(m+2)} d_B^{-(m+1)t} (\tau_2|_A \left[\frac{1}{d_A}\left(s\frac{1-d_E^{-n_s}}{d_E-1} + t-s n_s\right)|\tau_2)_A + |e)_A\right]\\
    &=  d_A^{-(m+2)} d_B^{-(m+1)t} \left[\frac{1}{d_A}\left(s\frac{1-d_E^{-n_s}}{d_E-1} + t-s n_s\right)d_A^{m+2} + d_A^{m+1}\right]\\
    &= \frac{d_B^{-(m+1)t}}{d_A} \left(s\frac{1-d_E^{-n_s}}{d_E-1} + t-s n_s + 1\right).
\end{align}
Similarly, for pseudopruities of $\rho_{A|\bfz}$ and $\rho_{RA|\bfz}$, we have
\begin{align}
    \E_{\rm Haar}\overline{\gamma}_A^{(m)} &= d_A^{-(m+2)} (e|_A \calT (1,0)^T \\
    &= d_A^{-(m+2)} d_B^{-(m+1)t} (e|_A \left[d_E^{-n_s}|\tau_2)_A + \frac{1}{d_A}\left(\frac{sd_E\left(1-d_E^{-n_s}\right)}{d_E-1} + (t-s n_s)d_E^{-n_s}\right)|e)_A\right]\\
    &= d_A^{-(m+2)} d_B^{-(m+1)t} \left[d_E^{-n_s}d_A^{m+1} + \frac{1}{d_A}\left(\frac{sd_E\left(1-d_E^{-n_s}\right)}{d_E-1} + (t-s n_s)d_E^{-n_s}\right)d_A^{m+2}\right]\\
    &= \frac{d_B^{-(m+1)t}}{d_A} \left(\frac{s d_E\left(1-d_E^{-n_s}\right)}{d_E-1} + (t-s n_s+1)d_E^{-n_s}\right),
\end{align}
and
\begin{align}
    \E_{\rm Haar}\overline{\gamma}_{RA}^{(m)} &= d_A^{-(m+2)} (\tau_2|_A \calT (1,0)^T \\
    &= d_A^{-(m+2)} d_B^{-(m+1)t} (\tau_2|_A \left[d_E^{-n_s}|\tau_2)_A + \frac{1}{d_A}\left(\frac{sd_E\left(1-d_E^{-n_s}\right)}{d_E-1} + (t-s n_s)d_E^{-n_s}\right)|e)_A\right]\\
    &= d_A^{-(m+2)} d_B^{-(m+1)t} \left[d_E^{-n_s}d_A^{m+2} + \frac{1}{d_A}\left(\frac{sd_E\left(1-d_E^{-n_s}\right)}{d_E-1} + (t-s n_s)d_E^{-n_s}\right)d_A^{m+1}\right]\\
    &= d_B^{-(m+1)t} \left[d_E^{-n_s} + \frac{1}{d_A^2}\left(\frac{s d_E\left(1-d_E^{-n_s}\right)}{d_E-1} + (t-s n_s)d_E^{-n_s}\right)\right]\\
    &= d_B^{-(m+1)t} d_E^{-n_s} + \mathcal{O}\left(\frac{1}{d_A^2}\right).
\end{align}

By taking the replica limit of $m \to -1$, and utilize the definition in Eq.~\eqref{eq:condQMI_partial_def}, we have the full expression for the R\'enyi-2 extended conditioned QMI as
\begin{align}
    &\E_{\rm Haar}\overline{I_2(R:A_t|\bfz)} = -\log_2 \overline{\gamma}_{R} -\log_2 \overline{\gamma}_{A} + \log_2 \overline{\gamma}_{RA} \\
    &= -\log_2 \left[\frac{1}{d_A} \left(s\frac{1-d_E^{-n_s}}{d_E-1} + t-s n_s + 1\right)\right] - \log_2\left[\frac{1}{d_A} \left(\frac{s d_E\left(1-d_E^{-n_s}\right)}{d_E-1} + (t-s n_s+1)d_E^{-n_s}\right)\right] + \log_2 d_E^{-n_s}\\
    &= 2N_A - n_sN_E - \log_2\left(s\frac{1-d_E^{-n_s}}{d_E-1} + t-s n_s + 1\right) - \log_2\left(\frac{s d_E\left(1-d_E^{-n_s}\right)}{d_E-1} + (t-s n_s+1)d_E^{-n_s}\right)\\
    &\simeq 2N_A - n_sN_E - \log_2\left[s\left(1-d_E^{-n_s}\right)d_E^{-1} + t-s n_s + 1\right]- \log_2\left[s \left(1-d_E^{-n_s}\right) + (t-s n_s+1)d_E^{-n_s}\right]\\
    &\simeq 2N_A - n_sN_E - \log_2\left(1+t-sn_s\right) - \log_2 s,
    \label{eq:avgQMI_partial_sol}
\end{align}
where the second to last line holds when $d_E \gg 1$. The last line is a further simplification for analysis convenience and leads to Eq.~\eqref{eq:avgQMI_partial_simplify} of Theorem~\ref{condMI_partial_theorem} in the main text.

\section{Additional numerical details}

\label{app:numeric_detail}

In this section, we provide additional numerical simulation results to support the theorems and estimation of lifetime in the main text. We simulate the dynamics of QMI with \texttt{TensorCircuit}~\cite{zhang2023tensorcircuit}.

\begin{figure}[t]
    \centering
    \includegraphics[width=0.65\textwidth]{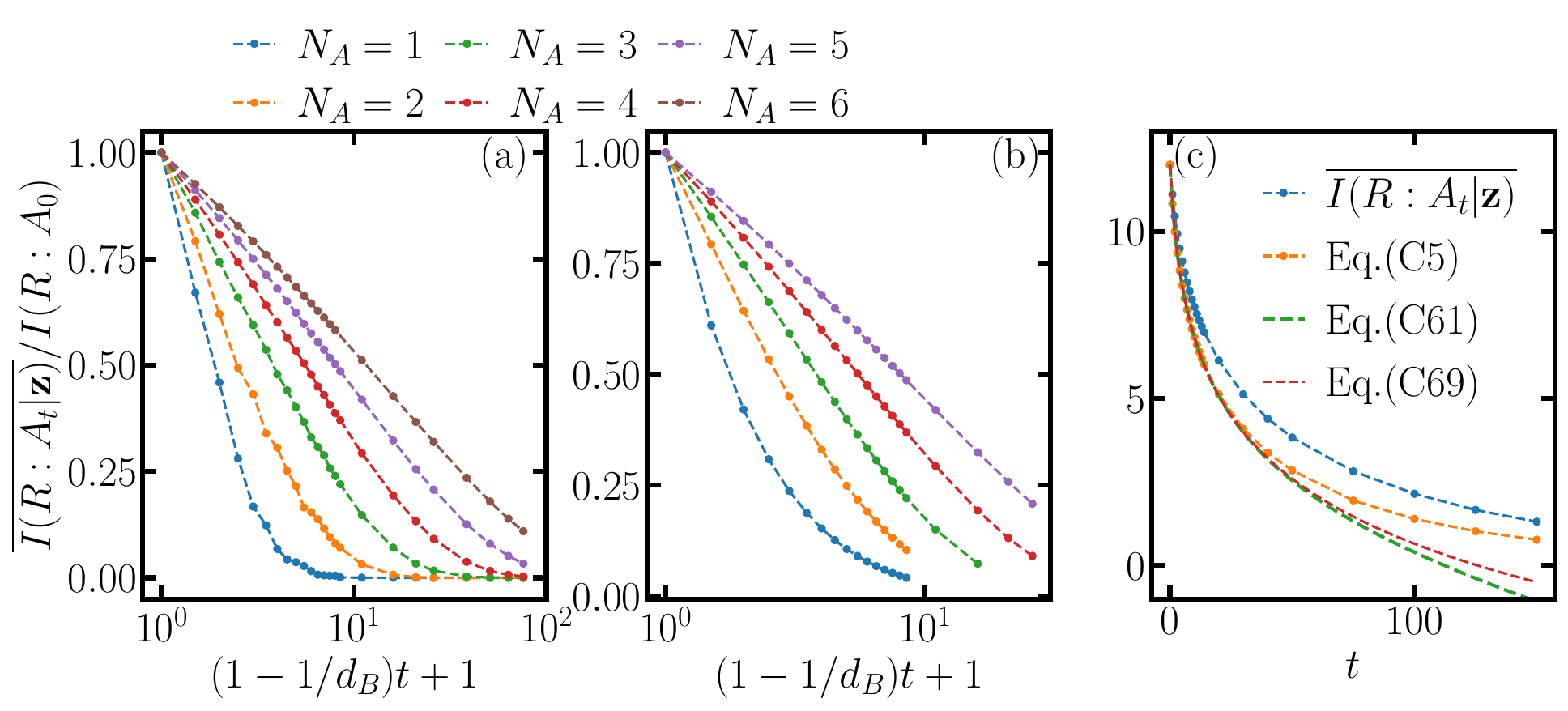}
    \caption{Dynamics of normalized measurement-conditioned QMI $\overline{I(R:A_t|\bfz)}$ with mid-circuit measurements and reset. In (a) and (b) we plot QMI with different data system $N_A$ and same bath $N_B = 1$ with randomly sampled unitary and identical unitary from Haar ensemble through different steps. In (c), dots are numerical simulation results of $\overline{I(R:A_t|\bfz)}$ with $N_A=5$ (same as purple in (a)).
    Orange dashed line represent numerical results of the lower bound in Eq.~\eqref{eq:avgMI_lb_sm}. Green and red dashed lines show
    the theoretical lower bounds Eq.~\eqref{eq:avgMI_sol_sm} and Eq.~\eqref{eq:cond_lb_asymp_sm}.}
    \label{fig:condQMI_detail_haar}
    \centering
    \includegraphics[width=0.45\textwidth]{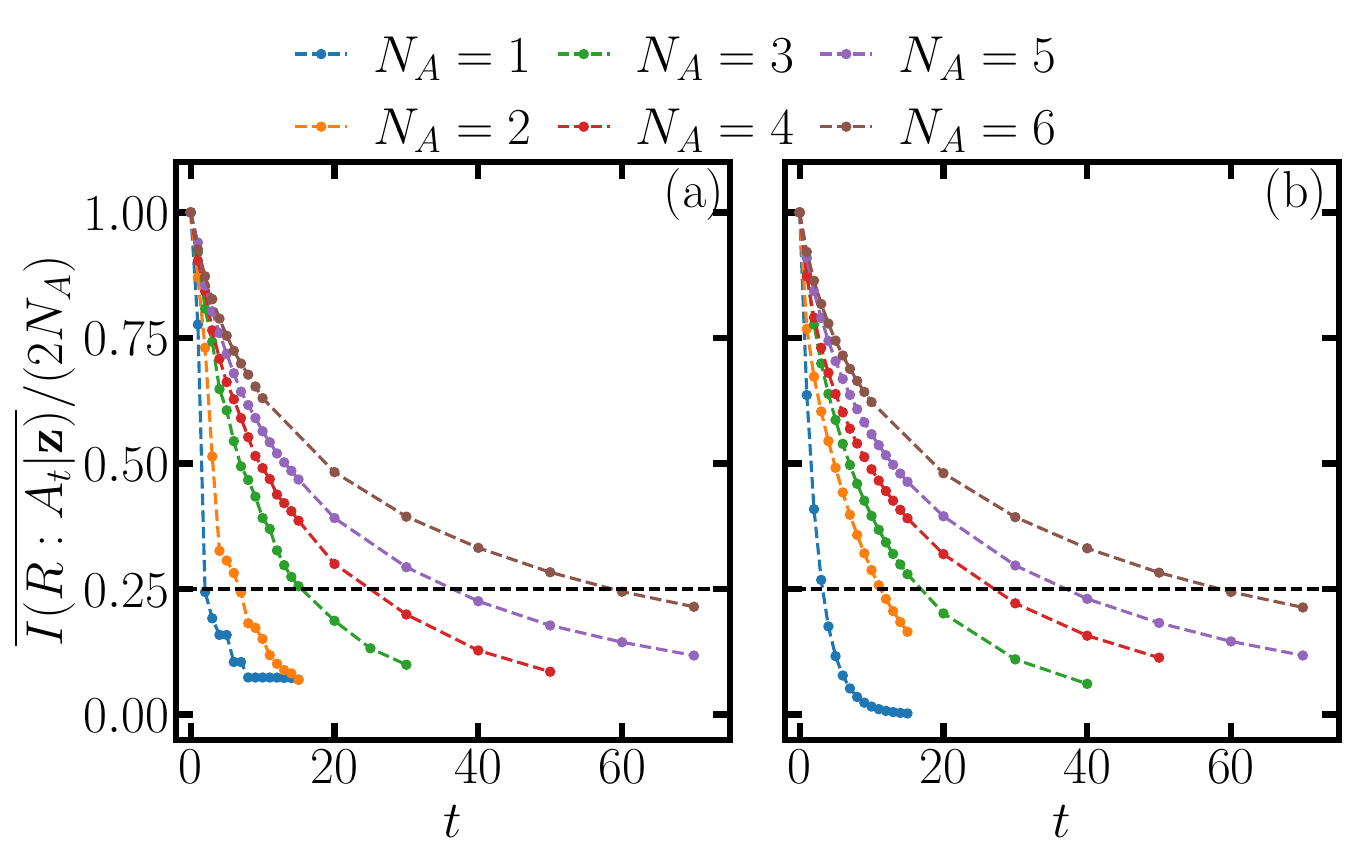}
    \caption{Dynamics of normalized measurement-conditioned QMI $\overline{I(R:A_t|\bfz)}$ with mid-circuit measurements and reset. In (a) and (b) we plot QMI with different data system $N_A$ and same bath $N_B = 1$ with Hamiltonian evolution with randomly sampled coefficients and identical Hamiltonian evolution.}
    \label{fig:condQMI_detail_ising}
\end{figure}

In Fig.~\ref{fig:condQMI_detail_haar}a-b, we show the decay of normalized measurement-conditioned QMI versus a rescaled time $(1-1/d_B)t + 1$ in various data system $N_A$ with either randomly sampled or identical unitary from Haar ensemble. With increasing $N_A$, we clearly see a logarithmic decay, and thus can estimate the corresponding QMI lifetime via a threshold (i.e. $\epsilon = 1/4$). Similar observations are also found for Hamiltonian evolution (shown in Fig.~\ref{fig:condQMI_detail_ising}).
In main text, we only show the asymptotic lower bound of Eq.~\eqref{eq:cond_lb_asymp_sm} (red), here we compare it to the complete form of Eq.~\eqref{eq:avgMI_sol_sm} (green) in Fig.~\ref{fig:condQMI_detail_haar}c. In fact, in the most valid range of decay, the two bounds overlap and align with the numerical results. Moreover, through a comparison to the numerical simulation of lower bound Eq.~\eqref{eq:avgMI_lb_sm} in derivation, we find that in early stage our theoretical results of either Eq.~\eqref{eq:avgMI_sol_sm} or Eq.~\eqref{eq:cond_lb_asymp_sm} agree well, and the deviation to the simulation of QMI (blue) is due to the difference between von Neumann entropy and its R\'enyi counterpart.

\begin{figure}[t]
    \centering
    \includegraphics[width=0.45\textwidth]{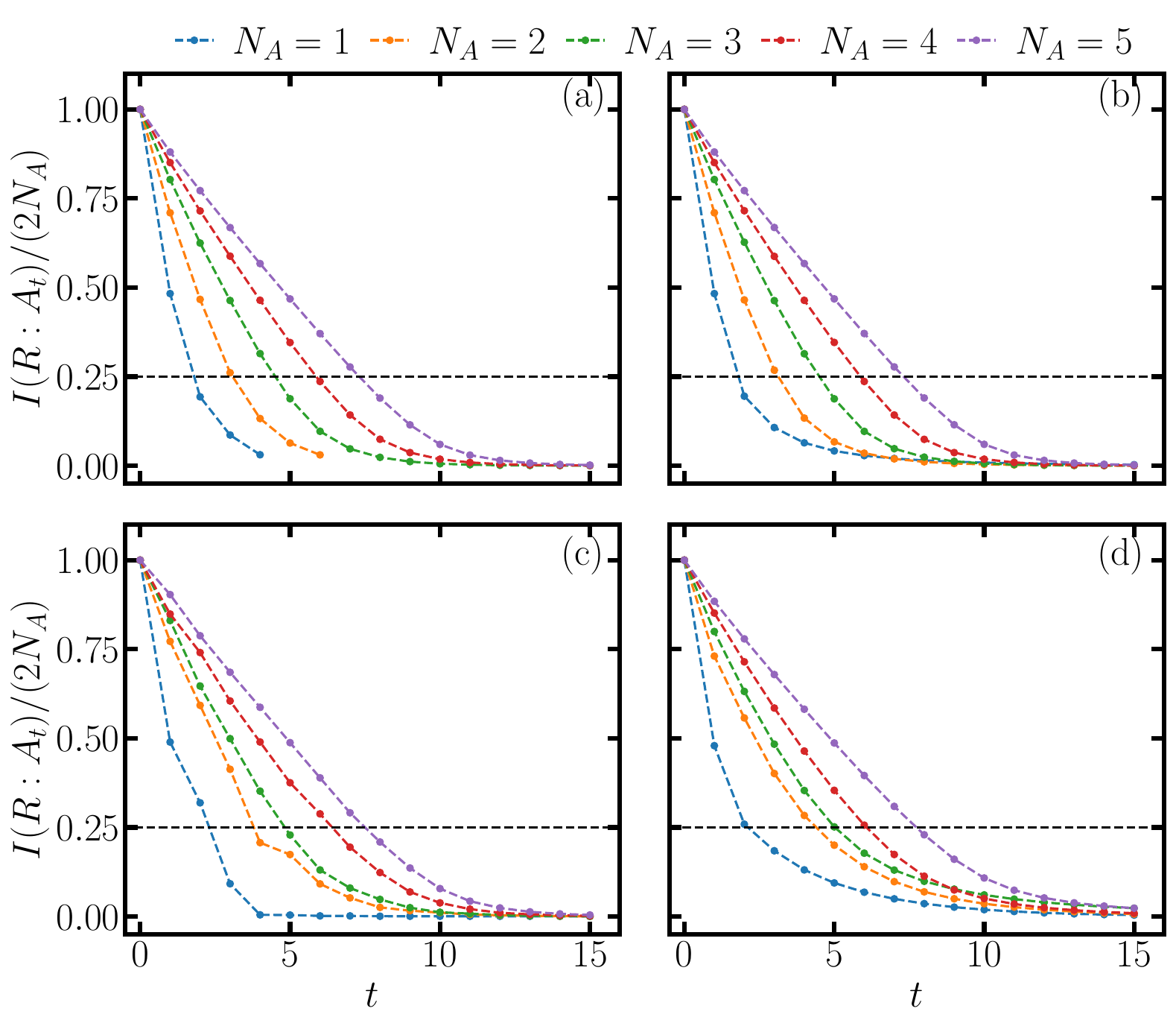}
    \caption{Dynamics of normalized measurement-unconditioned QMI $I(R:A_t)$ with mid-circuit measurements and reset. We plot QMI with different data system $N_A$ and same bath $N_B = 1$ with (a) randomly sampled Haar unitary (b) identical Haar unitary, (c) Hamiltonian evolution with random coefficients and (d) identical Hamiltonian evolution. }
    \label{fig:uncondQMI_detail}
\end{figure}

For measurement-unconditioned QMI, as shown in Fig.~\ref{fig:uncondQMI_detail} we still see that with increasing system size $N_A$, the dynamics of QMI behaves similarly as a linear decay for all possible four unitary settings under consideration. Therefore, we can estimate the corresponding QMI lifetime from numerical simulation of dynamics and we present the results in the main text and Appendix~\ref{app:indentical_or_not}.

\begin{figure}[t]
    \centering
    \includegraphics[width=0.7\textwidth]{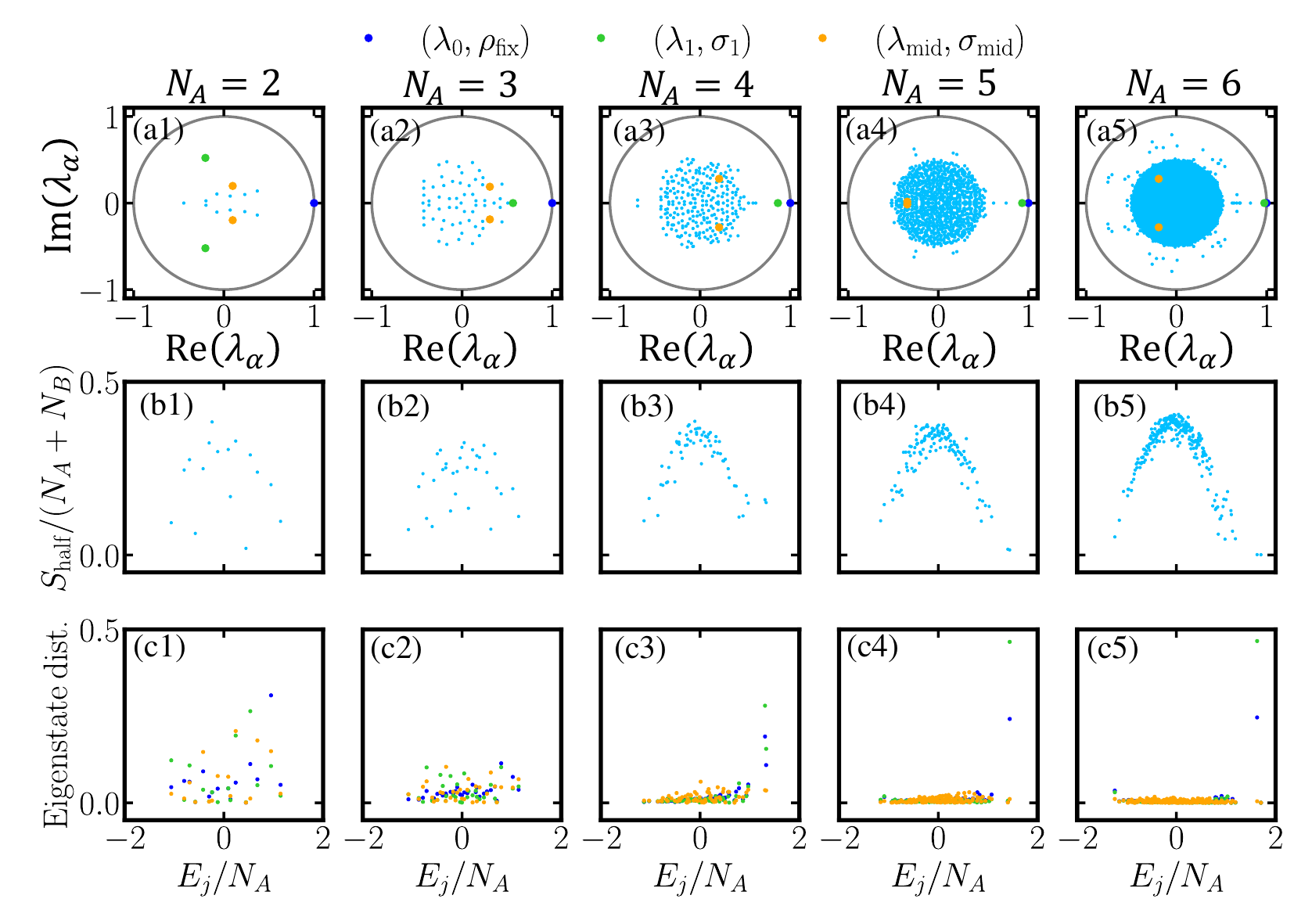}
    \caption{(a1)-(a5) Eigenspectrum of $\calP_0$. Blue, green and orange dots indicate the eigenstates $\rho_{\rm fix}, \sigma_1, \sigma_{\rm mid}$ to be considered. (b1)-(b5) Normalized bipartite entanglement entropy of Hamiltonian $H$ eigenstates versus normalized eigenenergies. (c1)-(c5) Normalized norm of overlap between eigenstates of $\calP_0$ and eigenstates of Hamiltonian versus normalized eigenenergies. In all cases, we have bath $N_B = 2$.}
    \label{fig:spectrum_detail}
\end{figure}

In Fig.~\ref{fig:spectrum_detail}b1-b5, we show the normalized bipartite entanglement entropy $S_{\rm half}/(N_A + N_B)$ of eigenstates of the Ising Hamiltonian in Eq.~\eqref{H_Ising} with different $N_A$ and $N_B = 2$. Since we only find the thermalized state entanglement spectrums, there does not exist ``many-body scar'' signature in the Hamiltonian considered here. Moreover, we evaluate the projection overlap between each eigenstate $\ket{E_j}$ and representative eigenstates of quantum channel $\calP_0$ appended with $\ketbra{0}{0}_B$ on bath. For a fair comparison, we normalize the norm of overlap and plot it versus normalized eigenenergies $E_j$ for different $N_A$ in Fig.~\ref{fig:spectrum_detail}c1-c5. For the fixed state (blue) and second largest eigenstate (green), the largest overlaps with Hamiltonian eigenstates lies at the boundaries, corresponding to eigenstates with nearly zero bipartite entanglement. In contrast, for $\sigma_{\mid}$ (orange), the operator with medium eigenvalue norm $|\lambda_{\rm mid}|$ from spectrum $\calP_0$, the major contribution on overlap comes from thermalized Hamiltonian eigenstates instead of the boundary parts. For visualization convenience, we mark the corresponding eigenvalues of $\rho_{\rm fix}, \sigma_1, \sigma_{\rm mid}$ in Fig.~\ref{fig:spectrum_detail}a1-a5.

\section{Details of noisy simulation and experiment}

\label{app:experiment_detail}

In this section, we provide details of the noisy simulation and experiments of mutual information dynamics on the IBM Quantum devices. 

For the experiment performed on IBM Quantum Torino presented in Fig.~\ref{fig:qmi_noisy} in the main text, we implement the brickwork circuit to mimic the Haar random unitaries considered in the theoretical studies. In a system of $N$ qubits, the unitary circuit is defined as
\begin{align}
    U = \prod_{\ell=1}^L \left(\prod_{i=1}^{\lfloor (N-1)/2\rfloor} {\rm CZ}_{2i,2i+1}\right) \left(\prod_{i=1}^{\lfloor N/2\rfloor} {\rm CZ}_{2i-1,2i}\right)\left(\otimes_{i=1}^N e^{-i\phi_i^{(\ell)} Y_i} e^{-i\theta_i^{(\ell)} X_i}\right),
\end{align}
where $\phi_i^{(\ell)}, \theta_i^{(\ell)}$ are random variables. $X_i, Y_i$ are Pauli-X and Y operators nontrivially support on $i$th qubit, and ${\rm CZ}_{i, i+1} = \ketbra{0}{0}_i \otimes \bI_{i+1} + \ketbra{1}{1}_i \otimes Z_{i+1}$ is the control-Z gate applied on nearest qubits $i, i+1$.

Additionally, we also consider the numerical simulation utilizing a variant of fast scrambling ansatz (FSA)~\cite{belyansky2020minimal} for IBM Quantum Sherbrooke. Compared to the noisy simulation 
of IBM Quantum Torino in the main text, the noisy simulation here does not consider the device shot noise, measurement error and idling decoherence.
The circuit ansatz is 
\be
    U_{\rm simu} = \prod_{\ell = 1}^{L} \left(\prod_{i<j} {\rm ECR}_{ij}\right) \left(\otimes_{i=1}^N e^{-i\phi_i^{(\ell)} Y_i} e^{-i\theta_i^{(\ell)} X_i}\right),
\ee
where ${\rm ECR}_{ij} \equiv \frac{1}{\sqrt{2}}(\bI_i \otimes X_j - X_i \otimes Y_j)$ with $X_i, Y_j$ denoting the Pauli-$X$ and $Y$ operator defined nontrivially on $i$th and $j$th qubit. ECR gate is indeed equivalent to the CNOT gate up to some single qubit rotations.


In Fig.~\ref{fig:purity_simu}, we show the decay of conditional state purity in our noisy simulation. Due to increasing number of gates in the noisy circuit, the purity of the post-measurement conditional state deviates from the ideal unity and undergo an approximately linear decay. Within the $15$ steps we simulated, the purity of conditional state remains $\gtrsim 0.75$, indicating the coherence of qubits is remained in our circuit of simulation.

\begin{figure}[t]
    \centering
    \includegraphics[width=0.35\textwidth]{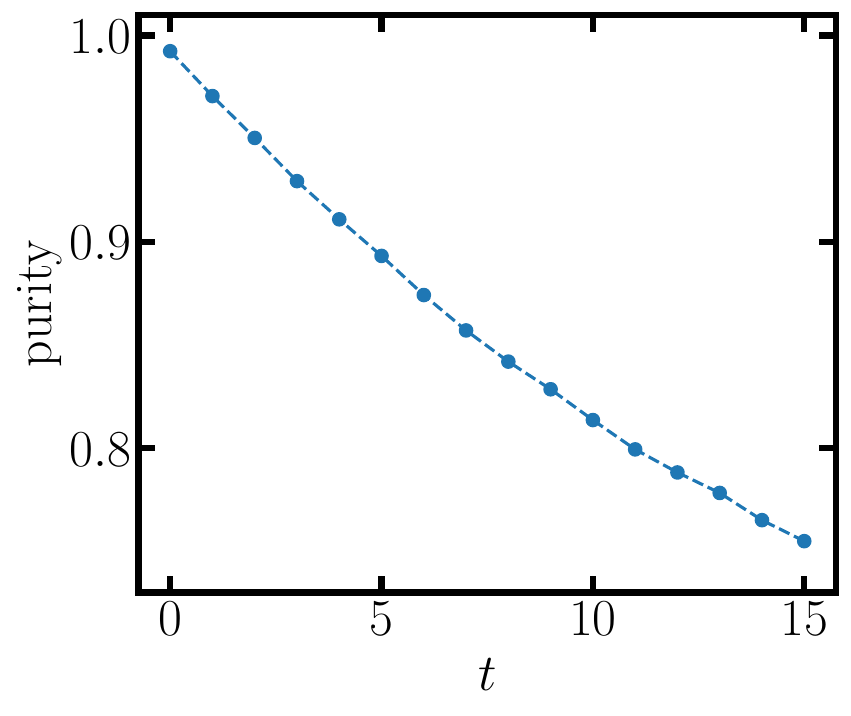}
    \caption{Purity of post-measurement conditional state at different time steps in noisy simulation. Here we perform simulation on a system of $N_A=3, N_B=1$ qubits with FSA consisting of $L=2$ layers, and $1024$ shots for mid-circuit measurements on the bath system. Each dot is averaged over $10$ randomized circuits.}
    \label{fig:purity_simu}
\end{figure}

\begin{figure}[t]
    \centering
    \includegraphics[width=0.45\textwidth]{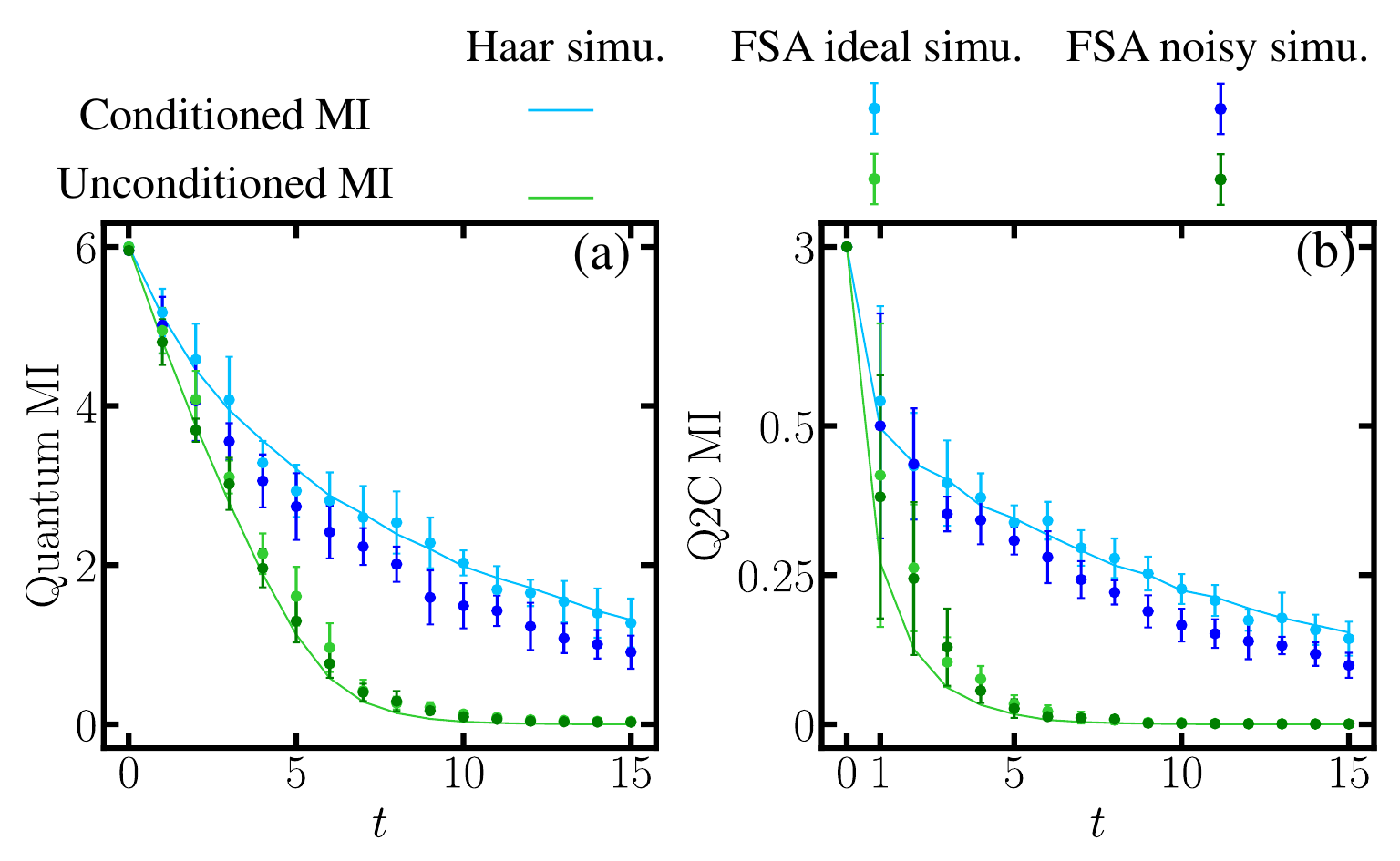}
    \caption{Dynamics of measurement-conditioned and unconditioned (a) QMI and (b) Q2C MI. Solid lines represent Haar simulation results. Dark and light colored dots with errorbars represent ideal and noisy simulation results separately with fast scrambling ansatz (FSA) for IBM Quantum, Sherbrooke. Here we perform simulation on a system of $N_A=3, N_B=1$ qubits with FSA consisting of $L=2$ layers, and $1024$ shots for mid-circuit measurements on the bath system. Each dot is averaged over $10$ randomized circuits.}
    \label{fig:qmi_noisy_app}
\end{figure}

For the IBM Quantum Sherbrooke we targeted at in Fig.~\ref{fig:qmi_noisy_app}, it has median $T_1$ as $275.73 \mu s$, median echoed cross-resonance (ECR) gate error $7.726\times10^{-3}$, median SX error $2.300\times10^{-4}$, median ECR gate length $5.333\times10^{-1} \mu s$, median readout (measurement) length $1.216 \mu s$, thus our simulation of $15$ steps is within the coherence time of qubits, which is also support by the conditional state purity $\gtrsim 0.75$ (see Fig.~\ref{fig:purity_simu}). Compared to the noisy simulation results presented in the main text, here the shot noise and idling decoherence are not considered in the noisy simulation.

In Fig.~\ref{fig:qmi_noisy_app}a, we evaluate the QMI dynamics implemented with FSA. For the ideal simulation, both measurement-conditioned (light blue dots) and unconditioned (light green dots) QMI align with the simulation results with Haar unitaries (solid lines), demonstrating its scrambling ability to mimic the dynamics with random unitaries. Next, we adopt the noisy circuit model (excluding measurement error) from Qiskit and simulate the QMI dynamics. In the presence of hardware noises, both conditioned (dark blue) and unconditioned (dark green) QMI show the same scaling of dynamics though with a small deviation from the ideal results. We thus expect that the exponential separation of QMI lifetime in conditional versus unconditional setup can be robust against circuit noise, suggesting a practical advantage on current-generation hardware. 

In Fig.~\ref{fig:qmi_noisy_app}b, we simulate the Q2C mutual information dynamics with FSA. Similar to the behavior of the QMI dynamics discussed above, the measurement-conditioned Q2C mutual information (blue) also decays logarithmically with steps, while the unconditioned one (green) simply decays linearly, and both of them present a small deviation in the noisy simulation.

\end{widetext}

\end{document}